%% file: main.tex
\pdfoutput=1
\documentclass[9pt,sigconf]{acmart}
\usepackage[utf8x]{inputenc}
\usepackage{balance}
\usepackage{float}
\usepackage{blindtext}
\usepackage{booktabs} 
\usepackage{xcolor,colortbl}
\usepackage{subfigure}
\usepackage{multirow}				
\usepackage{fixmath} 
\usepackage{graphicx}
\usepackage{dsfont}
\usepackage{balance}
\usepackage{marginnote}
\usepackage{multicol}
\usepackage[linesnumbered]{algorithm2e}
\usepackage{algpseudocode}
\usepackage{mathtools}
\usepackage{bm}
\usepackage{caption} 
\usepackage{enumerate}
\usepackage{lipsum}
\usepackage[most]{tcolorbox}
\usepackage{stfloats}
\usepackage{makecell}
\usepackage{enumitem}
\newcommand{\eat}[1]{}
\usepackage{cleveref}
\usepackage{listings}
\input{macros}

\copyrightyear{2021}
\acmYear{2021}
\setcopyright{acmcopyright}\acmConference[SIGMOD '21]{Proceedings of the 2021 International Conference on Management of Data}{June 20--25, 2021}{Virtual Event, China}
\acmBooktitle{Proceedings of the 2021 International Conference on Management of Data (SIGMOD '21), June 20--25, 2021, Virtual Event, China}
\acmPrice{15.00}
\acmDOI{10.1145/3448016.3457311}
\acmISBN{978-1-4503-8343-1/21/06}
\begin{document}
\title[\sys]{\sys: A Lightweight Approach for Optimizing Hybrid Complex Analytics Queries (Extended Version)}
\author{Rana Alotaibi}
\affiliation{%
  \institution{UC San Diego}}
\email{ralotaib@eng.ucsd.edu}
\author{Bogdan Cautis}
\affiliation{%
  \institution{University of Paris-Saclay}}
\email{bogdan.cautis@u-psud.fr}
\author{Alin Deutsch}
\affiliation{%
  \institution{UC San Diego}}
\email{deutsch@cs.ucsd.edu}
\author{Ioana Manolescu}
\affiliation{%
  \institution{Inria \& Institut Polytechnique de Paris}}
\email{ioana.manolescu@inria.fr}

%\author{Rana Alotaibi${\ssymbol{1}}$, Bogdan Cautis${\ssymbol{2}}$, Alin Deutsch${\ssymbol{1}}$, Ioana Manolescu${\ssymbol{3}}$}
%\affiliation{\vspace{0.1cm}
%\institution{${\ssymbol{1}}$UC San Diego, ${\ssymbol{2}}$University of Paris-Saclay,${\ssymbol{3}}$Inria \& Institut Polytechnique de
%     Paris, France}
%}
%\email{{ralotaib,deutsch}@eng.ucsd.edu;bogdan.cautis@u-psud.fr;ioana.manolescu@inria.fr }
 
\begin{abstract}		
\end{abstract}

\begin{abstract}
%\RA{Modified Abstract}
 Hybrid complex analytics workloads typically include $(i)$ data
management tasks (joins, selections, \etc), easily
expressed using relational algebra (RA)-based languages, and $(ii)$ complex analytics tasks (regressions, matrix decompositions, etc.), mostly expressed in linear algebra (LA) expressions. Such workloads are common in many
application areas, including scientific computing, web analytics, and business recommendation. 
Existing solutions for evaluating hybrid analytical tasks -- ranging from LA-oriented systems, to relational systems (extended to handle LA operations), to hybrid systems -- either optimize data management and complex tasks separately, exploit RA properties only  while leaving LA-specific optimization opportunities unexploited, or focus heavily on %low-level 
physical optimization, leaving semantic query optimization opportunities unexplored. Additionally, they are not able to exploit {\em precomputed (materialized) results} to avoid recomputing (part of) a given mixed (RA and/or LA) computation.

  In this paper, we take a major step towards filling this gap by proposing
  %by proposing %{\em a unified semantic query optimization framework} for such a hybrid setting}. We describe 
\sys, an extensible lightweight approach for optimizing hybrid complex analytics queries, based on a common abstraction that facilitates unified reasoning: {\em a relational model endowed with integrity constraints}. %which can be used to express the properties of the two computation formalisms.
%Our approach enables full exploration of LA properties and rewrites, as well as semantic query optimization. 
Our solution can be naturally and portably applied on top of pure LA and hybrid RA-LA platforms without 
modifying their internals. 
An extensive empirical evaluation shows that \sys\ yields significant performance gains on diverse workloads, 
ranging from LA-centered to hybrid.
\end{abstract}	
\settopmatter{printfolios=true}
\settopmatter{printacmref=true}
\fancyhead{}
\maketitle
\input{introduction}
\input{hadad-optimizations}
\input{problemstatement}
\input{preliminaries}
\input{overview}
\input{encoding}
\input{rewritingPruning}

\input{guarantees}

\input{experiments}
\input{related}
%%% Disabled now for double-blind constraint.
\begin{acks} 
This work is supported by a graduate fellowship from KACST.
\end{acks}
\bibliographystyle{abbrv}
\bibliography{main.bib}
\appendix
\input{appendixA}
\input{appendixB}
\input{appendixC}
\input{appendixD}
\input{appendixE}
\input{appendixF}
\input{appendixI}

\end{document}

%% file: macros.tex
\newcommand{\IM}[1]{\textcolor{red}{\reminder{IM:~#1}}}
\renewcommand{\IM}[1]{}
\newcommand{\RA}[1]{\textcolor{brown}{\reminder{RA:~#1}}}
\newcommand{\AD}[1]{\textcolor{green}{\reminder{AD:~#1}}}
\renewcommand{\AD}[1]{}
\newcommand{\CB}[1]{\textcolor{blue}{\reminder{CB:~#1}}}
\renewcommand{\CB}[1]{}
\newcommand {\sys}{HADAD}
\newcommand{\pacb}{PACB^{++}}
\newcommand{\calH}{\ensuremath{\boldsymbol{\mathcal {H}}}}

\newcommand{\calCC}{\ensuremath{\boldsymbol{\mathcal {C}}}}
\newcommand{\calVREM}{\ensuremath{\boldsymbol{\mathcal {VREM}}}}

\newcommand{\calMMC}{\ensuremath{\boldsymbol{\mathcal {MMC}}}}
\newcommand{\calView}{\ensuremath{\boldsymbol{\mathcal {V}}}}
\newcommand{\calQuery}{\ensuremath{\boldsymbol{\mathcal {Q}}}}
\newcommand{\calRW}{\ensuremath{\boldsymbol{\mathcal {RW}}}}

\newcommand{\calZero}{O}
\newcommand{\calIden}{I}
\newcommand{\calO}{R}

\newcommand{\calMI}{M}
\newcommand{\calMII}{N}
\newcommand{\calA}{A}
\newcommand{\calB}{B}
\newcommand{\calC}{C}
\newcommand{\calD}{D}
\newcommand{\calE}{E}
\newcommand{\calX}{X}
\newcommand{\calR}{R}
\newcommand{\calLL}{L}
\newcommand{\calQR}{Q}
\newcommand{\calU}{U}
\newcommand{\multiEOp}{\texttt{multi$_{E}$}}
\newcommand{\multiMSOp}{\texttt{multi$_{MS}$}}
\newcommand{\multiMOp}{\texttt{multi$_{M}$}}
\newcommand{\addMOp}{\texttt{add$_{M}$}}
\newcommand{\divMOp}{\texttt{div$_{M}$}}
\newcommand{\trOp}{\texttt{tr}}
\newcommand{\invMOp}{\texttt{inv$_{M}$}}
\newcommand{\detOp}{\texttt{det}}
\newcommand{\traceOp}{\texttt{trace}}
\newcommand{\diagOp}{\texttt{diag}}
\newcommand{\expOp}{\texttt{exp}}
\newcommand{\adjOp}{\texttt{adj}}
\newcommand{\sumDOp}{\texttt{sum$_{D}$}}
\newcommand{\dirctDOp}{\texttt{product$_{D}$}}
\newcommand{\sumOp}{\texttt{sum}}
\newcommand{\rowsumsOp}{\texttt{rowSums}}
\newcommand{\colsumsOp}{\texttt{colSums}}
\newcommand{\rowminOp}{\texttt{rowMin}}
\newcommand{\colminOp}{\texttt{colMin}}
\newcommand{\rowmaxOp}{\texttt{rowMax}}
\newcommand{\colmaxOp}{\texttt{colMax}}
\newcommand{\rowmeanOp}{\texttt{rowMean}}
\newcommand{\colmeanOp}{\texttt{colMean}}
\newcommand{\rowvarOp}{\texttt{rowVar}}
\newcommand{\colvarOp}{\texttt{colVar}}
\newcommand{\minOp}{\texttt{min}}
\newcommand{\maxOp}{\texttt{max}}
\newcommand{\meanOp}{\texttt{mean}}
\newcommand{\varOp}{\texttt{var}}
\newcommand{\QROp}{\texttt{QR}}
\newcommand{\CHOp}{\texttt{CHO}}
\newcommand{\LUOp}{\texttt{LU}}
\newcommand{\LUPOp}{\texttt{LUP}}
\newcommand{\calQ}{\ensuremath{\mathcal{Q}}}
\newcommand{\calI}{\ensuremath{\mathcal I}}
\newcommand{\calV}{\ensuremath{\mathcal V}}
\newcommand{\calVV}{V}
\newcommand{\calMM}{M}
\newcommand{\calNN}{N}
\newcommand{\query}{\mbox{:- }}
\newcommand{\mysection}[1]{\vspace{-3mm} \section{#1} \vspace{-.5mm}}
\newcommand{\mysubsection}[1]{\vspace{-2mm} \subsection{#1} \vspace{-.5mm}}
\newcommand{\mysubsubsection}[1]{\vspace{-1mm} \subsubsection{#1} \vspace{-0.5mm}}
\newcommand{\veat}{\vspace{-.5mm}}
\DeclareRobustCommand*\circled[1]{%
  \tikz[
    baseline=(char.base)
  ]{%
    \node[
      shape=circle,
      draw,
      inner sep=0.05pt,
      fill=black,
      black
     ] (char) {\color{white}\bfseries#1};%
   }%
}
\newcommand {\notopt}{\mathcal{P}^{\neg Opt}}
\newcommand {\opt}{\mathcal{P}^{Opt}}
\newcommand {\views}{\mathcal{P}^{Views}}

\newcommand{\cEQ}[3]{\ensuremath{\mbox{Eq}^{#3}\langle{#1},{#2}\rangle}}
\newcommand{\hadad}[3]{\ensuremath{HADAD\langle{#1},{#2},{#3}\rangle}}
\newcommand{\lprop}{\ensuremath{LA_{prop}}}
\newcommand{\lops}{\ensuremath{L_{ops}}}
\newcommand{\calL}{\ensuremath{\boldsymbol{\mathcal {L}}}}
\newcommand{\ie}{i.e.,~}
\newcommand{\eg}{e.g.,~}
\newcommand{\etc}{etc.~}

\newcommand{\st}{s.t.~}

\newcommand{\naive} {na\"ive}
\newcommand{\reminder}[1]{[\vadjust{\vbox to0pt{\vss\hbox to0pt{\hss{\Large $\Longrightarrow$}}}}{{\textsf{\small #1}}}]}
\newtheorem{example}{Example}[section]
\newtheorem{definition}{Definition}[section]
\newtheorem{theorem}{Theorem}[section]

\newtheorem{claim}{Claim}[section]

 \newcommand\rev[1]{%
  \protect\leavevmode
  \begingroup
    \color{blue}%
    #1%
  \endgroup
}
\crefformat{section}{\S#2#1#3}
\lstdefinestyle{mystyle}{
  numbers=left,
  stepnumber=0,
  numbersep=10pt,
  tabsize=4,
  showspaces=false,
  showstringspaces=false,
  mathescape=true,
  frame=single }
\newcommand{\figspa}{\vspace{-3mm}}
\newcommand{\figspb}{\vspace{-2mm}}
\newcommand{\figspc}{\vspace{-6mm}}
\makeatletter
\def\@fnsymbol#1{\ensuremath{\ifcase#1\or \dagger\or \ddagger\or
   \mathsection\or \mathparagraph\or \|\or **\or \dagger\dagger
   \or \ddagger\ddagger \else\@ctrerr\fi}}
\newcommand{\ssymbol}[1]{^{\@fnsymbol{#1}}}

%% file: introduction.tex
\mysection{Introduction}
\label{sec:Introduction}
Modern analytical tasks typically include $(i)$~data management tasks (\eg joins, filters) to perform pre-processing steps, including feature selection, transformation, and engineering~\cite{sculley2015hidden,kumar2016model,bose2017probabilistic,sparks2015automating,baylor2017tfx}, tasks that are easily expressed using RA%{\em relational algebra (RA)}
-based languages, as well as $(ii)$~complex analytics tasks (\eg regressions, matrices decompositions), which are mostly expressed using %{\em linear algebra (LA)} 
LA operations ~\cite{kumar2017data}. %Such workloads are common in several
%a number of 
%application domains including scientific computing, 
%\IM{idem abstract}  machine learning, and data mining~\cite{taft2014genbase}.
%web analytics, business recommendation, natural language processing~\cite{manning2003optimization}, or speech recognition~\cite{huang2014kernel}.
To perform such analytical tasks, data scientists can choose from a variety of systems, tools, and languages. Languages/libraries such as R~\cite{R} and NumPy~\cite{NumPy}, as well as LA %-oriented
systems such as SystemML~\cite{boehm2016systemml}, TensorFlow~\cite{abadi2016tensorflow} and MLlib ~\cite{sparkmlib} treat matrices and %linear algebra 
LA operations as first-class citizens: they offer a rich set of built-in LA operations. %and algorithms.
However, it can be difficult to express  pre-processing %and data transformation 
tasks in these systems. Further, \textit %{full-exploration} 
{\em expression rewrites}, based on equivalences that hold due to well-known LA properties, are not exploited in some of these systems,  %and rewrites in some of these systems are completely unexplored, which can lead
leading  to missed optimization opportunities. 

Many works %have sought to efficiently 
propose integrating RA and LA processing 
%in a hybrid environment
, where both algebraic styles can be used together~\cite{luo2018scalable,chen2017towards,montdb,kernert2013bringing,kunft2019intermediate,spores,boehm2018optimizing}. 
%\IM{We don't say anything about \cite{meng2016mllib}, can we to pack in an existing category?} \RA{I added it with LA-oriented systems}
 ~\cite{montdb} offers calling LA packages through user defined functions (UDFs), where libraries such as NumPy are embedded in the host language. Others suggest to extend RDBMS to treat LA objects as first-class citizens by using built-in functions to express LA operations~\cite{luo2018scalable,kernert2013bringing}. However, LA operations' semantics remain hidden behind these functions, where the optimizers treat as black-boxes. SPORES~\cite{spores} and SPOOF~\cite{boehm2018optimizing} optimize LA expressions by converting
them into RA, optimizing the latter, and then converting the result back to an (optimized) LA expression. They only focus on optimizing LA pipelines containing operations that can be expressed in RA. The restriction %of this approach 
is that LA properties of complex operations such as inverse, matrix-decompositions are entirely unexploited. Morpheus ~\cite{chen2017towards} speeds up LA pipelines over large joins by pushing computation into each joined table, thereby avoiding expensive materialization.
%Hybrid systems ~\cite{DugganESBHKMMMZ15,wang2017myria} provide an environment, where mixed RA and LA programs can be written and executed across different systems. 
LARA~\cite{kunft2019intermediate} focuses on {\em low-level }optimization by exploiting data layouts (\eg column-wise) to choose LA operators' physical implementations. A limitation of such approaches is that they lack \textit{high-level reasoning about LA properties and rewrites}, which can drastically enhance the pipelines' performance~\cite{thomas2018comparative}. 

Further, aforementioned solutions do not support
\textit{semantic query optimization}, which includes exploiting integrity constraints and materialized views and can bring enormous performance
advantages in hybrid RA-LA and even plain LA settings. 
%\RA{We already identified them above}
\eat{
We identify \textit{unexplored semantic optimization opportunities} in existing solutions for evaluating hybrid complex analytics queries. First,  they {\em do not fully exploit LA properties and rewrites}, whereas it has been shown that such rewrites can drastically enhance LA-based pipelines' performance~\cite{thomas2018comparative}. Second, they {\em do not support semantic query optimization}~\cite{AHV95}, which includes taking advantage of {\em partial materialized computation results}, \ie materialized views, known to improve the performance of a variety of queries.} %since they pre-compute partial query results.

% Further, as the aforementioned work do not reason with constraints, they cannot exploit  materialized views in an LA (or hybrid LA/RA) context; as shown in our experiments, such rewritings can bring large performance advantages.
%Our work can also complementing the optimizations of  LARA, RAVEN, SPORES or SPOOF, to extend to these platforms the benefits of views-based rewriting.  

 %\vspace{-2mm}	
 %\subsection{Motivation Scenario}

\eat{
\RA{Previous paragraph} 
%To enable such fruitful optimizations, 
%shown in the above scenario
We propose \sys, an extensible, lightweight, holistic optimizer for analytical queries, based on reasoning on a common abstraction: % about such a hybrid setting: 
\textit{relational model with integrity constraints}. This brings within reach {\em powerful cost-based optimizations across RA and LA}, without the need to modify the internals of the existing systems.} 

We propose \sys, an extensible lightweight framework for providing semantic query optimization (including views-based, integrity constraint-based, and LA property-based rewriting) on top of both pure
LA and hybrid RA-LA platforms, with no need to modify their internals.
At the core of \sys\ lies a common abstraction: \textit{relational model with integrity constraints}, which enables reasoning in hybrid settings.
Moreover, it makes it very easy to extend \sys's
semantic knowledge of LA operations by simply
declaring appropriate constraints, with no need to change \sys\ code.
As we show, constraints are sufficiently expressive to
declare (and thus allow \sys\ to exploit)
more properties of LA operations than previous work could consider.

%which brings within reach {\em a powerful cost-based optimization} without the need to modify the internals of the existing systems.}
\eat{
As we show, it is very easy to add within \sys\ knowledge about a wider range of LA operations than previous work could consider, while also enabling view-based rewriting and other semantic query optimizations using integrity constraints.
%which can be used to describe the properties of the two algebraic styles and to express view definitions.
}
\eat{\RA{redundant}\rev{The benefits of our optimized rewrites apply to both cross RA-LA platforms ~\cite{luo2018scalable,chen2017towards,montdb,kernert2013bringing,kunft2019intermediate,spores,boehm2018optimizing} and 
%systems that provide mixed programming interface, such as , 
LA %-oriented 
systems~\cite{R,NumPy,baylor2017tfx,sparkmlib}}.} %designed and built for matrix operations. %\IM{New:} 
Last but not least, our holistic, cost-based approach enables to judiciously apply for each query {\em the best available optimization}. For instance, given the computation $M(N P)$ for some matrices $M$, $N$ and $P$, we may rewrite it into $(M N) P$ if its estimated cost is smaller than that of the original expression, {\em or} we may turn it into $MV$ if a materialized view $V$ stores exactly the result of $(N P)$. %The rich and extensible set of constraints known to \sys\ covers a very large range of optimization opportunities. 
%\RA{From the problem statement by Ioana.}

\sys\ capitalizes on a framework previously introduced in~\cite{estocada-sigmod} for rewriting queries across many data models, using materialized views, in a polystore setting that does not include the LA model. The novelty of \sys\ is to extend the benefits of rewriting and views optimizations to pure LA and hybrid RA-LA computations, which are crucial for
%model analytics and 
ML workloads.
 
\vspace{2mm}
\noindent\textbf{Contributions}. %The technical contributions we make in this paper can be summarized as follows:	
The paper makes the following contributions: 
%\vspace{-2.6mm}
\begin{enumerate}[label=\circled{\arabic*},ref=\arabic*]
	\item We propose an {\em extensible lightweight approach to optimize hybrid complex analytics queries}. Our approach can be implemented on top of existing systems without %the need to 
modifying their internals; it is based on a powerful intermediate abstraction that supports reasoning in hybrid settings, namely {\em a relational model with integrity constraints}.
%	\item We propose to use a powerful intermediate abstraction to facilitate the reasoning about such a hybrid setting: \textit{the relational model with integrity constraints}. \IM{I thought there is no lightweight approach without the internal relational model, that is, it appeared to me that the first contribution didn't stand without the second, thus I unified them. We can rediscuss, of course.}
	\item We formalize the problem of {\em rewriting computations using previously materialized views in hybrid settings}.
		%combining relational and linear algebra. %constraints-based rewriting problem in the context of linear and relational algebra. 
To the best of our knowledge, %the concept of materialized views has not yet been adapted to LA-oriented systems. This 
ours is the first work that brings views-based rewriting under integrity constraints in the context of LA-based pipelines and hybrid %complex analytics 
analytical queries.
	\item We provide {\em formal guarantees} for our solution in terms of soundness and completeness.   
	\item We conduct an extensive set of empirical experiments on typical LA- and hybrid-based expressions, %(\ie realistic and synthetically generated pipelines), and
which show the benefits of \sys.
%In particular, LA-based pipeline experiments achieve speedups of {\em up to an order of magnitude}. 
\end{enumerate}
\vspace{2mm}
\noindent\textbf{Outline.}  The rest of this paper is organized as follows:  \S\ref{sec:hadad-opt} highlights \sys's
optimizations that go beyond the state of the art based on
{\em real-world} scenarios, \S\ref{sec:prob} formalizes the query
optimization problem %we solve 
in the context of a hybrid setting. %After some preliminaries (\S\ref{sec:pre}), 
\S\ref{sec:overview} provides an end-to-end overview of our approach.
%\IM{I shortened this one because it felt like the 5th time we said the same thing...}
%; at its core is a view-based query rewriting under integrity constraint algorithm based on an internal relational model,
%, invisible to users and applications
%previously introduced in~\cite{estocada-sigmod}.
%(RA and LA). 
%in the context of relational and linear algebra. 
\S\ref{sec:relencoding} presents our novel reduction of the rewriting problem into one that can be solved by existing techniques from the relational setting. %\IM{I don't think this is in that section:}\RA{I suggest to remove it, and show a decoding example in the technical report, we can't afford new figures:)} 
%we show how to transform rewritings obtained therein back into the native syntax of their respective underlying stores / engines. 
\S\ref{sec:opt} describes our extension to the query rewriting engine, integrating two different cost models, to help prune out inefficient rewritings as soon as they are enumerated. %t the early stages of the rewriting candidates' search. 
We formalize our solution's guarantees in \S\ref{sec:guaratees} and  present the experiments %our experimental evaluation 
in \S\ref{sec:exp}. We %then 
discuss related work and conclude in \S\ref{sec:related}.
% then conclude in Section ~\ref{sec:conclusion}.
%\vspace{1mm}

%% file: hadad-optimizations.tex
\mysection{HADAD %Semantic 
Optimizations}
\label{sec:hadad-opt}
We highlight below examples of performance-enhancing opportunities that are exploited by \sys\ and not being addressed by LA-oriented and cross RA-LA existing solutions.

\vspace{2mm}
\noindent\textbf{LA Pipeline Optimization.}
Consider the Ordinary Least Squares Regression (OLS) pipeline: $(X^TX)^{-1}(X^Ty)$, where $X$ is a square matrix of size 10K$\times$10K and $y$ is a vector of size 10K$\times$1. Suppose available a  materialized view $V=X^{-1}$.   \sys\ rewrites the pipeline to $(V(V^T(X^Ty))$, by exploiting  the LA properties  $(\calC\calD)^{-1}=\calD^{-1}\calC^{-1}$, $(\calC \calD) \calE = \calC (\calD \calE)$ and  $(\calD^{T})^{-1}=(\calD^{-1})^{T}$ as well as the view  $V$.
The rewriting is more efficient than the original pipeline
since it avoids computing the expensive inverse operation. Moreover, it optimizes the matrix chain multiplication order
to minimize the intermediate result size. This leads to a 150$\times$ speed-up on MLlib~\cite{meng2016mllib}.
Current popular LA-oriented systems ~\cite{R,NumPy,boehm2016systemml,abadi2016tensorflow,meng2016mllib} %~\cite{spores,boehm2018optimizing,kunft2019intermediate,aranasosIPSPPX20} 
are not capable of exploiting such rewrites, due to the lack of systematic exploration of standard LA properties and views.

\vspace{2mm}
\noindent\textbf{Hybrid RA-LA Optimization.}
Cross RA-LA platforms such as
Morpheus~\cite{chen2017towards}, SparkSQL~\cite{armbrust2015spark} and
others~\cite{kunft2019intermediate,aranasosIPSPPX20} can greatly benefit
from \sys's cross-model optimizations, which can find 
rewrites that they miss.

\vspace{2mm}
\textit{Factorization of LA Operations over Joins.}\
For instance, Morpheus implements a powerful optimization that
factorizes an LA operation on a matrix \textbf{M}
obtained by joining tables \textbf{R} and \textbf{S} and casting
the join result as a matrix.
Factorization pushes the LA operation on \textbf{M} %before the join,
to operate on \textbf{R} and \textbf{S}, cast as matrices.

Consider a specific instantiation of factorization:
\colsumsOp(\textbf{M}N), where matrix \textbf{M} has size
20M$\times$ 120 and N has size 120$\times$100;
both matrices are dense. The \colsumsOp\ operation 
sums up the elements in each column, returning the vector of these sums
(the operation is common in ML algorithms
such as K-means clustering ~\cite{macqueen1967some}).
%\AD{Rana, can you add a citation here?}
%\RA{Fixed}
On this pipeline, Morpheus applies a multiplication factorization rule
to push the multiplication by N down
to \textbf{R} and \textbf{S}: it computes
\textbf{R}N and \textbf{S}N, then concatenates the resulting matrices
to obtain \textbf{M}N. The size of this intermediate result is 20M$\times$100.
%This intermediate result is materialized as a dense 20M$\times$100 matrix of size $\approx$16GB (Morpheus is prototyped on R, which uses double precision).
Finally, \colsumsOp\ is applied to the intermediate result, reducing
%it to a vector of size 1$\times$100.
to a 1$\times$100 vector.

\sys\ can help Morpheus do much better, by pushing the \colsumsOp\
operator to \textbf{R} and \textbf{S} (instead of the multiplication
with N), then concatenating the resulting vectors.
This leads to much smaller intermediate results, since the
combined size of vectors
\colsumsOp$($\textbf{R}$)$ and \colsumsOp$($\textbf{S}$)$
is only 1$\times$120.

To this end, \sys\ rewrites the pipeline to \colsumsOp(\textbf{M})N by
exploiting the %LA%
property \colsumsOp$(AB)\ =$\colsumsOp$(A)B$
and applying its cost estimator, which favors
rewritings with a small intermediate result size.
Evaluating this \sys-produced rewriting, %\colsumsOp$($\textbf{M}$)N$,
Morpheus's multiplication pushdown rule no longer applies,
while the \colsumsOp\ pushdown rule is now enabled,
leading to 125$\times$ speed-up. %\\ %in our experiment.

\vspace{2mm}	
\textit{Pushing Selection from LA Analysis to RA Preprocessing.}\
Consider another hybrid example on a Twitter dataset ~\cite{twitter}.
The JSON dataset contains tweet ids, extended tweets, entities including
hashtags, filter-level, of %(\eg sensitive level),
media, URL, and tweet text, etc.
We implemented it on SparkSQL (with SystemML~\cite{boehm2016systemml}).

In the preprocessing stage, our SparkSQL
query constructs  a \textit{tweet-hashtag} filter-level 
matrix \textbf{N} of size 2M$\times$1000,
for all tweets posted from ``USA'' mentioning ``covid'',
where rows are tweets, columns are hashtags, and values are filter-levels.\footnote{Matrix $N$ is represented in MatrixMarket Format (MTX) since it is sparse.}.

\textbf{N} is then loaded into SystemML,
where rows with filter-level less than 4 are selected. The
result undergoes an Alternating Least Square (ALS) ~\cite{Mitchell97} computation. %in order to predict, given a set of tweets and hashtags, the sensitivity level of these tweets.
%\AD{Rana, do not cite spores for ALS, spores only uses it. Find ALS citation.}\\
%\RA{Fixed.}
A core building block of the ALS computation is the LA pipeline
$(uv^T-$ \textbf{N}$)v$. In our example,
$u$ is a tweet feature vector (of size 2M$\times$1)
and  $v$ is a hashtag feature vector (of size 1000$\times$1).

We have two materialized views available: $V_1$ stores the tweet id and text
as a {\em text datasource in Solr}, and $V_2$ stores tweet id, hashtag id,
and filter-level for all tweets posted from ``USA'', and is materialized on disk as CSV file. %\AD{Rana, where is V2 stored?} \RA{Fixed!}
The rewriting {\em modifies the preprocessing}
of \textbf{N} by introducing $V_1$ and $V_2$; it also
{\em pushes the filter-level selection} from the LA pipeline into
the preprocessing stage. To this end, it rewrites $(uv^T-$ \textbf{N}$)v$
to $uv^Tv-$\textbf{N}$v$, which is more efficient for two reasons.
First, \textbf{N} is ultra sparse (0.00018\% non-zero), which renders
the computation of \textbf{N}$v$ extremely efficient.
Second, SystemML evaluates the chain $uv^{T}v$ efficiently,
computing $v^{T}v$ first, which results in a scalar,
instead of computing $uv^{T}$, which results in a dense matrix
of size 2M$\times$1000 (\sys's cost model realizes this).
Without the rewriting help from \sys,
SystemML is unable to exploit its own efficient operations for
lack of awareness of the distributivity property of vector multiplication
over matrix addition,
$\calA v+\calB v=(\calA+\calB)v$.
The rewriting achieves 14$\times$ speed-up.

HADAD detects and applies all the above-mentioned optimizations combined. It captures RA-, LA-, and cross-model optimizations precisely because it reduces all rewrites to a single setting in which they can %interact and 
synergize: relational rewrites under integrity constraints.

\vspace{2mm}

%% file: problemstatement.tex
\mysection{Problem Statement}
\label{sec:prob}
We consider a set of {\em value domains} $\mathcal{D}_i$, e.g.,
$\mathcal{D}_1$ denotes integers, $\mathcal{D}_2$ denotes real
numbers, $\mathcal{D}_3$ strings, etc, %We consider
and two basic data
types: {\em relations (sets of tuples)} % as in classical database modeling, 
and {\em matrices} (bi-dimensional arrays). Any attribute in
a tuple or cell in a matrix is a value from some
$\mathcal{D}_i$. 
%In practical LA computations, matrices  typically only contain numeric types (usually real numbers). 
We assume {\em a matrix can be implicitly converted into a
relation} (the order among matrix rows is lost), and
{\em the opposite conversion} (each tuple becomes a matrix line, in
some order that is unknown, unless the relation was explicitly sorted
before the conversion).

We consider a hybrid language $\calL$, comprising a set $R_{ops}$ of (unary or binary) {\em  RA
  operators}; concretely, $R_{ops}$ comprises the standard relational selection, 
projection, and join. We also consider a set $L_{ops}$
of {\em LA operators}, comprising: unary  (e.g., inversion  and 
transposition) and binary (e.g., matrix product)  operators 
%unary operators which apply to a matrix and return a matrix (e.g., inversion  and  transposition), a number (e.g., the trace), or two matrices (e.g., the LU decomposition~\cite{kuttler2012linear}); unary operations applied to a matrix and a number and returning a matrix, such as the scalar-matrix multiplication; binary operations applied to two matrices, or a matrix and a number, and returning a matrix or a number, such as matrix sum and product, scalar product, etc. 
The full set $L_{ops}$ of  LA
operations we support is detailed in \S\ref{sec:matrix-al}. 
A {\em hybrid %(RA and LA) 
expression} in $\calL$ is defined as follows:
%\vspace{-4mm}
\begin{itemize}
\veat\item any value from a domain $\mathcal{D}_i$, any matrix, and any
  relation, is an expression;
\veat\item (RA operators): given some expressions $\calE, \calE'$, $ro_1(\calE)$ is
  also an expression, where $ro_1\in R_{ops}$ is a unary relational
  operator, and $\calE$'s type matches $ro_1$'s expected input type. The
  same holds for $ro_2(\calE,\calE')$, where $ro_2\in R_{ops}$ is a binary relational
operator (i.e., the join); 
\veat\item (LA operators): given some expressions $\calE,\calE'$ which are either
  numeric matrices or numbers (which can be seen as degenerate
  matrices of $1\times 1$),  and some real number $r$,  the following
  are also expressions: $lo_1(\calE)$ where $lo_1\in L_{ops}$ is a unary
  operator, and $lo_2(\calE, \calE')$ where $lo_2\in L_{ops}$ is a binary
  operator (again, provided that $\calE,\calE'$ match the expected input types
  of the operators).
\end{itemize}\veat

Clearly, an important set of {\em equivalence rules} hold over our hybrid
expressions, well-known respectively in the RA and the LA
literature. These equivalences lead to {\em alternative evaluation
  strategies} for each expression. Further, we assume given a
(possibly empty) set of {\em materialized views} $\calView \in \calL$, 
%$M$
%expressions 
which have been previously computed over some inputs (matrices
and/or relations), and whose results
are directly available (e.g., as a file on disk). Detecting when a
materialized view can be used instead of evaluating (part of) an
expression is another important source of alternative evaluation strategies. 

Given an expression $\calE$  and a {\em cost model} that assigns a cost
(a real number) to an expression, we consider the problem of
\textbf{identifying the most efficient %(lowest-cost) 
rewrite} 
%\textbf{efficiently generating all equivalent rewrites} 
derived from $\calE$ by: 
($i$)~exploiting RA and LA equivalence rules, and/or 
($ii$)~replacing part of an expression with a scan of a materialized
view equivalent to that expression. 

Below, we detail our approach, the equivalence rules we capture, and
two alternative cost models we devised for this hybrid %RA/LA
setting. Importantly, our solution (based on a {\em relational
  encoding with integrity constraints}) capitalizes on the framework
previously introduced in~\cite{estocada-sigmod}, where it was used to
{\em rewrite queries
using materialized views in a polystore setting}, where the data,
views, and query cover a variety of data models (relational,
JSON, XML, \etc). Those queries can be expressed in a combination of
%standard database 
query languages, including SQL, JSON query languages, XQuery, etc.
\textbf{The ability to rewrite such queries using heterogeneous
  views directly and fully
transfers to \sys}: thus, instead of a relation, we could have
the (tuple-structured) results of an XML or JSON query; views
materialized by joining an XML document with a JSON one and a
relational database could also be reused. The
novelty of our work is to \textbf{extend the benefits of rewriting
and view-based optimization to LA computations, crucial for
%modern analytics and 
ML workloads}. In %what follows 
\S\ref{sec:relencoding}, we focus on capturing matrix data and
LA computations in the relational framework, along with 
relational data naturally; this enables our novel, holistic optimization of
hybrid expressions.
%which includes  LA. 
\eat{\RA{I would suggest to state that from hereafter our focus on LA computations and we detail our contribution and architecture} }
\eat{We describe this on relational (only) and on matrix data, to be able to detail our novel treatment of hybrid expressions including LA. }
%\RA{I agreed with Ioana (in her email), that this work can be positioned as an extension to ESTOCADA by mainly focusing on matrix model and the reduction of LA-views based rewriting to the relational under integrity constraints. So, our problem statement still remains cross-model views based rewriting (hybrid scenarios which is ESTOCADA problem statement) with \textbf{extension} to matrix model  (main focus of this work)} 

%\IM{I guess this should go as follows:
%1.  \textbf{data} (relational tables +
%  matrices); 
%2.  \textbf{an expression language},
%  perhaps as a set of operators + a grammar showing how they can be
%  composed?
%3. recall what {\em equivalence} means in this context;
%4. \textbf{state problem as: given expression and set of materialized
%  views, find equivalent expressions (or, view-based rewritings)}...}

%\vspace{1mm}

%% file: preliminaries.tex
\mysection{PRELIMINARIES}
\label{sec:pre}
We recall conjunctive queries~\cite{chandra1977optimal}, integrity constraints~\cite{AHV95}, and query rewriting under constraints~\cite{pacb-paper}; these concepts are at the core of our approach. 

\mysubsection{Conjunctive Query and Constraints}
\label{sec:cq}
%Syntactically, 
A conjunctive query (or simply CQ) $\calQ$  is an expression 
of the form $\calQ (\overline x) \query \calR_1(\overline y_1),\dots,\calR_n(\overline y_n)$, where %$\overline x = <x_1, . . . , x_k>$,
each $R_i$ is a predicate (relation) of some finite arity, and $\overline x ,\overline y_1, \dots, \overline y_n$ are tuples of variables or constants. Each $\calR_i(\overline y_i)$ is called a relational atom. The expression $\calQ (\overline x)$ is  the \emph{head} of the query, while the conjunction of relational atoms $\calR_1(\overline y_1),\dots,\calR_n(\overline y_n)$ is its \emph{body}. All variables in the head are called \emph{distinguished}. Also, every variable in $\overline x $ must appear at least once in $\overline y_1,\dots, \overline y_n$.
%%%% Below the previous version %%%% 
\eat{\RA{This is the CQ previous version}
Syntactically, a conjunctive query $\calQ$ (or simply CQ) is an expression of
the form 
%\begin{center}
	$\calQ (x_1,\dots,x_k) \query \calR_1(\overline y_1),\dots,\calR_n(\overline y_n)$
%\end{center}
where $n \geq 0$, and each $R_i$, for $i = 1,\ldots,n$, is a relation. %and $\calQ$ is a fresh relation name. 
The expressions $\overline x,\overline y_1, \dots, \overline y_n$ are called {\em free tuples}, and contain either variables or constants. The tuples $\overline y_i$ must match the arities of the corresponding relation. Also, every variable in $\overline x = <x_1, . . . , x_k>$ must appear at least once in $\overline y_1,\dots, \overline y_n$. The expression $\calQ (x_1,\dots,x_k)$ is called the \emph{head} of the query, and $\calR_1(\overline y_1),\dots,\calR_n(\overline y_n)$ is called its
\emph{body}. Each expression $\calR_1(\overline y_i)$ is called an
\emph{atom}: notice that the atom is different from a relation, since
several atoms can refer to the same relation. 
}
Different forms of constraints have been studied in the literature~\cite{AHV95}. %They can be expressed using a fragment of first-order logic, known as the class of embedded dependencies.  
In this work, we use Tuple Generating Dependencies (\textbf{TGD}s) and Equality Generating Dependencies (\textbf{EGD}s), stated by formulas of the form $\forall x_1,\dots x_n$ $ \phi (x_1,\dots x_n) \rightarrow \exists z_1,\dots,z_k$ $\psi(y_1,\dots,y_m)$, where 
%\vspace{-3mm}
%\begin{center}
%$\forall$ $x_1,\dots x_n$ $ \phi (x_1,\dots x_n) \rightarrow \exists z_1,\dots,z_k$ $\psi(y_1,\dots,y_m)$
% \end{center}
% \vspace{-4mm}
 $\{z_1, \dots, z_k \} = \{y_1, \dots, y_m\} \char`\\ \{x_1,\dots, x_n\}$. %is a possibly empty variable set.
The constraint's \emph{premise} $\phi$  is a possibly empty conjunction of
relational atoms over variables $x_1,\dots, x_n$ and possibly constants. The constraint's \emph{conclusion} $\psi$  is a non-empty conjunction of atoms over variables $y_1,\dots, y_m$ and possibly constants, atoms that are relational ones in the case of \textbf{TGD}s or  equality atoms -- of the form  $w = w'$ --  in the case of \textbf{EGD}s. 
%The part of the constraint left of the $\rightarrow$ sign is called its {\em premise}.  
%A \emph{relational atom} has the form $R(w_1, \dots, w_l)$, and an \emph{equality atom} has the form $w = w'$,
%where each of $w,w',w_1, \dots, w_l$, are variables or constants. If  {\em all atoms are equalities}, the constraint is an  EGD; if {\em only relational atoms occur in $\psi$}, it is a TGD. 
For instance, consider a relation \emph{Review(paper, reviewer, track)} listing reviewers of papers submitted to a conference's tracks, and a relation \emph{PC (member, affiliation)} listing the affiliation of every program committee member~\cite{deutsch2009fol}.The fact that a paper can only be submitted to a single track is captured by the following EGD: $\forall p \forall r \forall t  \forall r' \forall t' Review(p,r,t) \wedge Review(p,r',t') \rightarrow t=t' $.  We can also express that papers can be reviewed only by PC members by the following TGD: $\forall p \forall r \forall t~ Review(p, r,t) \rightarrow \exists a ~PC (r, a)$. 

\mysubsection{\textbf{Provenance-Aware Chase \& Back-Chase}}
\label{sec:pacb}
A key ingredient leveraged in our approach is relational query rewriting using views, in the presence of constraints.  The state-of-the-art method for this task, called Chase \& Backchase, was introduced in~\cite{deutsch2006query} and  improved in~\cite{pacb-paper}, as the Provenance-Aware Chase \& Back-Chase (\textbf{PACB} in short). At the core of these methods is the idea to {\em model views as
  constraints}, in this way reducing the view-based rewriting problem to constraints-only
rewriting. Specifically, for a given view $V$ defined by a query, the constraint $V_{IO}$ states that {\em for every match of the view body against the input
data, there is a corresponding (head) tuple in the view output}, while the
constraint $V_{OI}$ states the converse inclusion, i.e., {\em each view
output tuple is due to a view body match}. From a set $\calV$ of view definitions, PACB therefore derives a set of view
constraints 
$C_\calV = \{ V_{IO}, V_{OI}\ |\ V \in \calV \}$.

Given a source schema $\sigma$ with a set of integrity constraints $\calI$,
a set $\calV$ of views defined over $\sigma$, %a target schema $\tau$ which includes $\calV$,
and a conjunctive query $Q$  over $\sigma$, 
the %constraints-only 
\textbf{rewriting problem} thus becomes:
find
every reformulation query $\rho$ 
over the schema of view names $\calV$ that is equivalent to $Q$ under the constraints $\calI \cup C_\calV$.
\begin{example}
For instance, if $\sigma = \{ R, S \}$, $\calI=\emptyset$, $\tau = \{
V \}$
and we have a view $V$ materializing the join of relations $R$ and $S$, 
$
V(x,y) \query R(x,z), S(z,y),
$
the pair of constraints capturing $V$ is the following:

\vspace{-4.5mm}
\begin{eqnarray*}
  V_{IO}: & & \forall x \forall z \forall y ~R(x,z) \wedge S(z,y) \rightarrow V(x,y)\\
  V_{OI}: & &  \forall x \forall y V(x,y) \rightarrow \exists z\ R(x,z) \wedge S(z,y).
\end{eqnarray*}
%\vspace{-4.5mm}

Given the query
$
Q(x,y) \query R(x,z), S(z,y),
$
 %over the schema $\sigma$, 
PACB finds the reformulation
$
\rho(x,y) \query V(x,y).
$
%over the schema $\tau$, comprising the view; 
Algorithmically, this is achieved by: 

($i$)~chasing $Q$ with the constraints $\calI \cup C_\calV^{IO}$, where
$C_\calV^{IO} = \{ V_{IO}\ |\ V \in \calV \}$; 
%\IM{I think we need to add an intuition for the chase here, as some readers may be lost. Maybe we can say:} 
 intuitively, this enriches (extends) $Q$ with all the consequences that follow from its atoms and the constraints $\calI \cup C_\calV^{IO}$.

($ii$)~restricting the chase result to only the $\calV$-atoms; the result is called the {\em universal plan}) $U$.

($iii$)~annotating each atom of the universal plan $U$ with a unique ID called a \emph{provenance term}.

($iv$)~chasing $U$ with the constraints in
$\calI \cup C_\calV^{OI}$, where $C_\calV^{OI} = \{ V_{OI}\ |\
V \in \calV \}$, and annotating each relational atom $a$
introduced by these chase steps with a \emph{provenance formula}\footnote{Provenance formulas are constructed from provenance terms using logical conjunction and disjunction.} $\pi(a)$, which gives the set of $U$-subqueries whose chasing led to the creation of $a$; the result of this phase, called the {\em
  backchase}, is denoted $B$.
    
%    provenance of a set of atoms is the logical conjunction of the provenances of its members
    
($v$)~matching $Q$ against $B$ and outputting as rewritings
the subsets of $U$  that are responsible for the introduction
(during the backchase) of the atoms in the image $h(Q)$ of $Q$; these rewritings are read off directly from the provenance formula $\pi(h(Q))$.
%\footnote{ Recall that a chase step $s$ with constraint $c$ matches $c$'s premise against existing atoms $e$ and adds new atoms $n$ corresponding to $c$'s conclusion.  To support fast detection of responsible atoms in Phase ($iv$), $s$ records that the $e$ atoms are responsible for the introduction of the $n$ atoms~\cite{DBLP:conf/sigmod/IleanaCDK14}.  Our optimization does not affect Phase ($iv$).  }

In our example, $\calI$ is empty, $C_\calV^{IO} = \{ V_{IO} \}$, and the result
of the chase in phase ($i$) is 
$
Q_1(x,y) \query R(x,z), S(z,y), V(x,y).
$
The universal plan obtained in ($ii$) by restricting $Q_1$ to the schema of view names is
$
U(x,y) \query V(x,y)^{p_{0}}
$, where $p_{0}$ denotes the provenance term of atom $V(x,y)$.
The result of backchasing $U$ with $C_\calV^{OI}$ in phase ($iv$) is
$
B(x,y) \query V(x,y)^{p_{0}},$ $R(x,z)^{p_{0}}, S(z,y)^{p_{0}}
$. Note that the provenance formulas
%formulas 
of the $R$ and $S$ atoms (a simple term, in this example) are introduced by chasing the view $V$. 
Finally, in phase ($v$) we find one %containment mapping image $h(Q)$ of $Q$ into the body of $B$, \IM{Trying to work without the containment mapping as some folks may not know what this is and we may get them as reviewers. Sigh.}
 match image given by $h$ from $Q$'s body into the $R$ and $S$ atoms from $B$'s body. %which were introduced during the backchase due to $U$'s atom $V(x,y)$,  
%Therefore, the provenance formula of $\pi(h(Q))$ = $p_{0}$ corresponds to the rewriting $\rho(x,y) \query V(x,y)$, which is an equivalent rewriting of $Q$ and 
The provenance formula $\pi(h(Q))$ of the image $h$ is $p_{0}$, which corresponds to 
%gives as 
an equivalent rewriting $\rho(x,y) \query V(x,y)$. 
\end{example}
%\IM{Does this belong before or after the pbm statement?}
%\RA{Which part you refer to ? you mean the preliminaries section?}
%\RA{Agreed moved, will switch the order}

%% file: overview.tex
%\vspace{1mm}
\mysection{\sys\ Overview}
\label{sec:overview}
We outline here our approach as an extension to~\cite{estocada-sigmod} for solving the rewriting problem introduced in \S\ref{sec:prob}.  

\vspace{2mm}
\noindent{\textbf{Hybrid Expressions and Views}.} 
%\IM{I turned it into Hybrid to connect with the pbm statement}
%\RA{The flow much better now. I put preliminary after pbm statement as well.}
A hybrid expression (whether asked as a query, or describing a materialized view) can be purely relational (RA), in which case we assume it is specified as a conjunctive query~\cite{chandra1977optimal}. Other expressions are purely LA ones; we assume that they are defined in a dedicated LA language such as R~\cite{R}, DML~\cite{boehm2016systemml}, \etc, using LA operators from our set $L_{ops}$ (see \S\ref{sec:matrix-al}), commonly used in real-world ML workloads. Finally, a hybrid expression can combine RA and LA, e.g., an RA expression (resulting in a relation) is treated as a matrix input by an LA operator, whose output may be converted again to a table and joined further, \etc

%\IM{Next, I justify focus on LA:}
Our approach is based on a reduction to a relational model. Below, we show how to bring our hybrid expressions - and, most specifically, their LA components - under a relational form (the RA part of each expression is already in the target formalism).
	
\begin{figure}[t!]
%\vspace{-6mm}
\begin{center}
\input{figure1}
\end{center}
\figspb
\caption{Outline of Our Reduction\label{fig:diagram}}
\vspace{-7mm}
\end{figure}
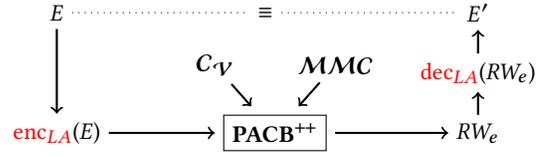

\vspace{2mm}
\noindent{\textbf{Encoding into a Relational Model}.}
Let $\calE$ be an LA expression (query) and $\calView$ be a set of materialized views. 
We reduce the LA-views based rewriting problem to the relational rewriting problem under integrity constraints, as follows (see Figure~\ref{fig:diagram}). 
First, we {\em encode relationally} $\calE$, $\calView$, and the set $L_{ops}$ of LA  operators. Note that the relations used in the encoding are {\em virtual} and {\em hidden}, i.e., invisible to both the application designers and users.  They only serve to support query rewriting via relational techniques. 

These virtual relations are accompanied by a set of relational {\em integrity constraints} $enc_{LA}(LA_{prop})$ that reflect a set $LA_{prop}$ of LA properties of the supported %LA operations 
$L_{ops}$ operations. For instance, we model the \emph{matrix addition} operation using a relation \addMOp$(\calMM,\calNN,\calR)$, denoting that $\calO$ is the result of $\calMM+\calNN$, together with a set of constraints stating that \addMOp\ is a \emph{functional} relation that is \emph{commutative}, \emph{associative}, \etc These constraints are Tuple Generating Dependencies (\textbf{TGD}s) or Equality Generating Dependencies (\textbf{EGD}s)~\cite{AHV95}, which are a generalization of key and foreign key dependencies. %(recall \cref{sec:pre}). 
We detail our relational encoding in \S\ref{sec:relencoding}.

\vspace{2mm}
\noindent{\textbf{Reduction from LA-based to Relational Rewriting}.} 
Our reduction translates the declaration of each view  $\calVV \in \calView$
to %additional 
constraints $enc_{LA}(\calVV)$ that reflect the correspondence
between $\calVV$'s input data and its output. Separately,  $\calE$  is also encoded as a relational query $enc_{LA}(\calE)$ over the relational encodings of $L_{ops}$ and its %basic ingredients 
matrices. %$enc_{LA}(\calE)$ is centered around conjunctive queries.

Now, the reformulation problem is reduced to a purely relational
setting, as follows. We are given a relational query $enc_{LA}(\calE)$ and a set $\calCC_{\calView}=enc_{LA}(\calVV_1) \cup \ldots \cup enc_{LA}(\calVV_n)$ of relational integrity constraints encoding the views $\calView$. We add as further input a set of relational constraints $enc_{LA}(LA_{prop})$, %which encodes relationally the $LA_{prop}$ %operators; 
we called them Matrix-Model Encoding constraints, or $\calMMC$ in short.  We must find the  rewritings $RW_r^i$ expressed over the relational views $\calCC_{\calView}$ and $\calMMC$,  for some integer $k$ and $1\leq i \leq k$, such that each  $RW_r^i$ is equivalent to $enc_{LA}(\calE)$ under these constraints $(\calCC_{\calView} \bigcup \calMMC)$.
Solving this problem yields a {\em relationally encoded rewriting} $RW_e$ expressed over the (virtual) relations used in the encoding; a final {\em decoding} step is needed to obtain $\calE'$, the rewriting of (LA or, more generally, hybrid) $\calE$ using the views $\calV$.

The challenge in coming up with the reduction consists in designing an encoding,  i.e., one in which rewritings found by ($i$) encoding relationally, ($ii$)~solving the resulting relational rewriting problem, and ($iii$)~decoding a resulting rewriting over the views, is guaranteed to produce an equivalent expression $\calE'$ (see \S\ref{sec:relencoding}).
%\IM{I replaced reformulation with rewriting because this is what we use above and below.}
%\IM{I felt the next phrase was "re-reading the figure" and not bringing much:}
% That is, $\calE'$ is a rewriting of $\calE$ given the views $\calView$ and $LA_{prop}$  if $\calE' = dec(RW_{e})$, where  $RW_{e}$ is a relational reformulation of $enc_{LA}(\calE)$ under the  constraints $\calCC_{\calView} \bigcup \calMMC$ obtained from encoding $\calView$ and $LA_{prop}$, respectively. 
%The reduction is detailed in \S\ref{sec:relencoding}.

\vspace{2mm}
\noindent{\textbf{Relational Rewriting Using Constraints}.} To solve the relational rewriting problem under constraints, the engine of choice is Provenance Aware Chase\&Backchase (PACB) ~\cite{pacb-paper}. The PACB engine ($\pacb$ hereafter) has been extended to utilize the $Pruned_{prov}$ algorithm discussed in~\cite{pacb-paper,ileana-thesis}, which prunes inefficient rewritings during the %rewritings 
search phase, based on a simple cost model (see \S\ref{sec:opt}).
% using two different matrix sparsity estimators. 
%\S\ref{sec:opt} details the choice of an efficient rewriting utilizing the $\pacb$ engine.

\vspace{2mm}
\noindent{\textbf{Decoding of the Relational Rewriting}.} For the selected relational reformulation $RW_{e}$ by $\pacb$, a \emph{decoding step} $dec(RW_{e})$ is performed to translate $RW_{e}$ into the native syntax of its respective underlying store/engine (\eg R, DML, etc.).
%\vspace{-1mm}

%% file: figure1.tex
\begin{tikzpicture}[node distance=16mm and 52mm]
\node (E) {$\calE$};
\node (Eprime) [right of=E, xshift=40mm] {${\calE}'$};
\draw [dotted] (E) -- (Eprime) node [midway, fill=white] {$\equiv$};
\node (encLAE)  [below of=E] {\textcolor{red}{enc$_{LA}$}($\calE$)}; 
\draw (E) -> (encLAE) [thick,->]; 
\node (RWe) [below of=Eprime] {$RW_e$};
\node (decRWe) [below of=Eprime, yshift=8mm]
{\textcolor{red}{dec$_{LA}$}($RW_e$)};
\draw (RWe) -> (decRWe) [thick,->]; 
\draw (decRWe) -> (Eprime) [thick,->]; 
\node (PACB) [right of=encLAE,xshift=13mm] {\fbox{\textbf{PACB$^{++}$}}}; 
\draw (encLAE) -> (PACB) [thick,->]; 
\draw (PACB) -> (RWe) [thick,->]; 
\node (Cv) [above of=PACB, xshift=-8mm, yshift=-7mm]
{$\calCC_{\calView}$}; 
\node (MMC) [above of=PACB, xshift=8mm, yshift=-7mm]
{$\calMMC$}; 
\draw (Cv) -> (PACB) [thick,->]; 
\draw (MMC) -> (PACB) [thick,->]; 
\end{tikzpicture}

%% file: encoding.tex
\mysection{Reduction to The Relational Model}
\label{sec:relencoding}
\begin{table*}[ht]
\centering
\begin{tabular}{||c|c||c|c||c|c||}
\rowcolor{lightgray}
\hline \hline
\textbf{Operation} & \textbf{Encoding} & \textbf{Operation}& \textbf{Encoding}  & \textbf{Operation} & \textbf{Encoding}\\
\hline
Matrix scan & $name(\calMI,n)$ & Inversion &\invMOp$(\calMI,\calO)$ & Cells sum & \sumOp$(\calMI,s)$ \\
\hline
Multiplication & \multiMOp$(\calMI,\calMII,\calO)$& \makecell{Scalar \\Multiplication} &\multiMSOp$(s,\calMI,\calO)$& Row sum &\rowsumsOp$(\calMI,\calO)$ \\
\hline
 Addition & \addMOp$(\calMI,\calMII,\calO)$& Determinant &\detOp$(\calMI,\calO)$ & Colsums &\colsumsOp$(\calMI,\calO)$ \\
\hline
Division & \divMOp$(\calMI,\calMII,\calO)$& Trace & \traceOp$(\calMI,s)$& Direct sum &\sumDOp$(\calMI,\calMII,\calO)$\\
\hline
 \makecell{Hadamard \\ product }& \multiEOp$(\calMI,\calMII,\calO)$& Exponential & \expOp$(\calMI,\calO)$ & Direct product& \dirctDOp$(\calMI,\calMII,\calO)$ \\
\hline
Transposition & \trOp$(\calMI,\calO)$& Adjoints &\adjOp$(\calMI,\calO)$& Diagonal &\diagOp$(\calMI,\calO)$\\
\hline
\end{tabular}
\caption{Snippet of the $\calVREM$ Schema}
\vspace{-5mm}
\label{tab:vrem}
\end{table*}
Our internal model is relational, and it makes prominent use of
expressive integrity constraints. This framework
suffices to describe the features and properties of most data models used today, notably including
relational, XML, JSON, graph, etc~\cite{estocada-sigmod,estocada-demo}. 

Going beyond, in this section, we present a novel way to {\em reason relationally about LA primitives/operations} by treating them as uninterpreted functions with black-box semantics, and adding {\em constraints that capture their important properties}. First, we give an overview of a wide range of LA operations that we consider in (\S\ref{sec:matrix-al}). Then, in (\S\ref{sec:vrem}), we show how matrices and their operations can be represented ({\em  encoded}) using a set of {\em virtual} relations, part of a schema
we call $\calVREM$ (for {\em Virtual Relational Encoding of
Matrices}), together with the integrity constraints $\calMMC$ that capture the LA properties 
%\RA{I think this redundant} (\eg associativity, commutativity, and distributivity, \etc) 
of these operations.  
%We emphasize that these relational instances are \emph{virtual} and \emph{invisible} to the application designers and users. 
Regardless
of matrix data’s physical storage, we only use $\calVREM$ to encode
LA expressions and views relationally to reason about them. (\S\ref{sec:rw-constraints})  exemplifies  relational rewritings obtained via our reduction.

%Going beyond, we now show how matrixes and their operations can also be represented ({\em
%  encoded}) using a set of {\em virtual} relations, part of a schema
%we call $\calVREM$ (for {\em Virtual Relational Encoding of
%Matrices}), as shown in Table~\ref{tab:vrem}, together with a set of
%integrity constraints that capture the LA properties (\eg associativity, commutativity, and distributivity, \etc) of these operations. We emphasize that these relational instances
%are \emph{virtual} and \emph{invisible} to the application designers and users. Regardless
%of matrix data’s physical storage, we only use $\calVREM$ to encode
%LA expressions and views relationally to reason about them. 

%In addition, we capture the intended meaning and the properties of the $\calVREM$ relations via integrity constraints.
%Next, we describe our internal virtual \textbf{VREM} schema relations.
%\vspace{-2mm}
\mysubsection{Matrix Algebra}
\label{sec:matrix-al}
We consider a wide range of matrix operations~\cite{kuttler2012linear,axler2015linear}, which are common in real-world machine learning algorithms~\cite{kaggle}:  element-wise multiplication (\ie Hadamard-product) (\multiEOp), matrix-scalar multiplication (\multiMSOp), matrix multiplication (\multiMOp), addition (\addMOp), division (\divMOp), transposition (\trOp), inversion (\invMOp), determinant (\detOp), trace (\traceOp), diagonal (\diagOp), exponential (\expOp), adjoints  (\adjOp), direct sum (\sumDOp), direct product (\dirctDOp), summation (\sumOp), rows and columns summation (\rowsumsOp, \colsumsOp, respectively), QR decomposition (\QROp), Cholesky decomposition {\CHOp), LU decomposition (\LUOp), and pivoted LU decomposition (\LUPOp).
%\vspace{-2mm}
\mysubsection{VREM Schema and Relational Encoding}
%\subsection{VREM Schema and Integrity Constraints}
\label{sec:vrem}

%We here describe in details our modeling of LA operations as a set of
%virtual relations ($\calVREM$). These virtual relations are
%accompanied by a set of integrity constraints. We refer to these
%constraints Matrix Model-Encoding Constraints $\calMMC$(s). 
To model LA operations on the $\calVREM$ relational schema (part of
which appears in Table~\ref{tab:vrem}), we also rely on a set of integrity constraints 
%called {\em Matrix Model-Encoding Constraints}, or  
$\calMMC$, which are encoded using relations in $\calVREM$.
%, as well as the key/uniqueness constraints 
%\IM{clarify: {\em the} key/uniqueness constraints - which ones?}.
%\RA{The EGDs constraints associated with each virtual relation. I removed the sentence above since we discuss them below + they are part of $\calMMC$} 
 We detail the encoding below.
%in short. 
%They mainly reflect the properties of LA operations, which are encoded using relations in $\calVREM$ as well as the key/uniqueness constraints. Next, we discuss the details of our relational encoding.

%\vspace{1mm}
%\noindent\textbf{Base Matrices and Dimensionality Modeling.} 
%\mysubsubsection{\textbf{Base Matrices and Dimensionality Modeling}.} 
%We denote by \calMM$_{i \times j}$($\mathcal{D}$)  a matrix of $i$ rows  $j$ columns, whose entries (values) come from a domain $\mathcal{D}$, e.g., the domain of real numbers $\mathbb{R}$. For brevity we just use \calMM$_{i \times j}$. 
%We define a virtual relation  $name(\calMM_1,n) \in \calVREM$ attaching a unique ID $\calMM_1$
%to any matrix identified by a name denoted $n$ (which may be e.g. of the form ``/data/\calMM.csv'').
%This relation (shown at the top left in Table~\ref{tab:vrem}) is accompanied by an EGD key constraint $\calI_{name} \in \calMMC_{m}$, where $\calMMC_{m}$ $\subset$ $\calMMC$, and $\calI_{name}$ states that {\em two matrices with the same  name $n$ have the same ID}: 
%\veat\begin{center}
%$\calI_{name}$: $\forall \calMM_1 \forall \calMM_2$ $name(\calMM_1,n) \wedge name(\calMM_2,n) \rightarrow \calMM_1 = \calMM_2$ 
%\end{center}\veat
	
\mysubsubsection{\textbf{Base Matrices and Dimensionality Modeling}.} 
We denote by \calMM$_{k \times z}$($\mathcal{D}$)  a matrix of $k$ rows  and $z$ columns, whose entries (values) come from a domain $\mathcal{D}$, e.g., the domain of real numbers $\mathbb{R}$. For brevity we just use \calMM$_{k \times z}$. 
We define a virtual relation  $name(\calMI,n) \in \calVREM$ attaching a unique ID $\calMI$
to any matrix identified by a name denoted $n$ (which may be e.g. of the form ``/data/\calMM.csv'').
This relation (shown at the top left in Table~\ref{tab:vrem}) is accompanied by an EGD key constraint $\calI_{name} \in \calMMC_{m}$, where $\calMMC_{m}$ $\subset$ $\calMMC$, and $\calI_{name}$ states that {\em two matrices with the same  name $n$ have the same ID}: 
\veat\begin{center}
$\calI_{name}$: $\forall \calMI \forall \calMII$ $name(\calMI,n) \wedge name(\calMII,n) \rightarrow \calMI = \calMII$
\end{center}\veat

Note that the matrix ID  in $name$ (and all the other virtual relations used in our encoding) are not IDs of individual matrix objects: rather, each identifies an {\em equivalence class} (induced by value equality) of expressions. That is, two expressions are assigned the same ID iff they yield
%in the same equivalence class, \ie iff they are 
value-based-equal matrices.  In Table ~\ref{tab:vrem}, we use $\calMI$ and $\calMII$ to denote an operation's input matrices' IDs and $\calO$ for the resulting matrix ID, and $s$ for scalar's input and output.

The dimensions of a matrix are captured by a $size(\calMI,k,z)$ relation, where $k$ and $z$ are the number of rows, resp. columns and $\calMI$ is an ID. An EGD constraint $\calI_{size} \in \calMMC_{m}$ holds on the $size$ relation, stating that the ID determines the dimensions: 

\begin{center}
$\calI_{size}$: $\forall \calMI  \forall k_{1} \forall z_{1} \forall k_{2} \forall z_{2} $ $size(\calMI,k_1,z_1) \wedge size(\calMI,k_2,z_2) \rightarrow k_1=k_2 \wedge z_1=z_2$
\end{center}

The identity $\calIden$ and zero $\calZero$ matrices are captured by the $Zero(\calZero)$ and $Identity(\calIden)$ relations, where $\calZero$ and $\calIden$ denote their IDs, respectively. They are accompanied by EGD constraints $\calI_{iden}$, $\calI_{zero} \in \calMMC_{m}$, stating that zero matrices with the same sizes have the same IDs (this also applies for identity matrices with the same size):

\begin{center}
$\calI_{zero}$: $\forall \calZero_1 \forall \calZero_2 \forall k \forall z $ $ Zero(\calZero_1) \wedge size(\calZero_1,k,z) \wedge Zero(\calZero_2) \wedge size(\calZero_2,k,z) \rightarrow \calZero_1=\calZero_2$
\end{center}
\begin{center}
$\calI_{iden}$: $\forall \calIden_1 \forall \calIden_2$ $ Identity(\calIden_1) \wedge size(\calIden_1,k,k) \wedge Identity(\calIden_2) \wedge size(\calIden_2,k,k) \rightarrow \calIden_1=\calIden_2$
\end{center}
%\vspace{1mm}\noindent\textbf{
\mysubsubsection{\textbf{Encoding Matrix Algebra Expressions}} 
LA operations are %treated as a black-box functions and 
encoded into dedicated relations, as shown in Table~\ref{tab:vrem}. %(see Appendix ~\ref{} for a complete list).
%shows a snippet of the relational encoding of matrix algebra operations. 
We  now illustrate the encoding of an LA expression on the $\calVREM$ schema.

%\vspace{-5mm}
\begin{example} Consider the LA expression $\calE$: $((\calMM\calNN)^T)$, where the two matrices \calMM$_{100 \times 1}$ and \calNN$_{1 \times 10}$ are stored as ``$\calMM$.csv'' and ``$\calNN$.csv'', respectively. 
%First, we assume a map from \textit{matrix names} to \textit{id} variables, \textbf{\textit{id(matrix-Name)}}s. For example, the scan of ```$\calMM$.csv'' as $\calMM$ is encoded as $name(\calMI$ $,``\calMM.csv")$, where $id(\calMM)=\calMI$. \IM{We don't need to "assume a map", we already have the "name" relation in the Table.}
The encoding function $enc_{LA}(\calE)$ takes as argument the LA expression $\calE$ and returns a conjunctive query whose: $(i)$ body is the relational encoding of $\calE$ using $\calVREM$ (see below), and $(ii)$ head has one distinguished variable,
    denoting the equivalence class of the result. For instance: 

%\vspace{1mm}
%\begin{tabular}{@{}l}
%$enc(((\calMM\calNN)^T)$ = \\
%$\qquad$  Let $enc(\calMM\calNN)$ =\\ 
%$\qquad$$\qquad$     	Let $enc(\calMM)$ = $\calQuery_0(\calMI) \query name(\calMI,``\calMM.csv")$;\\
%\phantom{$\qquad$$\qquad$     	Let }$enc(\calNN)$  = $\calQuery_1(\calNI) \query name(\calNI,``\calNN.csv")$;\\
%$\qquad$$\qquad$$\qquad$         $\calMOI$ = freshId()\\
%$\qquad$$\qquad$    	in \\
%$\qquad$         	$\calQuery_2(\calMOI) \query multi_{M}(\calMI,\calNI,\calMOI)$,$\calQuery_0(\calMI)$, $\calQuery_1(\calNI)$;\\
%$\qquad$  	  $\calMOII$ = freshId()\\
%$\qquad$  in \\
%$\calQuery(\calMOII) \query tr(\calMOI,\calMOII)$,  $\calQuery_2(\calMOI)$;\\
%\end{tabular}

\vspace{1mm}
\begin{tabular}{@{}l}
$enc(((\calMM\calNN)^T)$ = \\
$\qquad$  Let $enc(\calMM\calNN)$ =\\ 
$\qquad$$\qquad$     	Let $enc(\calMM)$ = $\calQuery_0(\calMI) \query name(\calMI,``\calMM.csv")$;\\
\phantom{$\qquad$$\qquad$     	Let }$enc(\calNN)$  = $\calQuery_1(\calMII) \query name(\calMII,``\calNN.csv")$; \\
$\qquad$$\qquad$$\quad$$\quad$$\calO_1$ = freshId()\\
$\qquad$$\qquad$    	in \\
$\qquad$  $\qquad$ $\quad$         	$\calQuery_2(\calO_1) \query$ \multiMOp$(\calMI,\calMII,\calO_1)$$\wedge$$\calQuery_0(\calMI)$$\wedge$ $\calQuery_1(\calMII)$;\\
$\qquad$$\qquad$   	  $\calO_2$ = freshId()\\
$\qquad$  in \\
$\qquad$ $\quad$ $\calQuery(\calO_2) \query$ \trOp$(\calO_1,\calO_2)$$\wedge$  $\calQuery_2(\calO_1)$;\\
\end{tabular}

\vspace{1mm}
\noindent
In the above, nesting is dictated by the syntax of $\calE$. From the inner (most indented) to the outer, we first encode $\calMM$ and $\calNN$ as small queries using the $name$ relation, then their product (to whom we assign the newly created identifier $\calO_1$
%$\calMOI$
), using the $multi_{M}$ relation and encoding the relationship  between this product and its inputs in the definition of $\calQuery_2(\calO_1)$. 
%$\calQuery_2(\calMOI)$. 
Next, we create a fresh ID $\calO_2$ 
%$\calMOII$ 
used to encode the full $\calE$ (the transposed of $\calQuery_2$) via relation $tr$, in the query $\calQuery(\calO_2)$. For brevity, we  omit the matrices' $size$ relations in this example and hereafter.
Unfolding %$\calQuery_2(\calMOI)$ 
$\calQuery_2(\calO_1)$ in the body of $\calQuery$ yields: 

%\vspace{.5mm}
%\begin{tabular}{l}
%$\calQuery(\calMOII)$ \query $tr(\calMOI,\calMOII)$, $multi_{M}(\calMI,\calNI,\calMOI)$,\\
%\phantom{$\calQuery(\calMOII)$ \query }$\calQuery_0(\calMI)$,$\calQuery_1(\calNI)$; \\
%\end{tabular}

\vspace{.5mm}
\begin{tabular}{l}
$\calQuery(\calO_2)$ \query \trOp$(\calO_1,\calO_2)$$\wedge$ \multiMOp$(\calMI,\calMII,\calO_1)$$\wedge$\\
\phantom{$\calQuery(\calO_2)$ \query }$\calQuery_0(\calMI)$$\wedge$$\calQuery_1(\calMII)$; \\
\end{tabular}
\vspace{1mm}

\noindent
Now, by unfolding $\calQuery_0$ and $\calQuery_1$ in $\calQuery$, we obtain the final encoding of $((\calMM\calNN)^T)$ as a conjunctive query $\calQuery$:  

%\vspace{.5mm}
%\begin{tabular}{l}
%$\calQuery(\calMOII)$ \query $tr(\calMOI,\calMOII)$,  $multi_{M}(\calMI,\calNI,\calMOI)$,\\
%\phantom{$\calQuery(\calMOII)$ \query} $name(\calMI,``\calMM.csv")$,$name(\calNI,``\calNN.csv")$;       	\\				  
%\end{tabular}
%\end{example}\veat

\vspace{.5mm}
\begin{tabular}{l}
$\calQuery(\calO_2)$ \query \trOp$(\calO_1,\calO_2)$$\wedge$  \multiMOp$(\calMI,\calMII,\calO_1)$$\wedge$\\
\phantom{$\calQuery(\calO_2)$ \query} $name(\calMI,``\calMM.csv")$$\wedge$$name(\calMII,``\calNN.csv")$;       	\\				  
\end{tabular}
\end{example}

\veat

\mysubsubsection{\textbf{Encoding LA Properties as Integrity Constraints}.} 
\label{sec:integrit-constraints}
 Figure~\ref{fig:snippet-mmc-1} shows some of the constraints $\calMMC_{LA_{prop}} \subset \calMMC$, which capture textbook LA properties~\cite{kuttler2012linear,axler2015linear} of our LA operations (\S\ref{sec:matrix-al}).
 %of matrix addition, transposition, multiplication, and inversion 
% \IM{Any reason to mention here 4 operators instead of the whole set?}.\RA{No. This just an example of how we encode some of LA properties, I rephrased it above.} 
 The TGDs (\ref{eq:com}), (\ref{eq:dis_trans}) and 
 %,(\ref{eq:dis_inv})
 (\ref{eq:inv_tr}) %(\ref{eq:asso}) 
 state that matrix addition is commutative, matrix transposition is distributive with respect to addition, %, matrix inversion is distributive with respect to multiplication
  and the transposition of the inverse of  matrix $\calMI$ is equivalent to the inverse
  of the transposition of $\calMI$, 
 %and matrix multiplication is associative, 
 respectively. We  also express that the {\em virtual  relations are functional} by using EGD key constraints. For example, the following  $\calI_{multi_{M}} \in \calMMC_{LA_{prop}}$ constraint states that $multi_M$ is functional, that is the products of pairwise equal matrices are equal.
% \begin{center}	
% \vspace{-5mm}
% \begin{align}
%& \calI_{multi_{M}}: \forall  \calMI  \forall \calMII  \forall \calMOI  \forall  \calMOII \nonumber\\
%& multi_{M}(\calMI, \calMII,\calMOI) \wedge multi_{M}(\calMI, \calMII,\calMOII) \rightarrow \nonumber\\ 
%&\calMOI=\calMOII \nonumber
% \end{align}
% \vspace{-4mm}
% \end{center}
 \begin{center}	
 \vspace{-6mm}
 \begin{align}
& \calI_{multi_{M}}: \forall  \calMI  \forall \calMII  \forall \calO_1  \forall  \calO_2 \nonumber\\
& \multiMOp(\calMI, \calMII,\calO_1) \wedge \multiMOp(\calMI, \calMII,\calO_2) \rightarrow \calO_1=\calO_2 \nonumber
 \end{align}

%\vspace{-4mm}
\end{center}
Other properties~\cite{axler2015linear,kuttler2012linear} of the LA operations we consider 
%(Section~\ref{sec:matrix-al}) 
are similarly encoded (see Appendix~\ref{appendixA}).

\begin{figure}[t!]
%\vspace{-4mm}
\begin{align}
& \forall  \calMI  \forall \calMII  \forall \calO ~ \addMOp(\calMI,\calMII,\calO) \rightarrow  \addMOp(\calMII,\calMI,\calO) \label{eq:com}\\
& \forall \calMI \forall \calMII  \forall \calO_1  \forall \calO_2  ~ \addMOp(\calMI,\calMII,\calO_1) \wedge \trOp(\calO_1,\calO_2) \rightarrow \nonumber\\
& \exists \calO_3  \exists \calO_4 ~ \trOp(\calMI,\calO_3) \wedge \trOp(\calMII,\calO_4) \wedge \addMOp(\calO_{3},\calO_{4},\calO_2) \label{eq:dis_trans}\\
%& \forall \calMI  \forall \calMII  \forall \calMOI  \nonumber\\
%&  multi_{M}(\calMI,\calMII,\calMOI)\wedge inv(\calMOI,\calMOII)\rightarrow \nonumber\\
%& \exists \calMOIII  \exists \calMOIIII inv(\calMI, \calMOIII) \wedge inv(\calMII, \calMOIIII) \wedge\nonumber \\
%& multi_{M}(\calMOIIII,\calMOIII,\calMOII) \label{eq:dis_inv}\\
& \forall \calMI  \forall \calO_1  \forall \calO_2 
~ \invMOp(\calMI,\calO_1) \wedge  \trOp(\calO_1,\calO_2) \rightarrow \nonumber \\
&  \exists \calO_3 ~ \trOp(\calMI,\calO_3) \wedge \invMOp(\calO_3,\calO_2)  \label{eq:inv_tr}
%&\forall \calMII  \forall \calMOI  \forall \calMIII  \forall \calMOII \nonumber\\
%& multi_{M}(\calMI,\calMII,\calMOI) \wedge multi_{M}(\calMOI,\calMIII,\calMOII) \rightarrow \nonumber\\
%& \exists \calMOIII multi_{M}(\calMII,\calMIII,\calMOIII) \wedge multi_{M}(\calMOI,\calMOIII,\calMOII) \label{eq:asso}
\end{align}
\vspace{-5mm}
\caption{$\calMMC$ Constraints Capturing Basic LA Properties}
\label{fig:snippet-mmc-1}
\vspace{-2mm}
\end{figure}

\vspace{0.5mm}
\mysubsubsection{\textbf{Encoding LA Views as Constraints.}}
\label{sec:view-enc}
We translate each view definition $\calVV$  (defined in LA language such as R, DML, etc) into relational constraints $enc_{LA}(\calVV) \in \calCC_{\calView}$, where $\calCC_{\calView}$ is the set of relational constraints used to capture the views $\calView$. These constraints show how the view's inputs are related to its output over the $\calVREM$ schema. Figure~\ref{fig:vcqexp} illustrates the encoding as a TGD constraint
of the view $\calVV:(\calNN)^{T} + (\calMM^{T})^{-1}$ stored in a file ``$\calVV$.csv'' and computed based on the matrices $\calNN$ and $\calMM$ (\eg stored as ``$\calNN$.csv'' and ``$\calMM$.csv'', respectively).
%for $V$ in R syntax). 
%\vspace{-5mm}
%\begin{figure}[t!] 
%$\forall \calM \forall \calN \forall \calMOI \forall \calMOII \forall \calMOIII \forall \calMOIII$\\
%$name(\calM,``\calMM.csv") \,\wedge\,$
%$name(\calN,``\calNN.csv") \,\wedge\,$
%$tr(\calN,\calMOI) \,\wedge\,$
%$tr(\calM,\calMOII) \,\wedge\,$
%$inv_M(\calMOII,\calMOIII) \,\wedge\,$
%$add_M(\calMOI,\calMOIII,\calMOIIII) \rightarrow name(\calMOIIII,``\calVV.csv")$%\IM{I turned into "V.csv" for uniformity}
%\vspace{-2.5mm}

\begin{figure}[t!] 
$\forall \calMI \forall \calMII \forall \calO_1 \forall \calO_2 \forall \calO_3 \forall \calO_4$\\
$name(\calMI,``\calMM.csv") \,\wedge\,$
$name(\calMII,``\calNN.csv") \,\wedge\,$
\trOp$(\calMII,\calO_1) \,\wedge\,$
\trOp$(\calMI,\calO_2) \,\wedge\,$
\invMOp$(\calO_2,\calO_3) \,\wedge\,$
\addMOp$(\calO_1,\calO_3,\calO_4) \rightarrow name(\calO_4,``\calVV.csv")$
%\IM{I turned into "V.csv" for uniformity}
\vspace{-2.5mm}

\caption{Relational Encoding of view $V$}
%\IM{Here, we have $name$ atoms. In previous examples, we don't have them. I guess this is because here we define a view...}
%\RA{Yes, and not all of $\calMMC$ constraints involve "name" atoms/relations, the views constraints $\calCC_{\calView}$ involve "name" atoms in the premise and in the conclusion to materialize the result matrix as a view} 
\vspace{-5mm}
\label{fig:vcqexp}
\end{figure}

\vspace{1mm}
\mysubsubsection{\textbf{Encoding Matrix Decompositions}.}
\label{sec:dec}
%Matrix decomposi\-tions play a crucial role in many LA computations. For instance, for every symmetric positive definite matrix $\calMM$ there exists a unique Cholesky Decomposition (CD) of the form $\calMM = \calL \calL^{T}$, where $\calL$ is a lower triangular matrix. 
%We model  CD, as well as other well-known decompositions (LU, QR, and Pivoted LU or PLU) as a set of virtual relations $\calVREM_{dec}$, which we add to  $\calVREM$. For instance, to CD we associate a relation $cho(\calMM_1,\calL)$, which denotes that $\calL$ is the output of the CD decomposition for a given matrix $\calMM$ whose ID is  $\calMM_1$. $cho$ is a functional relation, that is: every symmetric positive definite matrix has a unique CD decomposition. This functional aspect is captured by an EGD, conceptually similar to the constraint $\calI_{multi_{M}}$  (Section~\ref{sec:integrit-constraints}). The property $\calMM = \calL \calL^{T}$ is captured as a TGD constraint $I_{cho}\in \calMMC_{LA_{prop}}$: 
Matrix decomposi\-tions play a crucial role in many LA computations. For instance, for every symmetric positive definite matrix $\calMM$ there exists a unique Cholesky Decomposition (CD) of the form $\calMM = \calLL \calLL^{T}$, where $\calLL$ is a lower triangular matrix. 
We model  CD, as well as other well-known decompositions (LU, QR, and Pivoted LU or PLU) as a set of virtual relations $\calVREM_{dec}$, which we add to  $\calVREM$. For instance, to CD, we associate a relation \CHOp$(\calMI,L)$, which denotes that $L$ is the output of the CD decomposition for a given matrix $\calMM$ whose ID is  $\calMI$. \CHOp\ is a functional relation, meaning every symmetric positive definite matrix has a unique CD decomposition. This functional aspect is captured by an EGD, conceptually similar to the constraint $\calI_{multi_{M}}$  (\S\ref{sec:integrit-constraints}). The property $\calMM = \calLL \calLL^{T}$ is captured as a TGD constraint $I_{cho}\in \calMMC_{LA_{prop}}$:
\begin{center}
\vspace{-3mm}
\begin{align}
&\calI_{cho}: \forall \calMI ~ type(\calMI,``S") \rightarrow \exists ~ \calLL_1 \exists \calLL_2~ cho(\calMI,\calLL_1) \wedge \nonumber \\ 
&type(\calLL_1,``L") \wedge 
tr(\calLL_1,\calLL_2) \wedge multi_{M}(\calLL_1,\calLL_2,\calMI) \label{eq:cho}
\end{align}
\vspace{-2mm}
\end{center}
The atom $type(\calMI$,``S") indicates the type of matrix $\calMI$, where the constant ``S" denotes a matrix that is {\em symmetric positive definite}; similarly, $type(\calLL_1$,``L") denotes that the matrix $\calLL_1$ is a lower triangular matrix.    
%\eat{\RA{We assume this information is given for base matrices.}}
For each base matrix, its type (if available) (\eg symmetric, upper triangular, \etc) is specified as TGD constraint. For example, we state that a certain matrix $\calMM$ (and any other matrix value-equal to $\calMM$) is symmetric positive definite as follows: 

%The atom $type(\calMI$,``S") indicates the type of matrix $\calMI$, where the constant ``S" denotes a matrix that is {\em s}ymmetric positive definite; similarly, $type(\calL_{I1}$,``L") denotes that the matrix $\calL_{I1}$ is a lower triangular matrix.    
%%\eat{\RA{We assume this information is given for base matrices.}}
%For each base matrix, its type (if available) (\eg symmetric, upper triangular, \etc) is specified as TGD constraint. For example, we state that a certain matrix $\calMM$ (and any other matrix value-equal to $\calMM$) is symmetric positive definite as follows: 
%\begin{center}
%\vspace{-5mm}
%\begin{align}
% \forall \calMI ~name(\calMI,``\calMM.csv") \rightarrow type(\calMI,``S") \label{eq:symmetric}
%\end{align}
%\end{center}

\begin{center}
\vspace{-5mm}
\begin{align}
 \forall \calMI ~name(\calMI,``\calMM.csv") \rightarrow type(\calMI,``S") \label{eq:symmetric}
\end{align}
\end{center}

%Stating that $cho$ relation is a functional relation, in other words: every symmetric positive definite matrix has a unique CD decomposition is captured using the following EGD:  
%
%\begin{center}
%\vspace{-5mm}
%\begin{align}
%\forall \calMI \forall \calL_{1} \forall \calL_{2} ~ cho(\calMI, \calL_{1}) \wedge cho(\calMI, \calL_{2}) \rightarrow \calL_{1}= \calL_{2}
%\end{align}
%\vspace{-5mm}	
%\end{center}
%\eat{Also for every symmetric positive definite matrix $\calMM$ there exists a CD decomposition of the form $\calMM = \calL \calL^{T}$, where $\calL $ is a lower triangular matrix. This can be captured as a TGD constraint:
%\begin{center}
%\vspace{-4mm}
%\begin{align}
%&\forall \calMI ~ type(\calMI,``S") \rightarrow \exists ~ \calI \exists \calI_{o}~ cho(\calMI,\calI) \wedge type(\calI,``L") \wedge \nonumber \\ &tr(\calI,\calI_{o}) \wedge  multi_{M}(\calI,\calI_{o},\calMI) \label{eq:cho}
%\end{align}
%\end{center}
%where type($\calMI$,``S") indicates the type of a matrix $\calMI$.
%}

%\vspace{-5mm}
\begin{example} 
Consider a view $\calVV$=$\calNN + \calLL\calLL^{T}$, where $\calLL = cho(\calMM)$ and $\calMM$ is a symmetric positive definite matrix encoded as in (\ref{eq:symmetric}). Let $\calE$ be the  LA expression $\calMM+\calNN$. The reader realizes easily that $\calVV$ can be used to answer $\calE$ directly, thanks to the specific property of the CD decomposition  (\ref{eq:cho}), and since $\calMM +\calNN=\calNN+\calMM$, which is encoded in (\ref{eq:com}). However, at the syntactic level, $\calVV$ and $\calE$ are very dissimilar. Knowledge of (\ref{eq:com}) and (\ref{eq:cho}) and the ability to reason about them is crucial in order to efficiently answer $\calE$ based on $V$.
% Note  that view $V$ does not match into $\calE$, and in order to use the view $V$ to fully answer $\calE$, the property $\calMM=\calL\calL^{T}$, which is encoded in (\ref{eq:cho}) should be exploited together with $\calMM +\calNN=\calNN+\calMM$, which is encoded in (\ref{eq:com}). Without capturing such properties, the view $V$ can not be utilized to answer $\calE$.        
\end{example}

%The output matrix of CD decomposition is a lower triangular matrix $\calL$, which is not symmetric positive definite, meaning that CD decomposition can not be applied again on $\calL$. %; this is the fixed point of CD decomposition as it is only applied to a symmetric positive definite matrix, where we captured this fixed point as integrity constraint (\ref{eq:cho}). However, f
%For other decompositions, such as $QR(\calMM)=[\calQR,\calR]$ decomposition, where $\calMM$ is a square matrix, $\calQR$ is an orthogonal matrix~\cite{kuttler2012linear} and  $\calR$ is an upper triangular matrix, there exists a $QR$ decomposition for the orthogonal matrix $\calQR$ such that  $QR(\calQR)=[\calQR,I]$, where $I$ is an identity matrix and $QR(\calR)=[I,\calR]$. We say the {\em fixed point} of the QR decomposition is $QR(I)=[I,I]$. These properties of the $\calQR$ decompositions are captured with the following constraints, which are part of $\calMMC_{LA{prop}}$: 

The output matrix of CD decomposition is a lower triangular matrix $\calLL$, which is not necessary a symmetric positive definite matrix, meaning that CD decomposition can not be applied again on $\calLL$. %; this is the fixed point of CD decomposition as it is only applied to a symmetric positive definite matrix, where we captured this fixed point as integrity constraint (\ref{eq:cho}). However, f
For other decompositions, such as $QR(\calMM)=[\calQR,\calR]$ decomposition, where $\calMM$ is a square matrix, $\calQR$ is an orthogonal matrix~\cite{kuttler2012linear} and  $\calR$ is an upper triangular matrix, there exists a $QR$ decomposition for the orthogonal matrix $\calQR$ such that  $QR(\calQR)=[\calQR,I]$, where $I$ is an identity matrix and $QR(\calR)=[I,\calR]$. We say the {\em fixed point} of the QR decomposition is $QR(I)=[I,I]$. These properties of the $\calQR$ decompositions are captured with the following constraints, which are part of $\calMMC_{LA{prop}}$: 

%\begin{center}
%\vspace{-3mm}
%\begin{align}
%& \forall \calMI \forall n \forall i ~name(\calMI ,n) \wedge size(\calMI, i,i) \rightarrow ~ \exists \calQR_{O1} \exists \calR_{O1} \nonumber \\
%& QR(\calMI,\calQR_{O1},\calR_{O1}) \wedge  type(	\calQR_{O1}, ``O")  \wedge   type(	\calR_{O1}, ``U") \nonumber  \\
%& \wedge  multi_{M}(\calQR_{O1},\calR_{O1},\calMI)\\
%& \forall \calQR_{O1} ~ type(	\calQR_{O1}, ``O") \rightarrow \exists I ~  QR(\calQR_{O1},\calQR_{O1},I) \wedge identity(I) \nonumber \\
%& \wedge multi_{M}(\calQR_{O1},I,\calQR_{O1}) \\
%& \forall \calR_{O1}  ~ type( \calR_{O1}, ``U") \rightarrow \exists I ~  QR(\calR_{O1},I,\calR_{O1}) \wedge identity(I) \nonumber \\
%& \wedge multi_{M}(I,\calR_{O1}, \calR_{QI})\\
%& \forall I ~ identity(I) \rightarrow QR(I,I,I)
%\end{align}
%\end{center}

\begin{center}
\vspace{-3mm}
\begin{align}
& \forall \calMI \forall n \forall k ~name(\calMI ,n) \wedge size(\calMI, k,k) \rightarrow ~ \exists \calQR \exists \calR \nonumber \\
& QR(\calMI,\calQR,\calR) \wedge  type(\calQR, ``O")  \wedge   type(\calR, ``U") \nonumber  \\
& \wedge  multi_{M}(\calQR,\calR,\calMI)\\
& \forall \calQR ~ type(\calQR, ``O") \rightarrow \exists I ~  QR(\calQR,\calQR,I) \wedge identity(I) \nonumber \\
& \wedge multi_{M}(\calQR,I,\calQR) \\
& \forall \calR  ~ type(\calR, ``U") \rightarrow \exists I ~  QR(\calR,I,\calR) \wedge identity(I) \nonumber \\
& \wedge multi_{M}(I,\calR, \calR)\\
& \forall I ~ identity(I) \rightarrow QR(I,I,I)
\end{align}
\end{center}

Known LA properties of the other matrix decompositions (LU and PLU) are similarly encoded in Appendix~\ref{appendixA}.

\vspace{1mm}
\mysubsubsection{\textbf{Encoding LA-Oriented System Rewrite Rules}}
\label{sec:rwrules} 
Most  LA-oriented systems~\cite{R,NumPy} execute an incoming expression (LA pipeline) \emph{as-is}, that is:  run operations in a sequence, whose order is dictated by the expression syntax.
Such systems do not exploit basic LA properties, e.g., reordering a chain of multiplied matrices in order to reduce the intermediate size.
%the number of operations.
% o due to the lack of exploiting LA properties (\eg matrix chain multiplication optimal order can be achieved by exploiting the associative property of matrix multiplication) or supporting any rewrite rules/heuristics to further optimize LA pipelines. 
SystemML~\cite{boehm2016systemml} is the only system that models {\em some} LA properties as static rewrite rules. It also comprises a set of {\em rewrite rules} which modify the given expressions to avoid large intermediates for aggregation and statistical operations such as \rowsumsOp$(\calMM)$, \sumOp$(\calMM)$, etc. For example, SystemML uses rule:
$$
\sumOp(\calMM\calNN) =\sumOp(\colsumsOp(\calMM)^{T}\odot~\rowsumsOp(\calNN))\ \ \ \ (i)
$$
to rewrite \sumOp$(\calMM\calNN)$ (summing all cells in the matrix product)
%into $sum(colSums(\calMM)^{T}\odot~rowSums(\calNN))$,
where $\odot$ is a matrix element-wise multiplication, to avoid actually computing $\calMM\calNN$ and materializing it; similarly, it rewrites \sumOp$(\calMM^{T})$ into \sumOp$(\calMM)$, to avoid materializing  $\calMM^{T}$, etc. However, the performance benefits of rewriting depend on the rewriting power (or, in other words, on {\em how much the system understands  the semantics of the incoming expression}), as the following example shows.  
% has amount of understanding (knowledge, rwrwithout having enough reasoning about LA properties, potentially, several rules can not be fully triggered/explored which can lead to miss performance-saving opportunities. 

\begin{example}
Consider the LA expression $\calE$=$((\calMM^{T})^{k}(\calMM + \calNN)^{T})$, where $\calMM$ and $\calNN$ are square matrixes, and expression $\calE'$=\sumOp$(\calE)$, which computes the sum of all cells in $\calE$. The expression $\calE'$ can be rewritten to $RW_1:$ \sumOp$(\calE'')$, where $\calE''$ is:
	
	\begin{center}
	%\begin{align}
	\sumOp$($\colsumsOp$(\calMM+\calNN)^{T} \odot ~$ \rowsumsOp$(\calMM^{k}))$	
	 %\end{align}
	\end{center}
Failure to exploit the properties $LA_{prop1}: \calMM^{T}\calNN^{T}=(\calMM\calNN)^{T}$ and $LA_{prop2}:(\calMM^{n})^{T}=(\calMM^{T})^{n}$ prevents from finding rewriting $RW_1$.\\
$\calE'$ admits the alternative rewriting

$RW_2$: \sumOp$(($\colsumsOp$((\calMM^{T})^{k}))^T \,\odot\, ($\colsumsOp$(\calMM+\calNN))^T)$

\noindent
which can be obtained by directly applying the rewrite rule $(i)$ above and
\rowsumsOp$(\calMM^{T})$=\colsumsOp$(\calMM)^{T}$, without exploiting the properties $LA_{prop1}$
and $LA_{prop2}$.
%\RA{remove it already mentioned in details in experiments} 
%~\footnote{Interestingly, SystemML includes the rewrite rule $(i)$ but did not apply it during optimization. Thus, it does not find $RW_1$.}
However, $RW_2$ introduces more intermediate results than $RW_1$.
\end{example}

{\em To fully exploit the potential of rewrite rules (for statistical or aggregation operations), they should be accompanied by sufficient knowledge of, and reasoning on, known properties of LA operations}.

%In addition, in order
To bring such fruitful optimization to other LA-oriented systems lacking support of such 
%statistical or aggregation 
rewrite rules, we have incorporated SystemML's rewrite rules into our framework, encoding them as a set of integrity constraints over the virtual relations in the schema $\calVREM$, denoted $\calMMC_{StatAgg}$ ($\calMMC_{StatAgg}\subset \calMMC$).
%and which are part of our constraint set $\calMMC$. 
Thus, these rewrite rules can be exploited together with other LA properties.  
For instance, the rewrite rule $(i)$ is modeled by the following integrity constraint $ \calI_{sum} \in \calMMC_{StatAgg}$: 
%\vspace{-4mm}
%\begin{center}
%%$\calI_{sum}$: 
%\begin{align*}
%&\forall \calMI \forall \calNI \forall \calMO ~ multi_M(\calMI,\calNI,\calMO)\wedge sum(\calMO, s) \rightarrow \nonumber \\
%&\exists \calMOI \exists \calMOII \exists \calMOIII \exists \calMOIIII  colSums(\calMI,\calMOI) \wedge   tr(\calMOI,\calMOII) \nonumber \\
%& \wedge rowSums(\calNI,\calMOIII) \wedge
% multi_E(\calMOII,\calMOIII,\calMOIIII) \wedge sum(\calMOIIII,s)
%\end{align*}
%\end{center}

\vspace{-4mm}
\begin{center}
%$\calI_{sum}$: 
\begin{align*}
&\forall \calMI \forall \calMII \forall \calO ~ \multiMOp(\calMI,\calMII,\calO)\wedge \sumOp(\calO, s) \rightarrow \nonumber \\
&\exists \calO_1 \exists \calO_2 \exists \calO_3 \exists \calO_4  \colsumsOp(\calMI,\calO_1) \wedge   \trOp(\calO_1,\calO_2) \nonumber \\
& \wedge \rowsumsOp(\calMII,\calO_3) \wedge
 \multiEOp(\calO_2,\calO_3,\calO_4) \wedge \sumOp(\calO_4,s)
\end{align*}
\vspace{-5mm}
\end{center}
  We refer the reader to Appendix ~\ref{appendixB} for a full list of SystemML's encoded rewrite rules. 
\mysubsection{Relational Rewriting Using Constraints}
\label{sec:rw-constraints}
With the set of views constraints $\calCC_{\calView}$ and $\calMMC = \calMMC_{m} \cup \calMMC_{LA_{prop}} \cup \calMMC_{StatAgg}$,
%LA properties relationally encoded, 
 we rely on  \textbf{$\pacb$} %(see Section~\ref{sec:opt})  to identify the materialized views that may be leveraged to 
to rewrite a given expression under integrity constraints. We exemplify this below, and detail  \textbf{$\pacb$}'s inner workings in Section~\ref{sec:opt}. 

The view $\calVV$ shown in Figure~\ref{fig:vcqexp} can be used to \emph{fully} rewrite (return the answer for) the pipeline $\calQuery_{p}:(\calMM^{-1}+\calNN)^{T}$ by exploiting the TGDs (\ref{eq:com}), (\ref{eq:dis_trans}) and (\ref{eq:inv_tr}) listed in Figure~\ref{fig:snippet-mmc-1}, which describe the following three LA properties, denoted $LA_{prop_1}$: 
%
%\vspace{-2mm}
%\begin{center}
%$LA_{prop_{1}}:$\\
%$\calMI+\calMII=\calMII+\calMI$; 
%$((\calMI+\calMII))^{T}= (\calMI)^{T}+(\calMII)^{T}$  and 
%$((\calMI)^{-1})^{T}=((\calMI)^{T})^{-1}$. 
$\calMM+\calNN=\calMM+\calNN$; 
$((\calMM+\calNN))^{T}= (\calMM)^{T}+(\calNN)^{T}$  and 
$((\calMM)^{-1})^{T}=((\calMM)^{T})^{-1}$. 
%\end{center}
%\vspace{-5mm}
%\begin{figure}[h!]
%\begin{align}
% &\forall \calMI  \forall \calMII \forall \calMOI \nonumber \\
% & add_M(\calMI,\calMII,\calMOI) \rightarrow  add_M(\calMII,\calMI,\calMOI)\\
%& \IM{The one above was numbered (3) previously, refer to it instead?} \nonumber \\
% &\forall \calMI  \forall \calMII  \forall \calMOI  \forall\calMOII  \nonumber \\
% & add_M(\calMI,\calMII,\calMOI) \wedge tr(\calMOI,\calMOII) \rightarrow  \nonumber \\
% & \exists \calMOIII  \exists \calMOIIII tr(\calMI,\calMOIII) \wedge tr(\calMII,\calMOIIII) \wedge  \nonumber \\
%& \IM{The one above was numbered (5) previously, refer to it instead?} \nonumber \\
%& add_M(\calMOIII,\calMOIIII,\calMOII)\\
%&\forall \calMI  \forall \calMOI  \forall \calMOII  \nonumber \\
%& inv_M(\calMI,\calMOI) \wedge  tr(\calMOI,\calMOII) \rightarrow \nonumber \\
%&  \exists \calMOIII tr(\calMI,\calMOIII) \wedge inv_M(\calMOIII,\calMOII) 
%\end{align}
%\vspace{-5mm}
%\caption{Subset of $\calMMC$ Constraints to Exploit the View $V$}
%\label{fig:prop}
%\vspace{-3mm}
%\end{figure}
%
The relational rewriting $RW_0$ of $\calQuery_{p}$ using the view $\calVV$ is $RW_0(\calO_4) \query name (\calO_4, ``\calVV.csv")$.
%
%\begin{figure}[h!] 
%\vspace{-3mm}
%$RW_0(\calMOIIII):- name (\calMOIIII, ``V.csv")$
%\caption{Relational Rewriting $RW_0$}
%\label{fig:rwcqexp}
%\vspace{-3mm}
%\end{figure}
In this example, $RW_0$ is the only \textit{views-based} rewriting of $\calQuery_{p}$. However, \emph{five} other rewritings exist (shown in Figure~\ref{fig:equ-rws}), which reorder its operations just by exploiting the set $LA_{prop_{1}}$ of LA properties.
\begin{figure}[t!]
%\vspace{-7mm}
\begin{align}
&RW_1: (\calMM^{-1})^{T}+\calNN^{T}\ \ \ \ \ \   RW_2: (\calMM^{T})^{-1}+\calNN^{T} \nonumber\\
&RW_3: \calNN^T+(\calMM^{-1})^T\ \ \ \  \ \ RW_4: \calNN^{T}+(\calMM^{T})^{-1}\nonumber\\
&RW_5:(\calNN+\calMM^{-1})^{T}\nonumber
\end{align}
\vspace{-7.5mm}
\caption{Equivalent rewritings of the pipeline $\calQuery_{p}$.}
\label{fig:equ-rws}
\vspace{-4mm}
\end{figure}

%Lacking the views, 
Rewritings $RW_0$ to $RW_5$ have different evaluation costs. We discuss next
how we estimate which among these alternatives
(including evaluating $\calQuery_{p}$ directly) is likely the most efficient. 
\vspace{1mm}

%% file: rewritingPruning.tex
\mysection{Choice of an Efficient Rewriting}
\label{sec:cost-model}
We introduce our cost model (\S\ref{sec:cost-fun}), which can take two different sparsity estimators (\S\ref{sec:sparsity-estimator}). Then, we detail our extension to the PACB rewriting engine based on the  $Prune_{prov}$ algorithm (\S\ref{sec:prune}) to prune out inefficient rewritings.
\label{sec:opt}

\mysubsection{Cost Model}
\label{sec:cost-fun}
%\IM{I felt the previous was unclear when jumping from \emph{intermediate result size} to \emph{sparsity}. I added the connection: ``operations on 0s are very fast, this is why only non-zeros matter''. Please check!}\AD{This is really about the size of the intermediate result representation. added}
We estimate the cost of an expression $\calE$, denoted $\gamma(\calE)$, as the sum of the intermediate
result sizes if one evaluates  $\calE$  ``as stated'', in the syntactic order dictated by the expression.
Real-world matrices may be {\em dense} (most or all elements are non-zero) or {\em sparse}
(a majority of zero elements). The latter admit more economical representations that
do not store zero elements, which our intermediate result size measure excludes.
To estimate the number of non-zeros ({\em nnz}, in short), we incorporated two different
sparsity estimators from the literature (discussed in \S\ref{sec:sparsity-estimator}) into our framework. 

\begin{example}
Consider  $\calE_1=(\calMM\calNN)\calMM$ and $\calE_2=\calMM(\calNN\calMM)$, where we assume the matrices $\calMM_{50K \times 100}$ and $\calNN_{100 \times 50K}$ are dense. The total cost of $\calE_1$ is $\gamma(\calE_1)=50K\times50K$
%(the expected size of $\calMM\calNN$)
and  $\gamma(\calE_2)=100\times100$
%(the expected size of \calNN\calMM)
. 
\vspace{-2mm}
\end{example}

%We found this model to work well; devising a generic cost model for LA pipelines across different systems is beyond the scope of this paper. \IM{Not essential}

\mysubsection{LA-based Sparsity Estimators}
\label{sec:sparsity-estimator}
We outline below two existing {\em sparsity estimators}~\cite{spores,boehm2016systemml} that we have incorporated into our framework to estimate {\em nnz}.\footnote{Solving the problem of sparsity estimation is beyond the scope of this paper}. 
\mysubsubsection{\textbf{Naïve Metadata Estimator}} 
\label{sec:naiveEstimator}
The naïve metadata estimator~\cite{boehm2016declarative,spores} derives the sparsity of the output of LA expression solely from the base matrices' sparsity. %These incur \IM{I changed to "this incurs" because "this" refers to the Estimator (singular) in the subsection title. I moved it to singular (estimator, not estimators) becaus even if the two cited works do sth sligthly differently from each other, in our work we refer to A naive estimator (only one).}
This incurs no runtime overhead since metadata about the base matrices, including the {\em nnz}, columns and rows are available before runtime in a specific metadata file. The most common estimator is the {\em worst-case estimator}~\cite{boehm2016declarative}, which we use in our framework.
\mysubsubsection{\textbf{Matrix Non-zero Count (MNC) Estimator}.} 
\label{sec:MNCEstimator}
The MNC estimator ~\cite{sommer2019mnc} exploits matrix structural properties such as single non-zero per row, or columns with varying sparsity, for efficient, accurate, and general sparsity estimation; it relies on count-based histograms that exploit these properties. We have also adopted this framework into our approach, and 
%We refer the reader to  for more details about the MNC estimator.
%\footnot {MNC is open sourced at github.com/apache/systemml}. 
compute histograms about the \textit{base} matrices offline. However, the MNC framework still needs to derive and construct histograms for {\em intermediate results} online (during rewriting cost estimation).  We study this overhead in (\S\ref{sec:exp}). 

%MNC also lacks the support of estimating the sparsity of complex matrix operations. For this reason, we made the same assumption discussed in Section ~\ref{sec:naiveEstimator}.
\mysubsection{Rewriting Pruning: $\pacb$}
\label{sec:prune}
We extended the PACB rewriting engine with the $Prune_{prov}$  algorithm sketched and
discussed in ~\cite{ileana-thesis,pacb-paper}, to eliminate inefficient rewritings during the rewriting
search phase. The naïve PACB algorithm generates all minimal (by join count) rewritings %reformulations 
before choosing a \emph{minimum-cost} one. While this sufficies on the scenarios considered
in~\cite{pacb-paper,estocada-sigmod},
the settings we obtain from our LA encoding stress-test
the naïve algorithm, as commutativity, associativity, etc. blow up the space of alternate rewritings
exponentially. Scalability considerations forced us to further optimize naïve PACB
to find only \emph{minimum-cost %reformulation
rewritings}, aggressively pruning the others during the generation phase.
%Applying cost-based pruning of the provenance formula (Section~\ref{sec:pre}) while backchasing can reduce the search space.
%and speed up rewriting. 
%reformulations. 
We illustrate $Prune_{prov}$ and our improvements next.

%\vspace{1mm}
\noindent\textbf{$Prune_{prov}$ Minimum-Cost Rewriting. %Reformulation.
}Recall from (\S\ref{sec:pre}) that the minimal rewritings of a query $Q$ are
obtained by first finding the set $\calH$ of all matches (\ie containment mappings) from $Q$ to the
result $B$ of backchasing the universal plan $U$. Denoting with $\pi(S)$
the provenance formula of a set of atoms $S$, PACB computes the 
DNF form $D$ of $\bigvee_{h \in \calH} \pi(h(Q))$. Each conjunct $c$ of $D$
determines a subquery $sq(c)$ of $U$ which is guaranteed to be a rewriting of $Q$.
%yield all rewritings $\calRW$, where each conjuncg  $rw \in \calRW$, where $rw$ is a $U$-subquery.

\AD{confusing notation: $p$ used both for premise and for provenance, $C$ used both for constraint
and for provenance conjunct. Rewrote.}

The idea behind cost-based pruning is that, whenever the naive PACB backchase
would add a provenance conjunct $c$ to an existing atom $a$'s provenance formula
$\pi(a)$, $Prune_{prov}$ does so more conservatively:
if the cost $\gamma(sq(c))$ is larger than the minimum cost threshold $T$ found so far,
then $c$  will never participate in a minimum-cost rewriting and need not
be added to $\pi(a)$. Moreover, atom
$a$ itself need not be chased into $B$ in the first place if all its provenance conjuncts
have above-threshold cost.

\begin{example}
\vspace{-2mm}
Let $\calE$ = $\calMM(\calNN\calMM)$, where we assume for simplicity that $\calMM_{50K \times 100}$ and $\calNN_{100 \times 50K}$ are dense. Exploiting the associativity of matrix-multiplication $(\calMM\calNN)\calMM=\calMM(\calNN\calMM)$ during the chase leads to the following universal plan $U$ annotated with provenance terms: 
%\begin{center}
% %$U:$
% $U(\calMOII): 
% name(\calMI,``\calMM.csv")^{p_{0}} \wedge
% size(\calM,50000,100)^{p_{1}} \wedge
% name(\calNI,``\calNN.csv")^{p_{2}} \wedge
% size(\calNI,100,50000)^{p_{3}} \wedge
% multi_{M}(\calMI,\calNI,\calMOI)^{p_{4}}\wedge
% multi_{M}(\calMOI,\calMI,\calMOII)^{p_{5}}\wedge
% multi_{M}(\calNI,\calMI,\calMOIII)^{p_{6}}\wedge
% multi_{M}(\calMI,\calMOIII,\calMOII)^{p_{7}}$
%\end{center}
\begin{center}
 %$U:$
 $U(\calO_2): 
 name(\calMI,``\calMM.csv")^{p_{0}} \wedge
 size(\calMI,50000,100)^{p_{1}} \wedge
 name(\calMII,``\calNN.csv")^{p_{2}} \wedge
 size(\calMII,100,50000)^{p_{3}} \wedge$
\multiMOp$(\calMI,\calMII,\calO_1)^{p_{4}}\wedge$
\multiMOp$(\calO_1,\calMI,\calO_2)^{p_{5}}\wedge$
\multiMOp$(\calMII,\calMI,\calO_3)^{p_{6}}\wedge$
\multiMOp$\calMI,\calO_3,\calO_2)^{p_{7}}$
\end{center}

\AD{Excellent example! But nobody can follow even the one chase step with C, as you
are overloading variable names and every variable in the constraint does not bind
to the same variable in $U$.}
Now, consider in the back-chase the associativity constraint $C$:
%\vspace{-5mm}
%\begin{center}
%\begin{align*}
%&\forall \calMII  \forall \calMOI  \forall \calMIII  \forall \calMOII \nonumber\\
%& multi_{M}(\calMI,\calMII,\calMOI) \wedge multi_{M}(\calMOI,\calMIII,\calMOII) \rightarrow \nonumber\\
%& \exists \calMOIII multi_{M}(\calMII,\calMIII,\calMOIII) \wedge multi_{M}(\calMI,\calMOIII,\calMOII)
%\end{align*}
%\end{center
\vspace{-5mm}
\begin{center}
\begin{align*}
&\forall \calMI  \forall \calMII  \forall \calO_1  \forall \calO_2 \nonumber\\
& \multiMOp(\calMI,\calMII,\calO_1) \wedge \multiMOp(\calO_1,\calMI,\calO_2) \rightarrow \nonumber\\
& \exists \calO_4 \multiMOp(\calMII,\calMI,\calO_4) \wedge \multiMOp(\calMI,\calO_4,\calO_2)
\end{align*}
\end{center}
There exists a containment mapping $h$ embedding the two atoms in the premise $P$ of $C$ into the $U$
atoms whose provenance annotations are $p_4$ and $p_5$. The  provenance conjunct
collected from $P$'s image is $\pi(h(P))$=$p_{4} \wedge p_{5}$.

 %\IM{Below, I wanted to show \textbf{first} how pruning works and \textbf{second} when it leaves a (good) rewriting intact. The example was in the reverse order. But I'm not fully sure I understood it, please check?} \RA{It is correct. I did it in the reverse order, I just need to clarify how these formulas are bing introduced} 
{\em Without pruning}, the backchase would chase $U$ with the constraint $C$, yielding $U'$
which features additional  $\pi(h(P))$-annotated atoms
%$
%multi_{M}(\calNI,\calMI,\calMOIIII)^{p_4 \wedge p_5} \wedge multi_{M}(\calMI,\calMOIIII,\calMOII)^{p_4 \wedge p_5}.
%$

\multiMOp$(\calMII,\calMI,\calO_4)^{p_4 \wedge p_5} \wedge$ \multiMOp$(\calMI,\calO_4,\calO_2)^{p_4 \wedge p_5}$

$\calE$ has precisely two matches $h_1,h_2$ into $U'$.
$h_1(\calE)$ involves the newly added atoms as well as those annotated with $p_0, p_1, p_2, p_3$.
Collecting all their provenance annotations yields the
conjunct
$c_1 = p_{0} \wedge p_{1} \wedge p_{2} \wedge p_{3} \wedge  p_{4} \wedge p_{5}$. $c_1$ determines the
$U$-subquery $sq(c_1)$ corresponding to the rewriting $(\calMM\calNN)\calMM$, of cost $(50K)^{2}$.

$h_2(\calE)$'s image yields the provenance conjunct
$c_2 = p_{0} \wedge p_{1} \wedge p_{2} \wedge p_{3} \wedge  p_{6} \wedge p_{7}$,
which determines the rewriting $\calMM(\calNN\calMM)$
that happens to be the original expression $\calE$
of cost $100^2$.

The naive PACB would find both rewritings, cost them, and drop the former in favor of the latter.

\AD{this is not true, rewrote:
$h_1(\calE)$ comprises the atoms with 
$p_{0} \wedge p_{1} \wedge p_{2} \wedge p_{3} \wedge  p_{4} \wedge p_{5}$, %\IM{I think $U'$ and this provenance formula deserves more explanation} \RA{Agreed. I need to revisit this} 
which introduces a large intermediate result ($\gamma(sq(\pi(h(P)))$ = $(50K)^{2}$), above the threshold.
}

{\em With pruning}, the  threshold $T$ is the cost of the original expression
$100^2$. The chase step with $C$ is never applied, as it would introduce
the provenance conjunct $\pi(h(P))$ which determines U-subquery %$sq(\pi(h(P)) = 
%multi_{M}(\calMI,\calNI,\calMOI)^{p_{4}}$ $\wedge$ $multi_{M}(\calMOI,\calMI,\calMOII)^{p_{5}}$
$sq(\pi(h(P)) =$ 
\multiMOp$(\calMI,\calMII,\calO_1)^{p_{4}}$ $\wedge$ \multiMOp$(\calO_1,\calMI,\calO_2)^{p_{5}}$	

of cost $(50K)^2$ exceeding $T$.
The atoms needed as image of $\calE$ under $h_1$ are thus never produced while backchasing $U$,
so the expensive rewriting is never discovered.
This leaves only the match image $h_2(\calE)$, which corresponds to the efficient
rewriting $\calMM(\calNN\calMM)$.

%pruning removes the inefficient rewriting $(\calMM\calNN)\calMM$, leaving only the rewriting $\calE$.

%\IM{This was the original ``surviving example'', and it was introduced on $U$ (not $U'$). Do we need both $U$ and $U'$? Do we need both $h_2$ and $h$?} 
%The pruned chase step  (Definition~\ref{def:prune-chase-step}) does not apply since $\gamma (sq(\pi(h(p)))$ = $50K\times 50K$, which is greater than $T$. Thus, this results to having only one match %containment mapping 
%$h(\calE)$ from the expression $\calE$ into $U$, which gives us the original expression $\calE$ (where the intermediate size is 100x100) as an efficient rewriting, 
%and prunes-out non efficient alternative rewriting ($(\calMM\calNN)\calMM$), whose corresponding provenance conjunct is $p_{0} \wedge p_{1} \wedge p_{2} \wedge p_{3} \wedge  p_{4} \wedge p_{5}$, which introduces large intermediate (50Kx50K) compared to $\calE$. 
%Without applying the cost-based pruning while back-chasing, the back-chase would chase the constraint $C$. This results into $U'$, where $\calE$ has precisely two %containment mapping 
%match images into $U'$: one given by $h_1(\calE)$ yields the rewriting $(\calMM\calNN)\calMM$ and another given by $h_2(\calE)$, which gives the rewriting $\calMM(\calNN\calMM)$.  
\end{example}
\vspace{-1mm}
\noindent\textbf{Our improvements on $Prune_{prov}$.} Whenever the pruned chase step is applicable and applied for each TGD constraint, the original algorithm searches for all \emph{minimal-rewritings} $\calRW$ that can be found ``so far'', then it costs each $rw \in \calRW$ to find the ``so far'' \emph{minimum-cost one} $rw_e$  and adjusts the threshold $T$ to  the cost of $rw_e$. However, this strategy can cause \emph{redundant} costing of $rw \in \calRW$ whenever the pruned chase step is applied again for another constraint. Therefore, in our modified version of $Prune_{prov}$, we keep track of the rewriting costs already estimated, to prevent such redundant work. Additionally, the search for \emph{minimal-rewritings} ``so far'' (%homomorphic
matches of the query $Q$ into the evolving universal plan instance $U'$, see \S\ref{sec:pre}) whenever the pruned chase step is applied is modeled as a query evaluation of $Q$ against $U'$ (viewed as a symbolic/canonical database~\cite{AHV95}). This involves repeatedly evaluating the same query plan. However, the query is evaluated over evolutions of the same instance. Each pruned chase step {\em adds a few new tuples} to the evolving instance, corresponding to atoms introduced by the step, %leaving most of the instance unaffected by the step. 
while most of the instance is unchanged. Therefore, instead of evaluating the query plan from scratch, we employ incremental evaluation as in~\cite{pacb-paper}. The plan is kept in memory along with the populated hash tables, and whenever new tuples are added to the evolving instance, we push them to the plan.
\vspace{1mm}

%% file: guarantees.tex
\mysection{Guarantees on the reduction}
\label{sec:guaratees}
We detail the conditions under which we guarantee that our approach is
{\em sound} (\ie generates only equivalent, cost-optimal
rewritings), and 
{\em complete} (\ie finds all equivalent cost-optimal rewritings).

%We denote with $\calL$ the language of hybrid expressions described in \S\ref{sec:prob}.
\eat{
To abstract from the syntactic variations of LA programming languages,  
we define the language $\calL$ of {\em hybrid expressions}
in a surface-syntax-independent way:
given a set \lops\ of LA operations and a relational schema $\mathcal{R}$,
$\calL$ is the set of all expressions that admit an encoding to conjunctive queries over
$\mathcal{R} \cup \calVREM$, where $\calVREM$ denotes the virtual relations
introduced in (\S\ref{sec:relencoding}) for modeling LA
operations. \IM{Readers may be confused seeing a definition of $\calL$ here
 whereas ``hybrid expression'' were introduced in (\cref{sec:prob}); we
  meant the surface syntax independence there, too. Also: here we
  use $e$ as opposed to $E$ there. Should we move $\calL$ there, reuse
  it here, and turn $e$ into $E$?} \AD{good points. turned e into E.
  did not move \calH there as not needed there, and would need to refresh it here
  anyway. But shortened this section by not redefining \calL.}
}
Let $\calV \subseteq \calL$ be a set of materialized view definitions, where $\calL$ is the language of hybrid expressions described in
\S\ref{sec:prob}.

Let \lprop\ be a set of properties of the LA operations in \lops\ that admits
relational encoding over $\calVREM$. We say that \lprop\ is {\em terminating}
if it corresponds to a set of TGDs and EGDs with terminating chase (this holds for
our choice of \lprop).

Denote with $\gamma$ a cost model for expressions from $\calL$.
We say that $\gamma$ is {\em monotonic} if expressions are never assigned a lower
cost than their subexpressions (this is true for both models we used).

We call  $E \in \calL$ {\em $(\gamma,\lprop,\calV)$-optimal} if
for every $E' \in \calL$ that is (\lprop,\calV)-equivalent to $E$
we have $\gamma(E') \geq \gamma(E)$.

Let $\cEQ{\lprop}{\calV}{\gamma}(E)$ denote the set of all $(\gamma,\lprop,\calV)$-opti-mal
expressions that are (\lprop,\calV)-equivalent to $E$.

We denote with \hadad{\lprop}{\calV}{\gamma} our parameterized solution based on
relational encoding
followed by PACB++ rewriting and next by
decoding all the relational rewritings generated by the cost-based pruning PACB++
(recall Figure~\ref{fig:diagram}).
Given $E \in \calL$, \hadad{\lprop}{\calV}{\gamma}$(E)$ denotes all expressions
returned by \hadad{\lprop}{\calV}{\gamma} on input $E$.

%The following theorems refer to the notations introduced above.
	
\begin{theorem}[Soundness]
If the cost model $\gamma$ is monotonic,
then for every  $E \in \calL$ and every $rw \in  \hadad{\lprop}{\calV}{\gamma}(E)$, we have 
$rw \in \cEQ{\lprop}{\calV}{\gamma}(E)$. 
\end{theorem}

\begin{theorem}[Completeness]
If $\gamma$ is monotonic and $\lprop$ is terminating,  then
for every  $E \in \calL$ and every $rw \in \cEQ{\lprop}{\calV}{\gamma}(E)$,
we have $rw \in  \hadad{\lprop}{\calV}{\gamma}(E)$. 
\end{theorem}

%% file: experiments.tex
\mysection{Experimental Evaluation}
\label{sec:exp}
We evaluate \sys\ to answer research questions below about our approach:
\begin{itemize}
\item \textbf{\S\ref{sec:la-noviews}, \S\ref{sec:la-view}, \S\ref{sec:morpheus}} and \textbf{\S\ref{sec:micro-hybrid}}: Can HADAD find rewrites with/without views that lead to a greater performance improvement than original pipelines without modifying the internals of the existing systems?. Are the identified rewrites found by  state-of-the-art platforms?.   

\item \textbf{\S\ref{sec:la-opt}} and \textbf{\S\ref{sec:rw-overhead-hybrid}}: Is \sys’s optimization overhead compensated by the performance gains in execution?. 
\end{itemize}

We evaluate our approach, first on  \textbf{LA-based pipelines} (\S\ref{sec:exp-la}), then on 
%shown in Table~\ref{tab:pipelines-1} and  Table~\ref{tab:pipelines-2}, and then on 
\textbf{hybrid expressions} (\S\ref{sec:exp-hybrid}).
%and LA systems/tools listed in Table~\ref{tab:systems} and $(ii)$ real-world hybrid scenarios (discussed in Section~\ref{sec:exp-hybrid}), using SparkSQL~\cite{armbrust2015spark} as a polystore engine to execute the hybrid scenarios.
Due to space
constraints, we only discuss a subset of our results here and delegate other results to Appendix~\ref{appendixC},~\ref{appendixD},~\ref{appendixE}, and ~\ref{appendixF}.

\vspace{1mm}
\noindent\textbf{Experimental Environment.}
We used  a single node with
an Intel(R) Xeon(R) CPU E5-2640 v4 @ 2.40GHz, 20 physical cores ( 40 logical cores) %40 Cores (with hyper-threading)
 and 123GB
%RAM, disk read speed 616 MB/s, and disk write speed 455 MB/s.
 %(some systems spill intermediate results to disk). 
 We run on OpenJDK Java 8 VM 
 %\RA{I think the build is not important} (build 1.8.0\_242-b08).
. As for \textbf{LA systems/libraries}, we used \textbf{R 3.6.0}, \textbf{Numpy 1.16.6 (python 2.7)}, \textbf{TensorFlow r1.4.2}, \textbf{Spark 2.4.5 (MLlib)}, \textbf{SystemML 1.2.0}. For hybrid experiments, we use \textbf{MorpheusR}~\cite{morpheus} and \textbf{SparkSQL}~\cite{armbrust2015spark}.
%\IM{Maybe we can omit this:}
%We used Scala version 2.11.12  for SparkMLlib and SystemML.

\vspace{1mm}
\noindent\textbf{Systems Configuration Tuning.} We discuss here the most important installation and configuration details. We use a JVM-based linear algebra library for SystemML as recommended in~\cite{thomas2018comparative}, at the optimization level \textbf{4}. %, which includes all advanced rewrites. %\RA(This the recent optimization level I used. Previously, I ran it with level 2. However, I discovered level 4 introduced recently, and I re-ran the experiments with level 4. I noticed no difference from level 2 in terms of the rewrite rules that they apply.}. 
Additionally, we enable multi-threaded matrix operations in a single node. 
%(\textbf{sysml.cp.parall-el.ops=true}). 
We run Spark in a single node setting and %we configure Spark to import 
OpenBLAS (compiled from the source as detailed in~\cite{blas}) %as the LA back-end library 
to take advantage of its accelerations~\cite{thomas2018comparative}. %; we built OpenBLAS from the sources as detailed in~\cite{blas}. 
%\RA{Maybe this details is not necessary}
%The default OpenBLAS uses its internal threading rather than OpenMP, which can degrade performance when using SparkMLlib. Therefore, we use OpenMP as the threading implementation when compiling OpenBLAS. 
SparkMLlib’s
datatypes do not support many basic LA operations, such as scalar-matrix multiplication, Hadamard-product,
%(\ie Hadamard-product ), 
\etc To support them, we use the \texttt{Breeze} Scala library~\cite{Breeze}, convert MLlib’s
datatypes to Breeze types and express the basic LA operations. %in Spark. %We note that the conversion preserves the sparsity of the matrices. 
The driver memory allocated for Spark and SystemML is 115GB. To maximize TensorFlow performance, we compile it from the source to enable
architecture-specific optimizations. %to optimize it for the available CPU extensions/instructions (e.g., SMID). 
For all systems/libraries, we set the number of \textbf{cores} to \textbf{24} %to enable parallelism in a single node~\cite{thomas2018comparative}. All 
%all systems use \textbf{double precision} numbers.   
(\ie we use the command \texttt{\textbf{taskset -c 0-23}} when running R, NumPy and TensorFlow scripts).  All system use double precision numbers (\texttt{\textbf{double}}) by default, while TensorFlow uses single precision floating
point numbers (\texttt{\textbf{float}}). To enable fair companion with other systems, we use a double precision (\texttt{\textbf{tf.float64}}).%\texttt{\textbf{tf.placeholder}} to \texttt{\textbf{tf.float64}} (\ie double precision).
%\RA{SystemML recently added optimization 4, and I'm re-running the experiments by enabling level 4 optimization.} 
%\RA{I experimented Spark using JVM-based linear algebra library  and it is terrible. I did manage on to build Spark from source with BALS library in order to take advantage of native BLAS accelerations and it is much better that using JVM based library}.
\begin{table*}[t!]
\centering
\footnotesize                       % used font size
\setlength{\extrarowheight}{2pt}
\begin{tabular}{||c|c||c|c||c|c||}
\hline
\hline
\rowcolor{lightgray}
\textbf{No.} & \textbf{Expression} & \textbf{No.} & \textbf{Expression} & \textbf{No.} & \textbf{Expression}\\
\hline
P1.1 & \footnotesize $({\calMM} {\calNN})^T$ & P1.2 & \footnotesize ${\calA}^T + {\calB}^T $ & P1.3 & \footnotesize ${\calC}^{-1} {\calD}^{-1} $ \\
\hline
P1.4 &\footnotesize ${(\calA}+{\calB})v_1$ & P1.5 & \footnotesize  $(({\calD})^{-1})^{-1}$ & P1.6 & \footnotesize $\traceOp(s_1{\calD})$ \\
\hline
P1.7 & \footnotesize $(({\calA})^{T})^{T}$ & P1.8 &\footnotesize $s_1{\calA} + s_2{\calA}$ & P1.9 & \footnotesize  $\detOp({\calD}^{T})$\\
\hline 
P1.10 & \footnotesize  $\rowsumsOp({\calA}^T)$ & P1.11 & \footnotesize $\rowsumsOp({\calA}^T+{\calB}^T)$ & P1.12 &\footnotesize $\colsumsOp(({\calMM}{\calNN})$\\
\hline
P1.13 & \footnotesize  $\sumOp({\calMM}{\calNN})$ & P1.14 & \footnotesize $\sumOp(\colsumsOp(({\calNN}^{T}{\calMM}^{T}))$ & P1.15 & \footnotesize $({\calMM}{\calNN}){\calMM}$\\
\hline
P1.16 & \footnotesize  $\sumOp({\calA}^{T})$ & P1.17 &\footnotesize  $\detOp({\calC}{\calD}{\calC})$ & P1.18 &\footnotesize $\sumOp(\colsumsOp(({\calA}))$ \\
\hline
P1.19 & \footnotesize $({\calC}^{T})^{-1}$ & P1.20 & \footnotesize  \traceOp $({\calC}^{-1}) $ & P1.21 &\footnotesize  $({\calC}+{\calD}^{-1})^{T}$ \\
\hline
P1.22 & \footnotesize $\traceOp((\calC + \calD)^{-1})$ & P1.23 & \footnotesize $\detOp((\calC\calD)^{-1}) + \calD)$ & P1.24 &\footnotesize $\traceOp((\calC\calD)^{-1}))+ \traceOp(\calD)$ \\
\hline
P1.25 & \footnotesize $\calMM \odot (\calNN^{T} / (\calMM\calNN\calNN^{T}))$ & P1.26 &\footnotesize  $\calNN \odot (\calMM^{T} / (\calMM^T\calMM\calNN))$  & P1.27 & \footnotesize $\traceOp(\calD (\calC\calD)^{T})$\\
\hline
P1.28 & \footnotesize $ \calA \odot (\calA \odot \calB + \calA)$ & P1.29 &\footnotesize $ {\calD} \boldsymbol{\calC} {\calC} {\calC}$ & P1.30 & \footnotesize $ {\calNN}{\calMM} \odot {\calNN}{\calMM}{\calR}^{T}$ \\\hline
\end{tabular}
\caption{LA Benchmark Pipelines (Part 1)}
\vspace{-8mm}
\label{tab:LA-pipelines-1}
\end{table*}

\begin{table*}[t!]
\centering
\footnotesize                       % used font size
\setlength{\extrarowheight}{2pt}
\begin{tabular}{||c|c||c|c||c|c||}
%\hline
%\hline
%\multicolumn{6}{||c||}{\textbf{LA Pipelines}}\\
\hline
\hline
\rowcolor{lightgray}
\textbf{No.} & \textbf{Expression} & \textbf{No.} & \textbf{Expression} & \textbf{No.} & \textbf{Expression}\\
\hline
P2.1 & \footnotesize $\traceOp({\calC}+{\calD})$  & P2.2 & \footnotesize $\detOp({\calD}^{-1})$ & P2.3 & \footnotesize $\traceOp({\calD}^{T})$  \\
\hline
P2.4 &\footnotesize $s_1{\calA} + s_1{\calB}$ & P2.5 & \footnotesize $\detOp(({\calC}+{\calD})^{-1})$ & P2.6 & \footnotesize ${\calC}^T({\calD}^T)^{-1}$ \\
\hline
P2.7 & \footnotesize ${\calD} {\calD}^{-1} {\calC}$ & P2.8 &\footnotesize $\detOp({\calC}^T{\calD})$ & P2.9 & \footnotesize $\traceOp({\calC}^T{\calD}^T+{\calD})$\\
\hline 
P2.10 & \footnotesize   $\rowsumsOp({\calMM}{\calNN})$ & P2.11 & \footnotesize $\sumOp({\calA} + {\calB})$   & P2.12 &\footnotesize  $\sumOp(\rowsumsOp({\calNN}^{T}{\calMM}^{T}))$ \\
\hline
P2.13 & \footnotesize $(({\calMM}{\calNN}){\calMM})^T$  & P2.14 & \footnotesize  $(({\calMM}{\calNN}){\calMM}){\calNN}$ & P2.15 & \footnotesize  $\sumOp(\rowsumsOp({\calA}))$\\
\hline
P2.16 & \footnotesize $\traceOp({\calC}^{-1}{\calD}^{-1})+\traceOp(\calD)$   & P2.17 & \footnotesize $(((({\calC}+{\calD})^{-1})^{T})(({\calD}^{-1})^{-1}){\calC}^{-1}\calC$ & P2.18 & \footnotesize $\colsumsOp(({\calA}^T+{\calB}^T)$  \\
\hline
P2.19 & \footnotesize $({\calC}^{T}{\calD})^{-1}$ & P2.20 & \footnotesize  $({\calMM}({\calNN}{\calMM}))^T$ & P2.21 &\footnotesize $({\calD}^T{\calD})^{-1} ({\calD}^Tv_1)$ \\
\hline
P2.22 & \footnotesize $\expOp((\calC+\calD)^{T})$ & P2.23 & \footnotesize  $\detOp(\calC)*\detOp(\calD)*\detOp(\calC)$  & P2.24 &\footnotesize $({\calD}^{-1}{\calC})^{T}$ \\
\hline
P2.25 & \footnotesize $({u_1}v_2^{T}-{\calX})v_2$ & P2.26 &\footnotesize $\expOp((\calC+\calD)^{-1})$ & P2.27 & \footnotesize $((((\calC+\calD)^T)^{-1})D)C$ \\
\hline
\end{tabular}
\caption{LA Benchmark Pipelines (Part 2)}
\vspace{-6mm}
\label{tab:pipelines-2}
\end{table*}
\begin{table}[t!]
\footnotesize                       % used font size
\setlength{\extrarowheight}{2pt}
 \begin{tabular}{||c |r |r |r | r||} 
 \hline
 \rowcolor{lightgray}
 \textbf{Name} & \textbf{Rows} $_n$ & \textbf{Cols}$_m$ & \textbf{Nnz} ||X||$_0$ & \textbf{S}$_X$ \\ [0.5ex] 
 \hline 
 \hline 
 DFV & 1M & 100 & 8050 & 0.0080\%\\
 \hline
 2D\_54019 & 50K & 100 & 3700 & 0.0740\%\\
 \hline
 Amazon/($AS$) & 50K & 100 & 378 & 0.0075\%\\
 \hline
 Amazon/($AM$) & 100K & 100 & 673 & 0.0067\%\\
 \hline
 Amazon/($AL_{1}$) & 1M & 100 & 6539 & 0.0065\%\\
 \hline
 Amazon/($AL_{2}$) & 10M & 100 & 11897 & 0.0011\%\\
 \hline
 Amazon/($AL_{3}$) & 100K & 50K & 103557 & 0.0020\%\\
 \hline
 Netflix/($NS$) & 50K & 100 & 69559 & 1.3911\%\\
 \hline
 Netflix/($NM$) & 100K & 100 & 139344 & 1.3934\%\\
 \hline
 Netflix/($NL_{1}$) & 1M & 100 & 665445 & 0.6654\%\\
 \hline
 Netflix/($NL_{2}$) & 10M & 100 & 665445 & 0.0665\%\\
 \hline
 Netflix/($NL_{3}$) & 100K & 50K & 15357418 & 0.307\%\\
 \hline
\end{tabular}
\caption{Overview of Used Real Datasets.}
\label{tab:datasets-real}
\vspace{-7mm}
\end{table} 

\begin{table}[t!]
%\vspace{-2mm}
\footnotesize                       % used font size
\setlength{\extrarowheight}{2pt}
 \begin{tabular}{||c |r |r ||c | r|r|} 
 \hline
 \rowcolor{lightgray}
 \textbf{Name} & \textbf{Rows} $_n$ & \textbf{Cols}$_m$\\ [0.5ex] 
 \hline 
 \hline
 $Syn_1$ & 50K & 100\\
 \hline
 $Syn_2$ & 100 & 50K\\
 \hline
 $Syn_3$ & 1M & 100 \\
 \hline
 $Syn_4$ & 5M & 100 \\
 \hline
 $Syn_5$ & 10K &  10K\\
 \hline
\end{tabular}
\begin{tabular}{||c |c |c ||c | c|c|} 
 \hline
 \rowcolor{lightgray}
 \textbf{Name} & \textbf{Rows} $_n$ & \textbf{Cols}$_m$\\ [0.5ex]  
 \hline
 $Syn_6$ & 20K &  20K\\
 \hline
 $Syn_7$ & 100 &  1\\
 \hline
 $Syn_8$ & 50K & 1\\
 \hline
 $Syn_9$ & 100K &  1\\
 \hline
 $Syn_{10}$ & 100 &  100\\
 \hline
\end{tabular}
\caption{Syntactically Generated Dense Datasets}.
\label{tab:datasets-syn}
\vspace{-6mm}
\end{table}

\begin{table}[t!]
\footnotesize                       % used font size
\setlength{\extrarowheight}{2pt}
 \begin{tabular}{||c| c|| } 
 \hline
 \rowcolor{lightgray}
 \textbf{Matrix Name} & \textbf{Used Data} \\
 \hline 
 \hline  
 $\calA$ and $\calB$ & \makecell{AM, AL$_{1}$, AL$_{2}$, NM, NL$_{1}$, NL$_{2}$,\\ dielFilter, $Syn_3$ or $Syn_4$}\\
 \hline
 $\calC$ and  $\calD$  & $Syn_5$ or $Syn_6$ \\
 \hline 
 $\calMM$ & \makecell{AS, NS, $Syn_1$, or 2D\_54019} \\
 \hline
  $\calNN$ & $Syn_2$ \\
  \hline
  $\calR$ & $Syn_{10}$ \\
  \hline
 $\calX$ & AL$_{3}$ or NL$_{3}$\\
 \hline
 $v_1$,$v_2$ and $u_1$ & $Syn_7$, $Syn_8$ and $Syn_9$, respectively.\\
 \hline
\end{tabular}
\caption{Matrices used for each matrix name in a pipeline}.
\label{tab:matrices}
\vspace{-6mm}
\end{table}

\vspace{-2mm}
\mysubsection{LA-based Experiments}% Rewriting Time} 
\label{sec:exp-la}
In this experiments, we study the performance benefits of our approach on LA-based pipelines as well as our optimization overhead.

\vspace{1mm} 
\noindent\textbf{Datasets.}  
We used several \textbf{real-world, sparse matrices}, for which Table~\ref{tab:datasets-real} lists the dimensions and the sparsity (\textbf{S}$_X$)
%the number and percentage of non-zero elements (\textbf{S}$_X$)
 $(i)$ dielFilterV3real (\textbf{DFV} in short)  is an analysis of a microwave filter with different mesh qualities%results from the analysis of a 9-th order microwave combline filter with %second order (LT \char`\\ QN) 
%vector finite elements with different mesh qualities
~\cite{davis2011university};
$(ii)$ 2D\_54019\_highK (\textbf{2D\_54019} in short) is a 2D semiconductor device simulation~\cite{davis2011university}; 
$(iii)$ we used several subsets of an Amazon books review dataset~\cite{Amazon} (in JSON), and similarly 
$(iv)$ subsets of a Netflix movie rating dataset~\cite{Netflix}. The latter two were easily converted into matrices where columns are items and rows are customers~\cite{spores}; we extracted smaller subsets of all real datasets to ensure the various computations applied on them fit in memory (e.g., Amazon/(AS) denotes the small version of the Amazon dataset). 
We also used a set \textbf{synthetic, dense matrices}, described  in Table~\ref{tab:datasets-syn}.

\vspace{1mm}
\noindent\textbf{LA benchmark.}
We use a set  $\mathcal{P}$ of \textbf{57 LA pipelines} used in prior studies and/or frequently occurring in real-world LA computations%(where they are usually called {\em pipelines})
, as follows:
\begin{itemize}
\item 	\textbf{Real-world pipelines (10):} include:  a chain of matrix self products used for reachability queries and other graph analytics~\cite{sommer2019mnc} (\textbf{P1.29} in Table~\ref{tab:LA-pipelines-1}); 
expressions used in Alternating Least Square Factorization (ALS)~\cite{spores} (\textbf{P2.25} in Table~\ref{tab:pipelines-2}); 
Poisson Nonnegative Matrix Factorization (PNMF)(\textbf{P1.13} in Table~\ref{tab:pipelines-1})~\cite{spores}; 
Nonnegative Matrix Factorization (NMF)(\textbf{P1.25} and \textbf{P1.26} in Table ~\ref{tab:pipelines-1})~\cite{thomas2018comparative};
%complex predicate for image masking~\cite{sommer2019mnc} (\textbf{P1.30} in Table~\ref{tab:pipelines-1}); 
recommendation computation~\cite{sommer2019mnc} (\textbf{P1.30} in Table~\ref{tab:pipelines-1});
finally, Ordinary Least Squares Regression (OLS)~\cite{thomas2018comparative} (\textbf{P2.21} in Table~\ref{tab:pipelines-2}). 
\item \textbf{Synthetic pipelines (47):} were also generated, based on a set of basic matrix operations (inverse, multiplication, addition, etc.), and a set of combination templates,  written as a Rule-Iterated Context-Free Grammar (RI-CFG)~\cite{RICFG}. 
% which is similar to Context-Free Grammar, but it empowers the producer to have control over the generation of rules for different expressions. 
Expressions thus generated include \textbf{P2.16}, \textbf{P2.16}, \textbf{P2.23}, \textbf{P2.24} in Table~\ref{tab:pipelines-1}.
\end{itemize}

\vspace{1mm}
\noindent\textbf{Methodology.} In (\S\ref{sec:la-noviews}), we show the performance benefits of our approach to LA-oriented systems/tools mentioned above using a set $\notopt \subset \mathcal{P}$ of \textbf{38} pipelines in Table~\ref{tab:pipelines-1} and Table~\ref{tab:pipelines-2}, and the matrices in Table~\ref{tab:matrices}. The performance of these pipelines can be improved \emph{just by exploiting LA properties} (in the absence of views). For TensorFlow and NumPy, we present the results only for dense matrices, due to limited support for sparse matrices.  In (\S\ref{sec:la-view}), we show how our approach improves the performance of \textbf{30} pipelines from $\mathcal{P}$, denoted $\views$, using pre-materialized views. Finally, in (\S\ref{sec:la-opt}), we study our rewriting performance and \emph{optimization overhead} for the set $\opt = \mathcal{P}\setminus \notopt$ of \textbf{19} pipelines that are already optimized.

\mysubsubsection{\textbf{Effectiveness of LA Rewriting (No Views).}}
\label{sec:la-noviews}

%Figure~\ref{fig:rw-dis-naive} shows the distribution of the average time to select a rewriting for the $\notopt$ pipelines
% using the na\"ive cost model: 64\% of these need less than 25ms. The outlier is \textbf{P2.17}, which needs $\sim 200$ms. We run the same experiment using the MNC-based cost model (Figure~\ref{fig:rw-dis-mnc}): it has a higher optimization overhead due to the intermediate histograms construction.

% \begin{figure}[h!]
% \vspace{-5mm}
% \centering \captionsetup{format = hang}
% \subfigure[Rewriting Time Distribution (Naïve)\label{fig:rw-dis-naive}]{\includegraphics[width=0.49\linewidth, height=4.4cm]{charts/RewritingTimeDistribution/RewritingTime_NoViews_Naive.pdf}}
% \vspace{-3mm}
% \subfigure[Rewriting Time Distribution (MNC)\label{fig:rw-dis-mnc}]{\includegraphics[width=0.49\linewidth,height=3.7cm]{charts/RewritingTimeDistribution/RewritingTime_NoViews_MNC.pdf}}
% \vspace{-2mm}
% \caption{Rewriting time distribution}
% \vspace{-3.5mm}
% \label{fig:rw}
% \end{figure}

%\noindent\textbf{LA-Pipelines and Rewriting Evaluation.}
For each system, we run the original pipeline and our rewriting 5 times; we report the average of the last 4 running times.  We exclude the data loading time. For fairness, we ensured SparkMLib and SystemML compute the entire pipeline (despite their lazy evaluation mode). 
%Since SparkMLib and SystemML perform \textit{lazy evaluation} (computations are performed when the results are needed), 
So, we print a matrix cell value at random to force the systems to compute the entire pipeline. Additionally, for SystemML, we insert a break block (e.g., \texttt{\textbf{WHILE(FALSE){}}}) after each pipeline and before the cell print statement to prevent computing only the value of a single cell and force it to compute all the outputs. 
%We discuss a selection of these experiments below. %\RA{Stoped here}

%\vspace{1mm}
%\noindent\textbf{Discussion}. 
Figure~\ref{fig:la-batch1} illustrates the original pipeline  execution time $Q_{exec}$ and the selected rewriting execution time $RW_{exec}$ for \textbf{P1.1}, \textbf{P1.3}, {\textbf{P1.4}, and \textbf{P1.15}, including the rewriting time $RW_{find}$, using the MNC cost model. For each pipeline, 
the
%matrices' 
used datasets 
%(see Tables ~\ref{tab:datasets-real}  and ~\ref{tab:datasets-syn}
 are on top of the figure. For brevity in the figures, we use \textbf{SM} for SystemML, \textbf{NP} for NumPy, \textbf{TF} for Tensorflow, and \textbf{SP} for MLlib. 
  \begin{figure*}[!htbp]
\figspa
  \centering
  \subfigure[\textbf{P1.1}]{\includegraphics[scale=0.25,width=4.2cm,height=3.9cm]{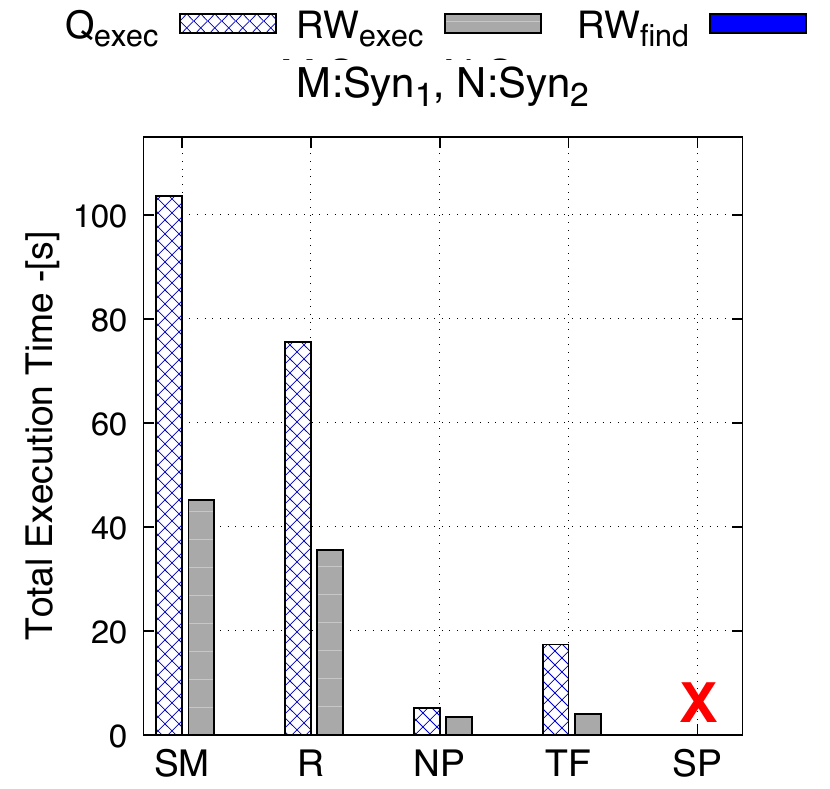}\label{fig:p1.1}}
  \subfigure[\textbf{P1.3}]{\includegraphics[scale=0.25,width=4cm,height=3.7cm]{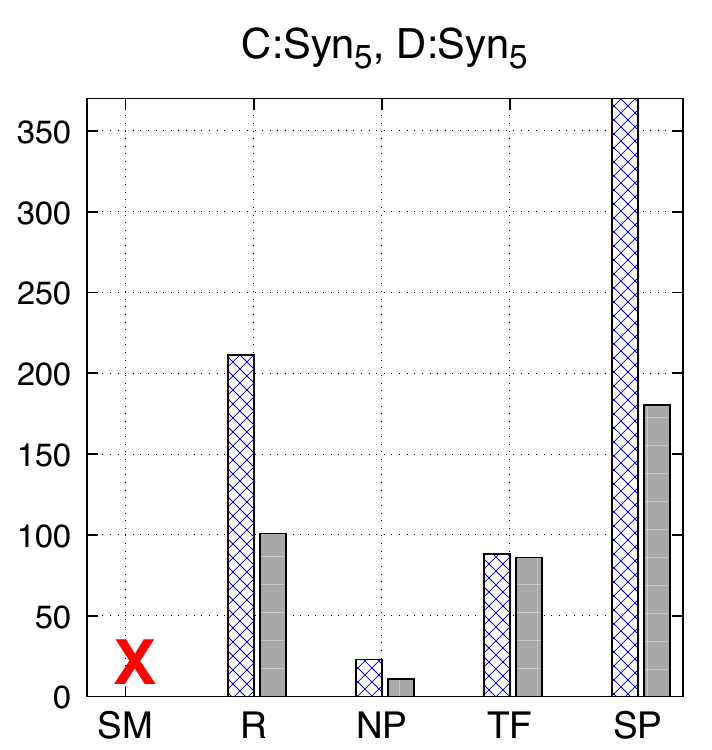}\label{fig:p1.3}}
    \subfigure[\textbf{P1.4}]{\includegraphics[scale=0.25,width=4cm,height=3.7cm]{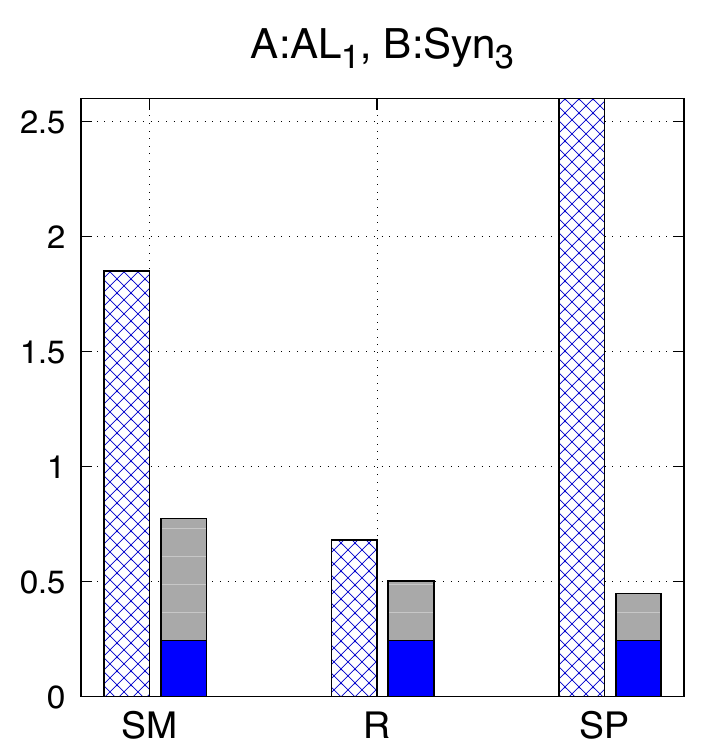}\label{fig:p1.4}}
        \subfigure[\textbf{P1.15}]{\includegraphics[scale=0.25,width=4cm,height=3.7cm]{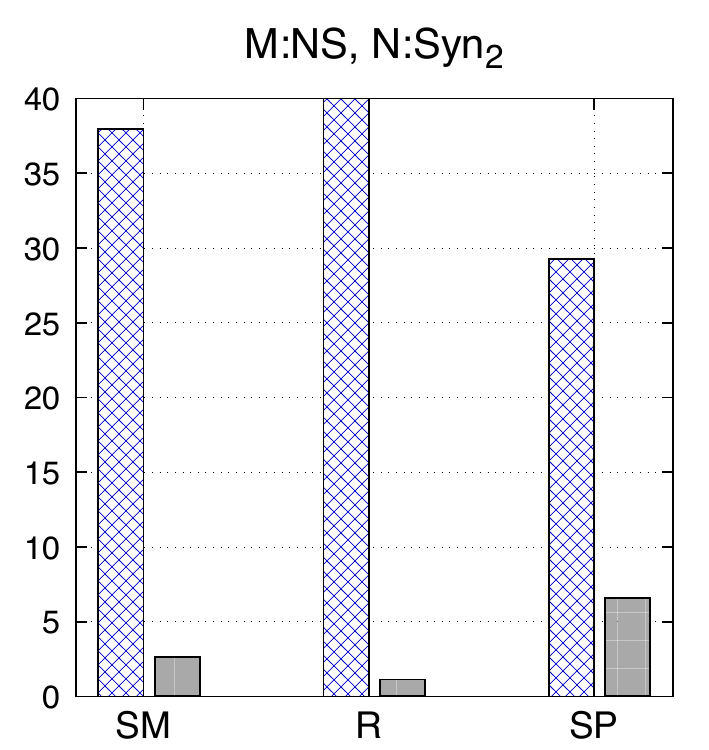}\label{fig:p1.15}}
\figspb\figspb
 \caption{P1.1, P1.3, P1.4, and P1.15 evaluation before and after rewrite}
 \label{fig:la-batch1}
\end{figure*}
\begin{figure*}[!htbp]
\figspa
  \centering
  \subfigure[\textbf{P1.13}]{\includegraphics[scale=0.25,width=4.2cm,height=3.9cm]{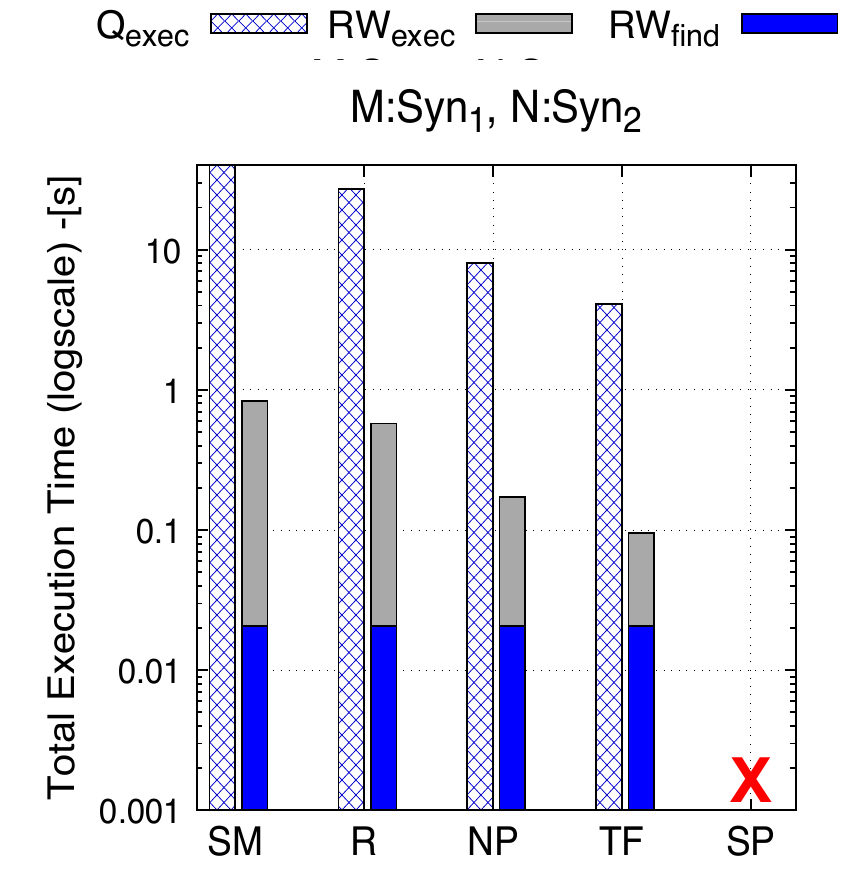}\label{fig:p1.13}}
  \subfigure[\textbf{P1.25}]{\includegraphics[scale=0.25,width=4.2cm,height=3.6cm]{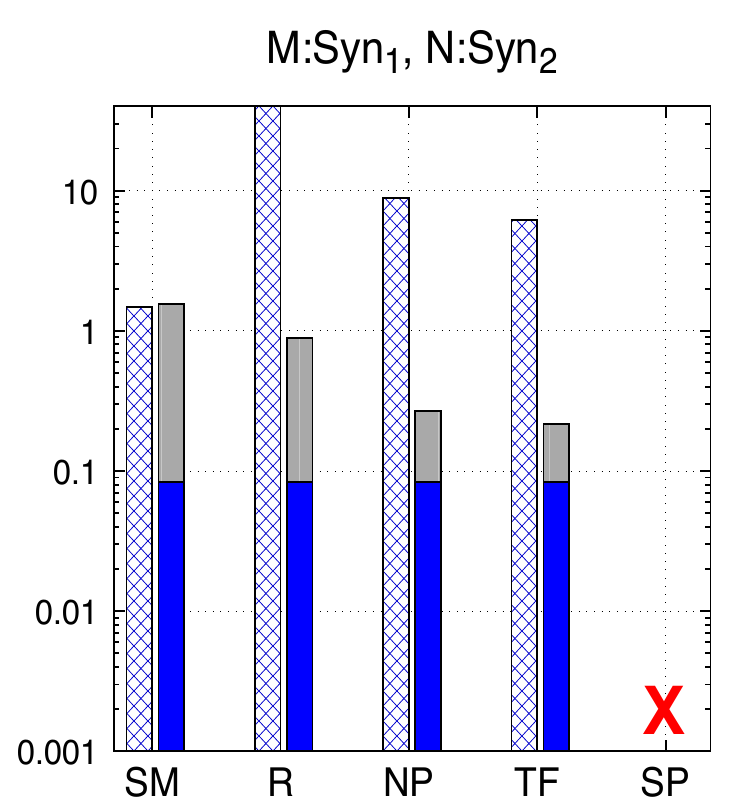}\label{fig:p1.25}}
    \subfigure[\textbf{P1.14}]{\includegraphics[scale=0.25,width=4.4cm,height=3.5cm]{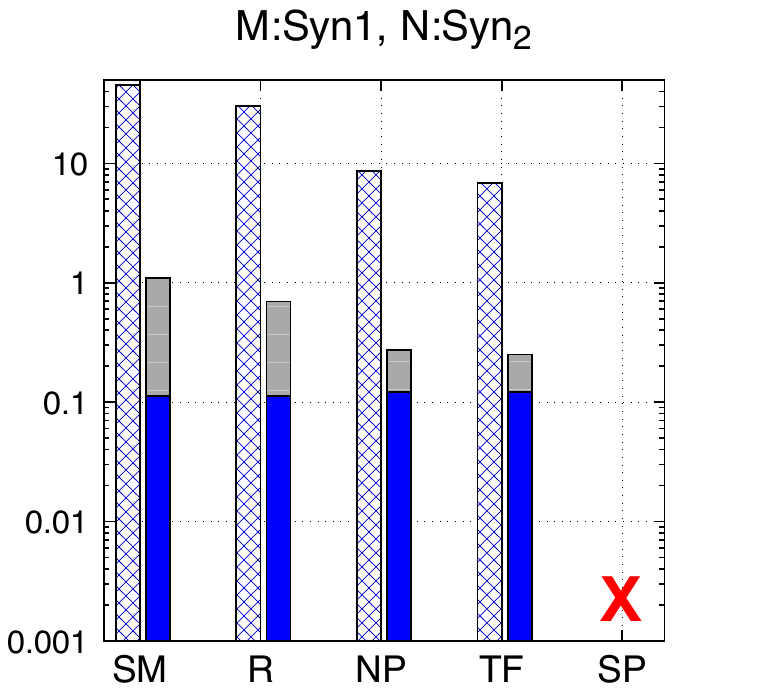}\label{fig:p1.14}}
        \subfigure[\textbf{P2.12}]{\includegraphics[scale=0.25,width=4.4cm,height=3.6cm]{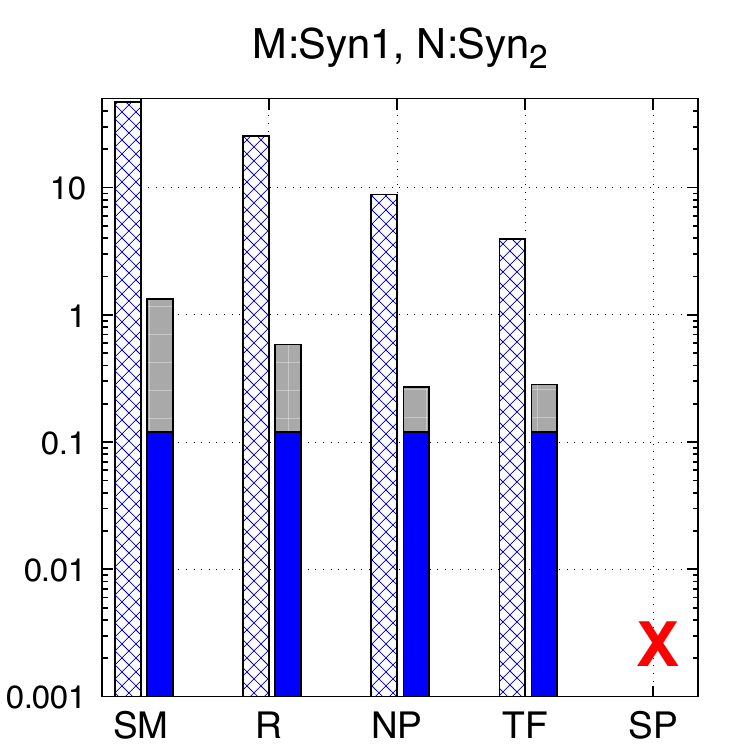}\label{fig:p2.12}}
\figspb
 \caption{P1.13, P1.25, P1.14 and P2.12 evaluation before and after rewrite}
 \label{fig:la-batch2}
\end{figure*}
For \textbf{P1.1} (see Figure~\ref{fig:p1.1}), both matrices are dense. The speed-up ($1.3\times$ to $4\times$) 
% our rewrite leads to speed-up of up to 4X  for TensorFlow, 2X for R and 2X for SystemML, and 1.5X speed-up for NumPy. The speed-up 
comes from rewriting $(\calMM\calNN)^{T}$ (intermediate result size to $(50K)^2$) into $\calNN^{T}\calMM^{T}$, much cheaper since both $\calNN^{T}$ and $\calMM^{T}$ are of size  $50K \times 100$.  
%Although the rewrite is straightforward, it is counterintuitive because one expects computing  $(\calMM\calNN)^{T}$ is more efficient than $\calNN^{T}\calMM^{T}$ since the former has %fewer number of one transpose operation. 
We exclude MLlib from this experiment since it failed to allocate memory for the intermediate matrix (Spark/MLLib limits the maximum size of a dense matrix).  
%\RA{I'm not sure if it is necessary to detail our tries. So, I commented this out}  
%As a workaround, we tried to use the distributed version of the matrix type (\textbf{BlockMatrix}, which we can run locally). The pipeline did not benefit from the rewrite; this is due to the slower version of the matrix-matrix multiplication provided by \textbf{BlockMatrix}, which dominates the execution time.  
As a variation (not plotted in the Figure), we ran the same pipeline with the ultra-sparse  $AS$ matrix (0.0075\% non-zeros) used as $\calMM$. 
The $Q_{exec}$ and $RW_{exec}$ time are very comparable using SystemML, because we avoid large dense intermediates.   In R, this scenario lead to a runtime exception and to avoid it,   %since the multiplication operator tries to densify the matrix $\calMM$. To avoid it,  
we cast $\calMM$ during load time to a dense matrix type. Thus, the speed-up achieved is the same as if $\calMM$ and $\calNN$ were both dense. 
If, instead, $NS$  (1.3911\% non-zeros) plays the role of $\calMM$, our rewrite achieves $\approx$ $1.8\times$ speed-up for SystemML. 

For  \textbf{P1.3} (Figure~\ref{fig:p1.3}), the speed-up comes from
rewriting $\calC^{-1} \calD^{-1}$ to $(\calD\calC)^{-1}$. %, thanks to the property $\calC^{-1}\calD^{-1} = (\calD\calC)^{-1}$. \IM{This explanation was trivial!}
Interestingly, for TensorFlow, the $Q_{exec}$ and $RW_{exec}$ time are very comparable.
%is the only system that applies this optimization by itself. 
%\RA{Maybe not necessary:}
%Note that the $RW_{find}$ time is negligible compared to the pipeline total execution time in TF (the $RW_{find}$ time is 0.0051\% of the pipeline's total execution time). 
SystemML timed-out ($>$1000 secs) for both original pipeline and its rewriting. 
%We observed that the inverse operation (\textbf{\texttt{inv}}) in SystemML performs poorly compared to other systems. 

For pipeline \textbf{P1.4} (Figure~\ref{fig:p1.4}), we rewrite $(\calA + \calB)v_1$ to $\calA v_1 + \calB v_1$. Adding a sparse matrix $\calA$ to a dense matrix $\calB$ results into materializing a dense intermediate of size $1M\times100$. Instead,   $\calA v_1 + \calB v_1$ has fewer non-zeros in the intermediate results, and $\calA v_1$ can be computed efficiently since $\calA$ is sparse. The MNC sparsity estimator has a noticeable overhead here. 
%during the rewriting time.
%to search for an efficient rewriting.
%and it is also due to the small size of the matrices ($1M \times 100$). 
%We observed that computing the matrix histogram for the intermediate of the addition operation is relatively costly compared to other operations. 
We run the same pipeline, where the dense $Syn_4$ matrix plays both $\calA$ and $\calB$ (not shown in the Figure). 
This leads to speed-up of up to %2.7X for SystemML, 1.6X for R, and 
$9\times$ for MLlib, which does not natively support matrix addition, thus we  
%the users have to implement these using RDD operations. 
convert its matrices to Breeze types in order to perform it (as in~\cite{thomas2018comparative}). 
%\RA{Maybe not necessary:}
%We observed that the addition of a sparse matrix of Breeze type to a dense matrix of Breeze type is slow comparing to the addition of two dense matrices or two sparse matrices. Thus, we densify the sparse matrix during the load time to speed up the execution for both the original pipeline and the rewriting.

\textbf{P1.15} (Figure~\ref{fig:p1.15}) is a matrix chain multiplication. The na\"ive left-to-right
evaluation plan  $(\calMM\calNN)\calMM$ computes an intermediate matrix of size $O(n^2)$, where $n$ is $50K$. Instead, the rewriting  $\calMM(\calNN\calMM)$ only needs an $O(m^2)$ intermediate matrix, where $m$ is $100$, and is much faster. To avoid MLLib memory failure on \textbf{P1.15}, 
we use the distributed matrix of type \textbf{BlockMatrix} for both matrices.  While $\calMM$ thus converted has the same sparsity, Spark views it as being of a dense type (
%and all operations 
multiplication on BlockMatrix is considered to produce dense matrices)~\cite{sparkmlib}. 
 %SystemML opts to densify the result of $\calMM\calNN$ given that $\calMM$ matrix, which uses the Netflix/NS dataset, is sparse, but it is not ultra sparse as $AS$ dataset. 
 SystemML does optimize the multiplication order if the user does not enforce it. % Additionally, if we specify the order of the  multiplication, SystemML does not optimize the chain. But, it optimizes the chain if no order is specified. \IM{I imagine it's hard to specify a mathematical expression without putting the symbols in some order. So, I prefer to say the user may "enforce" this (or not).}
Further (not shown in the Figure), we ran \textbf{P.15} with $AS$ in the role of $\calMM$. This is $4\times$ faster in SystemML since with an ultra sparse $\calMM$, multiplication is more efficient. This is not the case for  MLlib which views it as dense. 
  %matrix. 
  %For SystemML, %the $Q_{exec}$ and $RW_{exec}$ time are reduced by 4X since the multiplication becomes more efficient
%the ultra-sparse $\calMM$ reduces all times by 4, %\IM{Check:} 
%not for MLlib which still views $\calMM$ as dense. 
For R, we again had to densify $\calMM$ during loading to prevent crashes.
  %(discussed earlier). 
  
\vspace{1mm}
Figures~\ref{fig:p1.13} and ~\ref{fig:p1.25} study \textbf{P1.13} and \textbf{P1.25}, two real-world pipelines involved in
% the PNMF and NMF 
ML algorithms, %respectively, 
using the MNC cost model; note the log-scale $y$ axis. Rewriting  \textbf{P1.13} %$\sumOp(\calMM\calNN)$ 
to $\sumOp(t(\colsumsOp((\calMM))*\rowsumsOp(\calNN))$ yields a speed-up of $50\times$; %(it avoids  materializing a dense intermediate $\calMM\calNN$). 
while SystemML {\em has} this rewrite as a dynamic rewrite rule, it did not apply it. In addition, our rewrite allows other systems to benefit from it.
Not shown in the Figure, we re-ran this with $\calMM$ ultra sparse (using $AS$) and SystemML: the rewrite did not bring benefits, since $\calMM\calNN$ is already efficient. {\em In this experiment and subsequently, whenever MLlib is absent, this is due to its lack of support for LA operations} (here, sum of all cells in a matrix) on BlockMatrix. 
%\RA{Maybe not necessary:} 
%Also, in order to convert \textbf{BlockMatrix} type to Breeze type to be able to do the sum, this first requires casting the result of the multiplication from \textbf{BlockMatrix} to \textbf{LocalMatirx} type, which then leads memory allocation failure since the size of $\calMM\calNN$  >  INTEGER.MAX.
For \textbf{P1.25}, the important optimization is selecting the multiplication order in $\calMM \calNN \calNN^{T}$ (Figure~\ref{fig:p1.25}). SystemML is efficient here, due to its 
% can optimize this chain in a one go with no intermediates since the pipeline has transpose-self matrix multiplication pattern%the Gram matrix pattern, where the system has a dedicated operator \textbf{\texttt{tsmm} } that can efficiently compute this pattern. 
dedicated operator \texttt{tsmm} for {\em t}ranspose-\textit{s}elf {\em m}atrix {\em m}ultiplication and \texttt{mmchain} for matrix multiply chains.
% Note that the time it takes to search for this rewriting is 5.39\% of the pipeline execution time in SystemML.
%MLlib is excluded as it does not support divisions on BlockMatrixes. 

%\RA{Maybe not necessary:} 
% in order to perform the division, the result of the chain  $\calMM \calNN \calNN^{T}$ has to be casted from distributed \textbf{BlockMatirx} to \textbf{LocalMatirx} type since \textbf{BlockMatirx} does not support division operation. But, this again lead to memory allocation failur. 
	
Figures~\ref{fig:p1.14} and ~\ref{fig:p2.12} shows up to $42\times$ rewriting speed-up achieved by turning \textbf{P1.14} and \textbf{P2.12} 
%\boldsymbol{\calMM}^{T}))$  and $\sumOp(\colsumsOp((\boldsymbol{\calNN}^{T}\boldsymbol{\calMM}^{T}))$
%	
into $\sumOp(t(\colsumsOp((\calMM))*\rowsumsOp(\calNN))$. This exploits several properties: $(i)$ $(\calMM\calNN)^{T}=\calNN^{T}\calMM^{T}$, $(ii)$  $\sumOp(\calMM^{T})=\sumOp(\calMM)$, $(iii)$ $\sumOp(row/\colsumsOp((\calMM))= \sumOp(\calMM)$, 
 %$(iv)$ $\sumOp(\colsumsOp((\calMM)= \sumOp(\calMM)$ 
 and  $(iv)$ $\sumOp(\calMM\calNN)=\sumOp(t(\colsumsOp((\calMM))*\rowsumsOp(\calNN))$. SystemML captures $(ii)$, $(iii)$, and $(iv)$ as static rewrite rules, however, it is unable to exploit these performance-saving opportunities since it is unaware of $(i)$. Other systems lack support for more or all of these properties.
%We exclude MLlib from this experiment since sum operation is not supported by \textbf{BlockMatrix} type (same as \textbf{P1.13}). 
 %\RA{Maybe not necessary.}
 %We run the same experiments by making $\calMM$ ultra sparse using $AS$ dataset. This again make the computation of $\calNN^{T}\calMM^{T}$ more efficient in SystemML, and the pipeline did not benefit from the rewrite. 
 %However, the time to search for the rewrite is still negligible compared to the pipeline total execution time.  
 
Figure~\ref{fig:r-dis} shows the distribution of the significant rewriting speed-up %(speed-up = $Q_{exec}/(RW_{find}+RW_{exec})$) by our rewrites 
on $\notopt$ running on R, and using the MNC-based cost model. %(due to space constraints, 
%(we report other systems' results, and when using   the naive cost model in~\cite{technicalreport}). 
For clarity, we split the distribution into two figures: on the left, 25 $\notopt$ pipelines with speed-up lower than $10\times$; %\RA{Fixed}
%\IM{Discussed with Rana on Sat morning FR time (late Fri SD): the other cost model misses no rewritings; Rana will plot that instead and we'll mention (perhaps) that the naive model misses some.}
% \IM{Is rewriting making things a bit worse (slower)? Is this because it spends time finding yet it does not find?} \RA{This figure shows the distribution of the "selected rewriting execution" speed-up. Some pipelines, even after the rewrite, their speed-up is less than 10X; this is not because of the rewriting overhead. The time to search for a rewriting overall is very negligible compared to the "total execution time." E.g, P1.8 $s_1\calA + s_2 \calA$, the only alternative rewriting is $(s_1+s_2)\calA$, the speed-up achieved is $\sim$2X. Other example, P2.4 $\detOp(\calD^{T})$, the only alternative rewriting is $\detOp(\calD)$, the speed-up achieved is $\sim$ 1.2(not much), given that this pipeline doesn't introduce intermediate $\calD^{T}$, computing the determinant dominates the execution time. The time to search for P2.4's rewriting is 0.0004\% of the total execution time, which is $\sim$25sec (including rewriting time). Other example, P2.1 $tarce(\calC+\calD)$, the efficient alternative rewrite is $\traceOp(\calC) + \traceOp(\calD)$, where the achieved speed-up is 4.31X, the rewrite gets ride of all intermediates. So, in summary, this depends mostly on the pipelines, how much they benefit from the rewrite. Overall, searching for a rewrite is very small.},   
on the right, the remaining 13 with greater speed-up. 
%For the former, the naive cost model fails to identify efficient rewritings for 4 pipelines.  
%\IM{When reading this, the pesky reader will ask: ok, does the other cost model fare better?} \RA{Yes. The other cost-model will select the efficient rewritings for those 4 pipelines. I put the naive-cost model to show that we still did not miss many rewritings (just for 4 pipelines). The overhead of the other cost model can be noticeable for some pipelines. Do you suggest to re-plot this figure when using the other cost model.?. For sure, I will put it in the extended version that we cite.} 
% However, the time to search for a rewriting for these pipelines is very negligible.
%\RA{Maybe not necessary:}
%(the rewriting time is 0.023\% of the pipelines' total execution time on average) compared to the pipeline total execution time. 
Among the former, 87\%  achieved at least  $1.5\times$ speed-up. The latter are sped up by $10\times$ to $60\times$. \textbf{P1.5} is an extreme case here (not plotted): it is sped up by about $1000\times$, simply by rewriting $((\calD)^{-1})^{-1}$ into $D$. 
%\IM{I labeled it "extreme case" and left it here in the end}
%\RA{Maybe not necessary:}
 %The maximum speed-up in the figure (58X) comes from a rather simple optimization by rewriting  $\traceOp(\calD^{T})$ to $\traceOp(\calD)$, which avoids materializing large dense intermediate $\calD^{T}$.
\begin{figure*}[!htbp]
\figspa
  \centering	
  \subfigure[\textbf{P2.14}]{\includegraphics[scale=0.25,width=4.2cm,height=3.9cm]{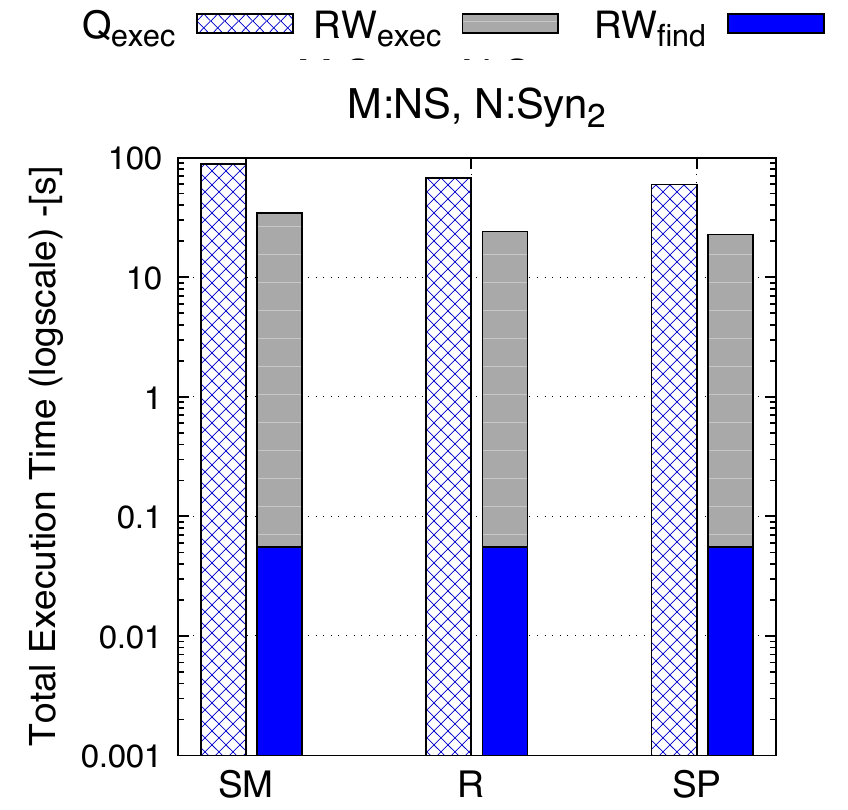}\label{fig:p2.14}}
  \subfigure[\textbf{P2.21}]{\includegraphics[scale=0.25,width=4.4cm,height=3.6cm]{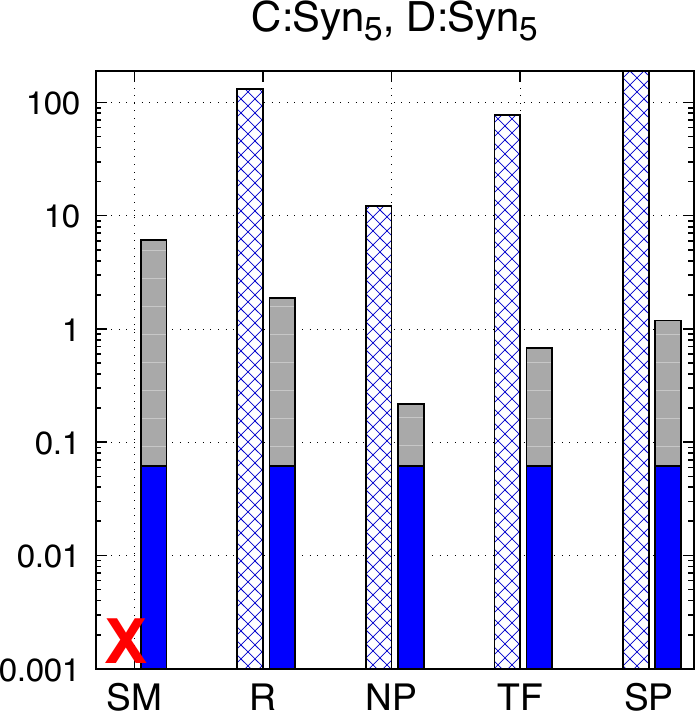}\label{fig:p2.21}}
    \subfigure[\textbf{P2.25}]{\includegraphics[scale=0.25,width=4.2cm,height=3.6cm]{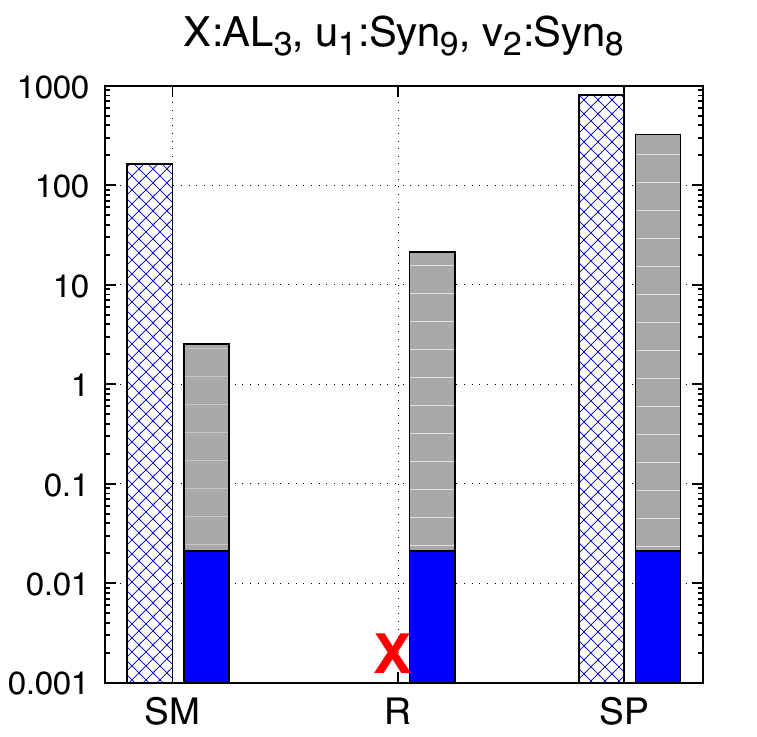}\label{fig:p2.25}}
        \subfigure[\textbf{P2.27}]{\includegraphics[scale=0.25,width=4.4cm,height=3.7cm]{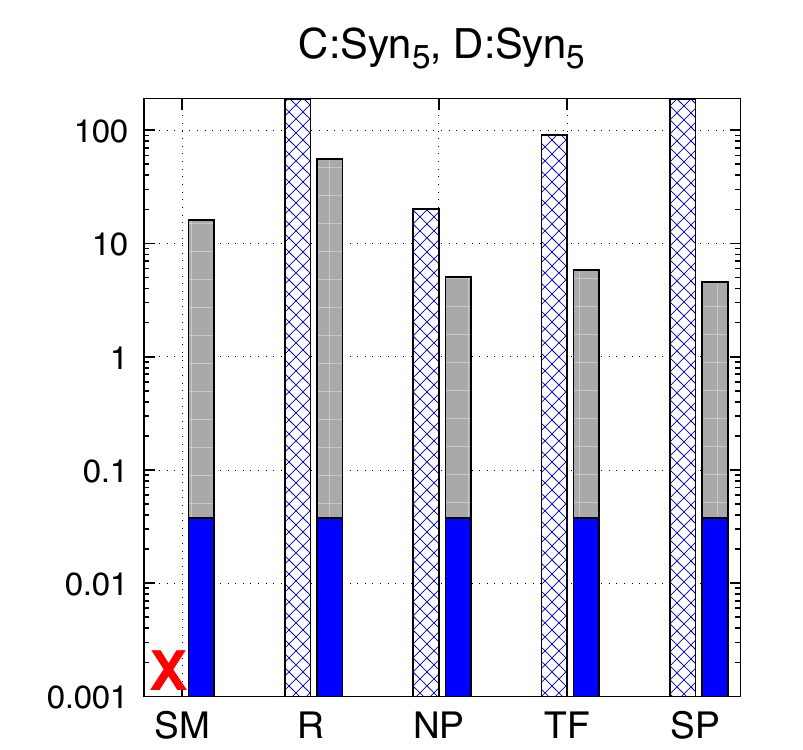}\label{fig:p2.27}}
 \figspb\figspb
 \caption{P2.14, P2.21, P2.25 and P2.27 evaluation before and after rewriting using the views $V_{exp}$}
 \label{fig:la-views}
\end{figure*} 

\begin{figure}[t!]
%\vspace{-1.5mm}
%  \includegraphics[width=\linewidth,height=4.4cm]{./charts/speedupDis/RSpeedupDis2.pdf}
    \includegraphics[width=\linewidth,height=4.4cm]{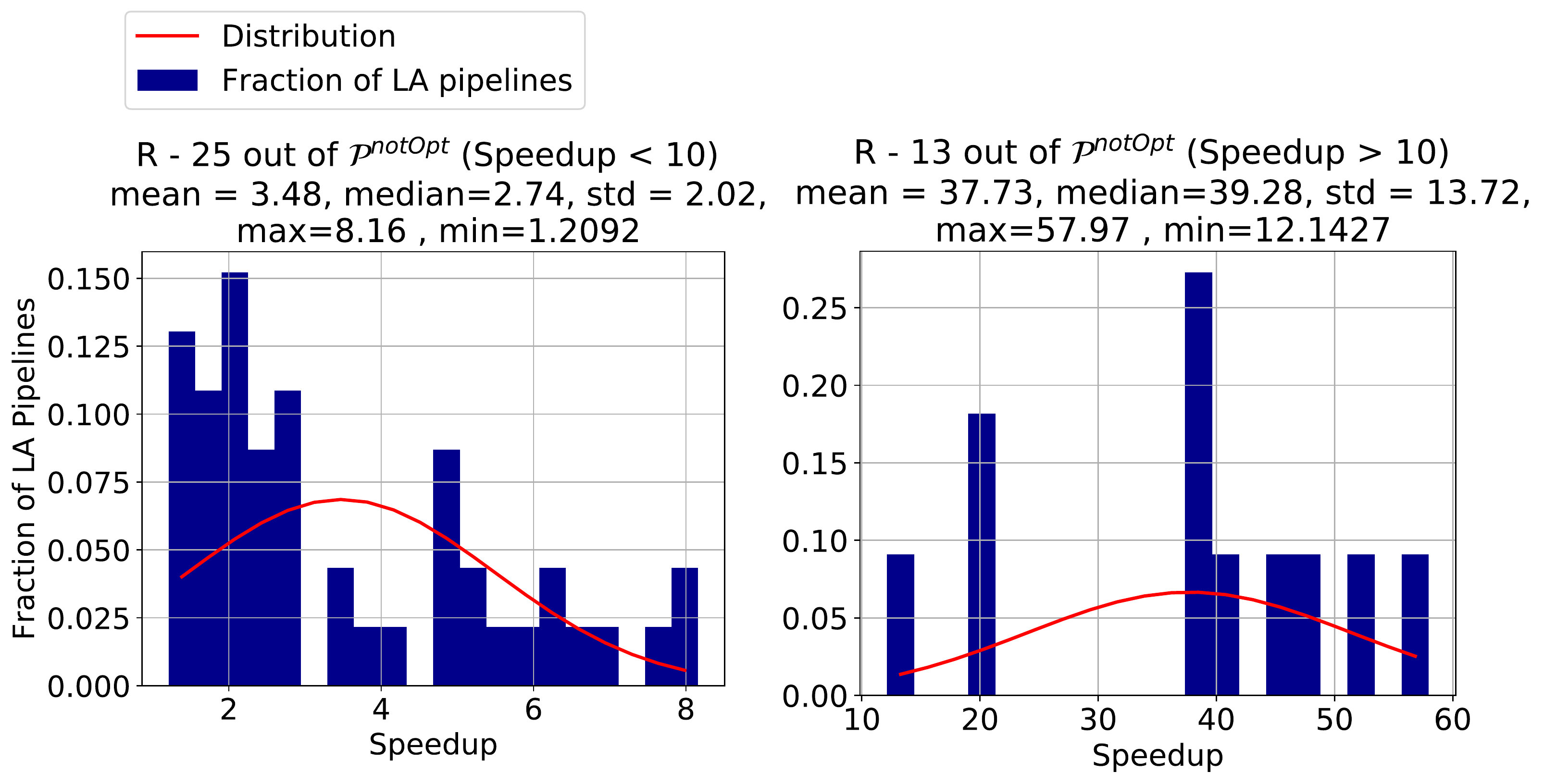}
  \vspace{-8mm}
  \caption{R speed-up on $\notopt$}
  \label{fig:r-dis}
\vspace{-6mm}
\end{figure}

\mysubsubsection{\textbf{Effectiveness of view-based LA rewriting}}
\label{sec:la-view}
We have %we enable the usage of the materialized views. 
defined a set  $V_{exp}$ of 12 views that pre-compute the result of some expensive operations (multiplication, inverse, determinant, etc.) which can be used to answer our   $\views$ pipelines, and materialized them on disk as CSV files. 
The experiments outlined below used the na\"ive cost model; all graphs have a log-scale $y$ axis.
\eat{More similar experiments are described to~\cite{technicalreport}.}   %which we find to work well.  
%30 pipelines $PIPE^{Views}$ out of 57 (listed in Table ~\ref{tab:pipelines-1} and Table ~\ref{tab:pipelines-2}) can be rewritten using the views $V_{exp}$. 

\vspace{1mm}
\noindent\textbf{Discussion}. For \textbf{P2.14} (Figure~\ref{fig:p2.14}), using the view $V_4=\calNN\calMM$ by and the multiplication associativity leads to up to $2.8\times$ speed-up.
 
Figure~\ref{fig:p2.21} shows the gain due to the view $V_1=\calD^{-1}$, %which pre-computes the result of an expensive operation (inverse),  
for the ordinary-least regression (OLS) pipeline \textbf{P2.21}. It has  8 rewritings, 4 of which use $V_1$; they are found thanks to the properties $(\calC\calD)^{-1}=\calD^{-1}\calC^{-1}$, $(\calC \calD) \calE = \calC (\calD \calE)$ and  $(\calD^{T})^{-1}=(\calD^{-1})^{T}$ among others. The cheapest rewriting is  $V (V^{T}(\calD^{T}v_1))$, since it introduces small intermediates due to the optimal matrix chain multiplication order. This rewrite leads to $70\times$, $55\times$ and $150\times$ speed-ups on R, NumPy and MLlib, respectively. %TensorFlow is omitted as \texttt{matmul} operator does not support matrix-vector multiplication.
%as it does not natively support matrix-vector multiplication. 
%It could run this pipeline by converting the matrices to NumPy, whose performance we already report separately. 
%\RA{One hack, would be to convert matrices in TensorFlow to NumPy's types, I honestly didn't try it since this will give us similar results as NumPy, which we already report}
%We omit TensorFlow from this experiment since the multiply operator (\textbf{\texttt{matmul}}) was coded for rank two or greater tensors, and it does not natively support 
 %\RA{Maybe not necessary:}
 %In SparkMLlib, the inverse operator is quite expensive compared to R, NumPy and TensorFlow. 
 On SystemML, the original pipeline timed out ($>1000$ seconds). 
 %\RA{We can omit this result}
 %For the pipeline \textbf{P2.6} (as shown in Figure ~\ref{fig:p2.6}), the speed-up attributes to rewriting $\calC^{T} (\calD^{T}) ^{-1}$ to $(V_1\calC)^{T}$. To get such a rewriting, the properties $(\calD^{T})^{-1}=(\calD^{-1})^{T}$ and $(\calC\calD)^{T}=\calC^{T}\calD^{T}$ are exploited.
  
Pipeline \textbf{P2.25} (Figure~\ref{fig:p2.25}) benefits from a view $V_5$, which pre-computes a dense intermediate vector multiplication result; then, rewriting based on the property $(\calA + \calB)v = \calA v + \calB v$ 
%$u_{1}v_{2}^{T}$. The rewrite of $({u_1}v_2^{T}-\boldsymbol{\calX})v_2$ to $V_5v_2 - \calX v_2$ leads to up to 
leads to a $65\times$ speed-up in SystemML. 
%This rewrites achieved by utilizing the view $V_5$ and exploiting the property $(\calA + \calB)v = \calA v + \calB v$, which exploits the sparsity of $\calX$, where $\calX v_2$ can be computed efficiently. 
For MLlib, as discussed before, to avoid memory failure, we used BlockMatrix types.  for all matrices and vectors, thus they were treated as dense. %matrix to avoid memory allocation failure of a matrix of \textbf{LocalMatirx} type. 
% Thus,  the rewrite leads to 2X speed-up, this is due to multiplying a \textbf{BlockMatirx} with another \textbf{BlockMatrix}, which always results in a dense matrix.  As for 
In R, the original pipeline triggers a memory allocation failure for the intermediate result, which the rewriting avoids. %\RA{Maybe not necessary:}
 %Another rewrite could be $(V_5-\calX)v_2$, but it is not the efficient one since it still introduces large intermediate $V_5-\calX$, and not exploiting the sparsity of $\calX$. 
%\vspace{-4mm}
%\begin{figure}[h!]
%\centering \captionsetup{format = hang}
%\includegraphics[width=0.54\linewidth]{charts/ViewsExp/P2-25_AL3.pdf}
%\centering
%\vspace{-4mm}
%\caption{P2.25 before- and after- rewrite using view $V_5$}
%\label{fig:p2.25}
%\vspace{-3mm}
%\end{figure}
Figure~\ref{fig:p2.27} shows that for \textbf{P2.27} exploiting the views $V_2=(\calD+\calC)^{-1}$ and $V_3=DC$ leads to speed-ups of 
 $4\times$ to $41\times$ on different systems. %for R and Numpy, $15\times$ and $41\times$ for TensorFlow and MLlib, respectively. 
Properties enabling rewriting here are $\calC+\calD=\calD+\calC$, $(\calD^{T})^{-1}=(\calD^{-1})^{T}$ and $(\calC \calD) \calE = \calC (\calD \calE)$. %The matrix multiplication order does not have an effect for this pipeline since all matrices are squares.  

\mysubsubsection{\textbf{Rewriting Performance and Overhead}} %Rewrite overhead.
 \label{sec:la-opt}
We now study the running time $RW_{find}$ of our rewriting algorithm, and the {\em rewrite overhead} defined as $RW_{find}$/($Q_{exec}$+ $RW_{find}$), where $Q_{exec}$ is the time to run the pipeline ``as stated''. We ran each experiment  $100$ times and report the average of the last $99$ times. The global trends are as follows. ($i$)~For a fixed pipeline and set of data matrices, {\em the overhead is slightly higher using the MNC cost model}, since histograms are built during optimization. ($ii$)~For a fixed pipeline and cost model, {\em sparse matrices lead to a higher overhead} simply because $Q_{exec}$ tends to be smaller. ($iii$)~{\em Some} (system, pipeline) pairs lead to a low $Q_{exec}$ when {\em the system applies internally the same optimization} that \sys\ finds ``outside'' of the system. 

Concretely, for the $\notopt$ pipelines,  on the dense and sparse matrices listed in Table~\ref{tab:matrices}, using the  na\"ive cost model, 64\% of the $RW_{find}$ times are under 25ms (50\% are under 20ms), and the longest is about 200m. Using the MNC estimator, 55\% took less than 20ms, and the longest (outlier) took about 300ms. 
Among the 38 $\notopt$ pipelines, SystemML finds efficient rewritings for a set of 9, denoted $\notopt_{SM}$, %\RA{These pipelines are 1.10,1.16,1.18,2.15,1.7,and 1.25.
while TensorFlow optimizes a different set of 11, denoted $\notopt_{TF}$. On these subsets, where \sys's optimization is redundant, using {\em dense} matrices, the overhead is very low: with the {\em MNC} model, 0.48\% to 1.12\% on $\notopt_{SM}$ (0.64\% on average), and 0.0051\% to 3.51\% on $\notopt_{TF}$ (1.38\% on average). Using the {\em na\"ive} estimator slightly reduces this overhead, but across $\notopt$, this model misses 4 efficient rewritings. On {\em sparse} matrices, the overhead is at most 4.86\% with the {\em na\"ive} estimator and up to 5.11\% with the {\em MNC} one.

Among the already-optimal pipelines $\opt$, 70\% involve expensive operations such as inverse, determinant, matrix exponential, leading to rather long $Q_{exec}$ times. Thus, the rewriting overhead is less than 1\% of the total time, on all systems, using sparse or dense matrices, and the na\"ive or the MNC-based cost models. For the other $\opt$ pipelines with short $Q_{exec}$, mostly matrix multiplications chains already in the optimal order, on {\em dense} matrices, the overhead reaches 0.143\% (SparkMlLib) to 9.8\% (TensorFlow)  using the na\"ive cost model, while the MNC cost model leads to an overhead of 0.45\% (SparkMlib) up to 10.26\% (TensorFlow). On {\em sparse} matrices, using the na\"ive and MNC cost models, the overhead reaches up to 0.18\% (SparkMLlib) to 1.94\% (SystemML), and  0.5\% (SparkMLlib) to 2.61\% (SystemML), respectively.

\begin{figure*}[!htbp]
  \centering
      \subfigure[\textbf{P1.12}]{\includegraphics[scale=0.25,width=4.2cm,height=3.6cm]{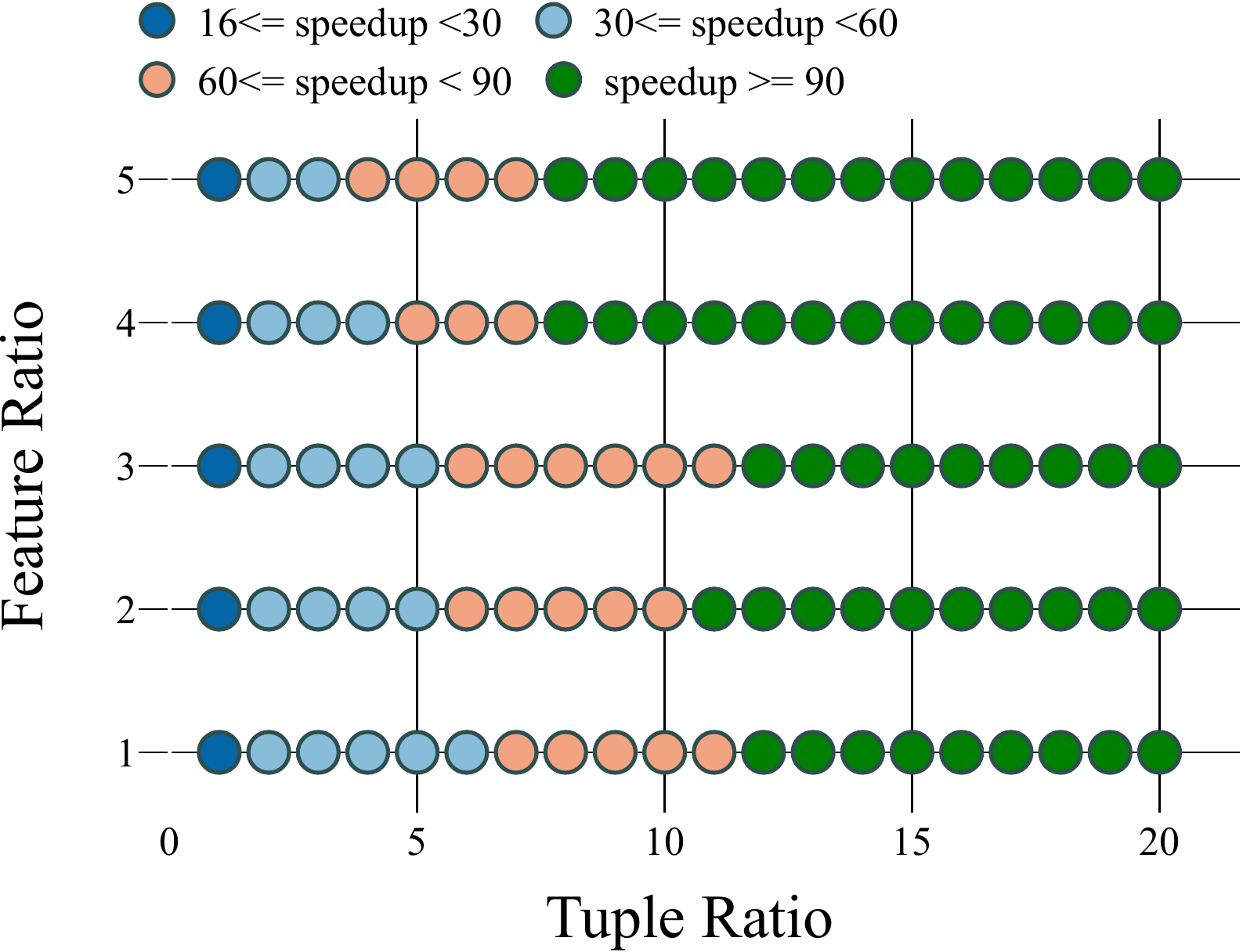}\label{fig:mor-p1.12}}
      \subfigure[\textbf{P2.10}]{\includegraphics[scale=0.25,width=4.2cm,height=3.6cm]{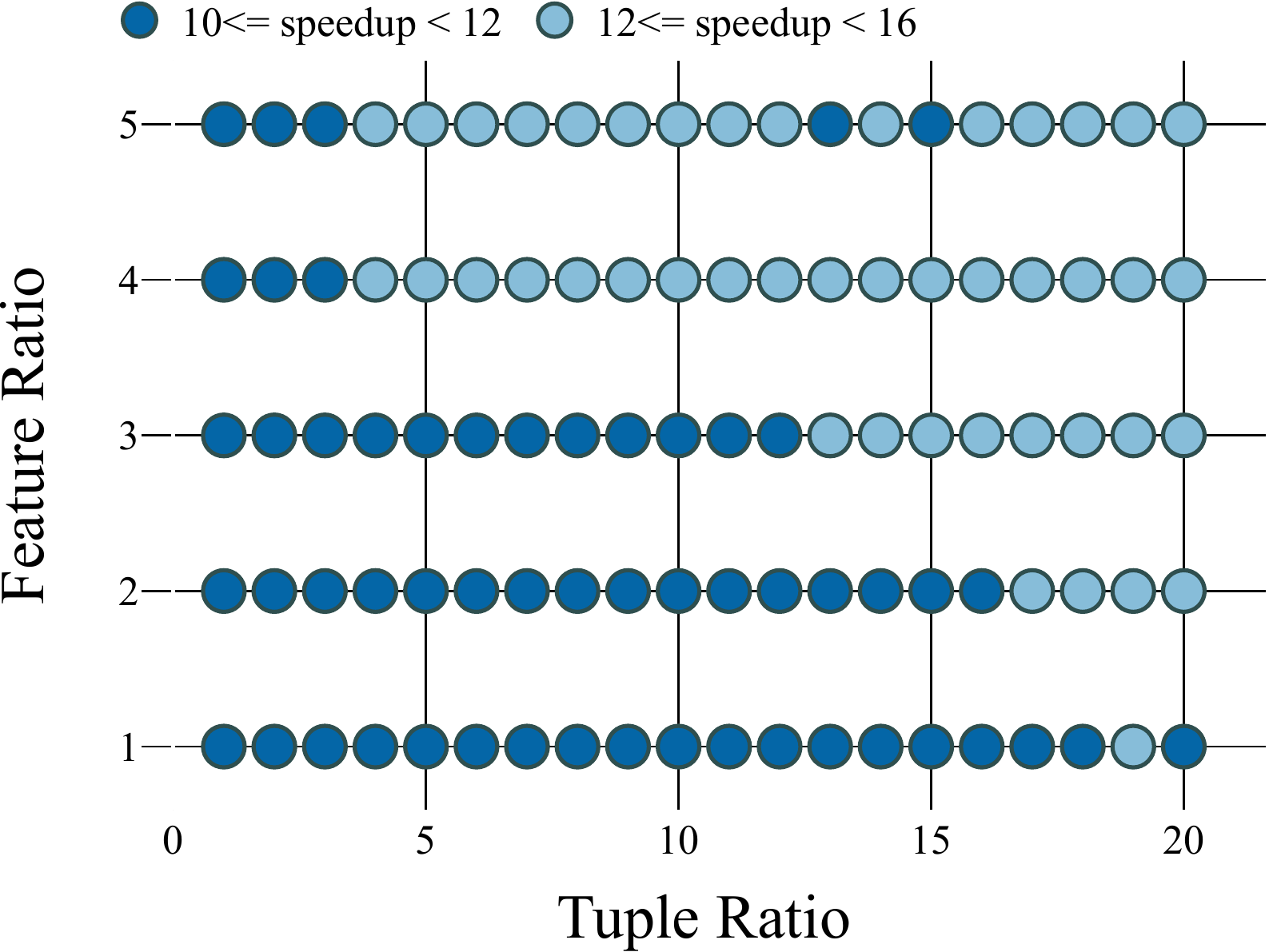}\label{fig:mor-p2.10}}
      \subfigure[\textbf{P2.11}]{\includegraphics[scale=0.25,width=4.2cm,height=3.6cm]{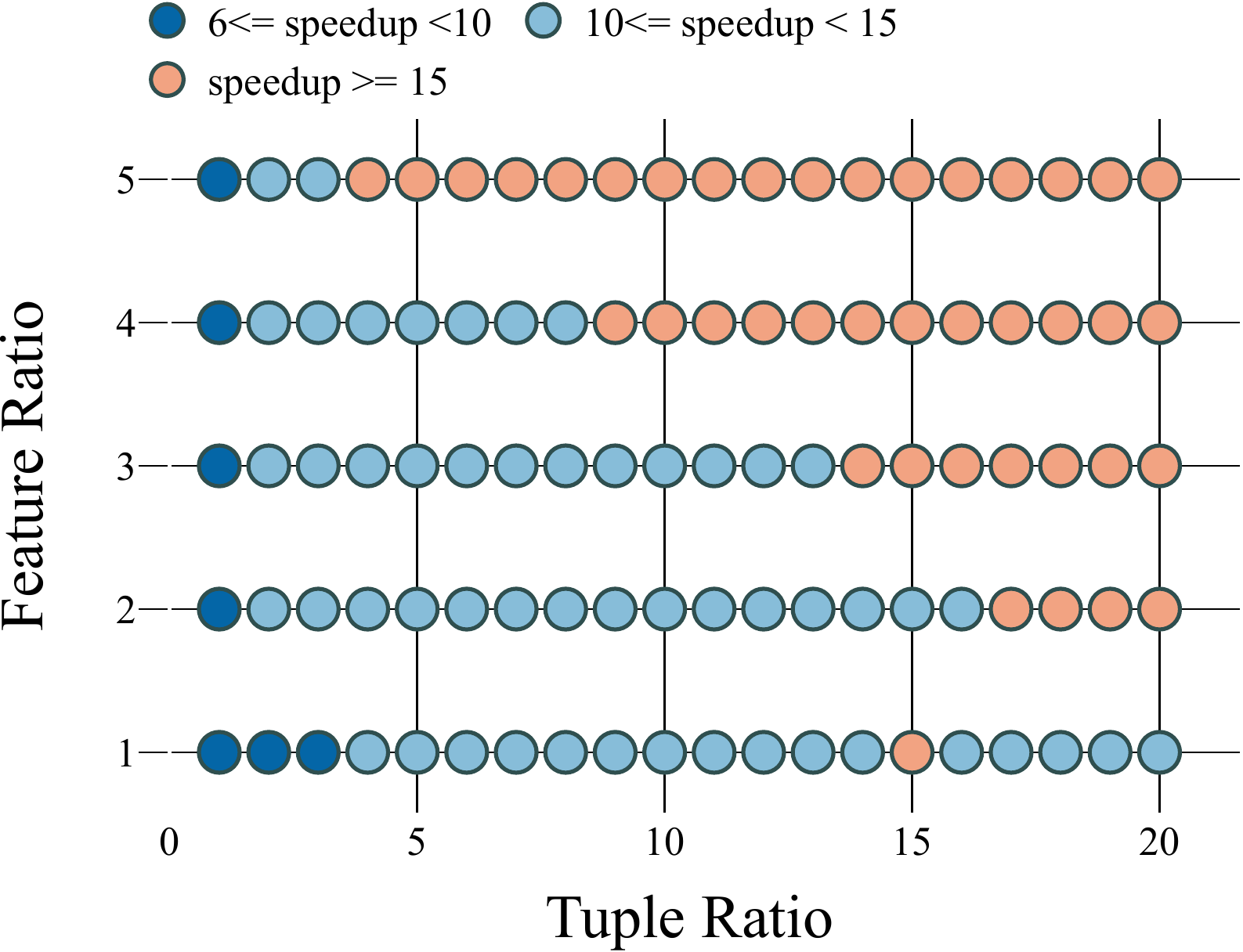}\label{fig:mor-p2.11}}
       \subfigure[\textbf{P2.15}]{\includegraphics[scale=0.25,width=4.2cm,height=3.6cm]{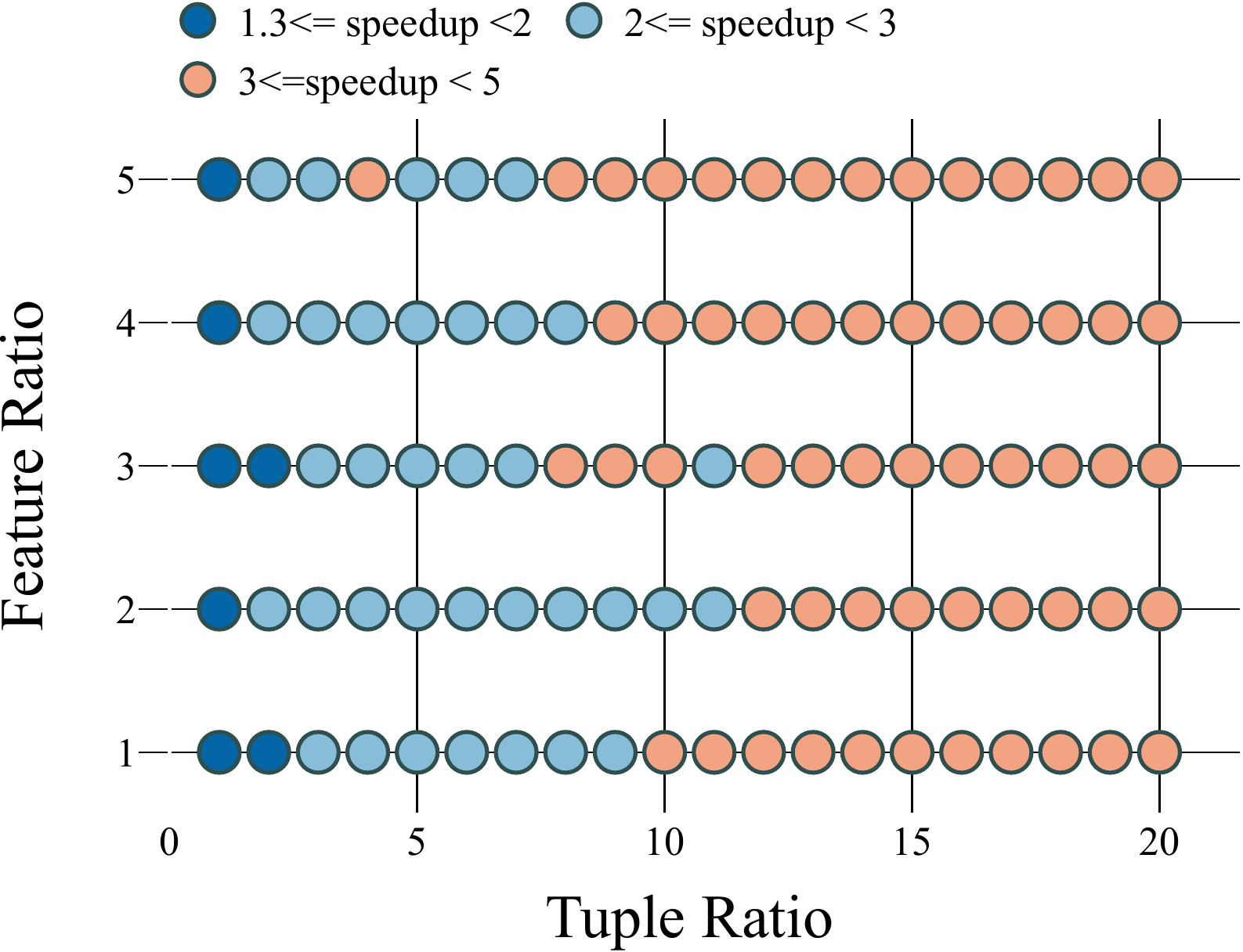}\label{fig:mor-p2.15}}
 \caption{Speed-ups of Morpheus (with HADAD rewrites) over  Morpheus (without HADAD rewrites) for pipelines P1.12, P2.10, P2.11 and P2.15 on synthetic data for a PK-FK join.}
 \label{fig:morphues-exp}
 \figspb
\end{figure*}
\mysubsection{Hybrid (LA and RA) Experiments}
\label{sec:exp-hybrid}
\newcommand {\MTabel}{\textbf{M}}
\newcommand {\NTabel}{\textbf{N}}
We now study the benefits of rewriting on hybrid scenarios combining RA and LA operations. In \S\ref{sec:morpheus}, we show the performance benefits of \sys\ to a cross RA-LA platform, MorpheusR~\cite{morpheus}. 
%, a cross RA-LA platform for pushing LA computations through joins.
We evaluate our hybrid micro-benchmark on SparkSQL+SystemML in \S\ref{sec:micro-hybrid}. (\S\ref{sec:rw-overhead-hybrid}) discusses our optimization overhead.

%\RA{We removed this section since the results exhibit similar trends as LA experiments } Finally, in  \S\ref{sec:rw-overhead-hybrid}, we discuss our optimization overhead.  %We discuss subset of the results here, and other results are discussed in~\cite{technicalreport} 

%\vspace{-4mm}
\mysubsubsection{\textbf{MorpheusR Experiments}}
\label{sec:morpheus}

We use the same experimental setup introduced in ~\cite{chen2017towards} for generating synthetic datasets for the PK-FK join of tables \textbf{R} and \textbf{S}. The quantities varied are the \textit{tuple ratio} ($n_S$/$n_R$) and feature ratio ($d_R$/$d_S$), where $n_S$ and $n_R$ are the number of rows  and $d_R$ and $d_S$ are the number of columns (features) in \textbf{R} and  \textbf{S}, respectively. We fix $n_R=1M$ and $d_S=20$. %, this helps quantify the amount of redundancy introduced by a PK-FK join on \textbf{R} and \textbf{S}. 
 The join of \textbf{R} and \textbf{S} outputs $n_s\times(d_R+d_S)$ matrix \MTabel, which is always dense.
We evaluate on Morpheus a set 
of 8 pipelines
and their rewritings found by \sys\ using the \naive\ cost model.
\eat{\RA{This info in the extended version}\footnote{\rev{We select a set of pipelines where Morpheus can rewrite (factorize) them. This includes pipelines that contain aggregate operations, transpose and matrix multiplication. For the inverse, we only evaluated (OLS), which is the realistic one}}} %Morpheus does not factorize trace, determinant, and element-wise operations.}

\vspace{2mm}
\noindent\textbf{Discussion}. 
\textbf{P1.12}: \colsumsOp$($\MTabel$N)$ is the example from
\S\ref{sec:hadad-opt}, with \MTabel\ the output (viewed
as matrix) of joining tables \textbf{R} and \textbf{S}
generated as described above.
$N$ is a $ncol($\MTabel$)\times 100$ %synthetic 
dense matrix. \sys's rewriting
yields up to 125$\times$ speed-up (see Figure~\ref{fig:mor-p1.12}).

Figure~\ref{fig:mor-p2.10} shows up to 15$\times$ speed-up
for \textbf{P2.10}: \rowsumsOp$(N$\MTabel$)$,
where the size of $N$ is  $100\times nrow($\MTabel$)$.
This is due to \sys's rewriting: $N$\rowsumsOp$($\MTabel$)$,
which enables Morpheus to push the \rowsumsOp\ operator to
\textbf{R} and \textbf{S} instead of computing
the large intermediate matrix multiplication. 

\textbf{P2.11}: \sumOp$(N+$\MTabel$)$ is run \textit{as-is}
by Morpheus since it does not factorize
element-wise operations, \eg addition.
However, \sys\ rewrites
\textbf{P2.11} into sum$(N)+$\sumOp$($\MTabel$)$, which
avoids the (large and dense) intermediate result of
the element-wise matrix addition. 
The \sys\ rewriting enables
Morpheus to execute \sumOp$($\MTabel$)$ by pushing
\sumOp\ to \textbf{R} and \textbf{S}, for up to
20$\times$ speed-up (see Figure~\ref{fig:mor-p2.11}). 

Morpheus evaluates
\textbf{P2.15}: \sumOp$($\rowsumsOp$($\MTabel$))$
by pushing the \rowsumsOp\ operator to
\textbf{R} and \textbf{S}.
\sys\ finds the rewriting $\sumOp($\MTabel$)$, which enables
Morpheus to push the \sumOp\ operation instead,
achieving up to 4.5$\times$ speed-up (see Figure~\ref{fig:mor-p2.15}).

%Figure~\ref{fig:mor-p1.12} shows up to 4.5$\times$ speed-up achieved by turning . The rewriting avoids the dene intermediate .  %However, we observed that the overhead of Morpheus's rewrite rules for \rowsumsOp\ and \sumOp operator are slightly comparable.

Since Morpheus does not exploit the associative property of
matrix multiplication, it cannot reorder multiplication
chains to avoid large intermediate results, which
lead to runtime exception in R (Morpheus' backend).
For example, for the chain $N^TN$\MTabel\ (in \textbf{P1.26}), %\AD{Why doesn't this one have a number like the others? It breaks the formula.} %\RA{The original pipeline failed (can not allocate memory for intermediate), So  I didn't repot the speed-up}. 
when the size of
\MTabel\ is 1M$\times$20 and the size of $N$ is 20$\times$1M,
the size of the $N^TN$ intermediate result is 1M$\times$1M,
which R cannot handle ((timed out >1000 seconds)). %\AD{that's what you wanted to say, rightt?}. \RA{Yes}
%\RA{Too much details ((1M$\times$1M)*8(size of double precision) = 8000G memory storage size). 
\sys\ exploits associativity and selects
the rewriting $N^T(N$\MTabel$)$ of intermediate result size
(100$\times$20).
 	
For pipelines \textbf{P1.14} and \textbf{P2.12}, that involve transpose operator, Morpheus applies its special rewrite rules that replace an operation on \MTabel$^T$ with an operation on \MTabel\ before pushing the operation to the base tables. %for pipelines \textbf{P1.14}: \sumOp$($\colsumsOp$($\MTabel$^TN^T))$ and \textbf{P2.12}: \sumOp$($\rowsumsOp$($\MTabel$^TN^T))$, Morpheus rewrites the pipelines to sum$(($\rowsumsOp$(N$\MTabel$)))$ and \sumOp$(($\colsumsOp$(N$\MTabel$)))$, respectively, and then applies its factorized matrix-multiplication rewrite rule on $N$\MTabel. 
\sys\ rewrites both pipelines to \sumOp$(N$\MTabel$)$, enabling again Morpheus to apply its factorized rewrite rule on $N$\MTabel\ and achieving speed-up ranging from 1.3 up to 1.5$\times$. 
%The reason is that for both rewritings (found by Morpheus and \sys), Morpheus's factorized matrix multiplication rewrite rule dominates the execution time.
%\RA{To replot -- too small}
\begin{figure*}[!htbp]
\figspa
  \centering
      \subfigure[\textbf{N matrix: 2M rows}]{\includegraphics[scale=0.27,width=5.8cm,height=3.9cm]{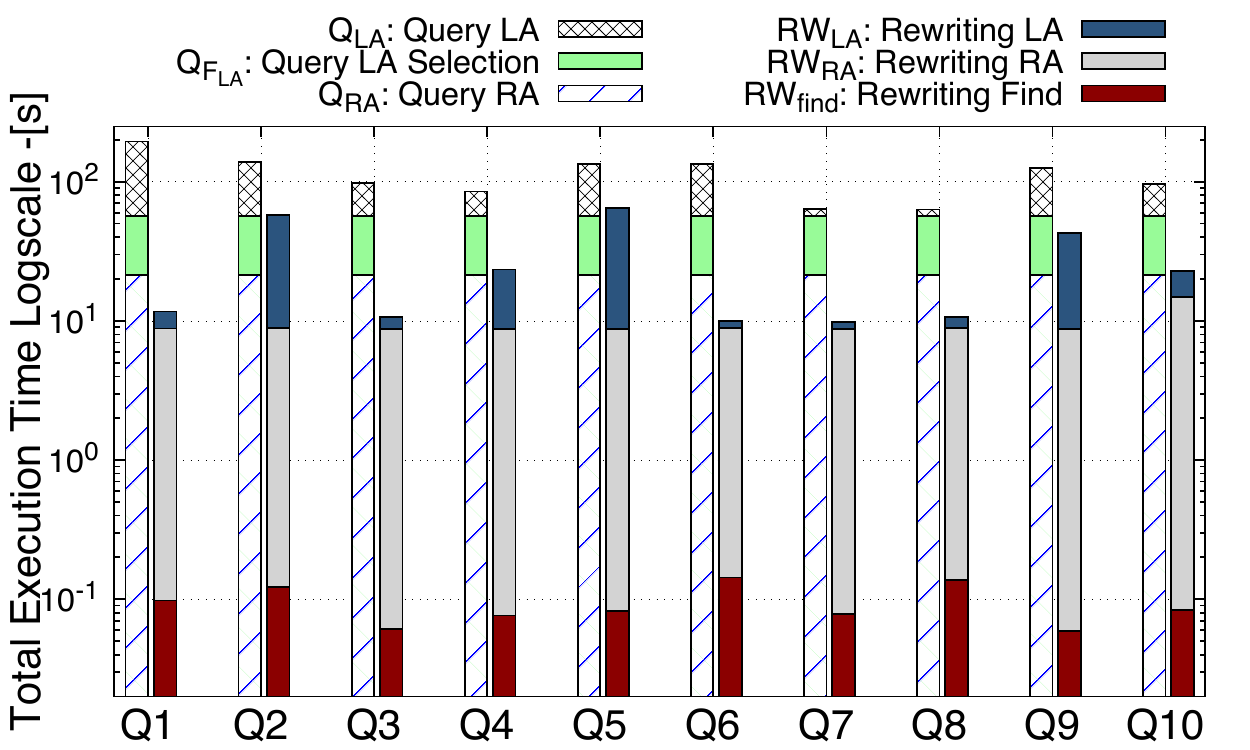}\label{fig:N2M}}
      \subfigure[\textbf{N matrix: 1M rows}]{\includegraphics[scale=0.27,width=5.8cm,height=3.26cm]{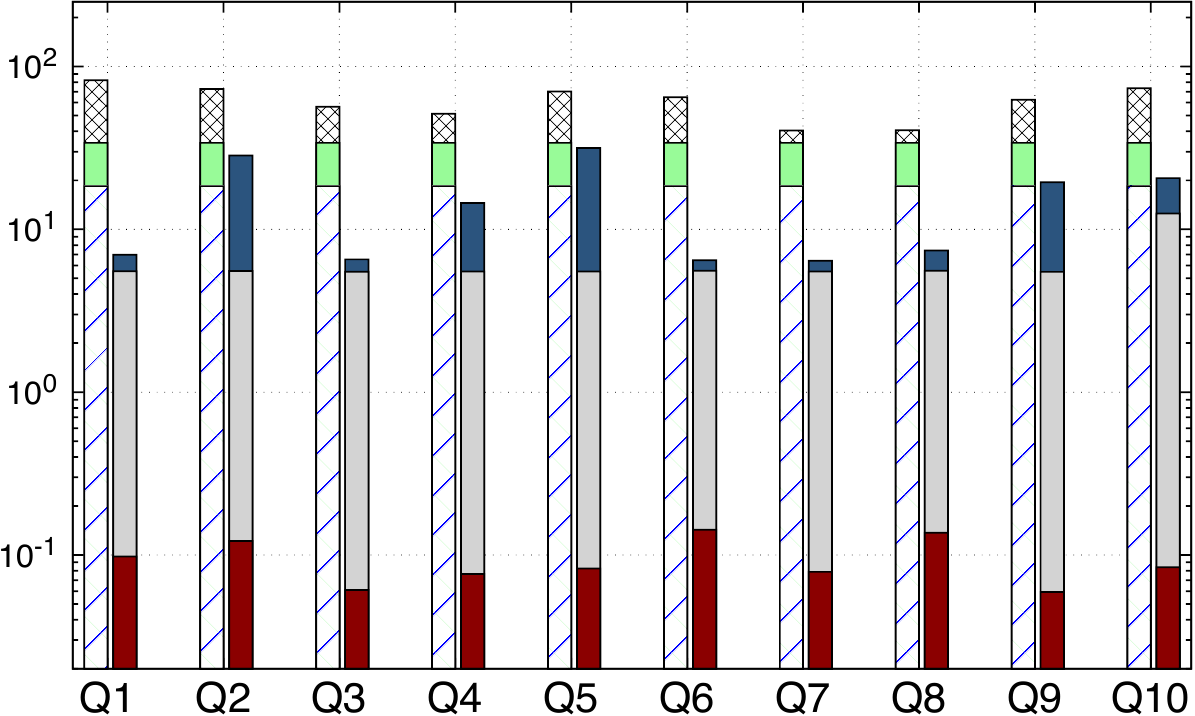}\label{fig:N1M}}
      \subfigure[\textbf{N matrix : 0.5 rows}]{\includegraphics[scale=0.27,width=5.8cm,height=3.26cm]{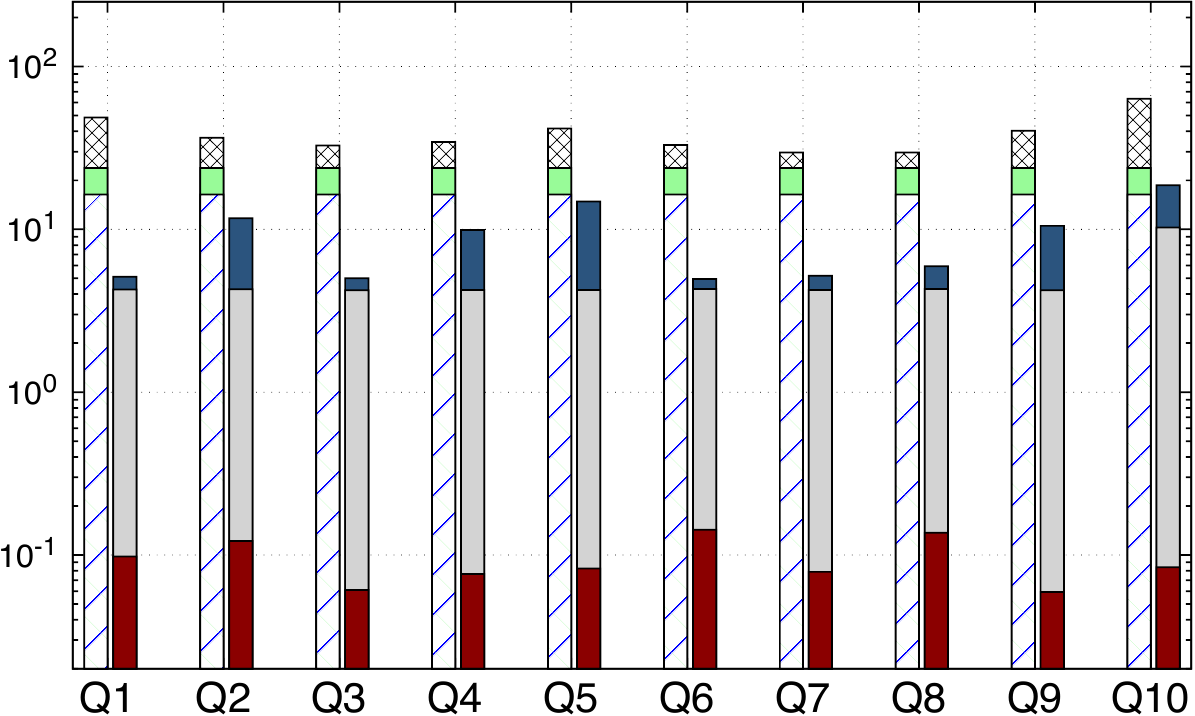}\label{fig:N5K}}
\figspb
 \caption{Micro-Hybrid Benchmark Twitter Dataset %\RA{TODO:Replot - font is too small}
 }
 \label{fig:micro-hybrid-twitter}
\figspb %\figspa
 \end{figure*}
 %\vspace{-1mm}
\mysubsubsection{\textbf{Micro-hybrid Benchmark Experiments}}
\label{sec:micro-hybrid}
In this experiment, we create a \textit{micro-hybrid benchmark} on Twitter\cite{twitter} and MIMIC \cite{MIMIC} datasets to empirically study \sys's rewriting benefits in a hybrid setting. The benchmark comprises ten different queries combining relational and linear algebra expressions. 
%We only discuss our Twitter dataset results here and present the other results on the MIMIC dataset in ~\cite{technicalreport}. 

\vspace{2mm}
\noindent\textbf{Twitter Dataset Preparation}. We obtain from Twitter API ~\cite{twitter} 16GB of tweets (in JSON). 
%worth of tweets in total. 
%The raw dataset is in a JSON format. 
We extract the structural parts of the dataset, which include user and tweet %\textit{structural} 
information, %such as the number of followers, likes, etc., 
and store them in tables \textbf{User (U)} and \textbf{Tweet (T)}, linked via PK-FK relationships.
The dataset %\textbf{TJ} 
is detailed in \S\ref{sec:hadad-opt}. 
The tables \textbf{User} and \textbf{Tweet} as well as \textbf{TweetJSON (TJ)} are stored in Parquet format. 
	
\vspace{1mm}
\noindent\textbf{Twitter Queries and Views}.
Queries consist of two parts:
($i$) RA preprocessing ($Q_{RA}$) and
($ii$) LA analysis ($Q_{LA}$).
In the $Q_{RA}$ part, queries construct two %main 
matrices: \MTabel\ and \NTabel. %The feature vectors in the base tables \textbf{Tweet} and \textbf{User} tables can be viewed as matrices, say \textbf{T} and \textbf{U}, respectively. 
The matrix \MTabel\ (2M$\times$12;\textit{dense})
is the output of joining \textbf{T} and \textbf{U}.
The construction of matrix \NTabel\ is described in
\S\ref{sec:hadad-opt}. %(\textit{ultra-sparse}) is a \textit{tweet-hashtag} filter level matrix constructed from \textbf{TJ}, where rows are tweets, columns are hashtags and values are filter level
%\footnote{\rev{Matrix $N$ is represented in Matrix Market Format (MTX) since it is sparse.}}.
We fix the $Q_{RA}$ part across all queries and vary the
$Q_{LA}$ part using a set of LA pipelines detailed below. %and illustrated in Table~\ref{tab:hybrid-la} 
% a well as \NTabel\  and \MTabel\ matrices. 

In addition to the views defined in \S\ref{sec:hadad-opt}, we
define three hybrid RA-LA materialized views:
$V_3$, $V_4$ and $V_5$, which store the result of applying
$\rowsumsOp$, $\colsumsOp$ and matrix multiplication
operations over base tables \textbf{T} and \textbf{U}
(viewed as matrices).

Importantly, rewritings based on these views can only be found
by {\em exploiting together LA properties} and
{\em Morpheus' rewrites rules}
(we incorporated them in our framework as a set of integrity
constraints).
The full list of queries and views is
in Appendix~\ref{appendixI}.
\begin{table}[ht]
\centering                     
\setlength{\extrarowheight}{2pt}
\begin{tabular}{||c|c||}
\hline
\hline
\rowcolor{lightgray}
\textbf{No.} & \textbf{Expression}\\
\hline
\textbf{P3.1}& \rowsumsOp$(X$\MTabel$)+(uv^T+$ \NTabel$^T)v$\\
\hline
\textbf{P3.2}& u\colsumsOp$((X$\MTabel$)^T)$+\NTabel \\
\hline
\textbf{P3.3}& $(($\NTabel$+X)v)$\colsumsOp$($\MTabel$)$\\
\hline
\textbf{P3.4}& $\sumOp(C+$\NTabel$$\rowsumsOp$(X$\MTabel$)v)$\\
\hline
\textbf{P3.5}& $u$\colsumsOp$($\MTabel$X)$+\NTabel\\
\hline
\textbf{P3.6}& \rowsumsOp$(($\MTabel$X)^T)+(uv^T+$ \NTabel$)v$\\
\hline
\textbf{P3.7}& $X$\NTabel$u+$\rowsumsOp$($\MTabel$)^T$\\
\hline
\textbf{P3.8}& \NTabel$\odot$\traceOp$(C+v$\colsumsOp$($\MTabel$X)C)$\\
\hline
\textbf{P3.9}& $X\odot $\sumOp$($\colsumsOp(C)$^T+$\rowsumsOp$($\MTabel$))$+\NTabel\\
\hline
\textbf{P3.10}& \NTabel$\odot$ \sumOp$((X+C)$\MTabel$)$\\
\hline\hline
\end{tabular}
\caption{LA pipelines used in micro-hybrid benchmark}
\vspace{-8mm}
\label{tab:hybrid-la}
\end{table}
\begin{figure*}[!htbp]
\figspa
  \centering
      \subfigure[\textbf{N matrix: 40K rows}]{\includegraphics[scale=0.27,width=5.8cm,height=3.9cm]{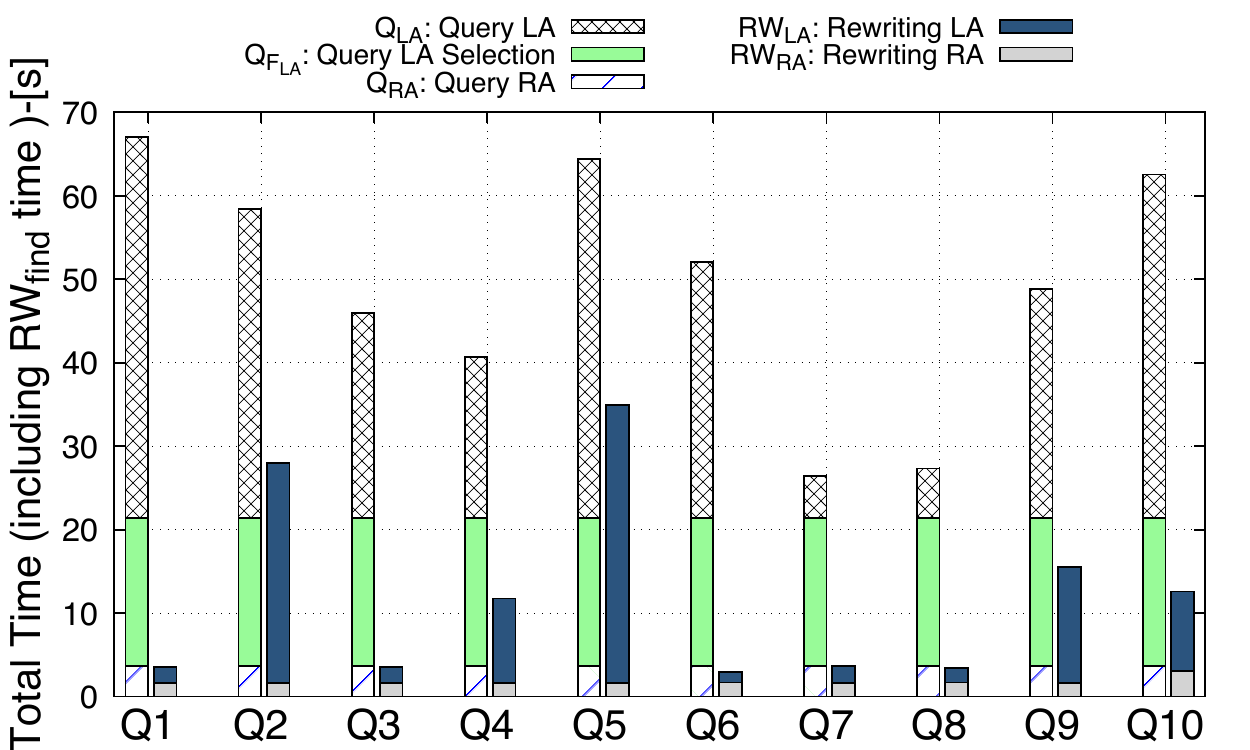}\label{fig:N40K}}
      \subfigure[\textbf{N matrix: 20K rows}]{\includegraphics[scale=0.27,width=5.8cm,height=3.26cm]{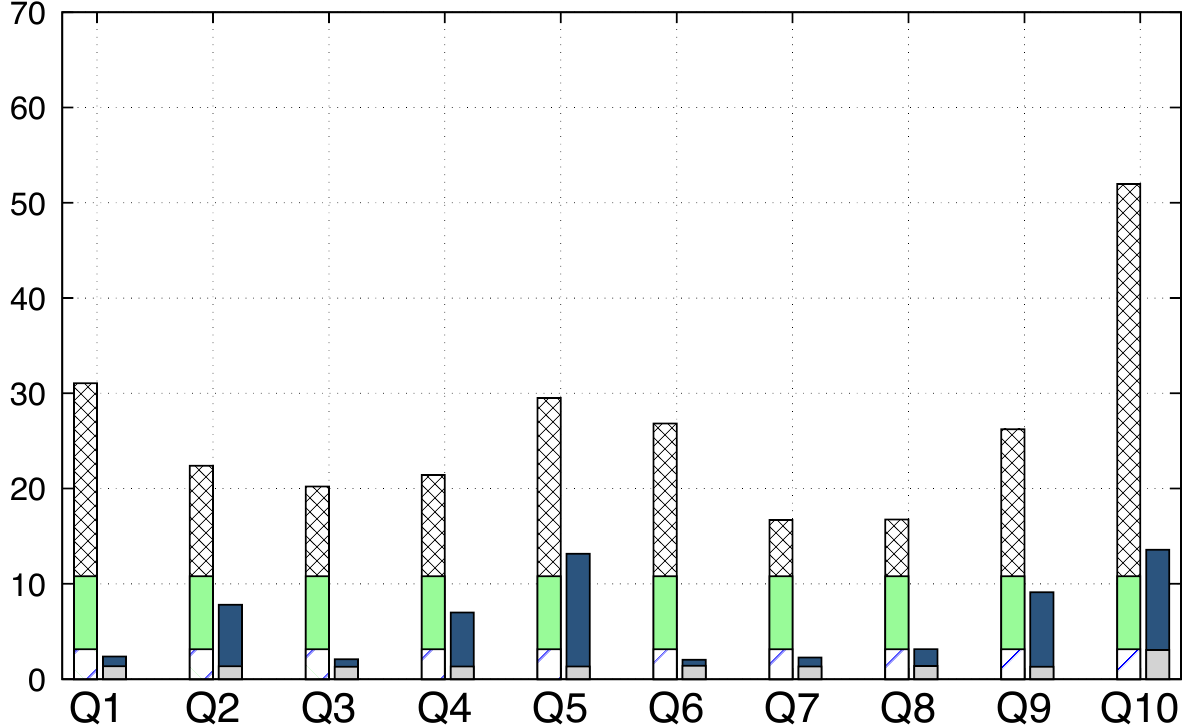}\label{fig:N20K}}
      \subfigure[\textbf{N matrix : 10K rows}]{\includegraphics[scale=0.27,width=5.8cm,height=3.26cm]{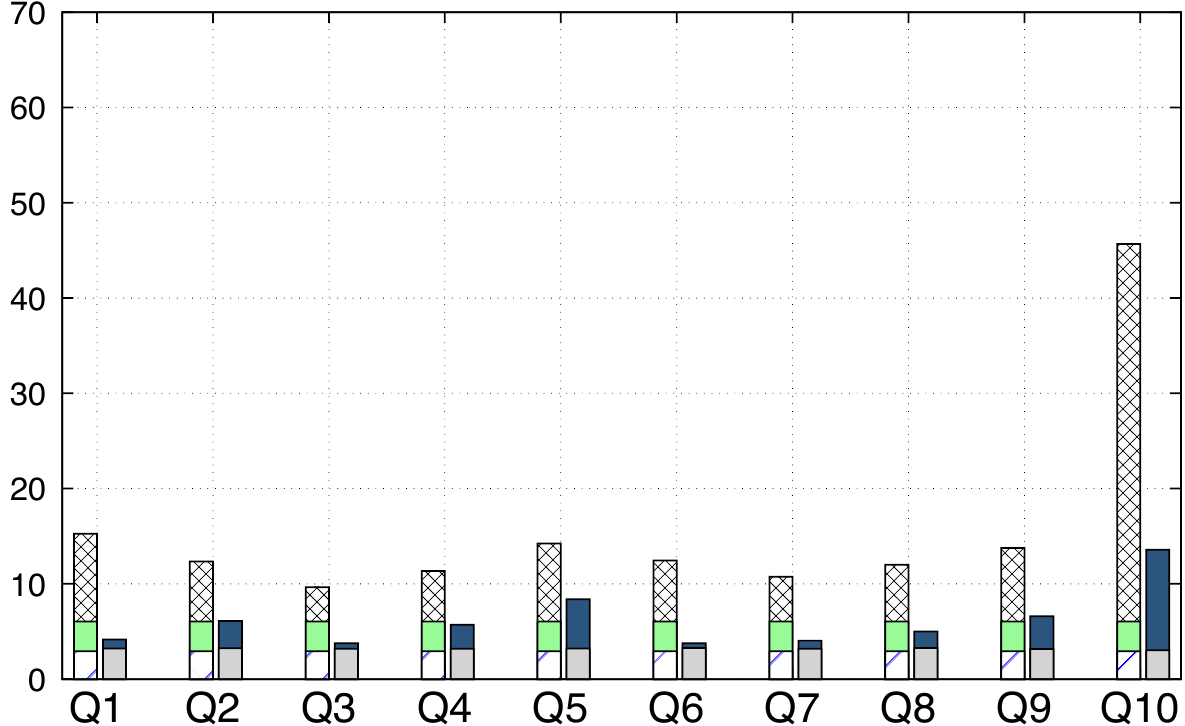}\label{fig:N10K}}
\figspa
 \caption{Micro-Hybrid Benchmark MIMIC}
 \label{fig:micro-hybrid-mimic}
\figspa
 \end{figure*}

\vspace{1mm}
\noindent\textbf{Discussion}. 
After construction of \MTabel\ and \NTabel\ by the $Q_{RA}$
part in SparkSQL, both matrices are loaded to SystemML to be
used in the $Q_{LA}$ part. %(varied using a set of LA pipelines, detailed below for each query).
%(varied using a set of pipelines in Table~\ref{tab:hybrid-la}). 
Before evaluating an LA pipeline on \MTabel\ and \NTabel, all queries select \NTabel\ 's rows ($Q_{F_{LA}}$) with \textit{filter-level} less than $4$ (medium).
For all of them, \sys\ rewrites the $Q_{RA}$ part of \NTabel\
as described in \S\ref{sec:hadad-opt}.
%For the first run of all queries (Figure~\ref{fig:N2M}), matrix \NTabel\  (2M$\times$1000) is constructed from \textbf{TJ} for all tweets that mention ``covid'' and are posted from ``USA'', and \MTabel\ (2M$\times$ 12) is constructed as described above. Both matrices are then loaded in SystemML to be used for the analysis part (varied using the set of pipelines in Table~\ref{tab:hybrid-la}). Before evaluating an LA pipeline, the queries filter \NTabel\ 's rows with \textit{filter-level} less than $4$ (medium). Now, we analyze  \sys's rewrites and optimizations for each query in detail in the following paragraphs: 

\vspace{1mm}
\noindent\textbf{Q1:} For the $Q_{LA}$ part , the query runs \textbf{P3.1} (see Table~\ref{tab:hybrid-la}). \sys\ applies several optimizations: ($i$) it rewrites $(uv^T+$ \NTabel\ $^T)v$ to $uv^Tv+$\NTabel$^Tv$, where $u$ and $v$ are synthetic vectors of size 	12$\times$1 and  2M$\times$1. First, \NTabel\  is ultra sparse, which makes the computation of \NTabel$^Tv$ extremely efficient. Second, SystemML evaluates $uv^{T}v$ efficiently in one go without intermediates, taking advantage of \texttt{\textbf{tsmm}} operator (discussed earlier) and \texttt{\textbf{mmchain}} for matrix multiply chains, where the best way to evaluate it computes $v^{T}v$ first, which results in a scalar, instead of computing $uv^{T}$, which results in a dense matrix of size 1000$\times$2M. Alone, SystemML is unable to exploit its own efficient operations for lack of awareness of the LA property $\calA v+\calB v=(\calA+\calB)v$; ($ii$) \sys\ also rewrites \rowsumsOp$(X$\MTabel$)$ into $XV_3$, where $V_3=$\rowsumsOp$($\textbf{T}$)$ $+ K$\rowsumsOp$($\textbf{U}$)$, by exploiting the property \rowsumsOp$(X$\MTabel$) = X$\rowsumsOp$($\MTabel$)$ together with Morpheus's rewrite rule: \rowsumsOp$($\MTabel$)\rightarrow$ \rowsumsOp$($\textbf{T}$)+K$\rowsumsOp$($\textbf{U}$)$ %where \MTabel\ is the join's output table (matrix) of \textbf{T} and \textbf{U} base tables (feature matrices). 
The rewriting achieves up to 16.5$\times$ speed-up. 

\vspace{1mm}
\noindent\textbf{Q2:} The speed-up of 2.5$\times$  comes from rewriting  the pre-processing part and turning \textbf{P3.2}: $u$\colsumsOp$((X$\MTabel$)^T)$+\NTabel\  to  $u(XV_4)^T$+\NTabel,  where $u$ and $X$ are synthetic matrices of size 2M$\times$1 and 1000$\times$2M, respectively. \sys\ exploits \colsumsOp$((X$\MTabel$)^T) = $\rowsumsOp$(X$\MTabel$)^T$ and \rowsumsOp$(X$\MTabel$) = X$\rowsumsOp$($\MTabel$)$ together with Morpheus's rewrite rule: \rowsumsOp$($\MTabel$)\rightarrow$ \rowsumsOp$($\textbf{T}$)+K$\rowsumsOp$($\textbf{U}$)$. Both the rewriting and the original LA pipeline  introduce an unavoidable large  dense intermediate %($u(XV_4)^T$ and $u$\\colsumsOp(Op$((X$\MTabel\ $)^T)$) 
of size 2M$\times$1000.  

\vspace{1mm}
\noindent\textbf{Q3:} %In addition to the rewriting of the pre-processing part, for \textbf{P3.3}, \sys\ ($i$) avoids introducing the large dense intermediate (\NTabel  $+ X$; 2M$\times$1000) by distributing the multiplication of $v$ (1000 $\times$ 1) and consider the sparsity of \NTabel, which again makes the multiplication with $v$ more efficient; ($ii$) directly rewrites \colsumsOp$($\MTabel$)$ to $V_3= [$\colsumsOp$($\textbf{T}$),$\colsumsOp$(K)$\textbf{U}$]$ by utilizing Morpheus's rewrite rule: \colsumsOp$($\MTabel$)\rightarrow [$\colsumsOp(\textbf{T})$, $\colsumsOp$(K)$\textbf{U}$]$. The obtained rewriting achieves 9.2$\times$ speed-up. 
The query runs $(($\NTabel$+X)v)$\colsumsOp$($\MTabel$)$ in the $Q_{LA}$ part, where dense matrices $X$ and $v$ are of size 2M$\times$1000 and 1000$\times$1, respectively. \sys\ avoids the dense intermediate (\NTabel  $+ X$) % 2M$\times$1000) 
by distributing the multiplication by $v$ and realizing that
the sparsity of \NTabel\ yields efficient multiplication. %\RA{SystemML doesn't exploit: (A+B)C=AC+BC} %with $v$. %which as discussed before it makes the multiplication with $v$ more efficient. 
It also directly rewrites \colsumsOp$($\MTabel$)$ to $V_4$, where $V_4= [$\colsumsOp$($\textbf{T}$),$\colsumsOp$(K)$\textbf{U}$]$ by utilizing
one of Morpheus's rewrite rules. The rewriting, including
the rewriting of the $Q_{RA}$ part of \NTabel, achieves 9.2$\times$ speed-up
(Figure~\ref{fig:N2M}-Q3).

\vspace{1mm}
\noindent\textbf{Q4:} In the $Q_{LA}$ part, \textbf{Q4} runs
\sumOp$(C+$\NTabel\rowsumsOp$(X$\MTabel$)v)$
(inspired by the COX proportional hazard regression model
used in SystemML's test suite~\cite{cox})%\AD{Rana, citation here}),\RA{Fixed}
, where synthetic dense matrices $C$,
$X$ and $v$ have size
2M$\times$1000,1000$\times$2M and 1$\times$1000, respectively. \sys\ ($i$) distributes the \sumOp\ operation %(\sumOp$(A+B)=$\sumOp$(A)+$\sumOp$(B)$) 
to avoid materializing the dense
addition (SystemML includes this rewrite rule but fails to
apply it);
($ii$) rewrites \rowsumsOp$(X$\MTabel$)$ to $XV_3$, where $V_3=$\rowsumsOp$($\textbf{T}$)$ $+ K$\rowsumsOp$($\textbf{U}$)$, by exploiting \rowsumsOp$(X$\MTabel$) = X$\rowsumsOp$($\MTabel$)$ together with Morpheus's rewrite rule
\rowsumsOp$($\MTabel$)\rightarrow$ \rowsumsOp$($\textbf{T}$)+K$\rowsumsOp$($\textbf{U}$)$\footnote{$K$ is the unique sparse indicator matrix that
captures the primary/foreign key dependencies between \textbf{T}
and \textbf{U}, introduced by Morpheus's rewrite rules~\cite{chen2017towards}.}.
The multiplication chain in the rewriting is efficient since \NTabel\ is
sparse. The rewriting of this query (including the rewriting of the $Q_{RA}$ part of \NTabel\ ) achieves 3.63$\times$ speed-up (Figure~\ref{fig:N2M}-Q4).
%Notably, both the rewriting and the original pipeline still pays the cost of materializing an intermediate of size 2M$\times$1000 ($N((XV_4)v)$ and  $N$\rowsumsOp$(X$\MTabel $)v$). 

\vspace{1mm}
\noindent\textbf{Q5:} \sys's rewriting speeds-up this query by 2.3$\times$. It rewrites $u$\colsumsOp$($\MTabel$X)$ in \textbf{P3.5} to $uV_4X$ (see \textbf{Q3} for $V_4$'s definition). The view is exploited by \sys\  due to utilizing \colsumsOp$($\MTabel$X)$ = \colsumsOp$($\MTabel$)X$ together with  the Morpheus's rewrite rule (shown in \textbf{Q3}) for pushing the \colsumsOp\ to the base tables (matrices) \textbf{U} and  \textbf{T}. The found rewriting enables SystemML to optimize the matrix-chain multiplication by computing $V_3X$ first, which results in 1$\times$1000 matrix instead of computing $uV_3$ which results in 2M$\times$12. The rewriting and the original pipeline still introduce unprevenbted dense intermediate of size 2M$\times$1000.         

\vspace{1mm}
\noindent\textbf{Q6:} The LA pipeline (\textbf{P3.6}) in this query is a variation of \textbf{P3.1}. In addition to distributing the multiplication of $v$, \sys\ rewrites \rowsumsOp$(($\MTabel\ $X)^T)$ to $(V_4X)^T$ (see \textbf{Q3} for $V_4$'s definition) by exploiting \rowsumsOp$(($\MTabel$X)^T)$ = \colsumsOp$($\MTabel$X)^T$  and \colsumsOp$($\MTabel$X)$ = \colsumsOp$($\MTabel$)X$, all together with Morpheus's rewrite rule as illustrated in \textbf{Q3}. The obtained rewriting (along with the rewriting of the pre-processing part) achieves speed-up of 13.4$\times$ .     

\vspace{1mm}
\noindent\textbf{Q8:} 
%In addition to rewriting the pre-processing part, \sys\ applies several optimizations for the analysis part (\textbf{P3.8}). First, it distributes the \traceOp\ operation to prevent materializing a dense intermediate, which SystemML does not apply. Second, it exploits the view $V_3$ (see \textbf{Q3}) by utilizing \colsumsOp$($\MTabel$X)$ = \colsumsOp$($\MTabel$)X$, which then enables SystemML to further optimize the matrix chain multiplication by computing $v((V_3X)C)$, where the size of $v$, $X$ and $C$ are 20K$\times$1, 12$\times$20K and 20K$\times$20K, receptively. The element-wise multiplication with \NTabel\ in the rewiring and the original pipeline is extremely efficient since  \NTabel\ is ultra-sparse. The rewriting speeds-up the query by 5.9$\times$.  
The $Q_{LA}$ part executes
\NTabel$\odot$\traceOp$(C+v$\colsumsOp$($\MTabel$X)C)$,
where the size of $v$, $X$ and $C$ are 20K$\times$1, 12$\times$20K and
20K$\times$20K, respectively. First, \sys\ distributes the \traceOp\
operation (which SystemML does not apply) to avoid the dense intermediate addition.
Second, \sys\ enables the exploitation of view $V_4$ (see \textbf{Q2})
by utilizing \colsumsOp$($\MTabel$X)$ = \colsumsOp$($\MTabel$)X$.
The resulting multiplication chain is optimized by the order
$v((V_4X)C)$.
The final element-wise multiplication with \NTabel\
is  efficient since
\NTabel\ is ultra-sparse. The combined $Q_{RA}$ and $Q_{LA}$
rewriting speeds up \textbf{Q8} by 5.94$\times$ (Figure~\ref{fig:N2M}-Q8). 

\vspace{1mm}
\noindent\textbf{Q9:} In addition to the rewriting of the $Q_{RA}$ part, the speed-up of 3$\times$  also attributes to turning \sumOp$($\colsumsOp$(C)^T+$\rowsumsOp$($\MTabel $))$ in \textbf{P3.9} to \sumOp$(V_5)$, where $V_5=[C$\textbf{T}$,CK$\textbf{U}$]$. The view $V_5$ is utilized by exploiting the property \sumOp$(C$\MTabel$)=$\sumOp$($\colsumsOp$(C)^T \odot$ \rowsumsOp$($\MTabel$))$ together with Morpheus's rewrite rule: $C$\MTabel $\rightarrow$ $[C$\textbf{T}$,CK$\textbf{U}$]$. The result of an element-wise multiplication with the dense matrix $X$ in the rewriting and the original pipelines is a dene intermediate (see Figure ~\ref{fig:N2M}-Q9).   
%where  $B$ is the join's output table (matrix) of $R$ and $S$ base tables (feature matrices). 

\vspace{1mm}
\noindent\textbf{Q10:} 
%\sys's rewrite leads to a speed-up of 3.9$\times$. In addition to rewriting the pre-processing part, \sys\ rewrites \sumOp$((X+C)$\MTabel$)$ \textbf{P3.10} to \sumOp$(X$\MTabel$)+$\sumOp$(V_5)$, where $V_5=[C$\textbf{T}$,CK$\textbf{U}$]$. The view $V_5$ is utilized by exploiting the property $(X+C)$\MTabel = $X$\MTabel$+C$\MTabel\ together with Morpheus's rewrite rule: $C$\MTabel $\rightarrow$ $[C$\textbf{T}$,CK$\textbf{U}$]$ (see \textbf{Q9}). This optimization goes beyond SystemML’s optimization since it does not consider distributing the multiplication of \MTabel, which then enables exploiting the view and distributing the \sumOp\ operation to avoid introducing a large dense intermediate.
\sys's rewrite the $Q_{LA}$: \NTabel$\odot$\sumOp$((X+C)$\MTabel$)$, where $X$ and $C$ are dense matrices of size 1000$\times$1M, to \NTabel$\odot$\sumOp$(X$\MTabel$)+$\sumOp$(V_5)$. The $V_5=[C$\textbf{T}$,CK$\textbf{U}$]$ is utilized by exploiting $(X+C)$\MTabel = $X$\MTabel$+C$\MTabel\ together with Morpheus's rewrite rule: $C$\MTabel $\rightarrow$ $[C$\textbf{T}$,CK$\textbf{U}$]$.  This optimization goes beyond SystemML’s optimization since it does not consider distributing the multiplication of \MTabel, which then enables exploiting the view and distributing the \sumOp\ operation to avoid a dense intermediate.  The obtained rewriting (with the rewriting of the $Q_{RA}$ part of \NTabel) achieves 3.91$\times$ (Figure~\ref{fig:N2M}-Q8).

\vspace{2mm}
\noindent\textbf{Twitter Varying Filter Selectivity}.
We repeat the benchmark for two different text-search selection conditions:
``Trump'' and ``US election'',
obtaining 1M and 0.5M rows for \NTabel,
respectively (we adjust the size of the synthetic matrices for 
dimensional compatibility).
As shown in Figures~\ref{fig:N1M} and ~\ref{fig:N5K}, the benefit
of the combined $Q_{RA}$ and $Q_{LA}$ stage rewriting increases
with data size, remaining significant across the spectrum.

\vspace{2mm}
\noindent\textbf{MIMIC Dataset Preparation}.  MIMIC dataset~\cite{MIMIC} comprises health data for patients. The total size of the dataset is 46.6 GB, and it consists of : (i) all charted data for all patients and their hospital admission information, ICU stays, laboratory measurements, caregivers’ notes, and prescriptions; (ii) the role of caregivers (e.g., MD stands for “medical doctor”), (iii) lab measurements (e.g., ABG stands for “arterial blood gas”) and (iv) diagnosis related groups (DRG) codes descriptions. We use subset of the dataset, which includes \textbf{Patients (P)}, \textbf{Admission (A)}, \textbf{Service (S)}, and \textbf{Callout (C)} tables. We convert tables' categorical features (columns) to  numeric using one-hot encoding.

\vspace{2mm}
\noindent\textbf{MIMIC Queries and Views}. Similar to the Twitter's dataset benchmark, queries consist of two parts: ($i$) preprocessing ($Q_{RA}$) and ($ii$) analysis ($Q_{LA}$). In the $Q_{RA}$ part, the queries construct two main matrices: \MTabel\ and \NTabel. The matrix \MTabel\ (40K$\times$82;\textit{dense}) is the join's output table (matrix) of \textbf{P} and \textbf{A}. The matrix \NTabel\ (40K$\times$30K;\textit{ultra-sparse}) is \textit{patient-service} outcome (e.g., cancelled (1),  serving (2), etc) matrix,  constructed from joining \textbf{C} and \textbf{S} for all patients who are in ``CCU'' care unit.  We fix the $Q_{RA}$ part across all queries and vary the $Q_{LA}$ part using a set of LA pipelines in Table ~\ref{tab:hybrid-la}, a well as \NTabel\  and \MTabel\ matrices.  For views, we define three cross RA-LA materialized views: $V_1$, $V_2$ and $V_3$, which store the result of applying $\rowsumsOp$, $\colsumsOp$ and matrix multiplication operations over \textbf{P} and \textbf{A} base tables (matrices), respectively. These views can only be found by {\em exploiting together LA properties} and {\em Morpheus rewrites' rules} in the same fashion as we detailed in Twitter's benchmark experiment.

\vspace{2mm}
\noindent\textbf{Discussion.}
The results  exhibit similar trends to the Twitter's benchmark. The first run of the benchmark is shown in Figure~\ref{fig:N40K}; both matrices  \MTabel\  and \NTabel\ are loaded in SystemML to be used for the analysis part (varied using the set of pipelines in Table~\ref{tab:hybrid-la}). Before evaluating an LA pipeline, all queries filter \NTabel\ 's rows with \textit{outcome} is equal to 2. \sys\ applies the same set of optimizations as described in Twitter's benchmark.  For the second and third runs of the queries (see Figures ~\ref{fig:N20K} and ~\ref{fig:N10K}), we construct the \NTabel\ matrix for patients who are in ``TSICU'' and ``MICU'' care units, where \NTabel's rows are 20K and 10K, respectively.

 \begin{figure*}[!htbp]
\figspa
  \centering
%      \subfigure[\textbf{P1.1}]{\includegraphics[scale=0.30,width=4.5cm,height=4cm]{charts/FactorizeLA/overhead/P1-1-FC-crop}}
      \subfigure[\textbf{P1.10}]{\includegraphics[scale=0.30,width=4.5cm,height=4cm]{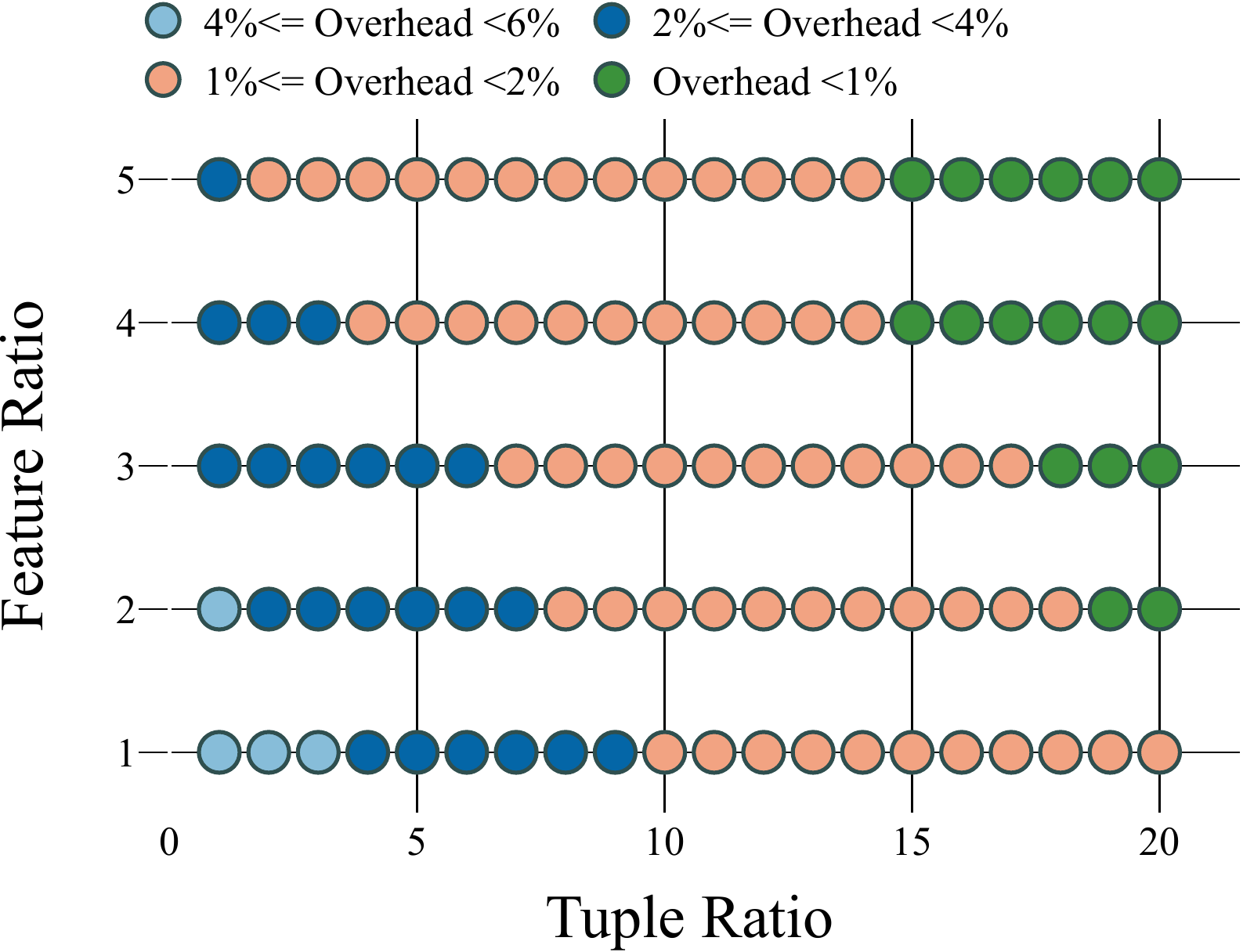}}
%      \subfigure[\textbf{P1.13}]{\includegraphics[scale=0.30,width=4.5cm,height=4cm]{charts/FactorizeLA/Overhead/P1-13-FC-crop}}
      \subfigure[\textbf{P1.16}]{\includegraphics[scale=0.30,width=4.5cm,height=4cm]{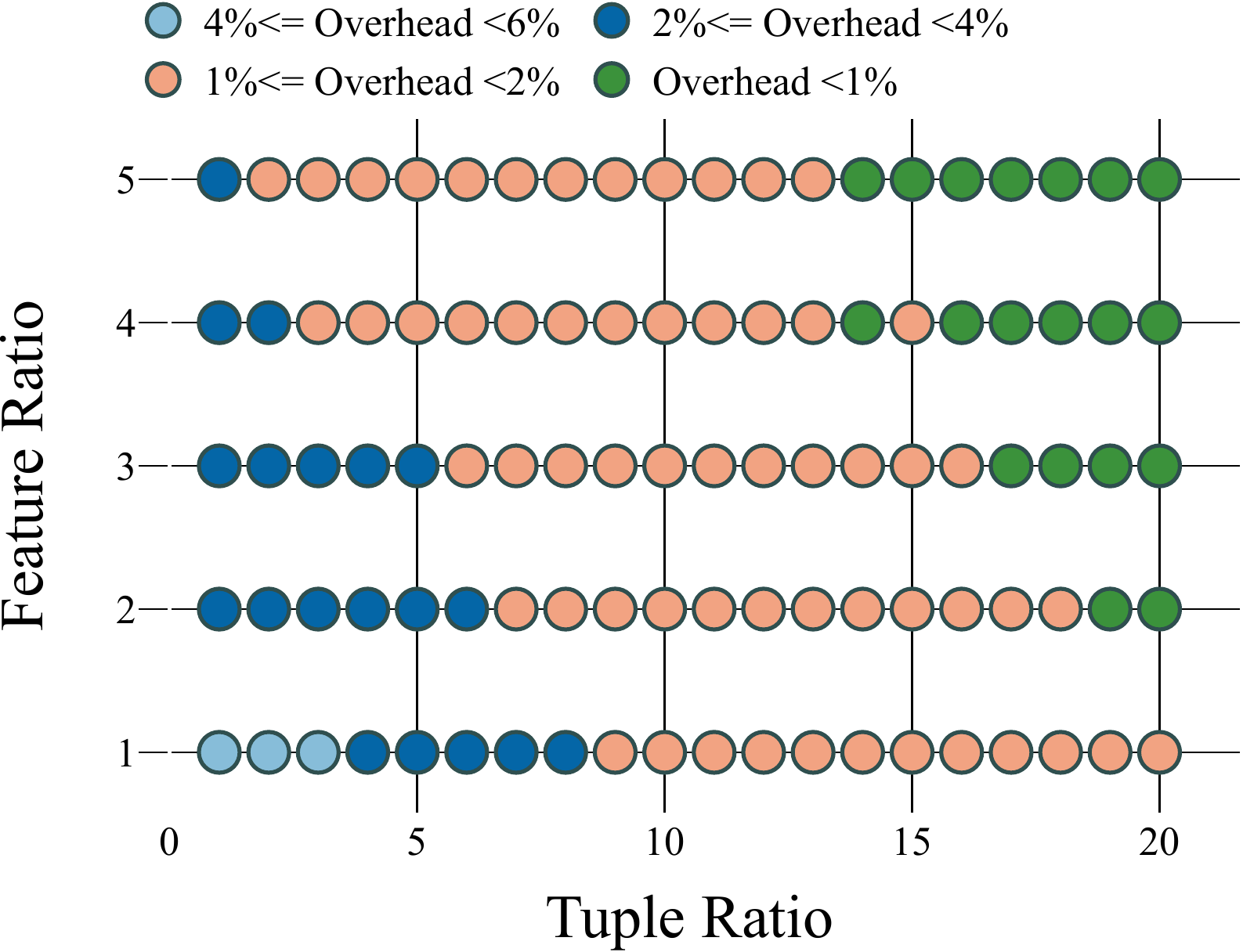}}
        \subfigure[\textbf{P1.18}]{\includegraphics[scale=0.30,width=4.5cm,height=4cm]{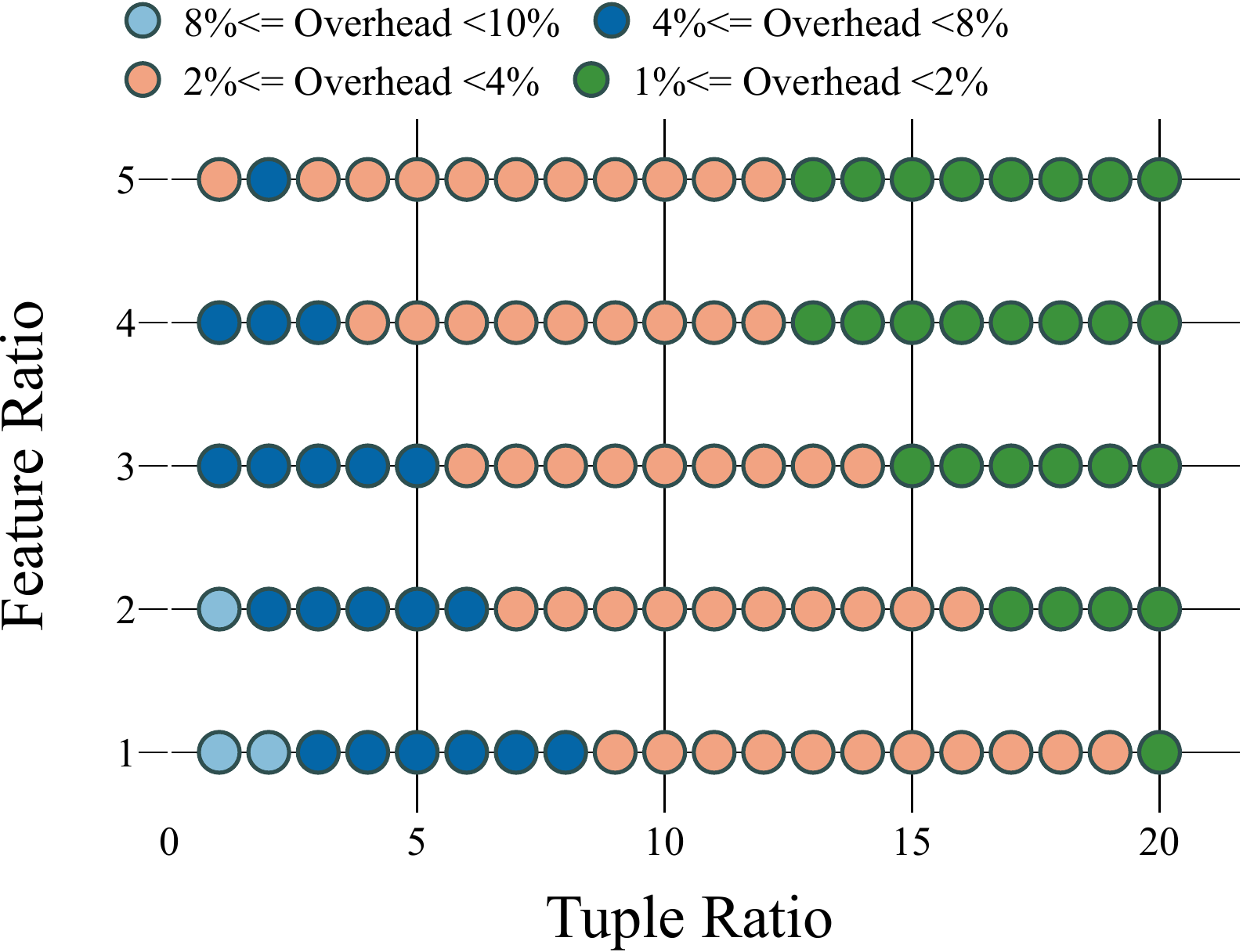}}
 %\figspb%\figspb\figspb
 \caption{HADAD $RW_{find}$ overhead as a percentage (\%) of the total time ($Q_{exec}+ RW_{fond}$) for pipelines P1.1, P1.10, P1.13, P1.16 and P1.18 running on Morpheus}
 %\figspc 
 \label{fig:la-overhead-1}
\end{figure*}

\mysubsubsection{\textbf{Rewriting Time Overhead}} 
\label{sec:rw-overhead-hybrid}
For MorpheusR, using the same experiment setup illustrated in \S\ref{sec:morpheus}, the rewriting time's overhead is very negligible compared to the pipelines' execution time for %a set  of %5
 pipelines that are already optimized or MorpheusR finds the same rewriting  found by \sys . For pipelines that contain matrix multiplication expressions, %such as \textbf{P1.1} and \textbf{P1.13}, 
 the rewriting time is generally less than 0.1\% of the total time. % as shown in Figure~\ref{fig:la-overhead-1}. 
 However, for the other pipelines such as \textbf{P1.10}, \textbf{P1.16}, and \textbf{P1.18}, which contain only aggregate operations, the rewriting time is up to 9\% of the total time (when the data size is very small (0.32GB) and the computation is extremely efficient) and less than 1\% (when the data size is large (19.2GB) and the computation is expensive) as shown in Figure ~\ref{fig:la-overhead-1}. 

%As for the micro-hybrid benchmark, the $RW_{find}$ reaches 78ms and up to 145 ms.

%\begin{figure*}[!htbp]
%\figspa
%  \centering
%      \subfigure[\textbf{P1.16}]{\includegraphics[scale=0.25,width=5.5cm,height=3.5cm]{charts/FactorizeLA/overhead/P1-16-FC-crop}}
%    
% \figspb\figspb\figspb
% \caption{$RW_{find}$ overhead as a percentage (\%) of the total time ($Q_{exec}+ RW_{fond}$) for pipelines 1.16 and 1.18}
% \label{fig:la-overhead-2}
%\end{figure*}
%\vspace{-2mm}
\mysubsection{Experiments Takeaway}
%\rev{We have shown that \sys\ brings significant performance-saving across LA-oriented and cross RA-LA platforms without the need to modify their internals. It improves their performance by order of magnitudes on typical LA-based and hybrid pipelines. Moreover, as we confirm experimentally, the time spent searching for rewritings is a small fraction of the query execution time hence a worthwhile investment. In addition, the rewriting overhead of pipelines that are already in the optimal form is very negligible compared to the original pipeline execution time in the presence of sparse/dense matrices and using na\"ive and MNC-based cost models.}
Our experiments with both real-life and synthetic datasets
show performance gains across the board, for small rewriting overhead,
in both pure LA and hybrid RA-LA settings. This is due to the fact
that \sys's rewriting power strictly subsumes that of
optimizers of reference platforms like R, Numpy, TensorFlow,
Spark (MLlib), SystemML and Morpheus.
Moreover, \sys\ enables optimization where it wasn't previously
feasible, such as across a cascade of unintegrated
tools, e.g. SparkSQL for preprocessing followed by SystemML for analytics.

%% file: related.tex
\mysection{Related Work and Conclusion}
\label{sec:related}
\eat{\RA{I think, we can further shorten the related work}}
\noindent\textbf{LA Systems/Libraries.} SystemML~\cite{boehm2016systemml} offers
% was initially  designed on top of Hadoop and then migrated to Spark. 
%The system offers 
high-level R-like LA language  
%abstractions, using a declarative language called Declarative Machine Learning (DML). 
%The language offers physical data independence, 
%where the users do not need to specify any data layout or formats. 
 %The system can perform local and distributed executions. 
 %The system 
 and applies some logical LA pattern-based rewrites and physical execution optimizations, based on cost estimates for the latter. 
 %The system also supports sparse and dense matrices. 
SparkMLlib~\cite{meng2016mllib} provides LA operations and built-in function implementations of  popular ML algorithms %such as linear regression, etc. 
on Spark RDDs.
 %and it can be executed in local or distributed settings.
% The library  supports  sparse and dense matrices, but the user has to select this type explicitly. 
R~\cite{R} and NumPy~\cite{NumPy} are two of the most popular computing environments for statistical data analysis, widely used in academia and industry. They provide a high-level abstraction that can simplify the programming of numerical and statistical computations, by treating matrices as first-class citizens and by providing a rich set of built-in LA operations. %and ML algorithms. 
However, LA properties in most of these systems remain unexploited, which makes them {\em miss opportunities to use their own highly efficient operators} (recall hybrid-scenario in \S\ref{sec:hadad-opt}).  
  %The pipeline execution in their setting is performed locally. 
Our experiments (\S\ref{sec:exp-la}) show that LA pipelines evaluation in  these systems can be sped up, by more than $10\times$, by our rewriting using ($i$)~LA properties and ($ii$)~materialized views. 
 %to be expressible and exploitable.    

%\RA{Need to do another pass on this. It has some redundancy from the introduction}
	
\vspace{1mm}	
\noindent\textbf{Bridging the Gap: RA and LA.} There has been a recent increase in research for unifying the execution of RA and LA expressions~\cite{luo2018scalable,montdb,kernert2013bringing,chen2017towards}. %As discussed in Section~\ref{sec:Introduction}, the 
A key limitation of these approaches is that {\em the semantics of LA operations remains hidden} behind built-in functions or UDFs, preventing performance-enhancing rewrites as shown in \S\ref{sec:morpheus}.
%,\ie LA routines, which the optimizers can only treat as black-boxes. 
\eat{	
\RA{Redundant}
Some of these works call LA packages through 
%user-defined functions (UDFs),
UDFs, where libraries such as R and NumPy are embedded in the host language~\cite{montdb}. Other works treat LA objects as first-class citizens and use built-in functions to express LA operations~\cite{luo2018scalable,kernert2013bringing,hellerstein2012madlib}.}

SPORES~\cite{spores}, SPOOF~\cite{boehm2018optimizing},  LARA~\cite{kunft2019intermediate} and RAVEN ~\cite{aranasosIPSPPX20} are closer to our work.
SPORES and SPOOF optimize LA expressions, by converting
them into RA, optimizing the latter, and then converting the result back to an (optimized) LA expression. 
%SPOOF operates on a restricted relational algebra, where 
%where every expression must have at most two free attributes. This to ensure that 
%every relational expression in every step of the optimization process can be expressed in LA. More general, SPORES only requires the optimized output to be expressible in LA. However, 
%\IM{Check this shorter phrase:} %\RA{done} 
%SPORES and SPOOF 
They are restricted to a small set of selected LA operations (the ones that can be expressed in RA), while we support significantly more (\S\ref{sec:matrix-al}), and model properties allowing to optimize with them. LARA relies on a declarative domain-specific language for collections and matrices, which can enable optimization  (\ie {\em selection pushdown}) across the two algebraic abstractions. It heavily focuses on low-level optimization such as exploiting the choice of data layouts for physical LA operators' optimization. 
 %\RA{RAVEN doesn't mention anything in their vision paper about exploiting LA properties, views and constraints} 
RAVEN takes a step forward by providing intermediate representation to % enable cross-optimization between ML and database operators and 
enhance in-database model inferencing performance. It  transforms  %many 
classical ML models %(e.g., decision tree)  
into equivalent neural networks %(NN) 
to leverage highly optimized ML engines on CPU/GPU.%However, both systems do not address semantic optimization under integrity constraints including RA/LA views-based and LA pure rewrites. %remain unexploited.
~\cite{BarceloH0S20, BrijderGBW19} study the expressive power of corss RA-LA \cite{BarceloH0S20} / LA \cite{BrijderGBW19} query languages. %but do not address semantic optimizations / rewriting under integrity constraints and/or views. All aforementioned works %do not reason with constraints, they cannot 
}
In contrast to \sys, as all aforementioned solutions do not reason with constraints, they provide no capabilities for {\em holistic} semantic query optimizations including RA/LA views-based and LA pure rewritings%lack exploiting  materialized views in an LA (or hybrid LA/RA) context; as shown in (\S\ref{sec:exp}),
; such optimizations can bring large performance saving as shown in \S\ref{sec:exp}.

We see \sys\ as complementary to all of these platforms, %providing an algorithm for RA/LA views- and constraint-based rewriting.  Our solution 
on top of which it can be naturally and portably applied.

\vspace{1mm}						
\noindent\textbf{Conclusion.}
\sys\ is an extensible lightweight framework for optimizing hybrid  analytics queries, based on the powerful intermediate abstraction of a \emph{a relational model with integrity constraints}. HADAD extends the capability of ~\cite{estocada-sigmod} with a reduction from LA (or LA view)-based rewriting to relational rewriting under constraints. It enables a full exploration of rewrites using a large set of LA operations, with no modification to the execution platform. Our experiments show significant performance gains on various LA and hybrid workloads across popular LA and cross RA-LA platforms.  
%Future work includes reasoning about  cell-wise operations, building upon the  FAQ~\cite{abo2016faq} framework.

 %it is not clear how both systems can fully exploit LA properties since they have show-cased their approach on LA operators that can only be expressed using RA. Additionally, both systems can lack exploiting materialized view which can drastically enhance the performance of LA-based pipelines as shown our experiments. In a complimentary fashion, we believe our approach can be adopted by SPORES and SPOOF.   s
%on a very limited number of LA operations (i.e., seven operations for SPORES). 
%In addition, it is not clear how their approach can be extended to support views-based rewriting for LA expressions. 
\vspace{-2mm}

%% file: appendixA.tex
\onecolumn
\
\section{$L_{ops}$ Operations Properties ($LA_{prop}$) Captured as Integrity Constraints}
\label{appendixA}
\begin{table*}[ht]	
\caption{$L_{ops}$ Operations Properties ($LA_{prop}$) Captured as Integrity Constraints}
\centering
\begin{tabular}{ |c|c| } 
\hline
\textbf{LA Property}  & \textbf{Relational Encoding as Integrity Constraints }\\
\hline
\multicolumn{2}{|c|}{\textbf{Addition of Matrices}}\\\hline
\hline
$M+N=N+M$ & $\forall \calMI,\calMII, \calO$ \addMOp$(\calMI, \calMII,\calO) \rightarrow$ \addMOp$(\calMII,\calMI,\calO)$\\\hline
$(M+N)+D =M+(N+D)$ & $\forall \calMI,\calMII,\calD, \calO_{1}, \calO_{2}$ ~ \addMOp$(\calMI,\calMII,\calO_{1}) \wedge$ \addMOp$(\calO_{1},\calD,\calO_{2}) \rightarrow$ \\ 
													   & $\exists \calO_3$ \addMOp$(\calMII,\calD,\calO_3) \wedge$ \addMOp$(\calMI,\calO_3,\calO_{2})$\\
\hline
$c(M+N)=cM+cN$ & $\forall c,\calMI,\calMII, \calO_{1}, \calO_{2}$ ~ \addMOp$(\calMI,\calMII,\calO_{1}) \wedge$ \multiMSOp$(c,\calO_{1},\calO_{2}) \rightarrow$\\
									 & $\exists \calO_{3}, \calO_{4} \multiMSOp(c,\calMI,\calO_{3}) \wedge \multiMSOp(c,\calMII,\calO_{4}) \wedge \addMOp(\calO_{3},\calO_{4},\calO_{2})$\\
\hline
$(c+d)M=cM+dM$ & $\forall c,d,s,\calMI,\calO_{1} add_S(c,d,s) \wedge \multiMSOp(s,\calMI,\calO_{1}) \rightarrow$ \\
							  & $\exists \calO_{2}, \calO_{3} \multiMSOp(c,\calMI,\calO_{2}) \wedge  \multiMSOp(d,\calMI,\calO_{3}) \wedge \addMOp(\calO_{2},\calO_{3},\calO_{1})$\\
\hline 
$M+0=M$ & $\rightarrow \exists Zero(\calZero)$ \\  
		&  $ \forall \calMI, n, \calZero$ ~ $name(\calMI,n) \wedge Zero(\calZero) \rightarrow \addMOp(\calMI,\calZero,\calMI)$\\
		& $ \forall \calZero$ ~ $Zero(\calZero) \rightarrow \addMOp(\calZero,\calZero,\calZero)$	\\													 
\hline
\hline	
\multicolumn{2}{|c|}{\textbf{Product of Matrices}}\\\hline
\hline
$(MN)D  = M(ND)$ & $\forall \calMI,\calMII,\calD , \calO_{1}, \calO_{2}$ ~$\multiMOp(\calMI,\calMII,\calO_{1}) \wedge \multiMOp(\calO_{1},\calD ,\calO_{2}) \rightarrow$ \\ 
													   & $\exists \calO_{3} \multiMOp(\calMII,\calD ,\calO_{3}) \wedge \multiMOp(\calMI,\calO_{3},\calO_{2})$\\
\hline	
$M(N+D) = MN+ MD $ & $\forall  \calMI,\calMII,\calD , \calO_{1}, \calO_{2}$ ~ $\addMOp(\calMII,\calD ,\calO_{1}) \wedge \multiMOp(\calMI,\calO_{1}, \calO_{2}) \rightarrow$ \\ 
															 & $\exists \calO_{3}, \calO_{4}$~$ \multiMOp(\calMI,\calMII,\calO_{3}) \wedge \multiMOp(\calMI,\calD ,\calO_{4}) \wedge \addMOp(\calO_{3},\calO_{4},\calO_{2})$\\
\hline	
$(M +N)D  = MD  + MD $ & $\forall  \calMI,\calMII,\calD , \calO_{1}, \calO_{2} \addMOp(\calMI,\calMII,\calO_{1}) \wedge \multiMOp(\calO_{1},\calD , \calO_{2}) \rightarrow$ \\ 
															 & $\exists \calO_{3}, \calO_{4} \multiMOp(\calMI,\calD ,\calO_{3}) \wedge \multiMOp(\calMII,\calD ,\calO_{4}) \wedge \addMOp(\calO_{3},\calO_{4},\calO_{2})$\\
\hline
$d(MN) = (dM)N$&  $\forall  d,\calMI,\calMII,\calO_{1},\calO_{2}$ ~ $\multiMOp(\calMI,\calMII,\calO_{1})\wedge \multiMSOp(d,\calO_{1},\calO_{2}) \rightarrow$\\
									  & $\exists \calO_{3} \multiMSOp(d,\calMI,\calO_{3}) \wedge  \multiMOp(\calO_{3},\calMII,\calO_{2})$\\
\hline	
$c(dM)=(cd)M$ & $\forall c,d \calMI,\calO_{1},\calO_{2} \multiMSOp(d,\calMI,\calO_{1}) \wedge \multiMSOp(c,\calO_{1},\calO_{2}) \rightarrow $\\
						& $\exists s $ $multi_S(c,d,s) \wedge \multiMSOp(s,\calMI,\calO_{2})$\\	
\hline 
$I_kM=M=MI_z$ & $\forall \calMI,n,k,z$ $name(\calMI,n),size(\calMI,k,z)\rightarrow  \exists  \calIden $ $ Identity(\calIden),size(\calIden,k,k)$ \\ 		
	& $\forall \calMI,n,k,z$ $name(\calMI,n),size(\calMI,k,z)\rightarrow  \exists  \calIden_1 $ $ Identity(\calIden_1),size(\calIden,z,z)$\\
	& $\forall \calMI,\calIden_1, n, k,z $ $name(\calMI,n),size(\calMI,k,z),Identity(\calIden),size(\calIden_1,k,k)\rightarrow \multiMOp(\calIden, \calMI, \calMI)$\\
	& $\forall \calMI,\calIden_1, n, k,z $ $name(\calMI,n),size(\calMI,k,z),Identity(\calIden),size(\calIden_1,z,z)\rightarrow \multiMOp(\calMI,\calIden,\calMI)$\\
\hline
\hline
\multicolumn{2}{|c|}{\textbf{Transposition of Matrices}}\\\hline		
\hline
$(MN)^T = N^TM^T$ & $\forall \calMI,\calMII,\calO_{1},\calO_{2} \multiMOp(\calMI,\calMII,\calO_{1})\wedge \trOp(\calO_{1},\calO_{2}) \rightarrow $ \\
										& $\exists \calO_{3}, \calO_{4} \trOp(\calMI,\calO_{3}) \wedge \trOp(\calMII,\calO_{4}) \wedge \multiMOp(\calO_{4},\calO_{3},\calO_{2})$\\
\hline
$(M+N)^T= M^T + N^T$ & $\forall \calMI,\calMII, \calO_{1}, \calO_{2} ~ \addMOp(\calMI,\calMII,\calO_{1}) \wedge \trOp(\calO_{1},\calO_{2}) \rightarrow $ \\ 
										   & $\exists \calO_{3}, \calO_{4} \trOp(\calMI,\calO_{3}) \wedge \trOp(\calMII,\calO_{4}) \wedge \addMOp(\calO_{3},\calO_{4},\calO_{2})$\\
\hline
$(cM)^T = c(M)^T$	&  $\forall c,\calMI,\calO_{1},\calO_{2} ~ \multiMSOp(c,\calMI,\calO_{1}) \wedge \trOp(\calO_{1},\calO_{2}) \rightarrow$\\
							& $\exists \calO_{3} \trOp(\calMI,\calO_{3}) \wedge \multiMSOp(c,\calO_{3},\calO_{2})$\\
\hline 
$((M)^T)^T = M$	& $\forall n,\calMI ~ name(\calMI,n) \rightarrow \exists \calO_{1} \trOp(\calMI,\calO_{1}) \wedge \trOp(\calO_{1},\calMI)$\\
\hline
$(I)^{T}=I$, where $I$ is identity matrix  &  $\forall \calIden   ~~Identity(\calIden) \rightarrow \trOp(\calIden,\calIden)$\\
$(O)^{T}=O$,  where $O$ is zero matrix &  $\forall O ~~Zero(O) \rightarrow \trOp(O,O)$\\
\hline
\hline	
\multicolumn{2}{|c|}{\textbf{Inverses of Matrices}}\\\hline	
\hline 
$((M)^{-1})^{-1} = M$	& $\forall n,\calMI ~name(\calMI,n) \rightarrow \exists \calO_{1} \invMOp(\calMI,\calO_{1}) \wedge \invMOp(\calO_{1},\calMI)$\\
\hline
$(MN)^{-1} = N^{-1}M^{-1}$ & $\forall \calMI,\calMII,\calO_{1},\calO_{2} ~~\multiMOp(\calMI,\calMII,\calO_{1})\wedge \invMOp(\calO_{1},\calO_{2}) \rightarrow $ \\
										& $\exists \calO_{3}, \calO_{4} \invMOp(\calMI,\calO_{3}) \wedge \invMOp(\calMII,\calO_{4}) \wedge \multiMOp(\calO_{4},\calO_{3},\calO_{2})$\\
\hline
$((M)^T)^{-1} = ((M)^{-1})^{T}$ & $\forall \calMI, \calO_{1}, \calO_{2} \trOp(\calMI,\calO_{1}) \wedge \invMOp(\calO_{1},\calO_{2}) \rightarrow$ \\
										  & $\exists \calO_{3} \invMOp(\calMI,\calO_{3}) \wedge \trOp(\calO_{3},\calO_{2})$\\
\hline
$((kM))^{-1}= k^{-1}M^{-1}$ & $\forall k,\calMI, \calO_{1}, \calO_{2} ~~\multiMSOp(k,\calMI,\calO_{1}) \wedge \invMOp(\calO_{1},\calO_{2} )\rightarrow$  \\
									 & $\exists \calO_{3},s $ $inv_S(k,s) \wedge \invMOp(\calMI,\calO_{3}) \wedge \multiMSOp (s,\calO_{3},\calO_{2})$\\
\hline
$M^{-1}M=I=MM^{-1}$ & $\forall \calMI,\calO_{1},\calO_{2}$ $\invMOp(\calMI, \calO_{1})  \wedge \multiMOp(\calO_{1},\calMI,\calO_{2}) \rightarrow Identity(\calO_{2})$\\
	& $\forall \calMI,\calO_{1},\calO_{2}$ $\invMOp(\calMI, \calO_{1})  \wedge \multiMOp(\calMI,\calO_{1},\calO_{2}) \rightarrow Identity(\calO_{2})$\\
\hline
\end{tabular}
\end{table*}

\begin{table*}[ht]	
\caption{$L_{ops}$ Operations Properties ($LA_{prop}$) Captured as Integrity Constraints}
\centering
\begin{tabular}{ |c|c| } 
\hline
\textbf{LA Property}  & \textbf{Relational Encoding as Integrity Constraints }\\
\hline
\multicolumn{2}{|c|}{\textbf{Determinant of Matrices}}\\\hline	
\hline
$\detOp(MN)=\detOp(M)*\detOp(N)$ & $\forall \calMI,\calMII, \calO_{1}, d $ $\multiMOp(\calMI,\calMII,\calO_{1}) \wedge \detOp(\calO_{1},d) \rightarrow$   \\
											& $\exists d_1,d_2 \detOp(\calMI,d_1) \wedge \detOp(\calMII,d_2) \wedge multi_S(d_1,d_2,d)$\\
\hline
$\detOp((M)^{T})=\detOp(M)$ & $\forall \calMI, \calO_{1}, d $ $\trOp(\calMI,\calO_{1}) \wedge \detOp(\calO_{1},d) \rightarrow \detOp(\calMI,d)$   \\
\hline
$\detOp((M)^{-1})=(\detOp(M))^{-1}$ & $\forall \calMI, \calO_{1}, d $ $\invMOp(\calMI,\calO_{1}) \wedge \detOp(\calO_{1},d) \rightarrow \exists d_1 \detOp(\calMI,d_1) \wedge inv_S(d_1,d)$   \\
\hline
$\detOp((cM))=c^k\detOp(M)$ & $\forall \calMI, c, k, d $ $size(\calMI,k,k) \wedge \multiMSOp(c,\calMI,d)\rightarrow $ $ \exists s_1, s_2 $\\
	& $ pow(c,k,s_1) \wedge \detOp (\calMI, s_2) \wedge multi_{S}(s_1,s_2,d)$\\
\hline
$\detOp((I))=1$ & $\forall \calIden_1, d Identity(\calIden_1)  \wedge \detOp(\calIden_1,d) \rightarrow  d=1$\\
\hline
\hline
\multicolumn{2}{|c|}{\textbf{ Adjoint of Matrices}}\\\hline
\hline
$\adjOp(M)=R^T$ & $\forall \calMI, \calO_{1} $ $ \adjOp(\calMI,\calO_{1})\rightarrow  \exists \calO_{2} $ $cof(\calMI,\calO_{2}) \wedge \trOp(\calO_{2},\calO_{1})$\\
\hline
$\adjOp(M)^T=\adjOp(M^T)$ & $\forall \calMI, \calO_{1},\calO_{2}$ $\adjOp(\calMI,\calO_{1})\wedge \trOp(\calO_{1},\calO_{2}) \rightarrow$ 
	 
	 					 $\exists \calO_{3}$ $\trOp(\calMI,\calO_{3})  \wedge \adjOp(\calO_{3}, \calO_{2})$\\
\hline
	$\adjOp(M)^{-1}=\adjOp(M^{-1})$ & $\forall \calMI, \calO_{1},\calO_{2}$ $\adjOp(\calMI,\calO_{1})\wedge \invMOp(\calO_{1},\calO_{2}) \rightarrow$ 
	 
	 					 $\exists \calO_{3}$ $\invMOp(\calMI,\calO_{3})\wedge \adjOp(\calO_{3},\calO_{2})$\\	
\hline	
	 $\adjOp(MN)=\adjOp(N)\adjOp(M)$ & $\forall \calMI, \calO_{1},\calO_{2}$ $\multiMOp(\calMI,\calMII,\calO_{1}) \wedge \adjOp(\calO_{1},\calO_{2}) \rightarrow$ 
	 
	 						$\exists \calO_{3}$ $\calO_{4}$\\ &$\multiMOp(\calO_{3},\calO_{4},\calO_{2}) \wedge \adjOp(\calMII,\calO_{3})\wedge \adjOp(\calMI,\calO_{4})$\\

\hline
\hline
\multicolumn{2}{|c|}{\textbf{Trace of Matrices}}\\\hline
\hline
$\traceOp(M+N)=\traceOp(M)+\traceOp(N)$ & $\forall \calMI, \calMII,\calO_{1}, s_1 $ $ \addMOp(\calMI,\calMII,\calO_{1}) \wedge $\\
	& $\traceOp(\calO_{1}, s_1) \rightarrow $ $\exists s_2,s_3 $ $\traceOp(\calMI,s_2)\wedge \traceOp(\calMII,s_3) \wedge add_s(s_2,s_3,s_1)$\\
\hline
$\traceOp(MN)=\traceOp(NM)$ & $\forall \calMI, \calMII,\calO_{1}, s_1 $ $ \multiMOp(\calMI,\calMII,\calO_{1}) \wedge $\\
	& $\traceOp(\calO_{1}, s_1) \rightarrow $ $\exists \calO_{2} $ $\multiMOp(\calMII,\calMI,\calO_{2})\wedge \traceOp(\calO_{2},s_1)$\\
\hline
$\traceOp(M^T)=\traceOp(M)$ & $\forall \calMI, \calO_{1}, s_1$ ~ $\trOp(\calMI,\calO_{1}) \wedge \traceOp(\calO_{1},s_1)$ $\rightarrow \traceOp( \calMI,s_1)$
\\
\hline
$\traceOp(cM)=c\traceOp(M)$ & $\forall \calMI, \calO_{1}, c, s_1$ ~ $\multiMSOp (c,\calMI, \calO_{1}) \wedge \traceOp(\calO_{1},s_1)$ $\rightarrow \exists s_2 \traceOp(\calMI,s_2) \wedge multi_s(c,s_2,s_1)$\\
\hline
$\traceOp(I_{k})=k$& $\forall \calIden,k,s_1 $ ~ $Identity(\calIden) \wedge  size(\calIden,k,k) \wedge \traceOp(\calIden,s_1)$ 

$\rightarrow s_1 =k $\\
\hline
\hline
\multicolumn{2}{|c|}{\textbf{Direct Sum}}\\\hline
\hline
$(M \oplus N)+(C \oplus D)= (M+C) \oplus (N+D)$ &  
$\forall \calMI,\calMII,\calO_{1},\calC,\calD,\calO_{2}, \calO_{3} $ $ \sumDOp(\calMI,\calMII,\calO_{1})$ \\& $\wedge \sumDOp(\calC,\calD,\calO_{2})\wedge \addMOp(\calO_{1},\calO_{2},\calO_{3})\rightarrow \exists \calO_{4},\calO_{5} \addMOp(\calMI,\calC,\calO_{4}) \wedge \addMOp(\calMII,\calD,\calO_{5})$\\& 
	$\wedge \sumDOp(\calO_{5},\calO_{3)}$
\\
\hline
$(M \oplus N)(C \oplus D)= (MC) \oplus (ND)$ & 
$\forall \calMI,\calMII,\calO_{1},\calC,\calD,\calO_{2}, \calO_{3} $ $ \sumDOp(\calMI,\calMII,\calO_{1})$ \\ 
& $\wedge \sumDOp(\calC,\calD,\calO_{2})\wedge \multiMOp(\calO_{1},\calO_{2},\calO_{3})\rightarrow \exists \calO_{4},\calO_{5} \multiMOp(\calMI,\calC,\calO_{4}) \wedge \multiMOp(\calMII,\calD,\calO_{5})$\\&
	$\wedge \sumDOp(\calO_{5},\calO_{3)}$

\\	
\hline
%$c(M \oplus N)= (cM \oplus cN)$ &  
%$\forall \calMI,\calMII,\calO_{1},c,\calO_{2} $ $ \sumDOp(\calMI,\calMII,\calO_{1})$ \\ 
%& $\wedge \multiMSOp(c,\calO_{1},\calO_{2}) \rightarrow \exists \calO_{3},\calO_{4} \multiMSOp(\calMI,c,\calO_{3}) \wedge
% \multiMSOp(\calMII,c,\calO_{4}) \wedge \sumDOp(\calO_{3},\calO_{4},\calO_{2}) $\\
\hline 
\multicolumn{2}{|c|}{\textbf{ Exponential of Matrices}}\\\hline
\hline
$\expOp(0)=I$ & $\forall \calZero, \calO_{1}$ ~$Zero(\calZero) \wedge \expOp(\calZero, \calO_{1}) \rightarrow Identity (\calO_{1})$\\
\hline
$\expOp(M^{T})=\expOp(M)^{T}$ & $\forall \calMI, \calO_{1},  \calO_{2}  \trOp(\calMI, \calO_{1}) \wedge  \expOp(\calO_{1}, \calO_{2}) \rightarrow \exists \calO_{3} \expOp(\calMI, \calO_{3}) \wedge \expOp(\calO_{3}, \calO_{2})$\\
\hline
\end{tabular}
\end{table*}

\begin{table*}[ht]	
\caption{Matrix Decompositions Properties Captured as Integrity Constraints}
\centering
\begin{tabular}{ |c|c| } 
\hline
\textbf{Decomposition Property}  & \textbf{Relational Encoding as Integrity Constraints }\\
\hline
\hline
\multicolumn{2}{|c|}{\textbf{Cholesky Decomposition (CD)}}\\\hline
$\CHOp(M)=L$ such that $M=LL^{T}$, where M is symmetric positive definite & 
 $\forall \calMI ~ type(\calMI,``S") \rightarrow \exists$ ~ $\calLL_1 \exists \calLL_2~ \CHOp(\calMI,\calLL_{1}) \wedge type(\calLL_1,``L") \wedge $ \\
 &$\trOp(\calLL_1,\calLL_2) \wedge \multiMOp(\calLL_1,\calLL_2,\calMI)$\\
\hline
\multicolumn{2}{|c|}{\textbf{QR Decomposition}}\\\hline
\hline
$QR(M)=[Q,R]$ such that $M=QR$ & $\forall \calMI \forall n \forall k ~name(\calMI ,n) \wedge size(\calMI, k,k) \rightarrow ~ \exists \calQR, \calR$ \\

	& $QR(\calMI,\calQR, \calR) \wedge  type(\calQR, ``O")  \wedge   type( \calR, ``U") \wedge$ \\
	& $ \multiMOp(\calQR, \calR,\calMI)$\\
	& $\forall \calQR ~ type(\calQR, ``O") \rightarrow \exists \calIden ~  QR(\calQR,\calQR,\calIden) \wedge identity(\calIden)$\\
	& $  \wedge \multiMOp(\calQR,\calIden,\calQR)$ \\
	& $\forall \calR  ~ type(\calR, ``U") \rightarrow \exists \calIden ~  QR(\calR,\calIden,\calR) \wedge identity(\calIden)$\\
	& $ \wedge \multiMOp(\calIden,\calR,\calR)$ \\
	& $\forall \calIden ~ identity(\calIden) \rightarrow QR(\calIden,\calIden,\calIden)$\\
\hline
\multicolumn{2}{|c|}{\textbf{LU Decomposition}}\\\hline
\hline
$\LUOp(M)=[L,U]$ such that $M=LU$ & $\forall \calMI \forall n \forall k ~name(\calMI ,n) \wedge size(\calMI, k,k) \rightarrow ~ \exists \calLL, \calU$ \\

	& $\LUOp(\calMI,\calLL,\calU) \wedge  type(\calLL, ``L") \wedge type(\calU, ``U") \wedge $\\
	
	& $\multiMOp(\calLL,\calU,\calMI)$\\
	& $\forall \calLL ~ type( \calLL, ``L") \rightarrow \exists \calIden ~  \LUOp( \calLL, \calLL,\calIden) \wedge identity(\calIden)$\\
	& $\wedge \multiMOp( \calLL,\calIden, \calLL) $ \\
	& $\forall \calU  ~ type(\calU, ``U")\rightarrow \exists \calIden ~  \LUOp(\calU,\calIden,\calU) \wedge identity(\calIden)$\\
	& $\wedge \multiMOp(\calIden,\calU,\calU)$ \\
	& $\forall \calIden ~ identity(\calIden) \rightarrow \LUOp(\calIden,\calIden,\calIden)$\\
\hline
\multicolumn{2}{|c|}{\textbf{Pivoted LU Decomposition}}\\\hline
\hline
$\LUPOp(M)=[L,U,P]$ such that $PM=LU$, where $M$ is a square matrix & $\forall \calMI \forall n \forall k ~name(\calMI ,n) \wedge size(\calMI, k,z) \rightarrow ~ \exists \calLL, \calU, P,\calO $  \\

	& $\LUPOp(\calMI,\calLL,\calU, P) \wedge  type(\calLL, ``L")  \wedge   type(\calU, ``U")  \wedge type( P, ``P")  $\\
	& $\wedge  \multiMOp(\calLL,\calU,\calO) \wedge \multiMOp(P,\calMI,\calO)$\\
		& $\forall \calLL ~ type( \calLL, ``L") \wedge size( \calLL, k,z) \rightarrow \exists \calIden ~  \LUPOp( \calLL, \calLL,\calIden,\calIden) \wedge $\\
		& $identity(\calIden) $
		$\wedge \multiMOp( \calLL,\calIden, \calLL) \wedge  \multiMOp(\calIden,\calLL,\calLL)$ \\
			& $\forall \calU ~ type( \calU, ``U") \rightarrow \exists \calIden ~  \LUPOp( \calU, \calIden,\calU, \calIden) \wedge$ \\
	& $ identity(\calIden) \wedge \multiMOp( \calIden,\calU, \calU)$\\
	& $\forall \calIden ~ identity(\calIden) \rightarrow \LUOp(\calIden,\calIden,\calIden)$\\

\hline
\end{tabular}
\end{table*}

%% file: appendixB.tex
\onecolumn
\section{SystemML Rewrite Rules Encoded As Integrity Constraints}
\label{appendixB}
\begin{table*}[ht]	
\caption{SystemML Algebraic Aggregate Rewrite Rules Captured as Integrity Constraints}
\centering	
\begin{tabular}{ |c|c| } 
\hline
\textbf{SystemML Algebraic Simplification Rule}  & \textbf{Integrity Constraints $\calMMC_{StatAgg}$}\\
\hline
\multicolumn{2}{|c|}{\textbf{UnnecessaryAggregates}}\\\hline
\hline
\sumOp(t(M))-> \sumOp(M) & $\forall \calMI,\calR_{1}, s$ $\trOp(\calMI,\calR_{1}),\sumOp(\calR_{1},s) \rightarrow \sumOp(\calMI,s) $\\
\hline
\sumOp(rev(M))-> \sumOp(M) & $\forall \calMI,\calR_{1}, s$ $rev(\calMI,\calR_{1}),\sumOp(\calR_{1},s) \rightarrow \sumOp(\calMI,s) $\\
\hline
\sumOp(\rowsumsOp(M))-> \sumOp(M) & $\forall \calMI,\calR_{1}, s$ $\rowsumsOp(\calMI,\calR_{1}),\sumOp(\calR_{1},s) \rightarrow \sumOp(\calMI,s) $\\
\hline
\sumOp(\colsumsOp(M))-> \sumOp(M) & $\forall \calMI,\calR_{1}, s$ $\colsumsOp(\calMI,\calR_{1}),\sumOp(\calR_{1},s) \rightarrow \sumOp(\calMI,s) $\\
\hline
\minOp(\rowminOp(M))-> \minOp(M) & $\forall \calMI,\calR_{1}, s$ $\rowminOp(\calMI,\calR_{1}),\minOp(\calR_{1},s) \rightarrow \minOp(\calMI,s) $\\
\hline
\minOp(\colminOp(M))-> \minOp(M) &  $\forall \calMI,\calR_{1}, s$ $\colminOp(\calMI,\calR_{1}),\minOp(\calR_{1},s) \rightarrow \minOp(\calMI,s) $\\
\hline
\maxOp(col\maxOp(M))-> \maxOp(M) &$\forall \calMI,\calR_{1}, s$ $\colmaxOp(\calMI,\calR_{1}),\maxOp(\calR_{1},s) \rightarrow \maxOp(\calMI,s) $\\
\hline
\maxOp(row\maxOp(M))-> \maxOp(M) &$\forall \calMI,\calR_{1}, s$ $\rowmaxOp(\calMI,\calR_{1}),\maxOp(\calR_{1},s) \rightarrow \maxOp(\calMI,s) $\\
\hline
\multicolumn{2}{|c|}{\textbf{pushdownUnaryAggTransposeOp}}\\\hline
\hline
\rowsumsOp(t(M))->t(\colsumsOp(M)) & $\forall \calMI,\calR_{1},\calR_{2}$ $\trOp(\calMI,\calR_{1})\wedge \rowsumsOp(\calR_{1},\calR_{2}) \rightarrow \exists \calR_{3}$ $ \colsumsOp(\calMI,\calR_{3})\wedge \trOp(\calR_{3},\calR_{2})$\\
\hline
\colsumsOp(t(M))->t(\rowsumsOp(M)) & $\forall \calMI,\calR_{1},\calR_{2}$ $\trOp(\calMI,\calR_{1})\wedge  \colsumsOp(\calR_{1},\calR_{2}) \rightarrow \exists \calR_{3}$ $ \rowsumsOp(\calMI,\calR_{3}) \wedge \trOp(\calR_{3},\calR_{2})$\\
\hline
\rowmeanOp(t(M))->t(\colmeanOp(M)) & $\forall \calMI,\calR_{1},\calR_{2}$ $\trOp(\calMI,\calR_{1})\wedge \colmeanOp(\calR_{1},\calR_{2}) \rightarrow \exists \calR_{3}$ $ \rowmeanOp(\calMI,\calR_{3})\wedge \trOp(\calR_{3},\calR_{2})$\\
\hline
\colmeanOp(t(M))->t(\rowmeanOp(M)) & $\forall \calMI,\calR_{1},\calR_{2}$ $\trOp(\calMI,\calR_{1})\wedge \rowmeanOp(\calR_{1},\calR_{2}) \rightarrow \exists \calR_{3}$ $ \colmeanOp(\calMI,\calR_{3} \wedge \trOp(\calR_{3},\calR_{2})$\\
\hline
\rowvarOp(t(M))->t(\colvarOp(M)) &  $\forall \calMI,\calR_{1},\calR_{2}$ $\trOp(\calMI,\calR_{1})\wedge \rowvarOp(\calR_{1},\calR_{2}) \rightarrow \exists \calR_{3}$ $ \colvarOp(\calMI,\calR_{3})\wedge \trOp(\calR_{3},\calR_{2})$\\
\hline
\colvarOp(t(X))->t(\rowvarOp(X)) & $\forall \calMI,\calR_{1},\calR_{2}$ $\trOp(\calMI,\calR_{1})\wedge \colvarOp(\calR_{1},\calR_{2}) \rightarrow \exists \calR_{3}$ $ \rowvarOp(\calMI,\calR_{3})\wedge \trOp(\calR_{3},\calR_{2})$\\
\hline
\rowmaxOp(t(M))->t(\colmaxOp(M)) & $\forall \calMI,\calR_{1},\calR_{2}$ $\trOp(\calMI,\calR_{1})\wedge \rowmaxOp(\calR_{1},\calR_{2}) \rightarrow \exists \calR_{3}$ $ \colmaxOp(\calMI,\calR_{3})\wedge \trOp(\calR_{3},\calR_{2})$\\
\hline
\colmaxOp(t(M))->t(\rowmaxOp(M)) & $\forall \calMI,\calR_{1},\calR_{2}$ $\trOp(\calMI,\calR_{1})\wedge \colmaxOp(\calR_{1},\calR_{2}) \rightarrow \exists \calR_{3}$ $ \rowmaxOp(\calMI,\calR_{3})\wedge \trOp(\calR_{3},\calR_{2})$\\
\hline
\rowminOp(t(M))->t(\colminOp(M)) & $\forall \calMI,\calR_{1},\calR_{2}$ $\trOp(\calMI,\calR_{1}) \wedge \rowminOp(\calR_{1},\calR_{2}) \rightarrow \exists \calR_{3}$ $ \colminOp(\calMI,\calR_{3}) \wedge \trOp(\calR_{3},\calR_{2})$\\
\hline
\colminOp(t(M))->t(\rowminOp(M)) & $\forall \calMI,\calR_{1},\calR_{2}$ $\trOp(\calMI,\calR_{1})\wedge  \colminOp(\calR_{1},\calR_{2}) \rightarrow \exists \calR_{3}$ $ \rowminOp(\calMI,\calR_{3})\wedge \trOp(\calR_{3},\calR_{2})$\\
\hline
\multicolumn{2}{|c|}{\textbf{simplifyTraceMatrixMult}}\\\hline
\traceOp(MN)->\sumOp(M$\odot$t(N)) & $\forall \calMI, \calMII, \calR_{1}, r $ $\multiMOp(\calMI,\calMII,\calR_{1})\wedge \traceOp(\calR_{1},r) \rightarrow$ \\
							 & $\exists \calR_{3},\calR_{4} \trOp(\calMII,\calR_{3})\wedge \multiEOp(\calMI,\calR_{3},\calR_{4}) \wedge  \sumOp(\calR_{4},r)$\\
\hline
%\multicolumn{2}{|c|}{\textbf{DistributiveBinaryOperation}}\\\hline
%(Y$\odot$X+X)->(Y+1)$\odot$X & $\forall idY,idX, idR1,idR2 $ $multi_s(idY,idX,idR1),add(idR1,idX,idR2) \rightarrow $ \\
%							& $\exists idR3 $ $add_s(idY,"1",idR3), multi_s(idR3,idX,idR2)$\\
%\hline
%(Y$\odot$X-X)->(Y-1)$\odot$X & $\forall idY,idX, idR1,idR2 $ $multi_s(idY,idX,idR1),sub(idR1,idX,idR2) \rightarrow $ \\
%							& $\exists idR3 $ $sub_s(idY,"1",idR3), multi_s(idR3,idX,idR2)$\\
%\hline
%(X+Y$\odot$X)->(1+Y)$\odot$X & $\forall idY,idX, idR1,idR2 $ $add(idY,idX,idR1),multi_s(idR1,idX,idR2) \rightarrow $ \\
%							& $\exists idR3 $ $add_s("1",idY,idR3), multi_s(idR3,idX,idR2)$\\
%\hline
%(X-Y$\odot$X)->(1-Y)$\odot$X & $\forall idY,idX, idR1,idR2 $ $sub(idY,idX,idR1),multi_s(idR1,idX,idR2) \rightarrow $ \\
%							& $\exists idR3 $ $sub_s("1",idY,idR3), multi_s(idR3,idX,idR2)$\\
\hline
\multicolumn{2}{|c|}{\textbf{simplifySumMatrixMult}}\\\hline
\hline
\sumOp(MN) -> \sumOp(t(\colsumsOp(M))$\odot$\rowsumsOp(N))& $\forall \calMI,\calMII, \calR_{1}, r$ $\multiMOp(\calMI,\calMII,\calR_{1}) \wedge \sumOp(\calR_{1},r)\rightarrow$ \\
											  & $\exists \calR_{2},\calR_{3},\calR_{4},\calR_{5}$ $ \colsumsOp(\calMI,\calR_{2}) \wedge \trOp(\calR_{2},\calR_{3}) \wedge \rowsumsOp(\calMII,\calR_{4}) \wedge$\\
											  & $\multiEOp(\calR_{3},\calR_{4},\calR_{5}), \wedge \sumOp(\calR_{5},r)$\\
										  
\hline
\colsumsOp(MN) -> \colsumsOp(M)N 					  & $\forall \calMI,\calMII,\calR_{1},\calR_{2}$ $\multiMOp(\calMI,\calMII,\calR_{1}) \wedge \colsumsOp(\calR_{1},\calR_{2})\rightarrow$ \\
											  & $\exists \calR_{3} $ $ \colsumsOp(\calMI,\calR_{3}) \wedge \multiMOp(\calR_{3},\calMII,\calR_{2})$\\
\hline	
\hline
\rowsumsOp(MN) -> M\rowsumsOp(N)				  	  & $\forall \calMI,\calMII,\calR_{1},\calR_{2}$ $\multiMOp(\calMI,\calMII,\calR_{1}) \wedge \rowsumsOp(\calR_{1},\calR_{2})\rightarrow$ \\
											  & $\exists \calR_{3} $ $ \rowsumsOp(\calMII,\calR_{3}) \wedge \multiMOp(\calMI,\calR_{3},\calR_{2})$\\
\hline	
%\multicolumn{2}{|c|}{\textbf{simplifyColWiseAgg}}\\\hline
\hline
\colsumsOp(M)->M if x is row vector & $\forall \calMI,n,i $ $name(\calMI,n) \wedge size(\calMI,``1",j)\rightarrow \colsumsOp(\calMI,\calMI)$\\
\hline
\colmeanOp(M)->M if x is row vector & $\forall \calMI,n,j $ $name(\calMI,n)  \wedge  size(\calMI,``1",j)\rightarrow \colsumsOp(\calMI,\calMI)$\\
\hline
\colvarOp(M)->M if x is row vector & $\forall \calMI,n,j $ $name(\calMI,n)  \wedge  size(v``1",j)\rightarrow \colvarOp(\calMI,\calMI)$\\
\hline
\colmaxOp(M)->M if x is row vector & $\forall \calMI,n,j $ $name(\calMI,n) \wedge  size(\calMI,``1",j)\rightarrow \colmaxOp(\calMI,\calMI)$\\
\hline
\colminOp(M)->M if x is row vector & $\forall \calMI,n,j $ $name(\calMI,n) \wedge  size(\calMI,``1",j)\rightarrow \colminOp(\calMI,\calMI)$\\
\hline
\colsumsOp(M)->\sumOp(M) if x is col vector & $\forall \calMI,i, \calR_{1} $ $\colsumsOp(\calMI, \calR_{1}) \wedge   size(\calMI,i,``1")\rightarrow \sumOp(\calMI,\calR_{1})$\\
\hline	
\colmeanOp(M)->\meanOp(M) if x is col vector & $\forall \calMI,i, \calR_{1} $ $\colmeanOp(\calMI, \calR_{1}) \wedge size(\calMI,i,``1")\rightarrow \meanOp(\calMI,\calR_{1})$\\
\hline
\colmaxOp(X)->\maxOp(M) if x is col vector & $\forall \calMI,i, \calR_{1} $ $\colmaxOp(\calMI, \calR_{1})\wedge   size(\calMI,i,``1")\rightarrow \maxOp(\calMI,\calR_{1})$\\
\hline
\colminOp(M)->\minOp(X) if x is col vector &$\forall \calMI,i, \calR_{1} $ $\colminOp(\calMI, \calR_{1}) \wedge   size(\calMI,i,``1")\rightarrow \minOp(\calMI,\calR_{1})$\\
\hline
\colvarOp(M)->\varOp(M) if x is col vector & $\forall \calMI,i, \calR_{1} $ $\colvarOp(\calMI, \calR_{1})\wedge  size(\calMI,i,``1")\rightarrow \varOp(\calMI,\calR_{1})$\\
\hline						 
\end{tabular}
\end{table*}

\begin{table*}[ht]	
\centering	
\begin{tabular}{ |c|c| } 
\hline
\textbf{SystemML Algebraic Simplification Rule}  & \textbf{Integrity Constraints $\calMMC_{StatAgg}$}\\
\hline	
\multicolumn{2}{|c|}{\textbf{simplifyRowWiseAgg}}\\\hline
\hline
\rowsumsOp(M)->M if x is col vector & $\forall \calMI,n,i $ $name(\calMI,n) \wedge size(\calMI,i,``1")\rightarrow \rowsumsOp(\calMI,\calMI)$\\
\hline
\rowmeanOp(M)->M if x is col vector &$\forall \calMI,n,i $ $name(\calMI,n) \wedge size(\calMI,i,``1")\rightarrow \rowmeanOp(\calMI,\calMI)$\\
\hline
\rowvarOp(M)->M if x is col vector & $\forall \calMI,n,i $ $name(\calMI,n) \wedge size(\calMI,i,``1")\rightarrow \rowvarOp(\calMI,\calMI)$\\
\hline
\rowmaxOp(M)->M if x is col vector & $\forall \calMI,n,i $ $name(\calMI,n) \wedge size(\calMI,i,``1")\rightarrow \rowmaxOp(\calMI,\calMI)$\\
\hline
\rowminOp(M)->M if x is col vector &  $\forall \calMI,n,i $ $name(\calMI,n) \wedge size(\calMI,i,``1")\rightarrow \rowmaxOp(\calMI,\calMI)$\\
\hline
\rowsumsOp(M)->\sumOp(M) if x is row vector & $\forall \calMI,j, \calR_{1} $ $\rowsumsOp(\calMI, \calR_{1}) \wedge   size(\calMI,``1",j)\rightarrow \sumOp(\calMI,\calR_{1})$\\
\hline
\rowmeanOp(M)->\meanOp(M) if x is row vector & $\forall \calMI,j, \calR_{1} $ $\rowmeanOp(\calMI, \calR_{1}) \wedge   size(\calMI,``1",j)\rightarrow \meanOp(\calMI,\calR_{1})$\\
\hline
\rowmaxOp(M)->\maxOp(M) if x is row vector & $\forall \calMI,j, \calR_{1} $ $\rowmaxOp(\calMI, \calR_{1}) \wedge   size(\calMI,``1",j)\rightarrow \maxOp(\calMI,\calR_{1})$\\
\hline
\rowminOp(X)->\minOp(M) if x is row vector &  $\forall \calMI,j, \calR_{1} $ $\rowminOp(\calMI, \calR_{1}) \wedge   size(\calMI,``1",j)\rightarrow \minOp(\calMI,\calR_{1})$\\
\hline
\rowvarOp(X)->\varOp(M) if x is row vector & $\forall \calMI,j, \calR_{1} $ $\rowvarOp(\calMI, \calR_{1}) \wedge   size(\calMI,``1",j)\rightarrow \varOp(\calMI,\calR_{1})$\\
\hline
\hline	
\multicolumn{2}{|c|}{\textbf{pushdownSumOnAdd}}\\\hline	 
\hline
\sumOp(M+N) -> \sumOp(A)+\sumOp(B) & $\forall \calMI,\calMII,s $ $ \addMOp\calMI,\calMII,s_1) \wedge \sumOp(\calMI,s_1) \rightarrow$\\
						  & $\exists s_2, s_3$ $\sumOp(\calMI,s_2)  \wedge \sumOp(\calMII,s_3) \wedge \wedge add_s(s_2,s_3,s_1)$\\
\hline
\hline
\multicolumn{2}{|c|}{\textbf{ColSumsMVMult}}\\\hline
\hline
\colsumsOp(M*N) -> t(M)N & $\forall \calMI,\calMII,\calR_{1},\calR_{2},i$ $size(\calMII,i,``1") \wedge \multiEOp(\calMI,\calMII,\calR_{1}) \wedge  \colsumsOp(\calR_{1},\calR_{2})$\\
							 & $\exists \calR_{3} \trOp(\calMI,\calR_{3}) \wedge \multiMOp(\calR_{3},\calMII,\calR_{2}) $\\
\hline 
\rowsumsOp(M*M) -> Mt(N)& $\forall \calMI,\calMII,\calR_{1},\calR_{2},j$ $size(\calMII,``1",j) \wedge \multiEOp(\calMI,\calMII,\calR_{1}) \wedge  \rowsumsOp(\calR_{1},\calR_{2})$\\
							 & $\exists \calR_{3} \trOp(\calMII,\calR_{3}) \wedge \multiMOp(\calMI,\calR_{3},\calR_{2}) $\\
\hline 
\hline
\end{tabular}
\end{table*}

%% file: appendixC.tex
\onecolumn
\section{$\notopt$ and $\views$ pipelines Rewrites}
\label{appendixC}
\begin{table*}[ht]
\centering
\setlength{\extrarowheight}{2pt}
\begin{tabular}{||c|c||c|c||c|c||}
\hline
\hline
\rowcolor{lightgray}
\textbf{No.} & \textbf{Rewrite} & \textbf{No.} & \textbf{Rewrite} & \textbf{No.} & \textbf{Rewrite}\\
\hline
P1.1 & $N^TM^T$ & P1.2 &  $(A+B)^T $ & P1.3 &  $(DC)^{-1} $ \\
\hline
P1.4 &  $\calA v_1+\calB v_1$ & P1.5 &  $D$ & P1.6 &  $s_1\traceOp(D)$ \\
\hline
P1.7 &  $A$ & P1.8 &  $(s_1+s_2)A$ & P1.9 & $\detOp({\calD})$\\
\hline 
P1.10 &  $\colsumsOp({\calA})^T$ & P1.11 &  $\colsumsOp(A+B)^{T}$ & P1.12 & $\colsumsOp({\calMM})\calNN$\\
\hline
P1.13 &  $\sumOp(\colsumsOp(\calMM)^{T}*\rowsumsOp(\calNN)$) & P1.14 & $\sumOp(\colsumsOp(\calMM)^{T}*\rowsumsOp(\calNN)$) & P1.15 &  ${\calMM}({\calNN}{\calMM})$\\
\hline
P1.16 &   $\sumOp({\calA})$ & P1.17 & $\detOp(\calC)*\detOp(\calD)*\detOp(\calC)$ & P1.18 & $\sumOp(A)$ \\
\hline
P1.25 & \multicolumn{5}{|c|}{$\calMM \odot (\calNN^{T} / (\calMM(\calNN\calNN^{T})))$}\\ 
\hline
\end{tabular}
\caption{$\notopt$ Pipelines (Part 1) Rewrites}
\vspace{-8mm}
\label{tab:pipelines-1}
\end{table*}

\begin{table*}[ht]
\centering                     
\setlength{\extrarowheight}{2pt}
\begin{tabular}{||c|c||c|c||c|c||}
\hline
\hline
\rowcolor{lightgray}
\textbf{No.} & \textbf{Rewrite} & \textbf{No.} & \textbf{Rewrite} & \textbf{No.} & \textbf{Rewrite}\\
\hline
P2.1 &  $\traceOp(\calC)+\traceOp(\calD)$  & P2.2 &  $1/\detOp({\calD})$ & P2.3 &  $\traceOp({\calD})$  \\
\hline
P2.4 & $s_1({\calA} +{\calB})$ & P2.5 &  $1/\detOp(({\calC}+{\calD}))$ & P2.6 &  $(D^{-1}C)^T$ \\
\hline
P2.7 &  ${\calC}$ & P2.8 & $\detOp({\calC})* \detOp({\calD})$ & P2.9 &  $\traceOp(DC)+\traceOp(D)$\\
\hline 
P2.10 &    ${\calMM}rowSums{\calNN})$ & P2.11 &  $\sumOp({\calA}) + \sumOp({\calB})$   & P2.12 &  $\sumOp(\colsumsOp(\calMM)^{T}*\rowsumsOp(\calNN)$\\
\hline
P2.13 &  $({\calMM}({\calNN}{\calMM}))^T$  & P2.14 &   $({\calMM}({\calNN}{\calMM})){\calNN}$ & P2.15 &   $\sumOp({\calA})$\\
\hline
P2.16 &  $\traceOp((DC)^{-1})+trace{\calD})$   & P2.17 &  $(((({\calC}+{\calD})^{-1})^{T})D$ & P2.18 &  $\rowsumsOp({\calA}+B)^T$  \\
\hline
P2.25 &  \multicolumn{5}{|c|}{$u_1v_2^{T} v_2 - \calX v_2$}\\ 
\hline
\end{tabular}
\caption{$\notopt$ Pipelines (Part 2) Rewrites}
\vspace{-6mm}
\label{tab:pipelines-2}
\end{table*}

\begin{table*}[ht]
\centering                     
\setlength{\extrarowheight}{2pt}
\begin{tabular}{||c|c||c|c||c|c||}
\hline
\hline
\rowcolor{lightgray}
\textbf{No.} & \textbf{Expression} & \textbf{No.} & \textbf{Expression} & \textbf{No.} & \textbf{Expression}\\
\hline
$V_1$ &  $(D)^{-1}$   & $V_2$ &  $(C^{T})^{-1}$ & $V_3$ &  $NM$ \\
\hline
$V_4$ & $u_1v_{2}^{T}$ & $V_5$ &  $DC$ & $V_6$ &  $A+B$ \\
\hline
 $V_7$ &  $C^{-1}$ &  $V_8$ & $C^{T}D$ &  $V_9$ &  $(D+C)^{-1}$\\
\hline 
 $V_{10}$ &  $\detOp(CD)$ & $V_{11}$ &  $\detOp(DC)$  & $V_{12}$ & $(DC)^{T}$\\
\hline
\end{tabular}
\caption{The set of views $V_{exp}$}
\vspace{-6mm}
\label{tab:pipelines-2}
\end{table*}

\begin{table*}[ht]
\centering
                       % used font size
\setlength{\extrarowheight}{2pt}
\begin{tabular}{||c|c||c|c||c|c||}
%\hline
%\hline
%\multicolumn{6}{||c||}{\textbf{LA Pipelines}}\\
\hline
\hline
\rowcolor{lightgray}
\textbf{No.} & \textbf{Rewrite} & \textbf{No.} & \textbf{Rewrite} & \textbf{No.} & \textbf{Rewrite}\\
\hline
P1.2 &  $(V_6)^T$ & P1.3 &  $V_7V_1$ & P1.4 &  $(V_6)v_1$ \\
\hline
P1.11 & $\colsumsOp(V_6)^{T}$  & P1.15 & ${\calMM}(V_3)$& P1.17 &  $V_{10}*\detOp(C)$ \\
\hline
P1.19 &  $V_2$ & P1.20 & $\traceOp(V_7)$ & P1.21 &   $(C+V_1)^{T}$\\
\hline 
P1.22 &  $\traceOp(V_9)$ & P1.24 &  $\traceOp(V_1V_7)+\traceOp(D)$ & P1.29 & $V_5CCC$\\
\hline
P1.30 &  $ V_3\odot V_3{\calR}^{T}$  & P2.2  & $\detOp(V_1)$  & P2.4&  $s_1(V6)$\\
\hline
P2.5 &   $\detOp(V_9)$ & P2.6 &  $(V_1C)^T$ & P2.9 & $\traceOp(V_{12})+\traceOp(D)$ \\
\hline
P2.11 &  $\sumOp(V_6)$ & P2.13 &  $(MV_3)^T$ & P2.14 &  $MV_3N$ \\
\hline
P2.16 &  $\traceOp(V_7V_1)+trace{\calD})$ & P2.17 & $(V_9^{T})D$  & P2.18 & $\rowsumsOp(V_6)^T$  \\
\hline
P2.20 &  $(MV_3)^T$ & P2.21 & $V_1 (V_1^{T}(\calD^{T}v_1))$  & P2.25 &  $V_4v_1-Xv_1$\\
\hline
P1.23 &  $\detOp((V_7V_1) + \calD)$ & P2.26 & $exp(V_9)$ & P2.27 &  $V_9^TV_5$ \\\hline
\end{tabular}
\caption{$\views$ Pipelines Rewrites}
\vspace{-8mm}
\label{tab:pipelines-1}
\end{table*}

%% file: appendixD.tex
\onecolumn
\section{Additional Results: $\notopt$ Pipelines - Na\"ive-based Cost Model}
\label{appendixD}
\newcommand{\includeCroppedPdf}[2][]{%
    \immediate\write18{pdfcrop #2}%
    \includegraphics[#1]{#2-crop}}

\begin{figure*}[!htbp]
\figspa
  \centering
  \subfigure[\textbf{P1.2}]{\includegraphics[scale=0.25,width=4.2cm,height=3.6cm]{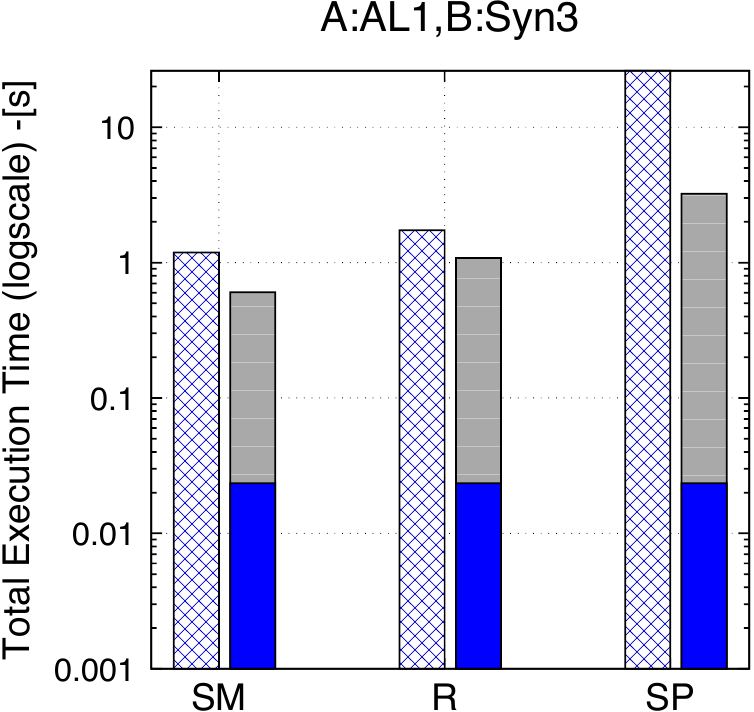}}
  \subfigure[\textbf{P1.2}]{\includegraphics[scale=0.25,width=4.4cm,height=3.6cm]{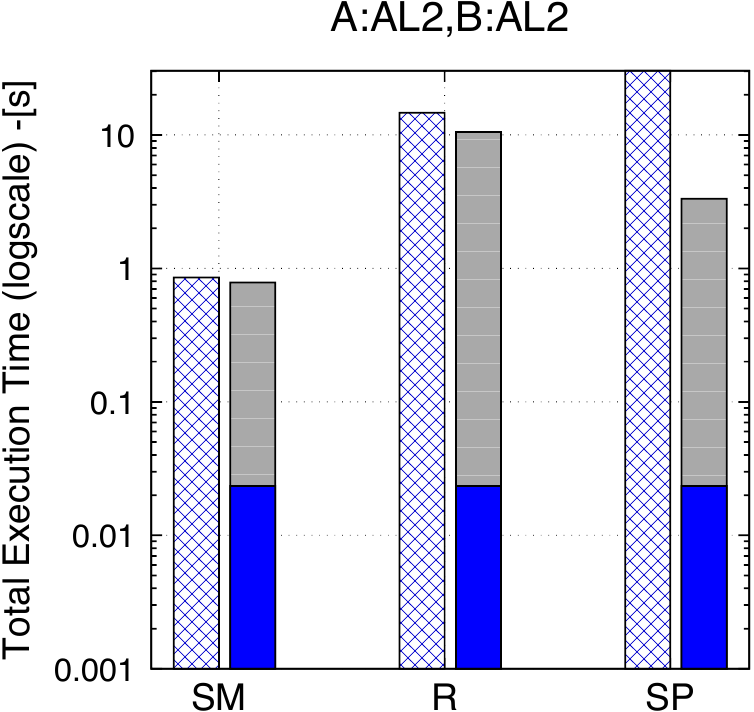}}
    \subfigure[\textbf{P1.2}]{\includegraphics[scale=0.25,width=4.2cm,height=3.6cm]{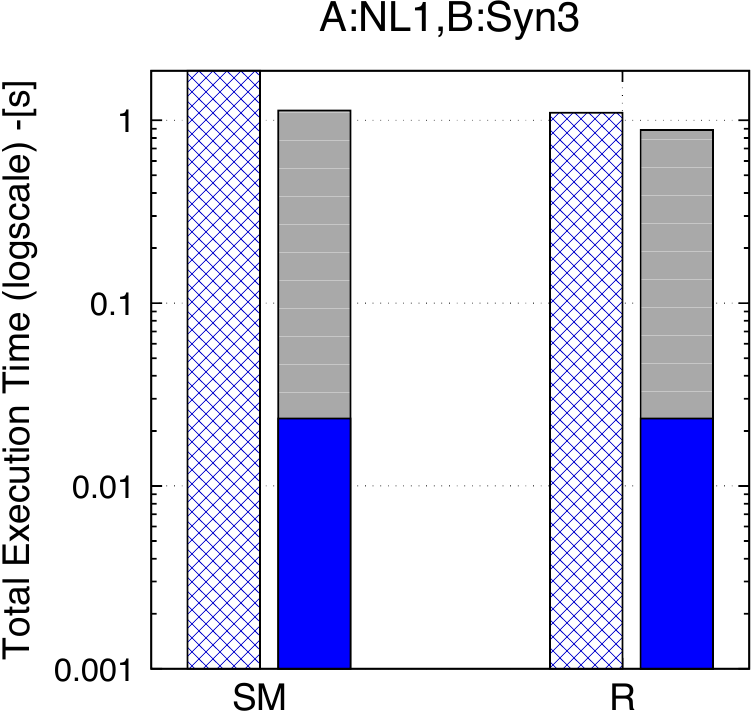}}
 \figspb\figspb\figspb
 \caption{P1.2 evaluation time with and without rewriting}
 \figspc
 \label{fig:la-p1.2}
\end{figure*}

\begin{figure*}[!htbp]
\figspa
  \centering
  \subfigure[\textbf{P1.2}]{\includegraphics[scale=0.25,width=4.2cm,height=3.6cm]{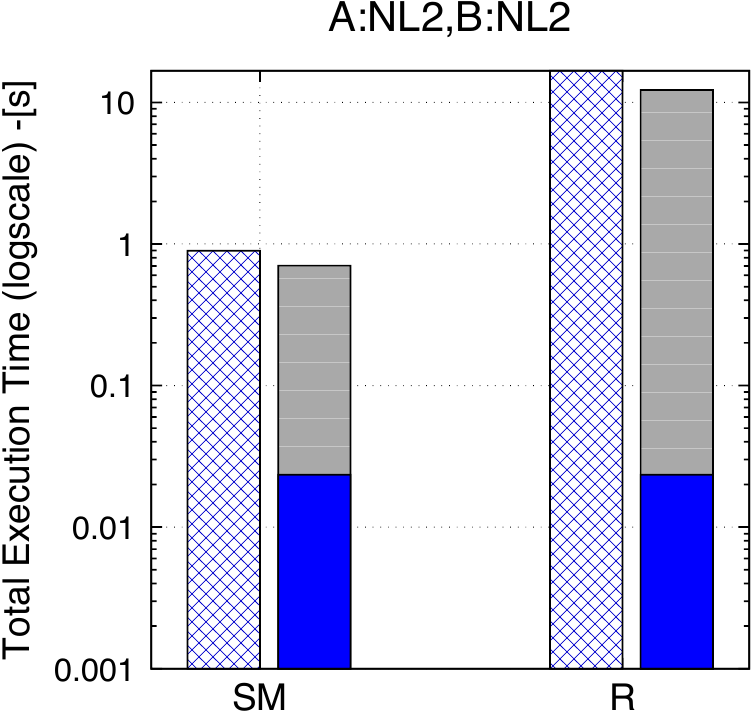}}
  \subfigure[\textbf{P1.2}]{\includegraphics[scale=0.25,width=4.4cm,height=3.6cm]{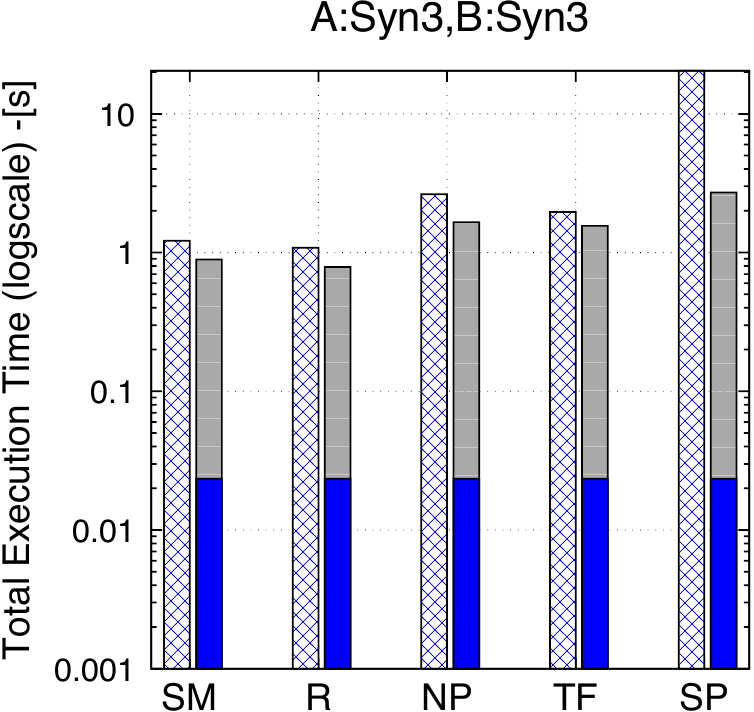}}
    \subfigure[\textbf{P1.2}]{\includegraphics[scale=0.25,width=4.2cm,height=3.6cm]{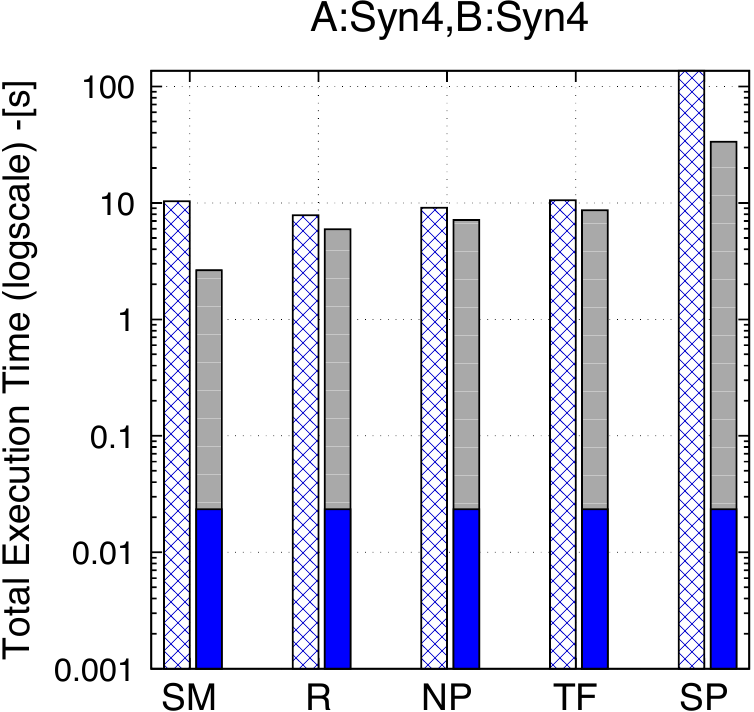}}
 \figspb\figspb\figspb
 \caption{P1.2 evaluation time with and without rewriting}
 \figspc
 \label{fig:la-p1.2}
\end{figure*}

\begin{figure*}[!htbp]
\figspa
  \centering
  \subfigure[\textbf{P1.6}]{\includegraphics[scale=0.25,width=4.2cm,height=3.6cm]{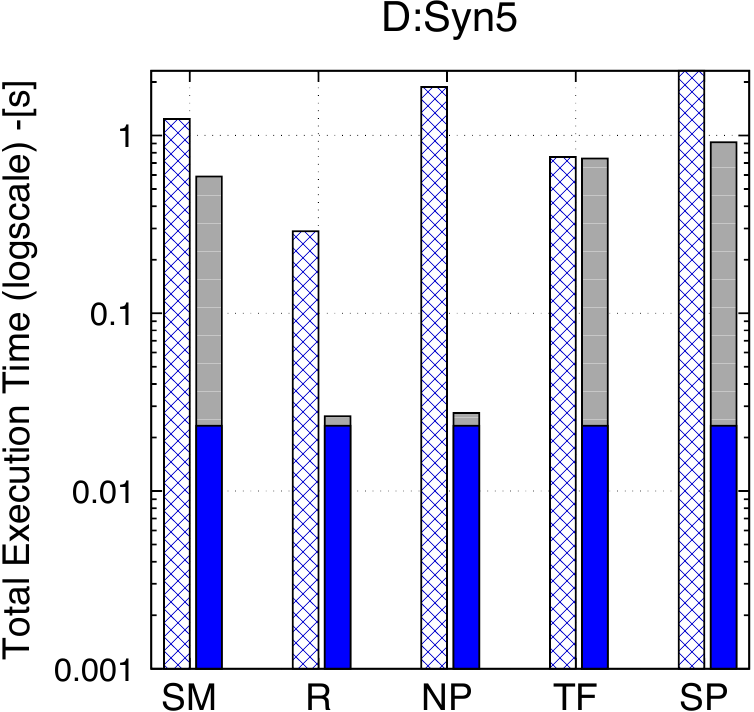}}
  \subfigure[\textbf{P1.6}]{\includegraphics[scale=0.25,width=4.4cm,height=3.6cm]{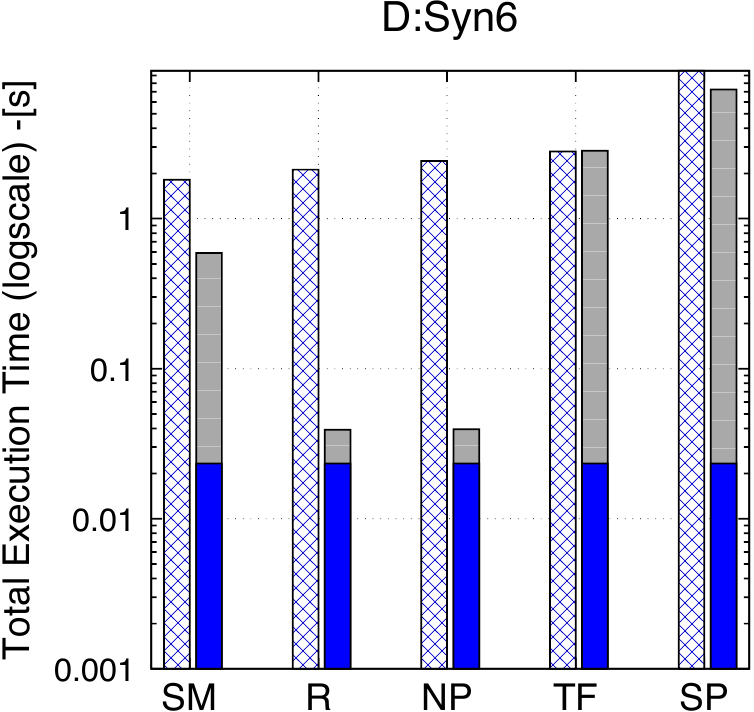}}
 \figspb\figspb\figspb
 \caption{P1.6 evaluation time with and without rewriting}
 \figspc
 \label{fig:la-p1.6}
\end{figure*}

\begin{figure*}[!htbp]
\figspa
  \centering
  \subfigure[\textbf{P1.8}]{\includegraphics[scale=0.25,width=4.2cm,height=3.6cm]{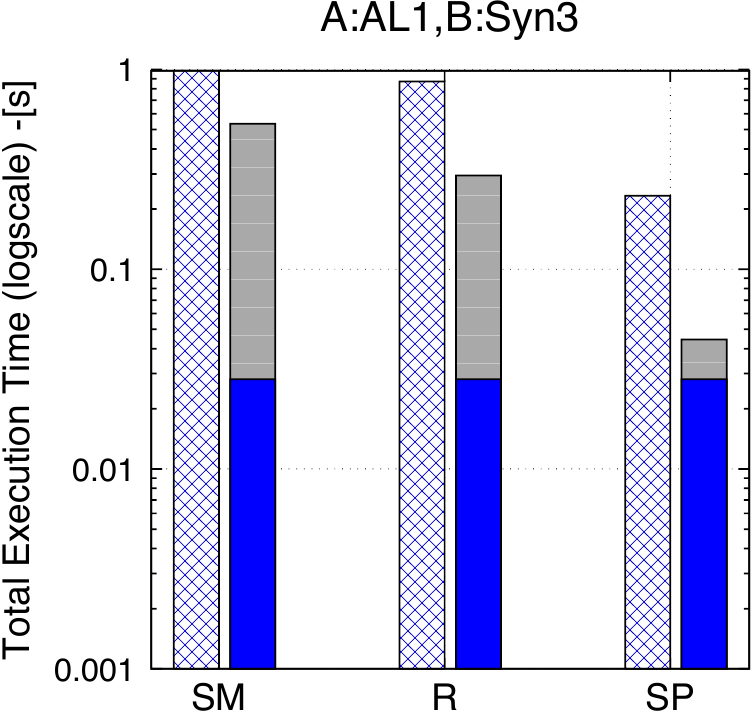}}
  \subfigure[\textbf{P1.8}]{\includegraphics[scale=0.25,width=4.4cm,height=3.6cm]{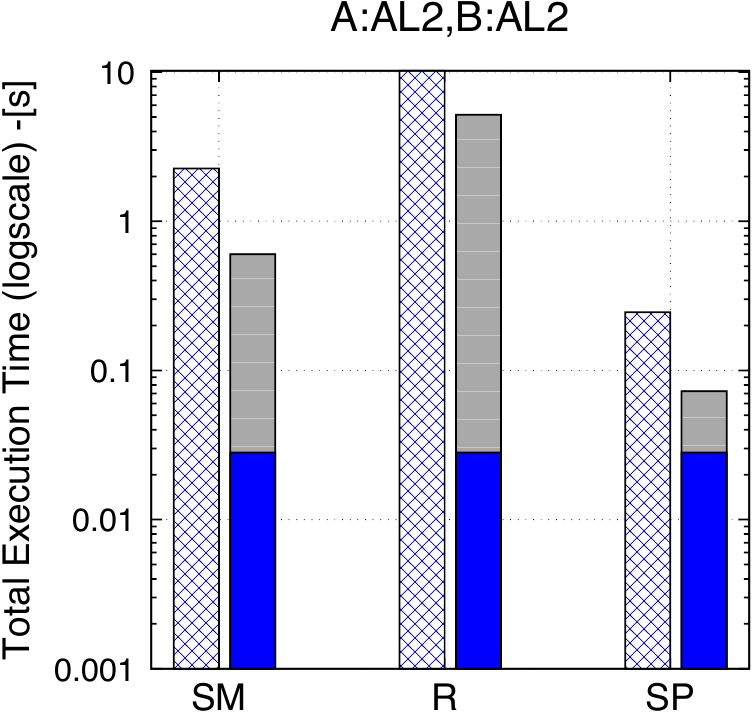}}
    \subfigure[\textbf{P1.8}]{\includegraphics[scale=0.25,width=4.2cm,height=3.6cm]{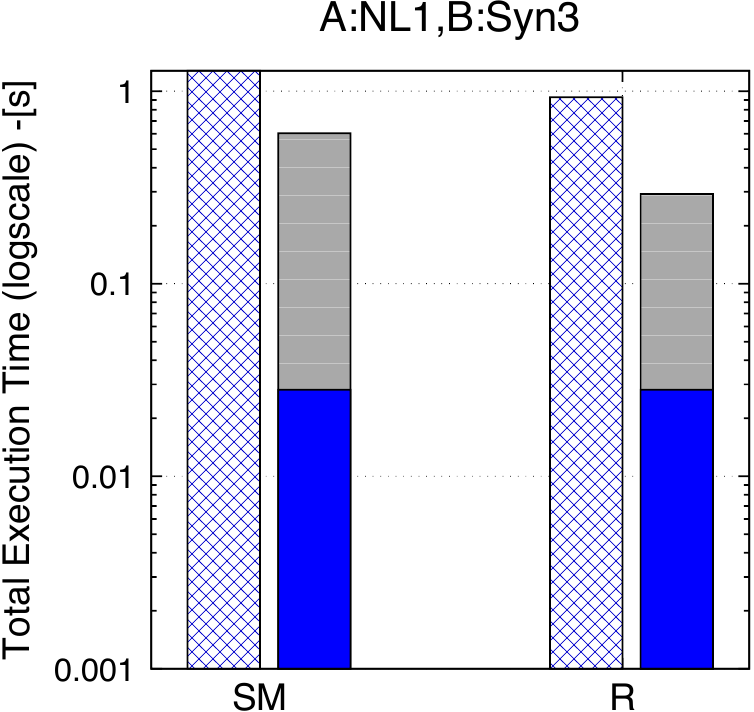}}
 \figspb\figspb\figspb
 \caption{P1.8 evaluation time with and without rewriting}
 \figspc
 \label{fig:la-p1.8}
\end{figure*}

\begin{figure*}[!htbp]
\figspa
  \centering
  \subfigure[\textbf{P1.8}]{\includegraphics[scale=0.25,width=4.2cm,height=3.6cm]{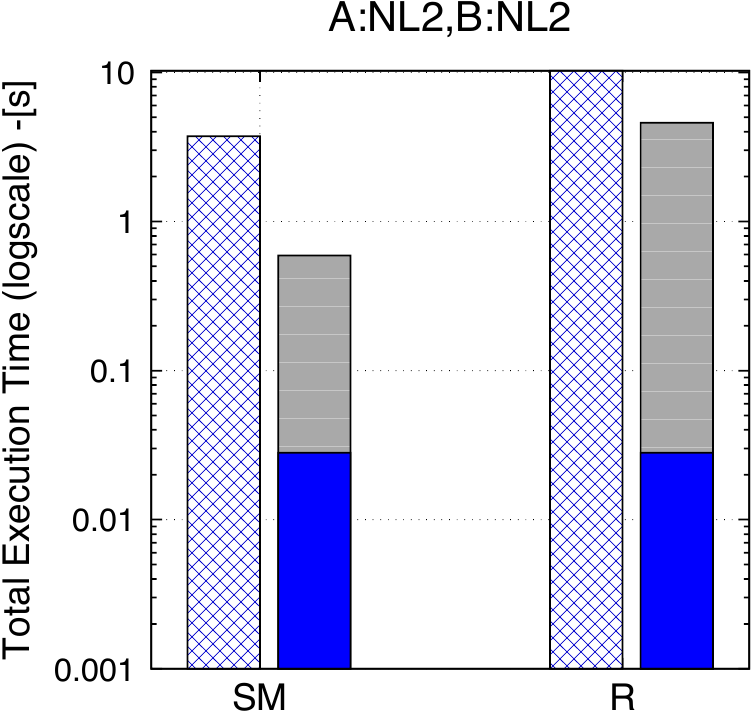}}
  \subfigure[\textbf{P1.8}]{\includegraphics[scale=0.25,width=4.4cm,height=3.6cm]{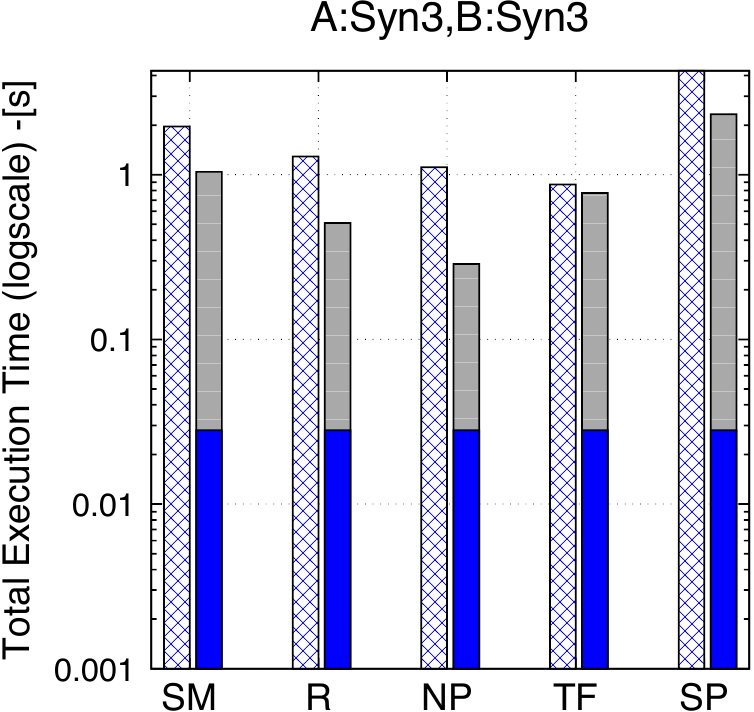}}
    \subfigure[\textbf{P1.8}]{\includegraphics[scale=0.25,width=4.2cm,height=3.6cm]{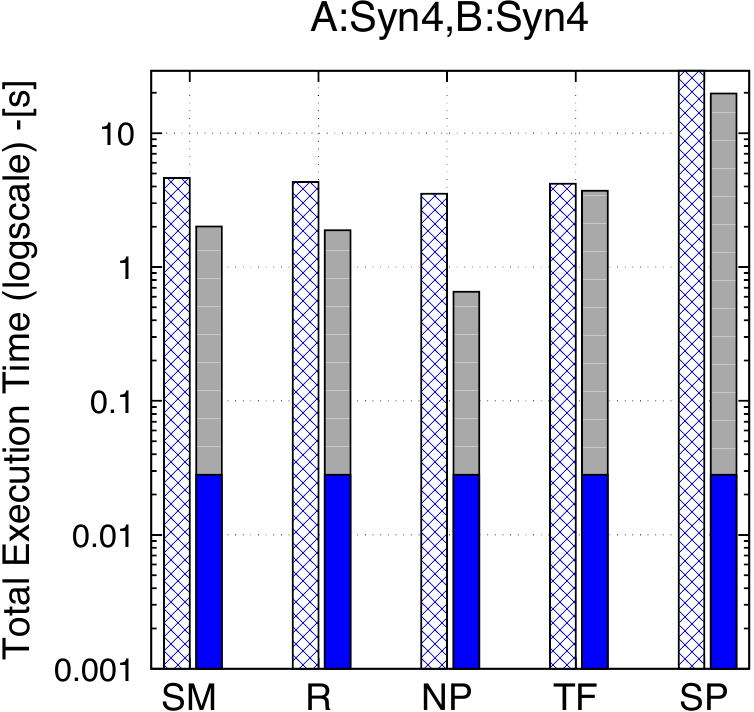}}
 \figspb\figspb\figspb
 \caption{P1.8 evaluation time with and without rewriting}
 \label{fig:la-p1.8}
\end{figure*}

\begin{figure*}[!htbp]
\figspa
  \centering
  \subfigure[\textbf{P1.9}]{\includegraphics[scale=0.25,width=4.2cm,height=3.6cm]{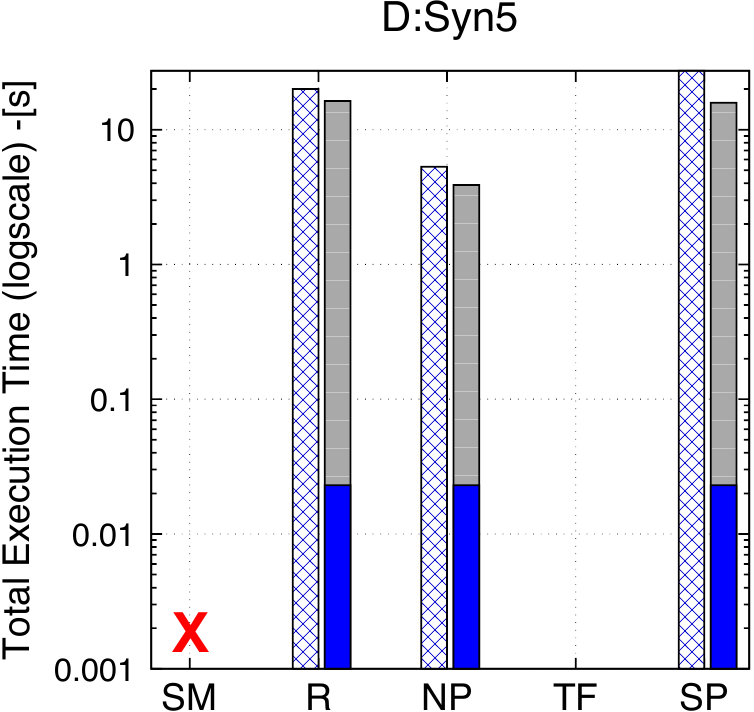}}
 \figspb\figspb\figspb
 \caption{P1.9 evaluation time with and without rewriting}
 \label{fig:la-p1.9}
\end{figure*}

\begin{figure*}[!htbp]
\figspa
  \centering
  \subfigure[\textbf{P1.10}]{\includegraphics[scale=0.25,width=4.2cm,height=3.6cm]{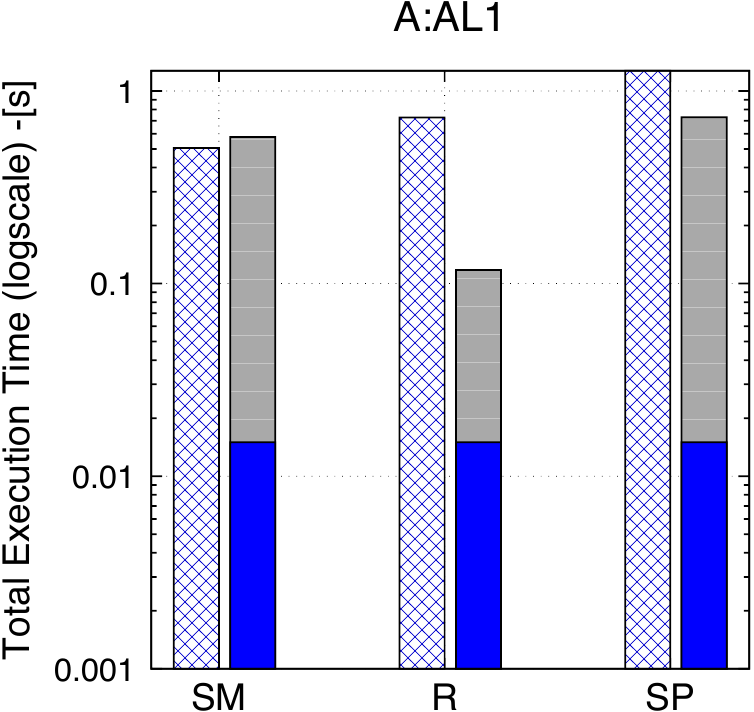}}
  \subfigure[\textbf{P1.10}]{\includegraphics[scale=0.25,width=4.4cm,height=3.6cm]{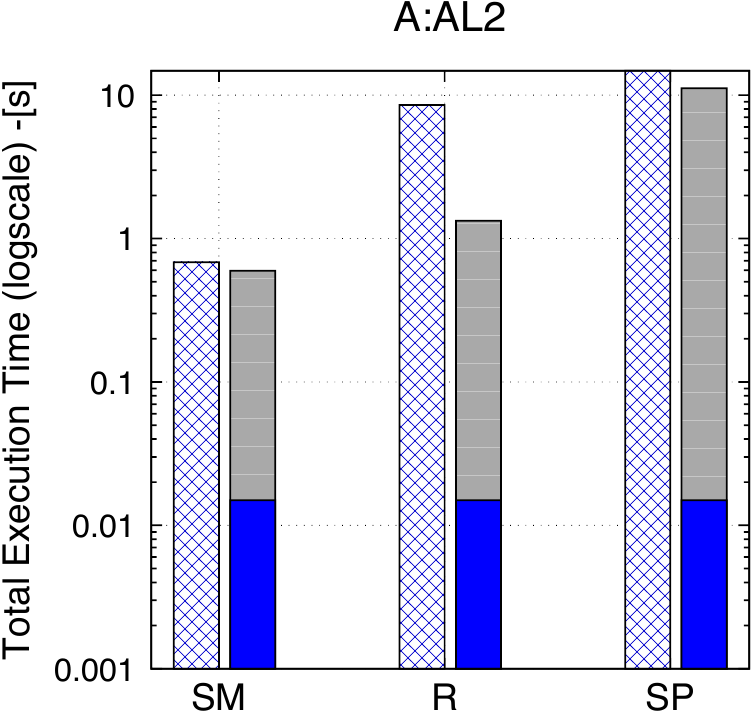}}
    \subfigure[\textbf{P1.10}]{\includegraphics[scale=0.25,width=4.2cm,height=3.6cm]{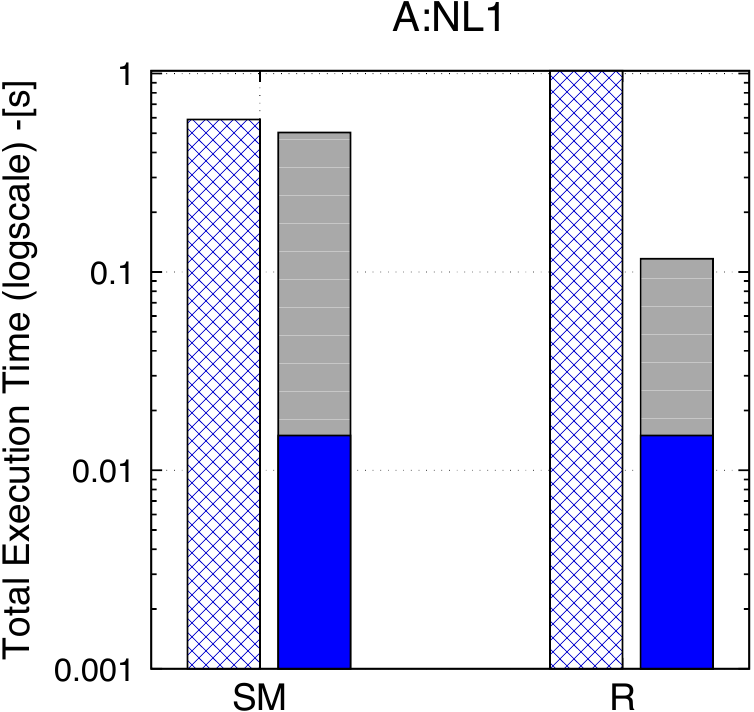}}
 \figspb\figspb\figspb
 \caption{P1.10 evaluation time with and without rewriting}
 \label{fig:la-p1.10}
\end{figure*}

\begin{figure*}[!htbp]
\figspa
  \centering
  \subfigure[\textbf{P1.10}]{\includegraphics[scale=0.25,width=4.2cm,height=3.6cm]{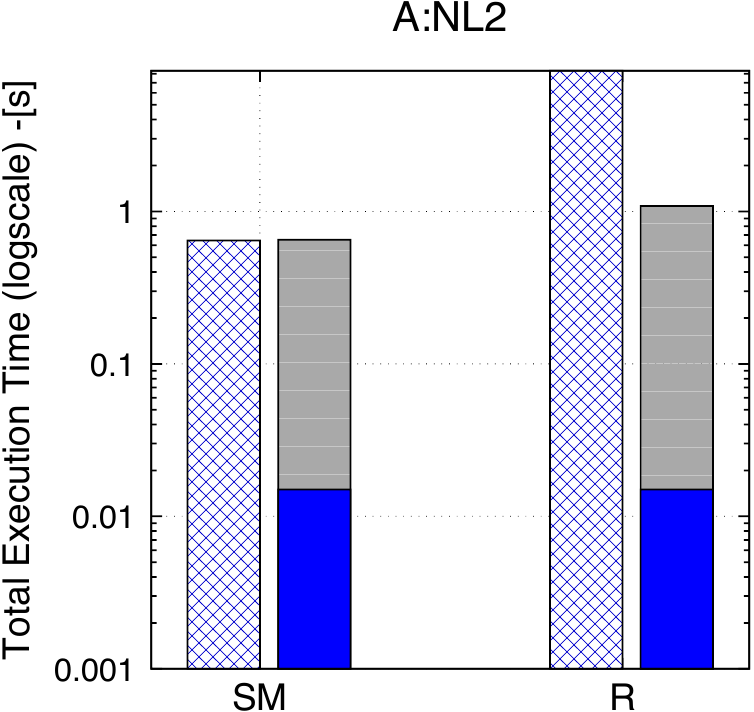}}
  \subfigure[\textbf{P1.10}]{\includegraphics[scale=0.25,width=4.4cm,height=3.6cm]{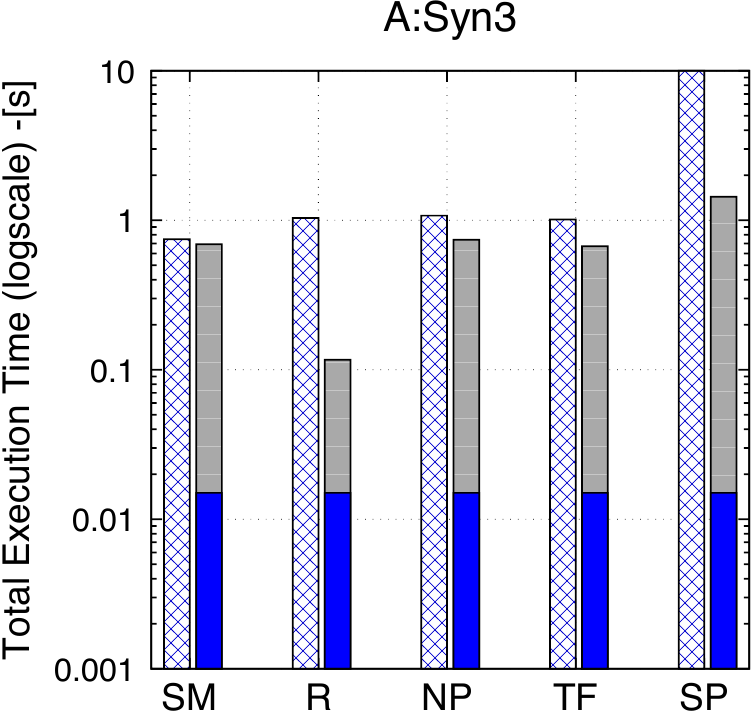}}
    \subfigure[\textbf{P1.10}]{\includegraphics[scale=0.25,width=4.2cm,height=3.6cm]{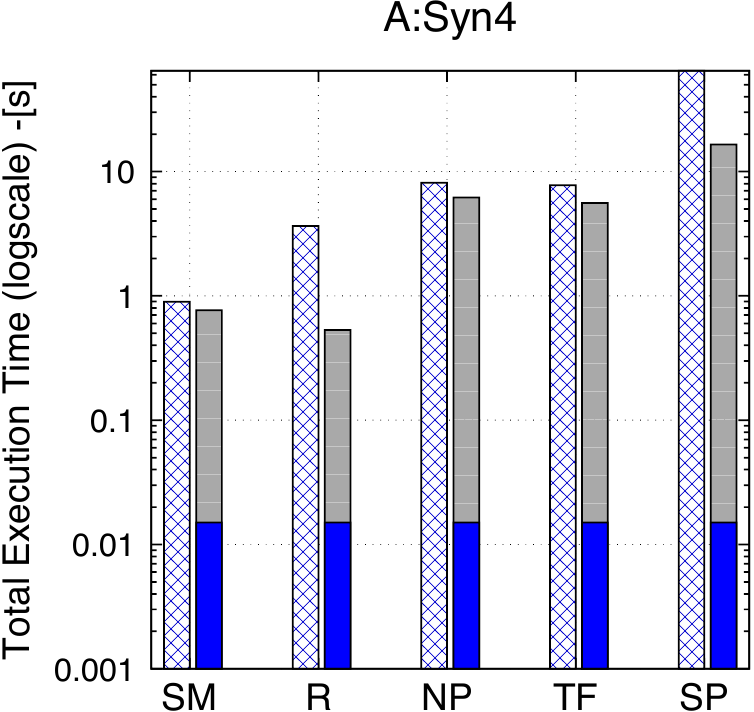}}
 \figspb\figspb\figspb
 \caption{P1.10 evaluation time with and without rewriting}
 \label{fig:la-p1.10}
\end{figure*}

\begin{figure*}[!htbp]
\figspa
  \centering
  \subfigure[\textbf{P1.11}]{\includegraphics[scale=0.25,width=4.2cm,height=3.6cm]{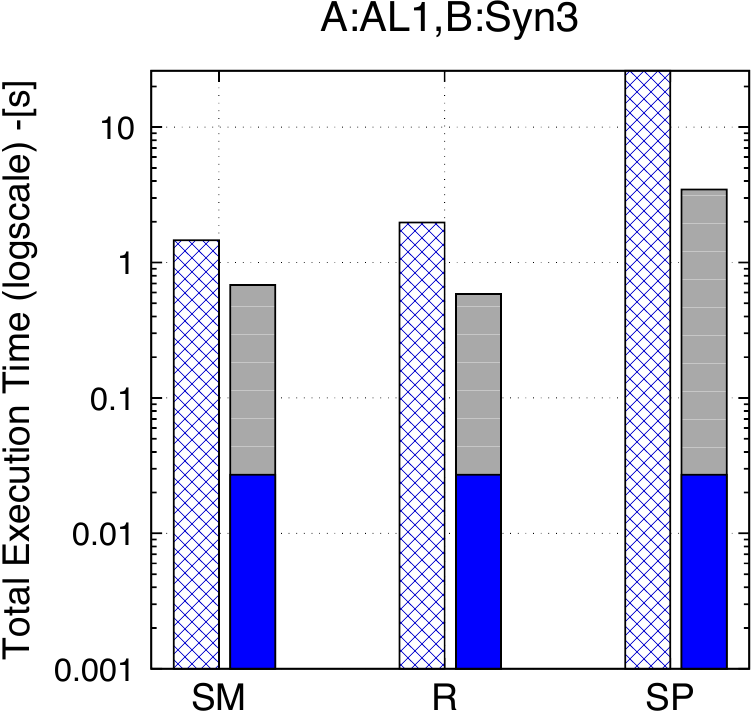}}
  \subfigure[\textbf{P1.11}]{\includegraphics[scale=0.25,width=4.4cm,height=3.6cm]{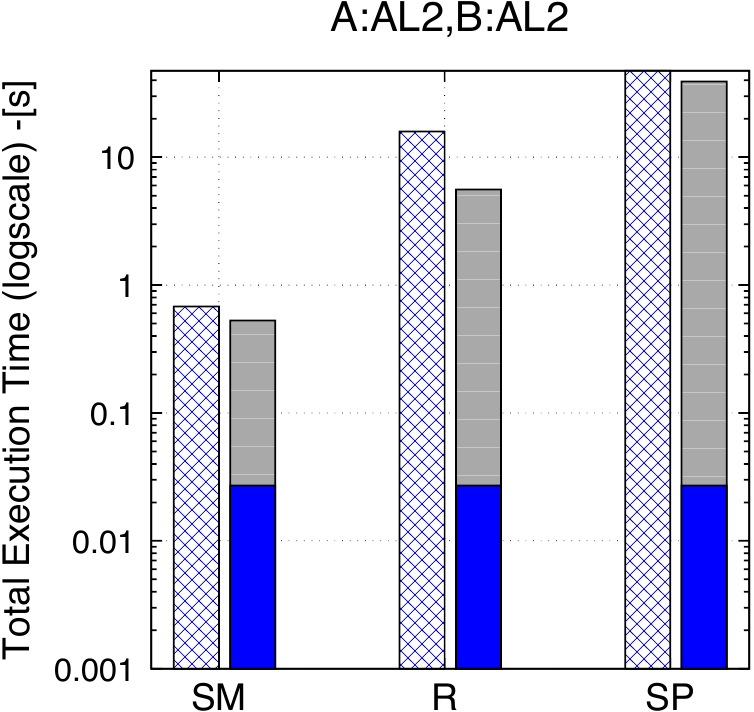}}
    \subfigure[\textbf{P1.11}]{\includegraphics[scale=0.25,width=4.2cm,height=3.6cm]{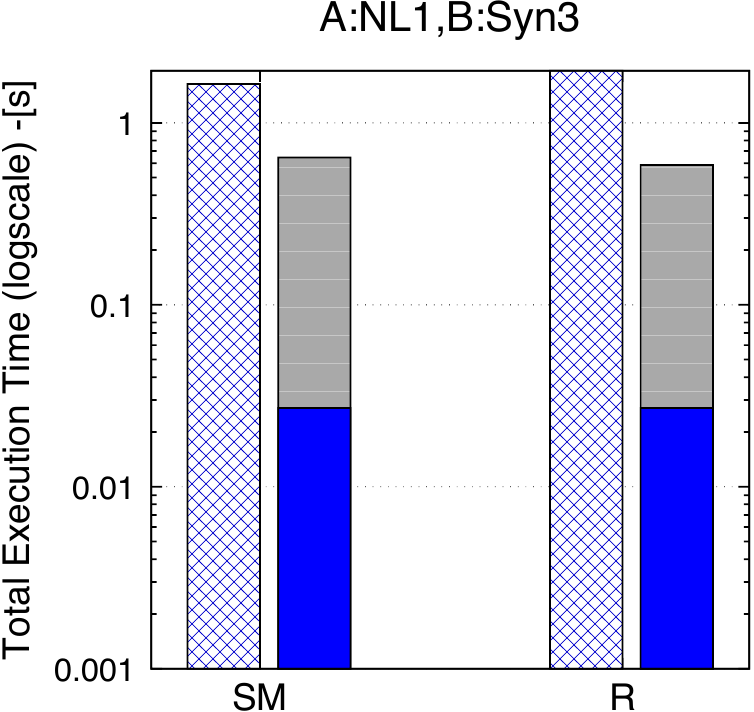}}
 \figspb\figspb\figspb
 \caption{P1.11 evaluation time with and without rewriting}
 \label{fig:la-p1.11}
\end{figure*}

\begin{figure*}[!htbp]
\figspa
  \centering
  \subfigure[\textbf{P1.11}]{\includegraphics[scale=0.25,width=4.2cm,height=3.6cm]{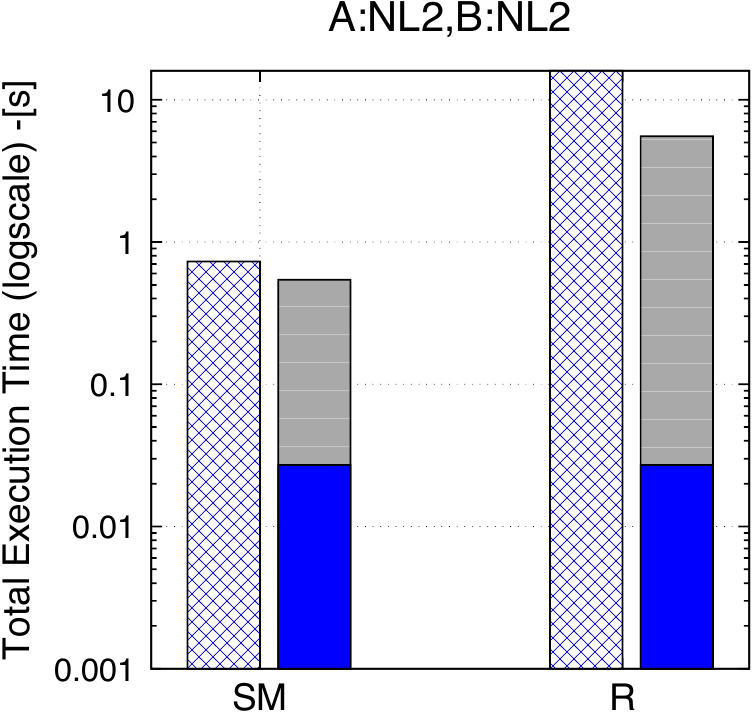}}
  \subfigure[\textbf{P1.11}]{\includegraphics[scale=0.25,width=4.4cm,height=3.6cm]{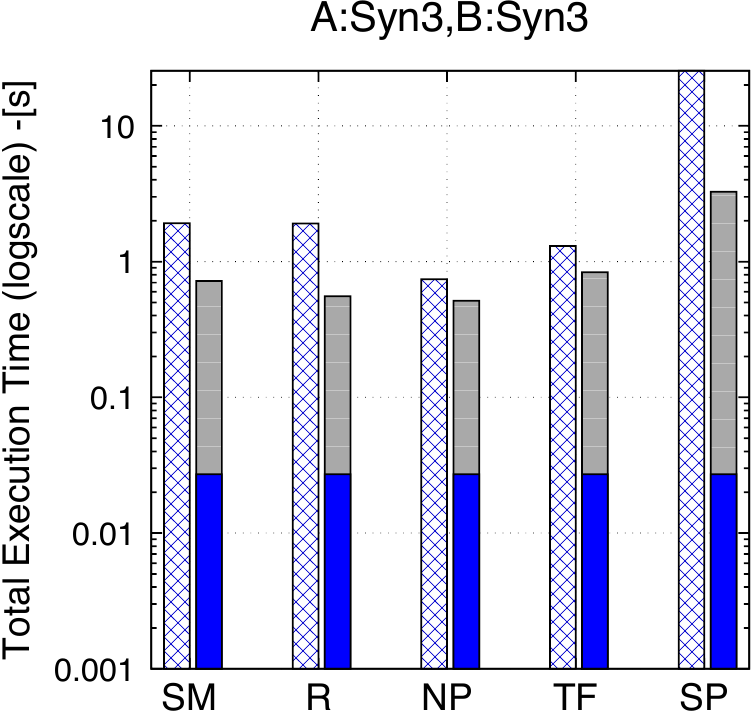}}
    \subfigure[\textbf{P1.11}]{\includegraphics[scale=0.25,width=4.2cm,height=3.6cm]{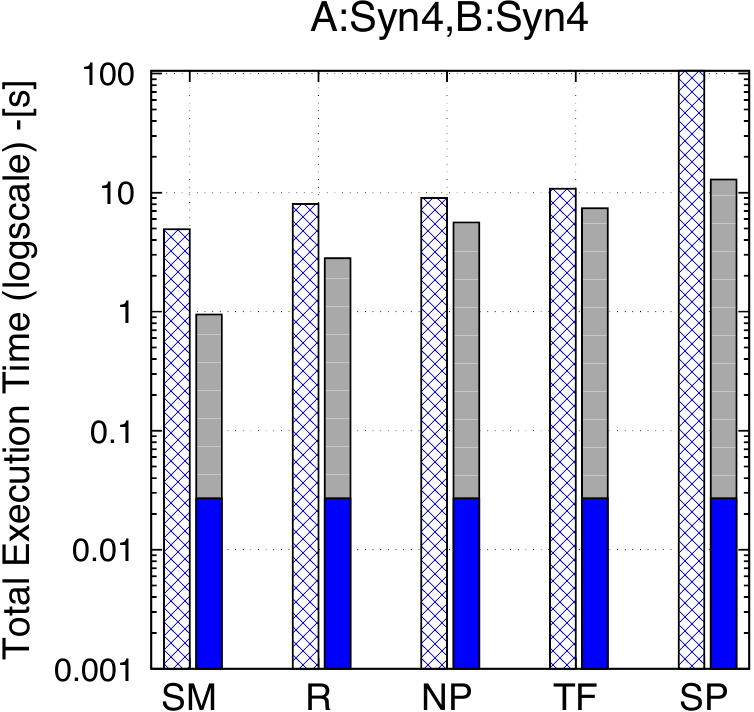}}
 \figspb\figspb\figspb
 \caption{P1.11 evaluation time with and without rewriting}
 \label{fig:la-p1.11}
\end{figure*}

\begin{figure*}[!htbp]
\figspa
  \centering
  \subfigure[\textbf{P1.12}]{\includegraphics[scale=0.25,width=4.2cm,height=3.6cm]{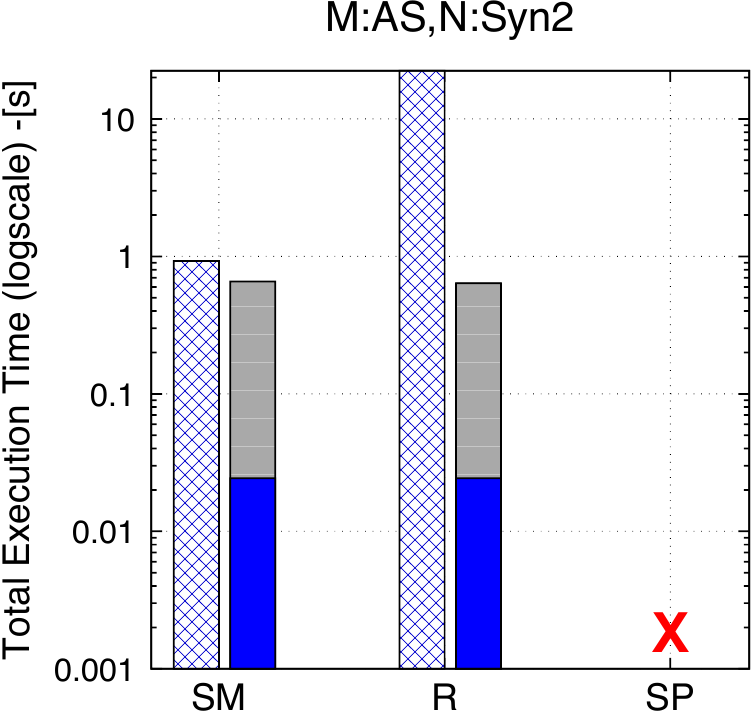}}
  \subfigure[\textbf{P1.12}]{\includegraphics[scale=0.25,width=4.4cm,height=3.6cm]{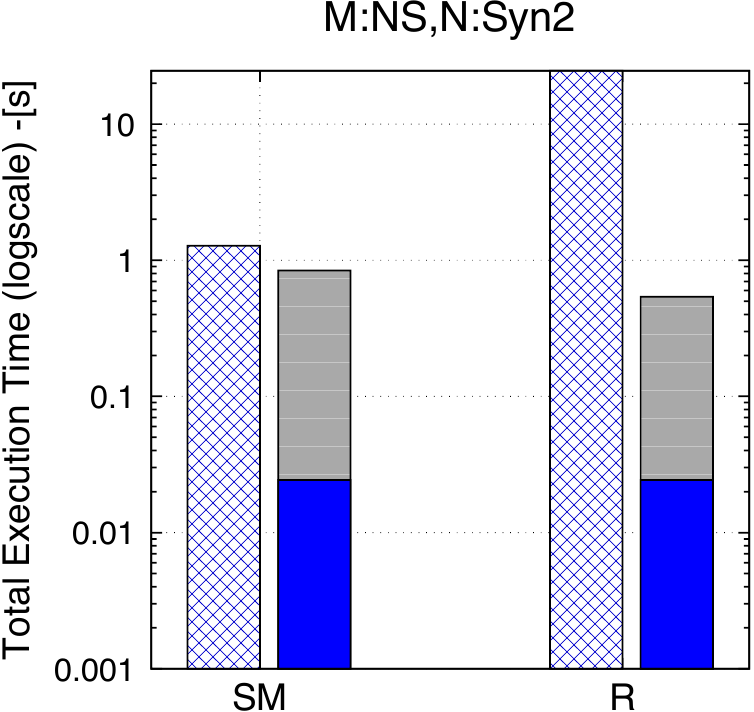}}
    \subfigure[\textbf{P1.12}]{\includegraphics[scale=0.25,width=4.2cm,height=3.6cm]{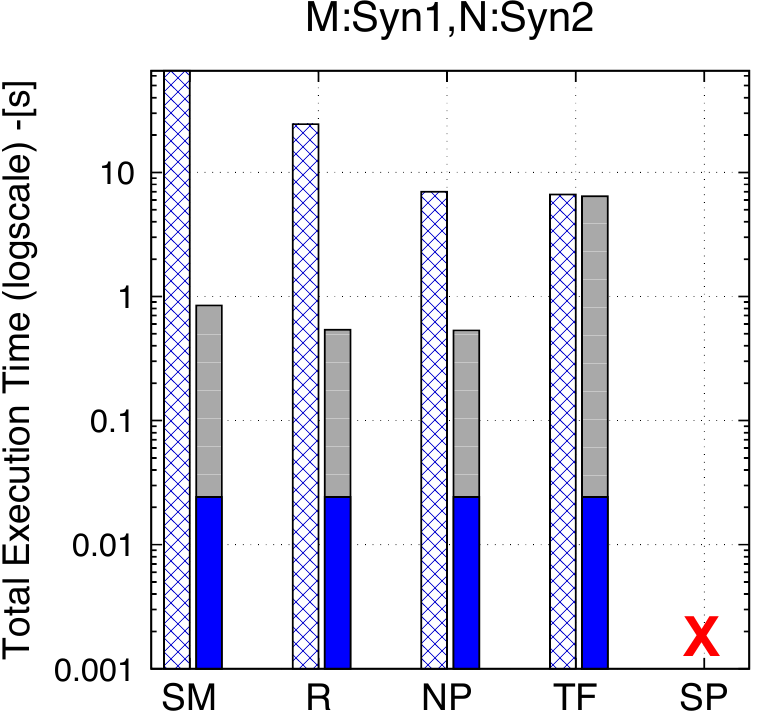}}
 \figspb\figspb\figspb
 \caption{P1.12 evaluation time with and without rewriting}
 \label{fig:la-p1.12}
\end{figure*}

\begin{figure*}[!htbp]
\figspa
  \centering
  \subfigure[\textbf{P1.14}]{\includegraphics[scale=0.25,width=4.2cm,height=3.6cm]{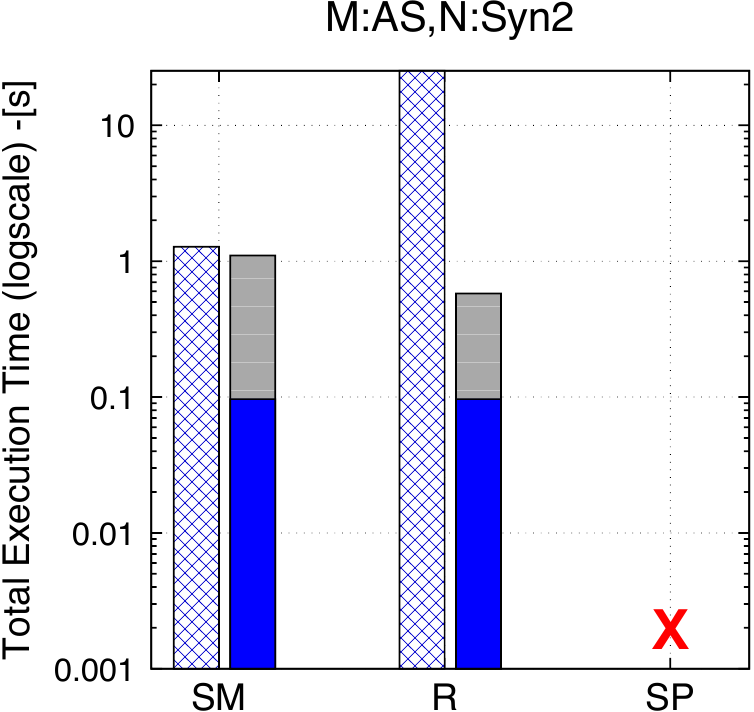}}
  \subfigure[\textbf{P1.14}]{\includegraphics[scale=0.25,width=4.4cm,height=3.6cm]{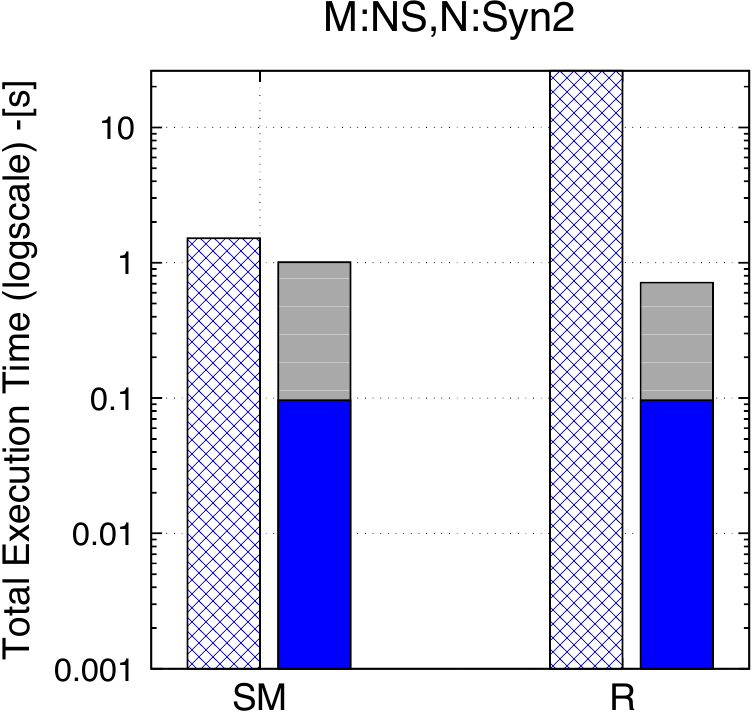}}
 \figspb\figspb\figspb
 \caption{P1.14 evaluation time with and without rewriting}
 \label{fig:la-p1.14}
\end{figure*}

\begin{figure*}[!htbp]
\figspa
  \centering
  \subfigure[\textbf{P1.15}]{\includegraphics[scale=0.25,width=4.2cm,height=3.6cm]{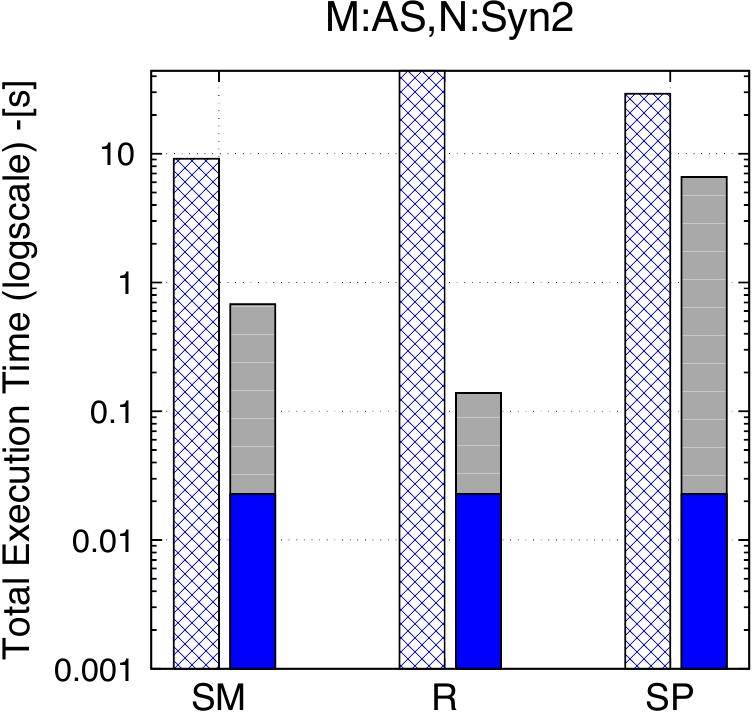}}
  \subfigure[\textbf{P1.15}]{\includegraphics[scale=0.25,width=4.4cm,height=3.6cm]{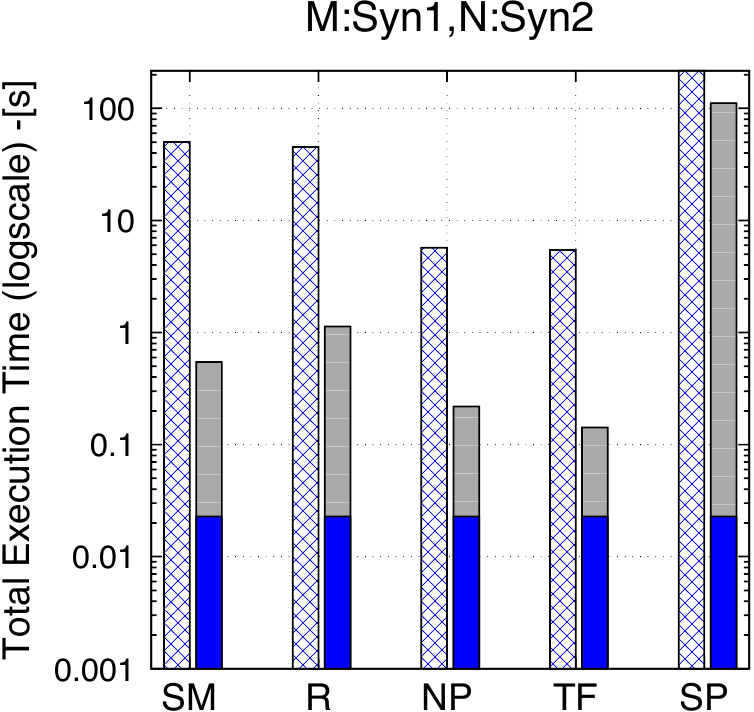}}
 \figspb\figspb\figspb
 \caption{P1.15 evaluation time with and without rewriting}
 \label{fig:la-p1.15}
\end{figure*}

\begin{figure*}[!htbp]
\figspa
  \centering
  \subfigure[\textbf{P1.16}]{\includegraphics[scale=0.25,width=4.2cm,height=3.6cm]{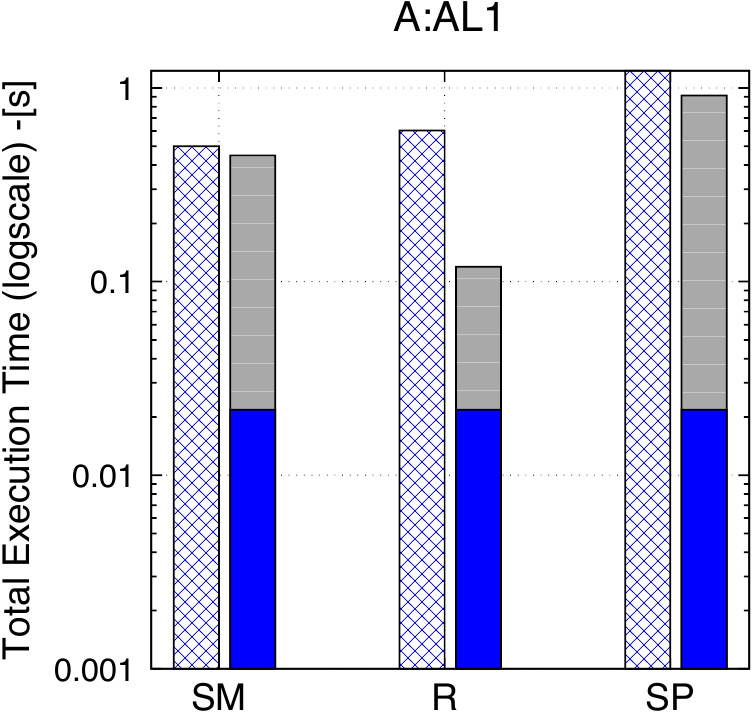}}
  \subfigure[\textbf{P1.16}]{\includegraphics[scale=0.25,width=4.4cm,height=3.6cm]{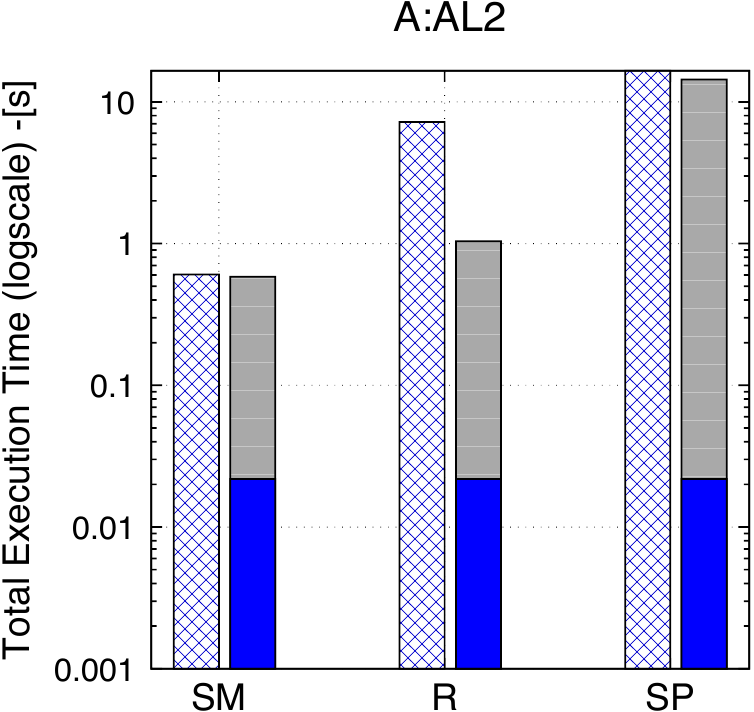}}
    \subfigure[\textbf{P1.16}]{\includegraphics[scale=0.25,width=4.2cm,height=3.6cm]{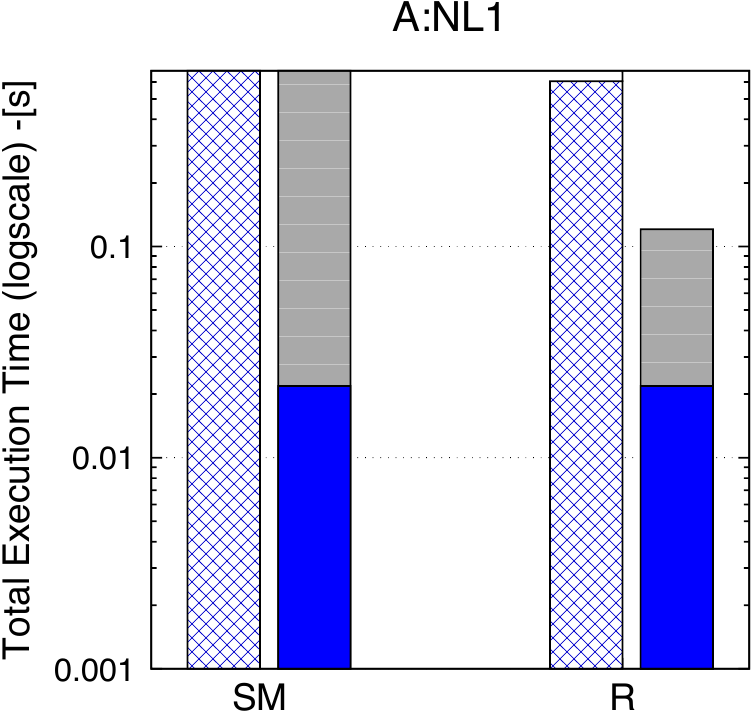}}
 \figspb\figspb\figspb
 \caption{P1.16 evaluation time with and without rewriting}
 \label{fig:la-p1.16}
\end{figure*}

\begin{figure*}[!htbp]
\figspa
  \centering
  \subfigure[\textbf{P1.16}]{\includegraphics[scale=0.25,width=4.2cm,height=3.6cm]{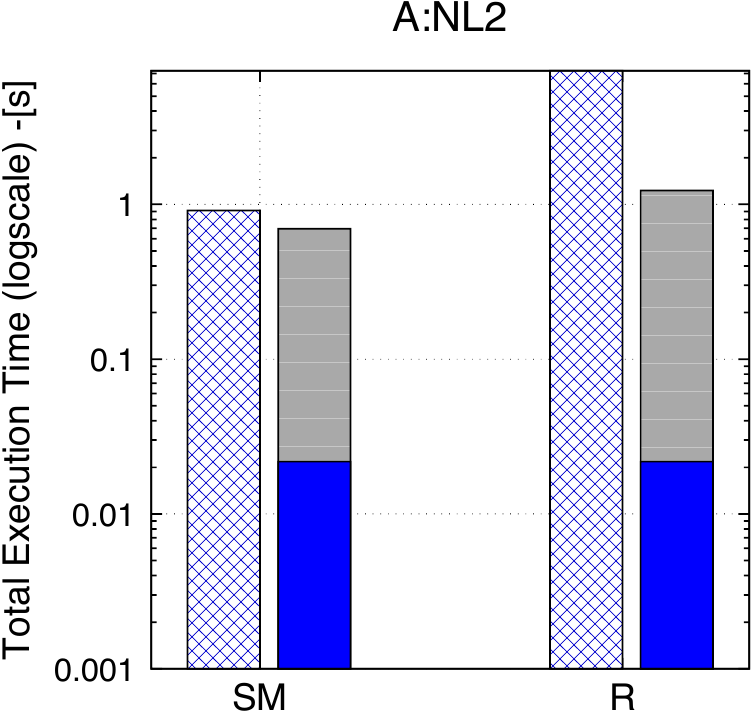}}
  \subfigure[\textbf{P1.16}]{\includegraphics[scale=0.25,width=4.4cm,height=3.6cm]{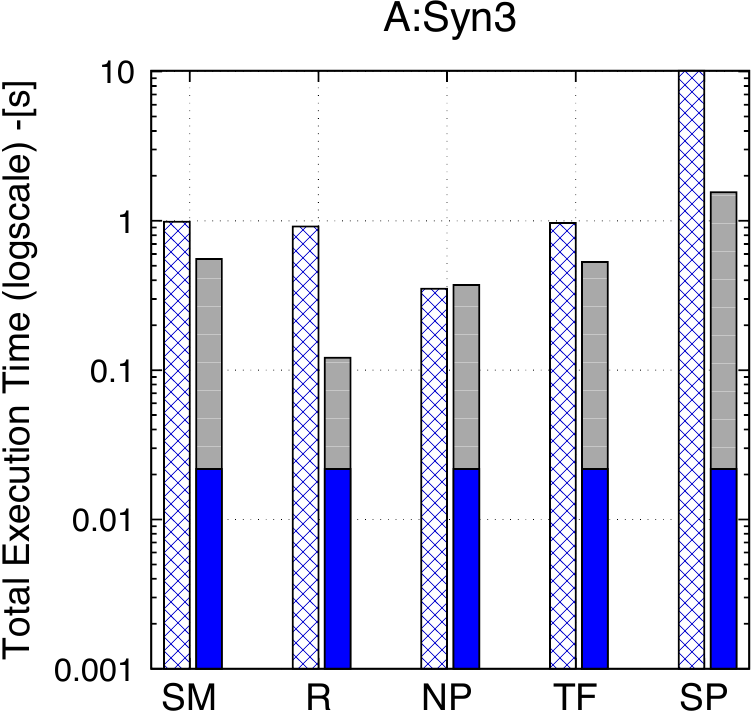}}
    \subfigure[\textbf{P1.16}]{\includegraphics[scale=0.25,width=4.2cm,height=3.6cm]{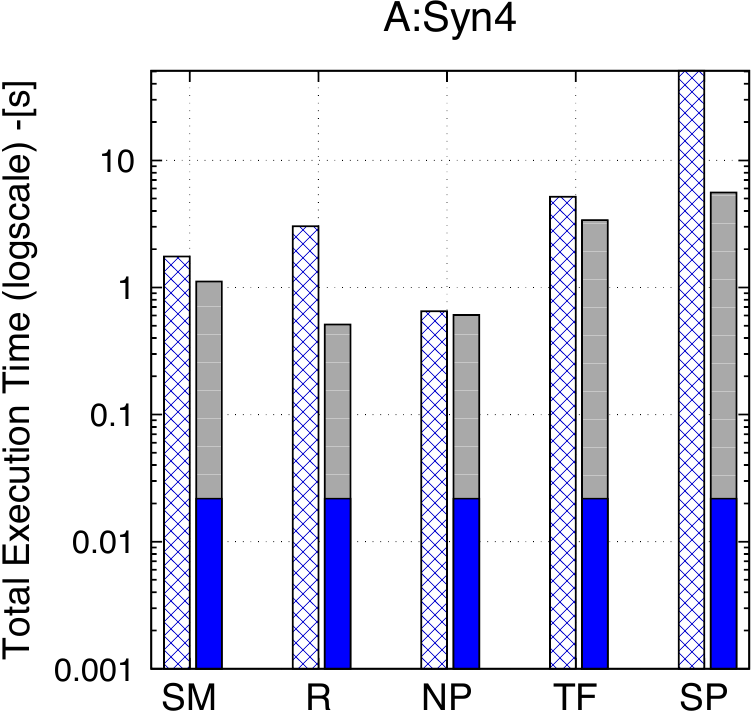}}
 \figspb\figspb\figspb
 \caption{P1.16 evaluation time with and without rewriting}
 \label{fig:la-p1.16}
\end{figure*}

\begin{figure*}[!htbp]
\figspa
  \centering
  \subfigure[\textbf{P1.17}]{\includegraphics[scale=0.25,width=4.2cm,height=3.6cm]{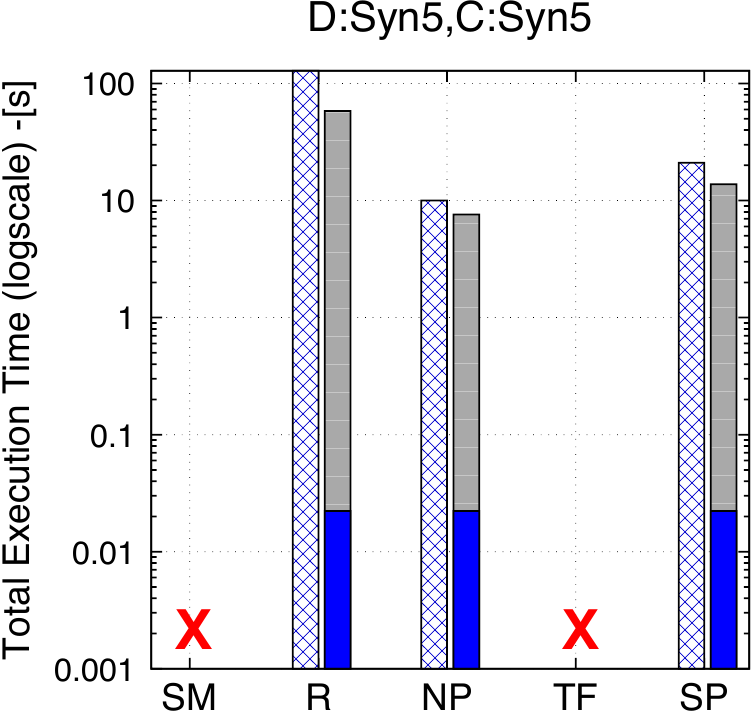}}
  \figspb\figspb\figspb
 \caption{P1.17 evaluation time with and without rewriting}
 \label{fig:la-p1.17}
\end{figure*}

\begin{figure*}[!htbp]
\figspa
  \centering
  \subfigure[\textbf{P1.18}]{\includegraphics[scale=0.25,width=4.2cm,height=3.6cm]{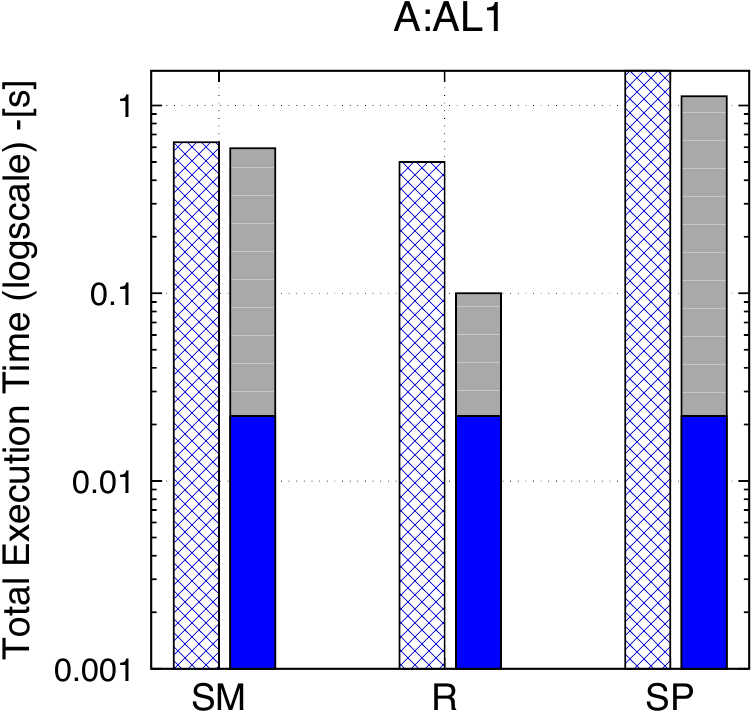}}
  \subfigure[\textbf{P1.18}]{\includegraphics[scale=0.25,width=4.4cm,height=3.6cm]{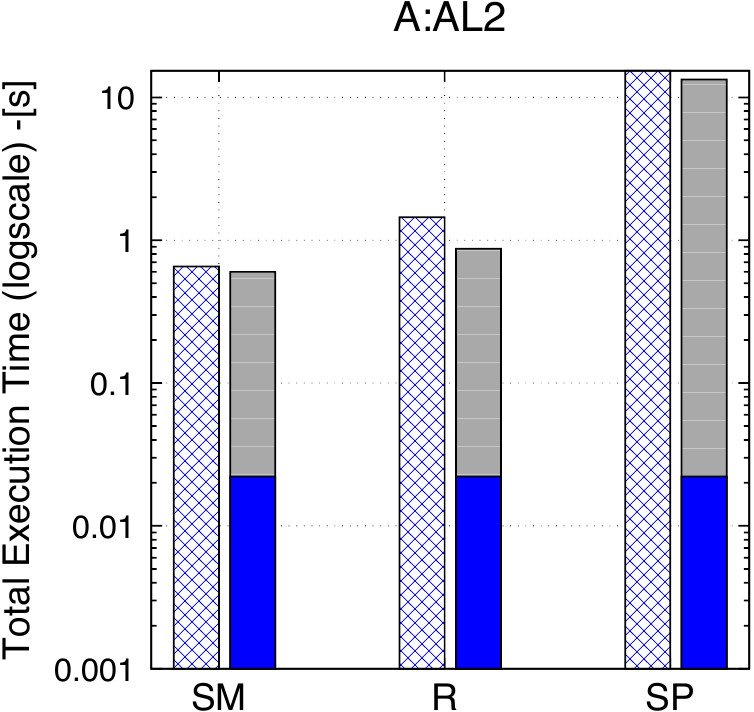}}
    \subfigure[\textbf{P1.18}]{\includegraphics[scale=0.25,width=4.2cm,height=3.6cm]{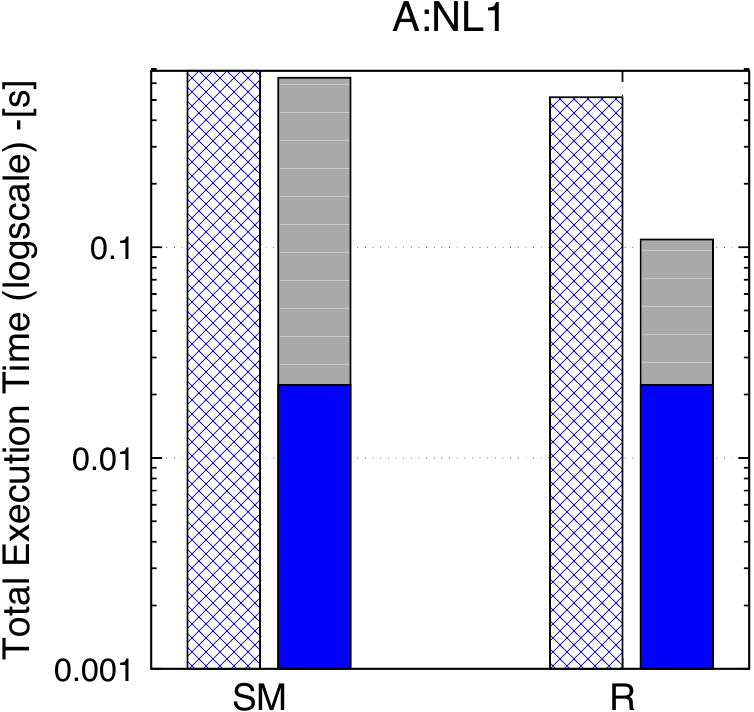}}
 \figspb\figspb\figspb
 \caption{P1.18 evaluation time with and without rewriting}
 \label{fig:la-p1.18}
\end{figure*}

\begin{figure*}[!htbp]
\figspa
  \centering
  \subfigure[\textbf{P1.18}]{\includegraphics[scale=0.25,width=4.2cm,height=3.6cm]{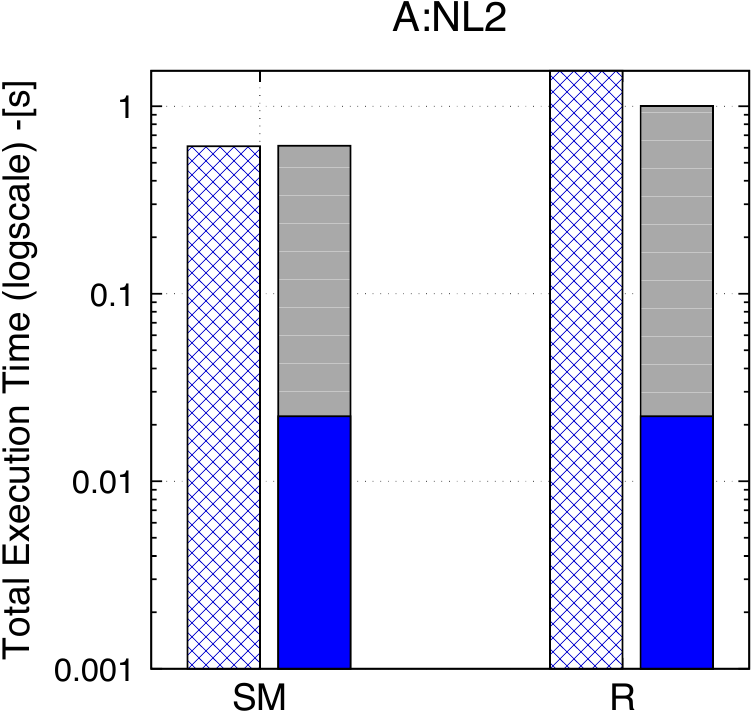}}
  \subfigure[\textbf{P1.18}]{\includegraphics[scale=0.25,width=4.4cm,height=3.6cm]{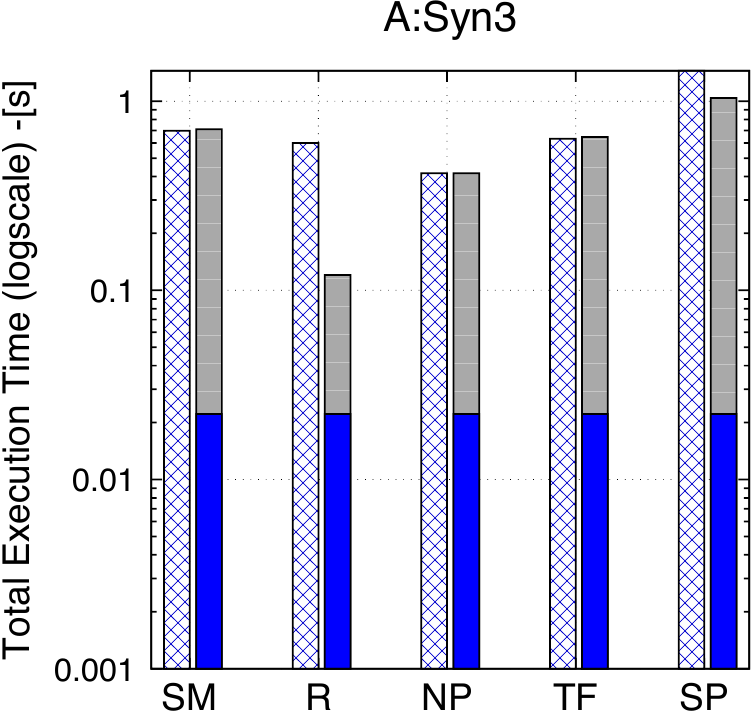}}
    \subfigure[\textbf{P1.18}]{\includegraphics[scale=0.25,width=4.2cm,height=3.6cm]{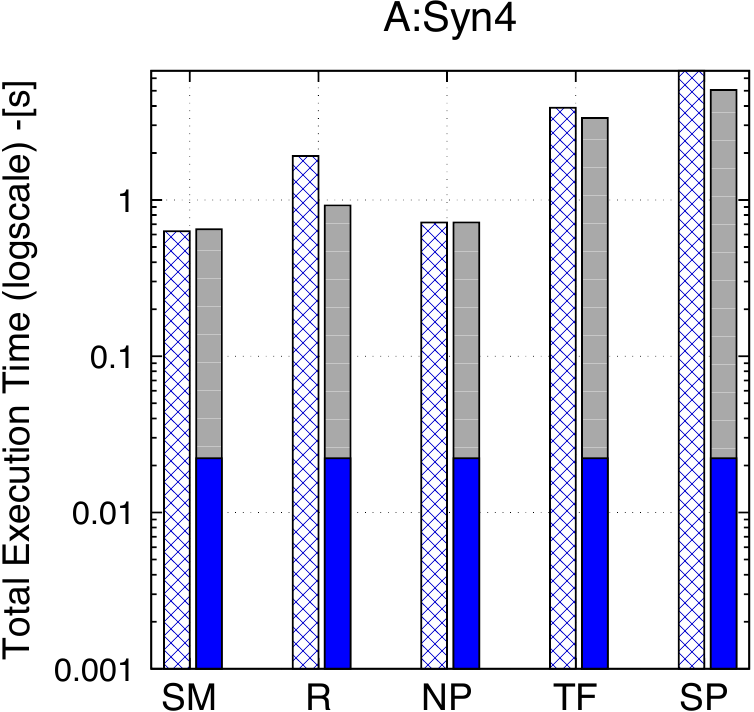}}
 \figspb\figspb\figspb
 \caption{P1.18 evaluation time with and without rewriting}
 \label{fig:la-p1.18}
\end{figure*}

\begin{figure*}[!htbp]
\figspa
  \centering
  \subfigure[\textbf{P1.25}]{\includegraphics[scale=0.25,width=4.2cm,height=3.6cm]{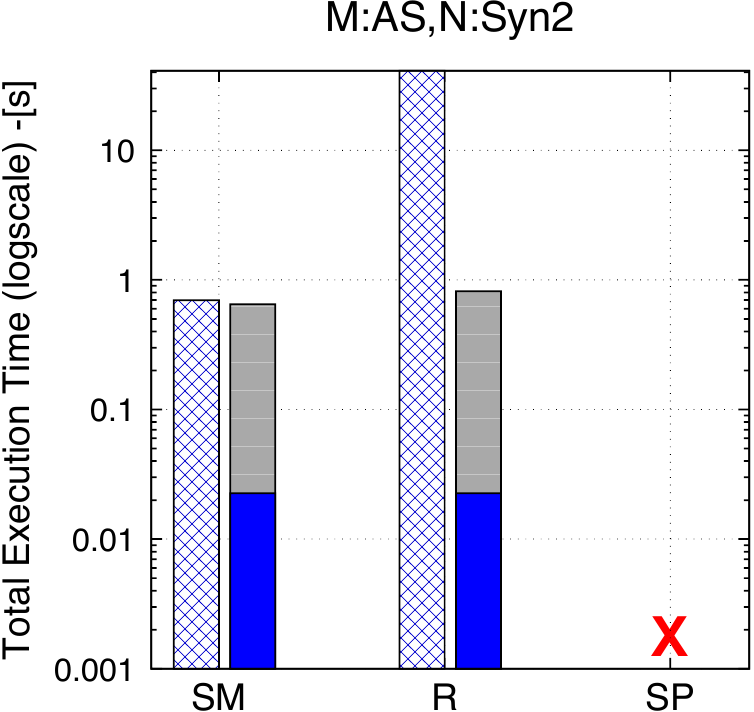}}
  \subfigure[\textbf{P1.25}]{\includegraphics[scale=0.25,width=4.4cm,height=3.6cm]{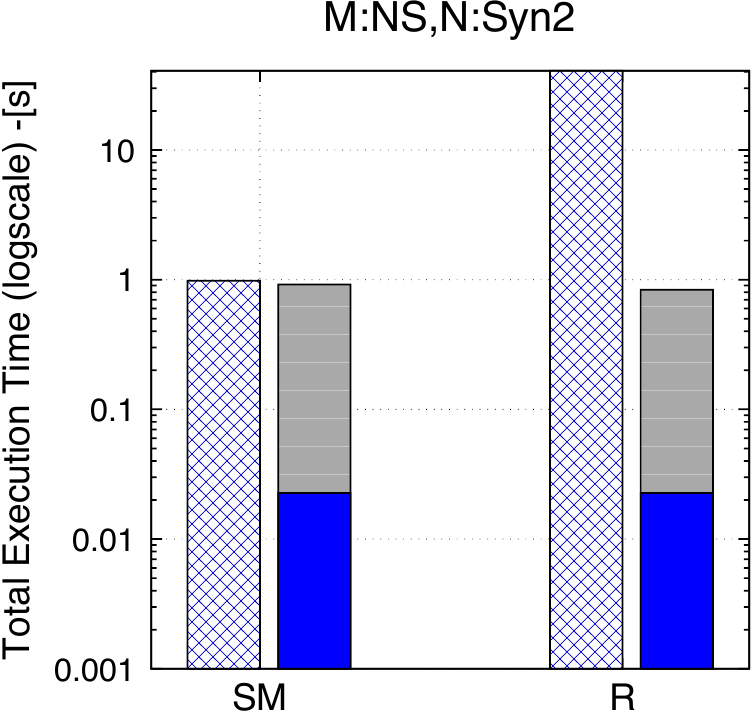}}
 \figspb\figspb\figspb
 \caption{P1.25 evaluation time with and without rewriting}
 \label{fig:la-p1.25}
\end{figure*}

\begin{figure*}[!htbp]
\figspa
  \centering
  \subfigure[\textbf{P2.1}]{\includegraphics[scale=0.25,width=4.2cm,height=3.6cm]{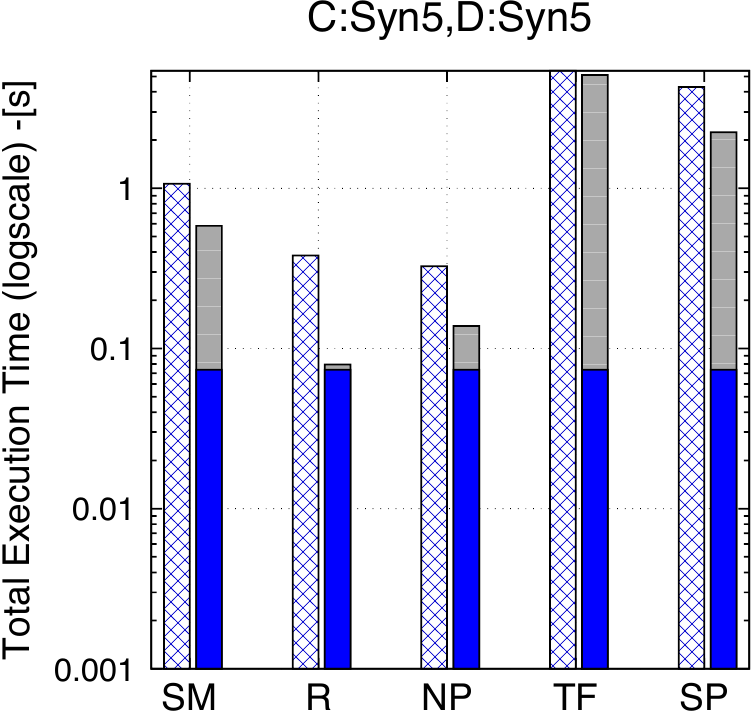}}
   \subfigure[\textbf{P2.1}]{\includegraphics[scale=0.25,width=4.2cm,height=3.6cm]{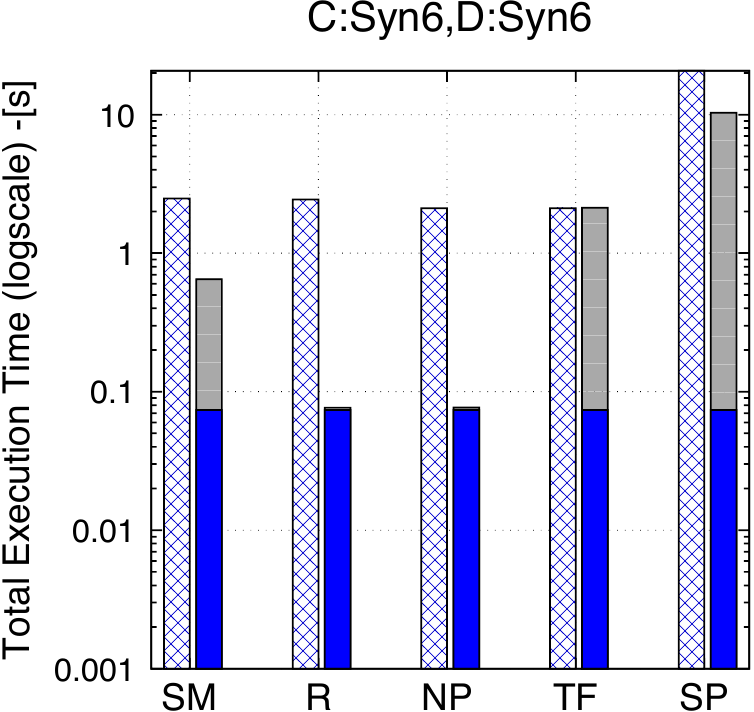}}
  \figspb\figspb\figspb
 \caption{P2.1 evaluation time with and without rewriting}
 \label{fig:la-p2.1}
\end{figure*}

\begin{figure*}[!htbp]
\figspa
  \centering
  \subfigure[\textbf{P2.2}]{\includegraphics[scale=0.25,width=4.2cm,height=3.6cm]{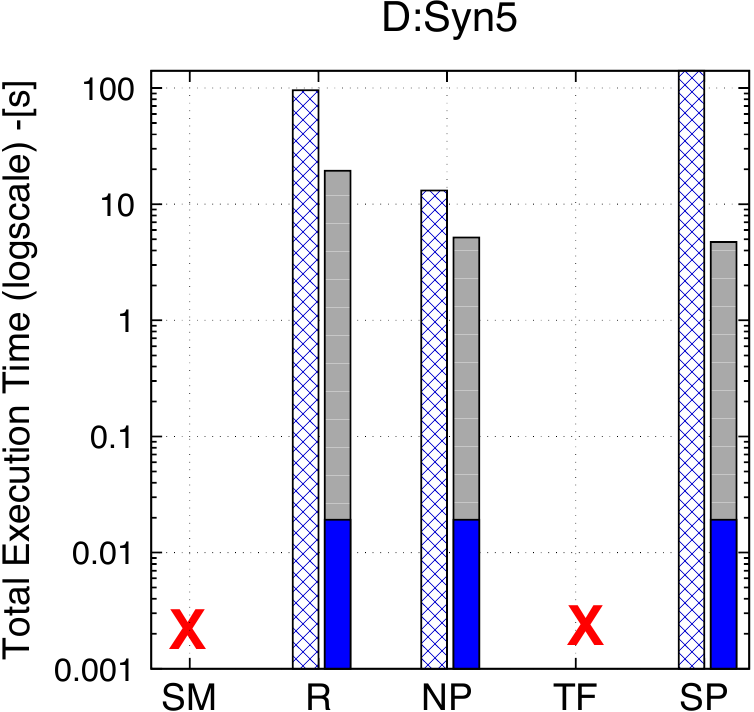}}
  \figspb\figspb\figspb
 \caption{P2.2 evaluation time with and without rewriting}
 \label{fig:la-p2.2}
\end{figure*}

\begin{figure*}[!htbp]
\figspa
  \centering
  \subfigure[\textbf{P2.3}]{\includegraphics[scale=0.25,width=4.2cm,height=3.6cm]{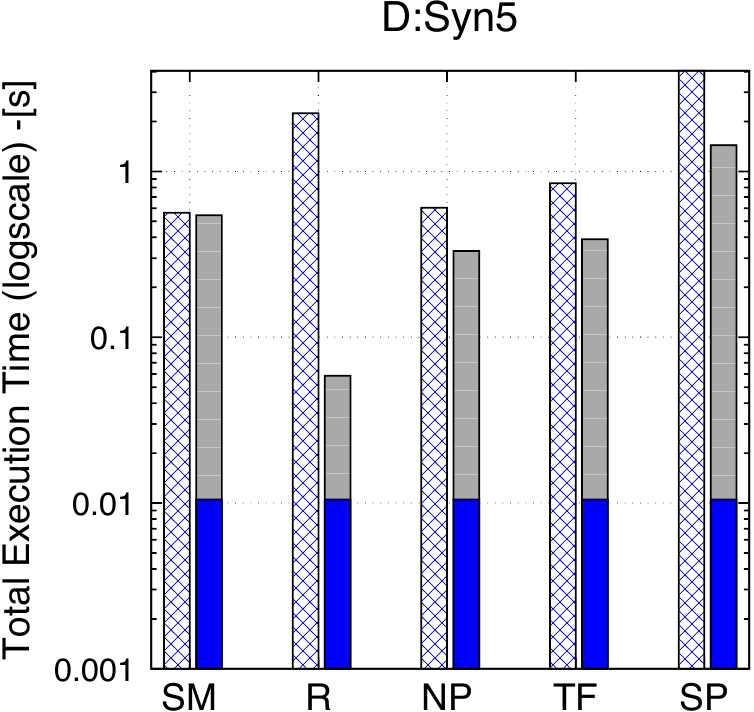}}
   \subfigure[\textbf{P2.3}]{\includegraphics[scale=0.25,width=4.2cm,height=3.6cm]{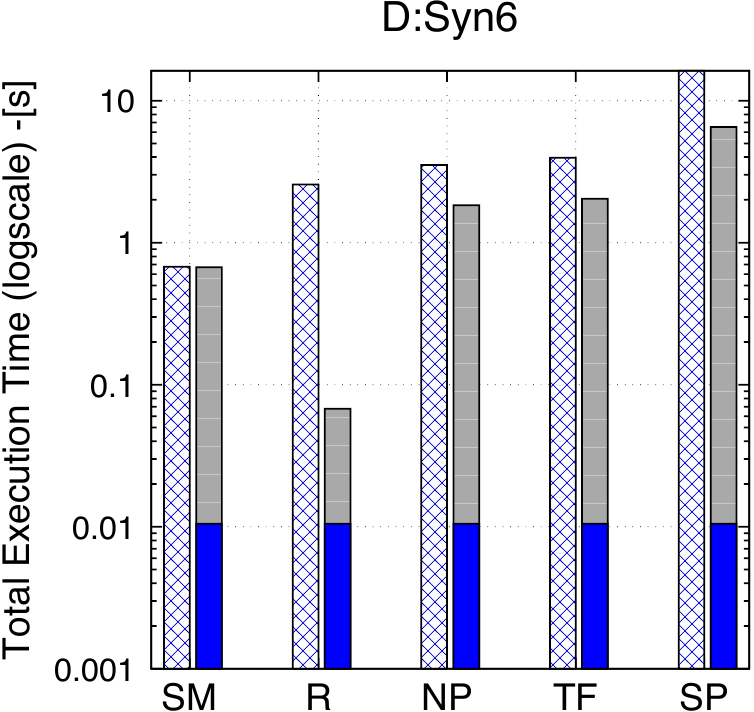}}
  \figspb\figspb\figspb
 \caption{P2.3 evaluation time with and without rewriting}
 \label{fig:la-p2.3}
\end{figure*}

\begin{figure*}[!htbp]
\figspa
  \centering
  \subfigure[\textbf{P2.4}]{\includegraphics[scale=0.25,width=4.2cm,height=3.6cm]{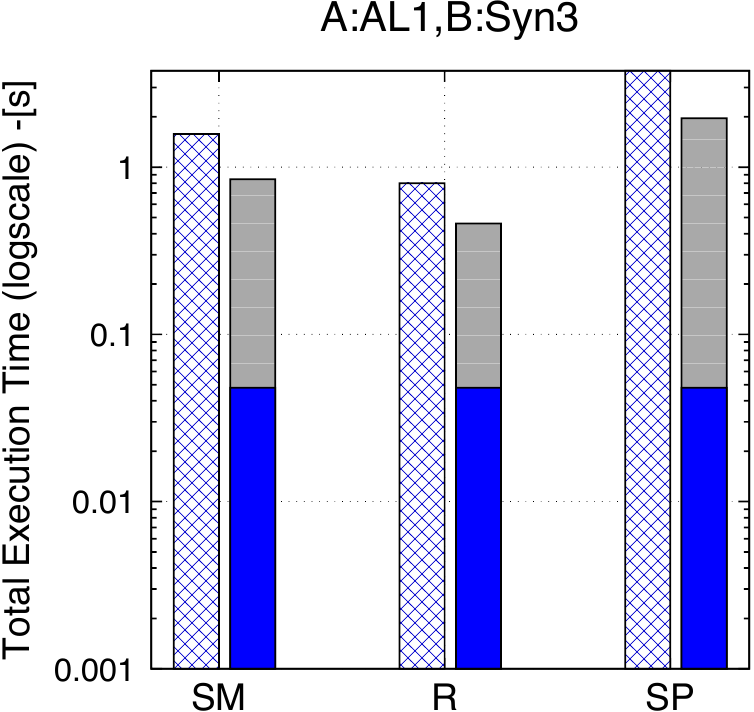}}
  \subfigure[\textbf{P2.4}]{\includegraphics[scale=0.25,width=4.4cm,height=3.6cm]{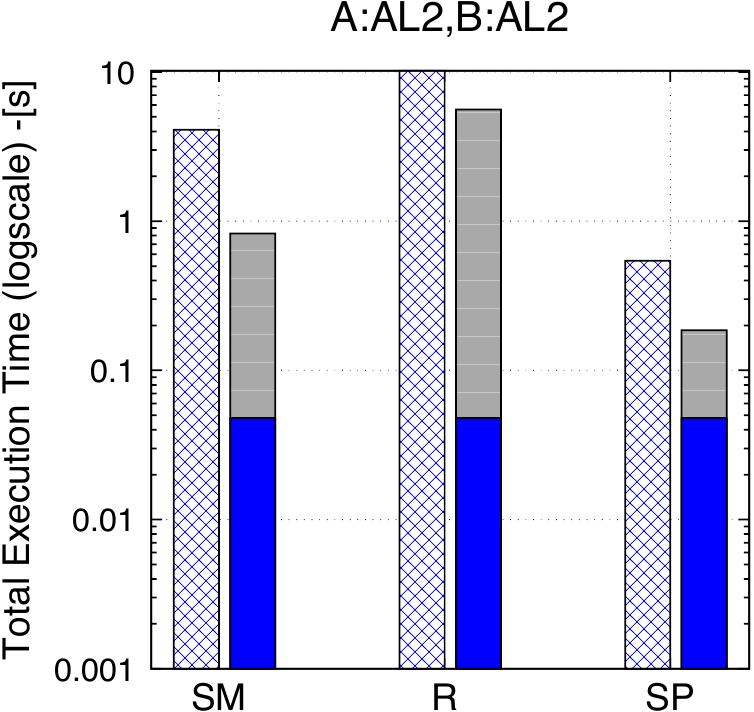}}
    \subfigure[\textbf{P2.4}]{\includegraphics[scale=0.25,width=4.2cm,height=3.6cm]{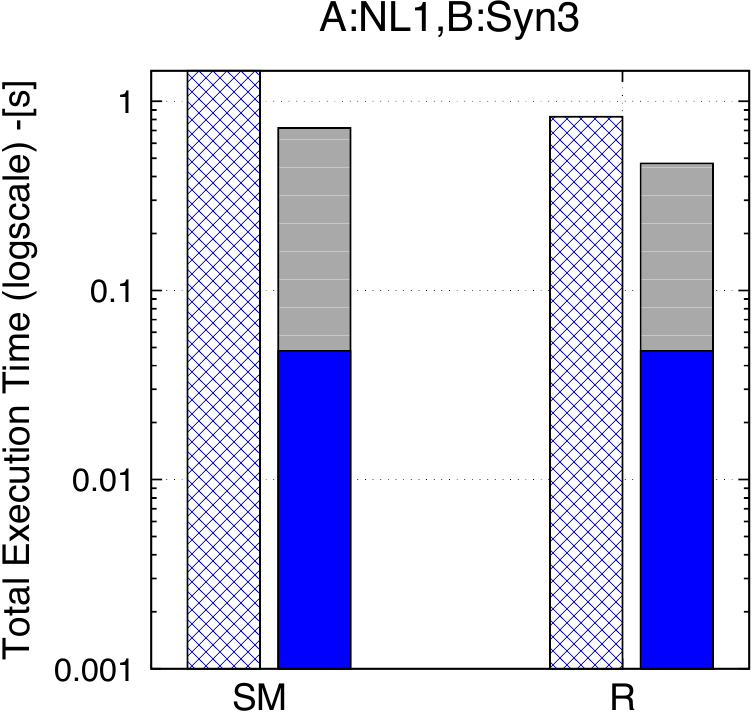}}
 \figspb\figspb\figspb
 \caption{P2.4 evaluation time with and without rewriting}
 \label{fig:la-p2.4}
\end{figure*}

\begin{figure*}[!htbp]
\figspa
  \centering
  \subfigure[\textbf{P2.4}]{\includegraphics[scale=0.25,width=4.2cm,height=3.6cm]{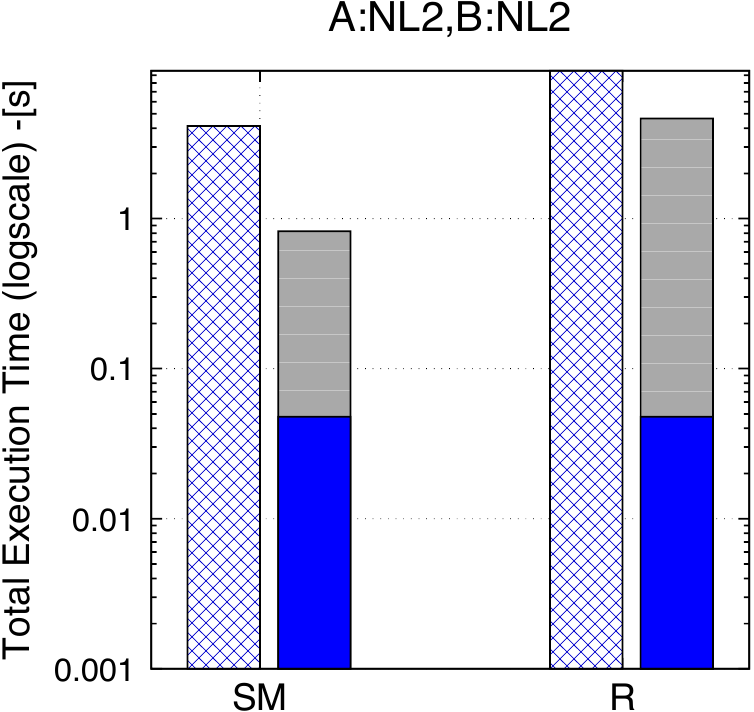}}
  \subfigure[\textbf{P2.4}]{\includegraphics[scale=0.25,width=4.4cm,height=3.6cm]{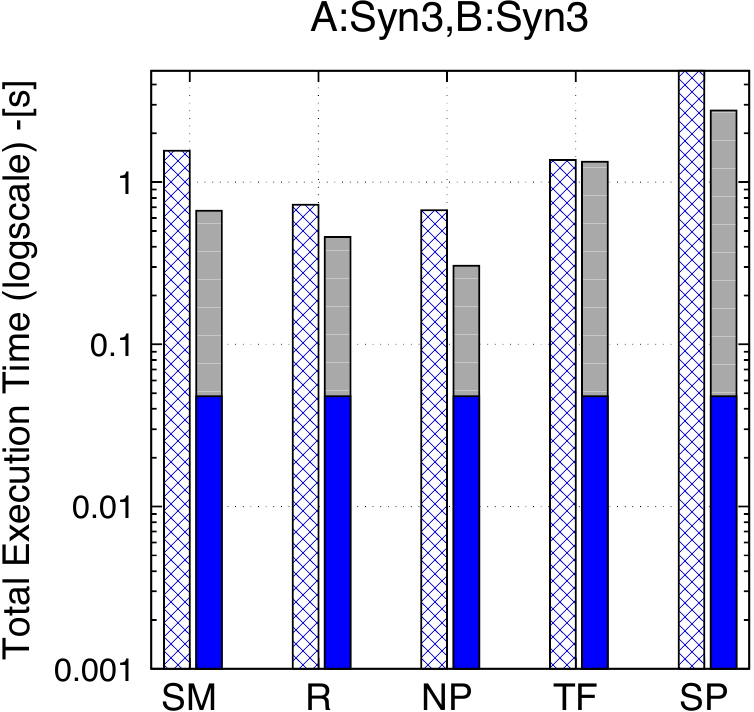}}
    \subfigure[\textbf{P2.4}]{\includegraphics[scale=0.25,width=4.2cm,height=3.6cm]{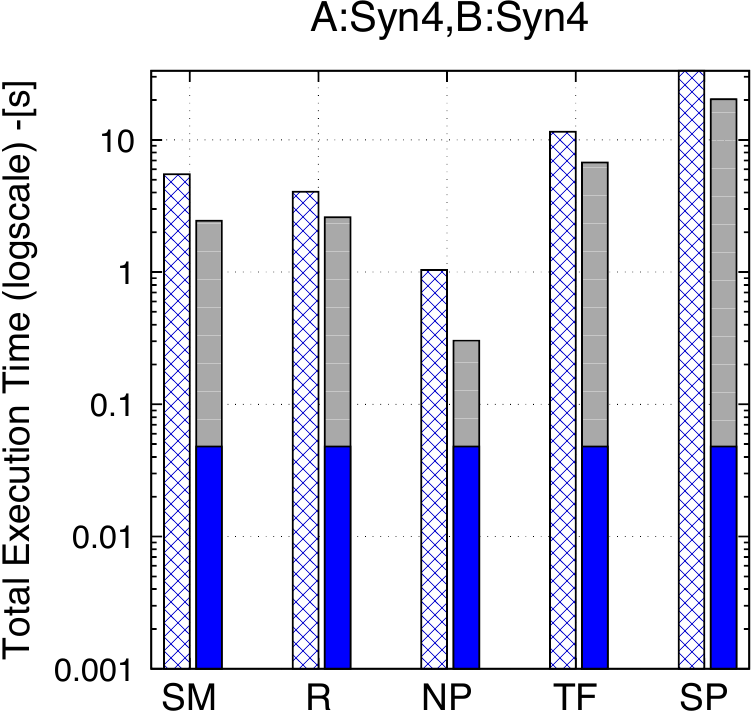}}
 \figspb\figspb\figspb
 \caption{P2.4 evaluation time with and without rewriting}
 \figspc
 \label{fig:la-p2.4}
\end{figure*}

\begin{figure*}[!htbp]
\figspa
  \centering
  \subfigure[\textbf{P2.5}]{\includegraphics[scale=0.25,width=4.2cm,height=3.6cm]{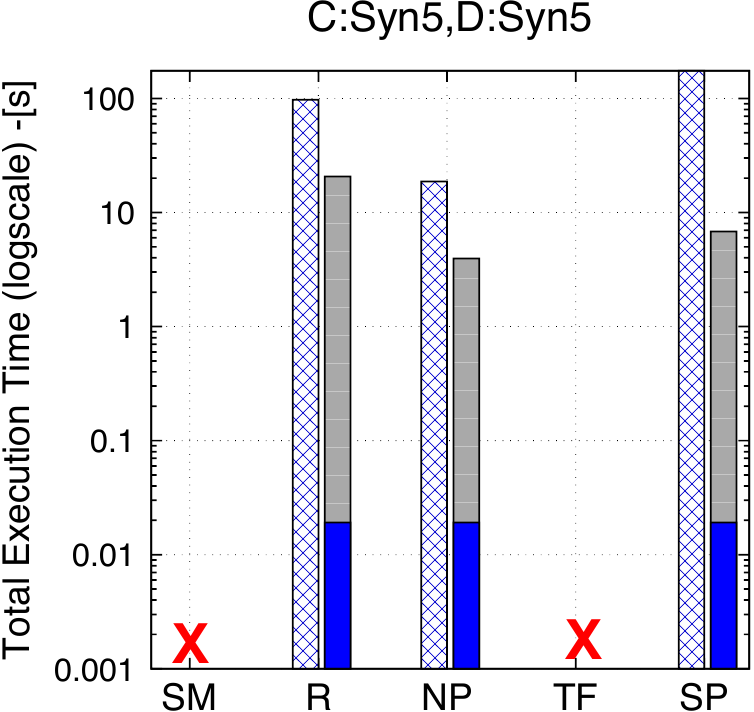}}
  \subfigure[\textbf{P2.6}]{\includegraphics[scale=0.25,width=4.2cm,height=3.6cm]{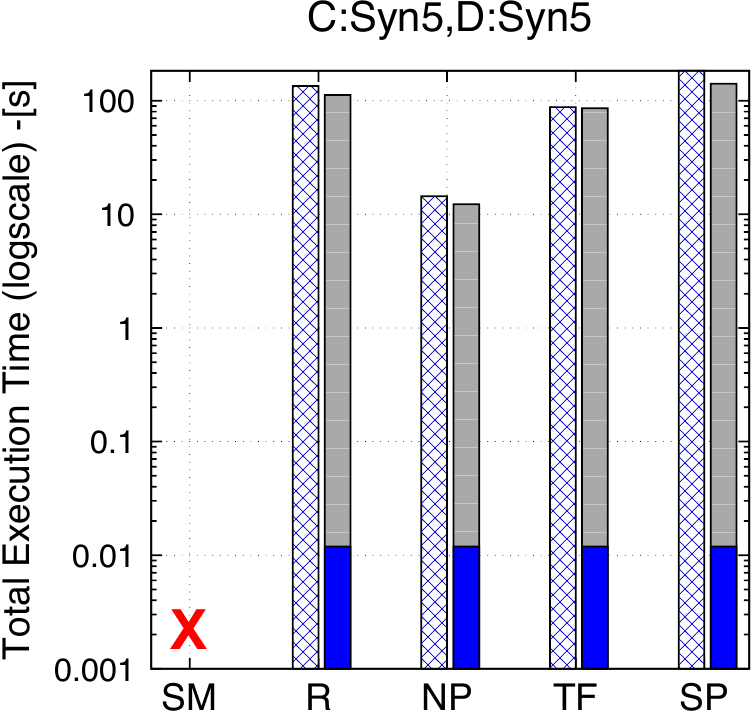}}
  \subfigure[\textbf{P2.8}]{\includegraphics[scale=0.25,width=4.2cm,height=3.6cm]{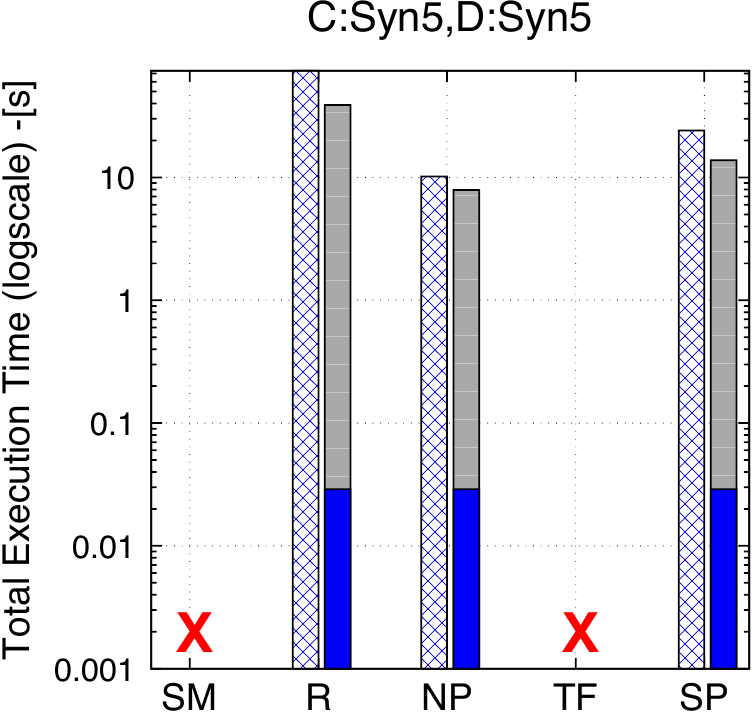}}
  \figspb\figspb\figspb
 \caption{P2.5, P2.6 and P2.8 evaluation time with and without rewriting}
 \label{fig:la-p2.568}
\end{figure*}

\begin{figure*}[!htbp]
\figspa
  \centering
  \subfigure[\textbf{P2.9}]{\includegraphics[scale=0.25,width=4.2cm,height=3.6cm]{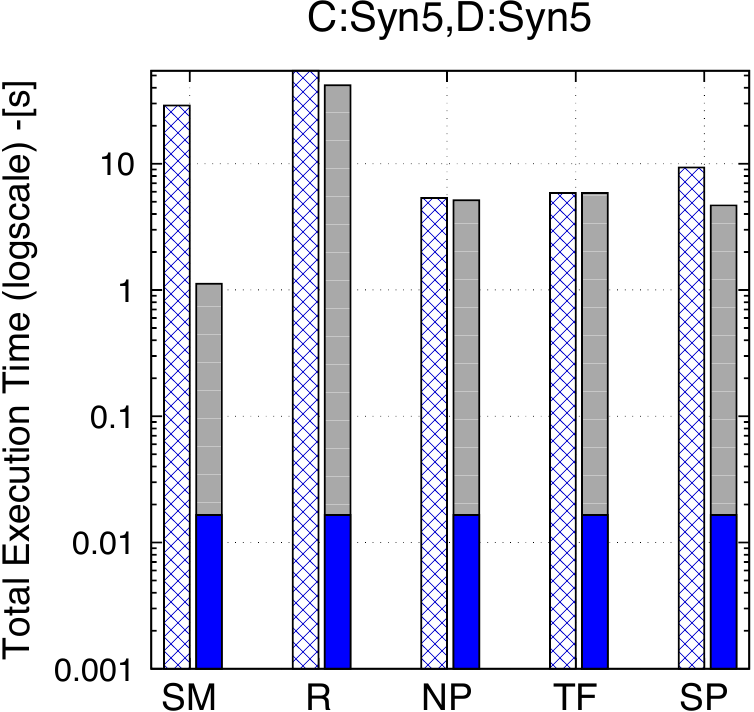}}
  \subfigure[\textbf{P2.9S}]{\includegraphics[scale=0.25,width=4.2cm,height=3.6cm]{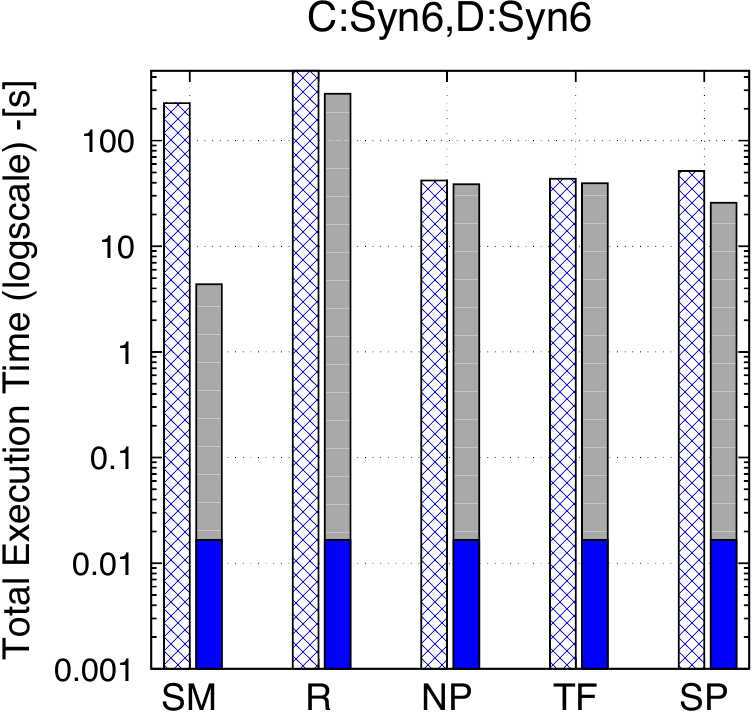}}
   \figspb\figspb\figspb
 \caption{P2.9 evaluation time with and without rewriting}
 \label{fig:la-p2.9}
\end{figure*}

\begin{figure*}[!htbp]
\figspa
  \centering
  \subfigure[\textbf{P2.10}]{\includegraphics[scale=0.25,width=4.2cm,height=3.6cm]{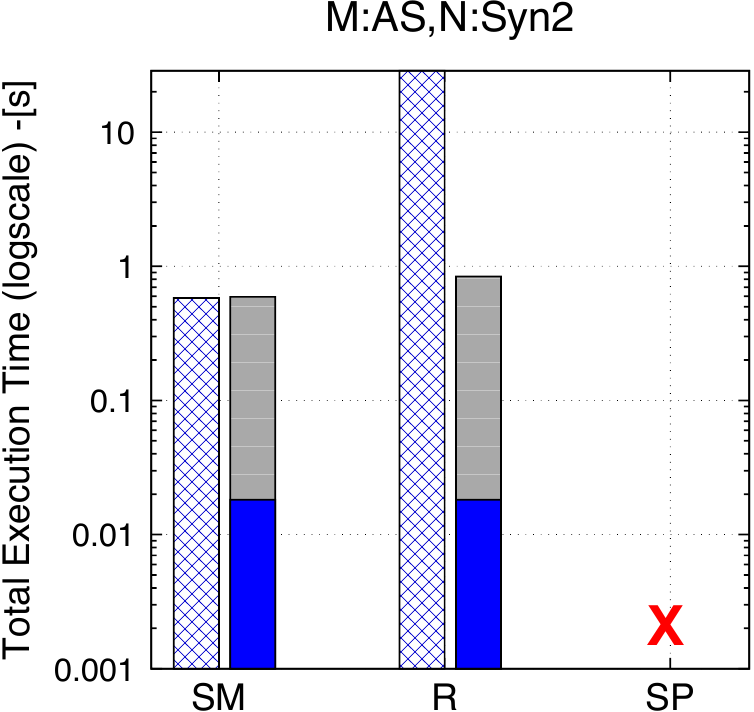}}
  \subfigure[\textbf{P2.10}]{\includegraphics[scale=0.25,width=4.4cm,height=3.6cm]{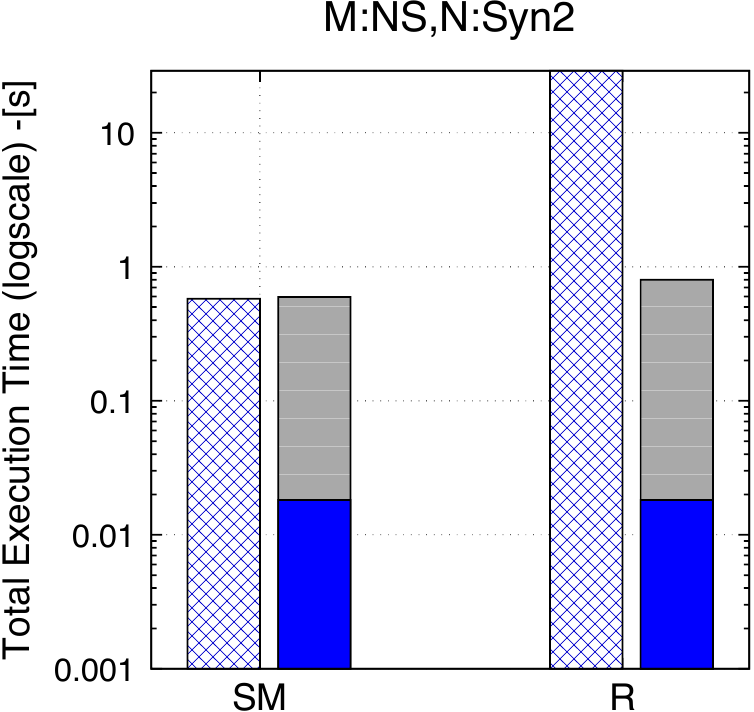}}
    \subfigure[\textbf{P2.10}]{\includegraphics[scale=0.25,width=4.2cm,height=3.6cm]{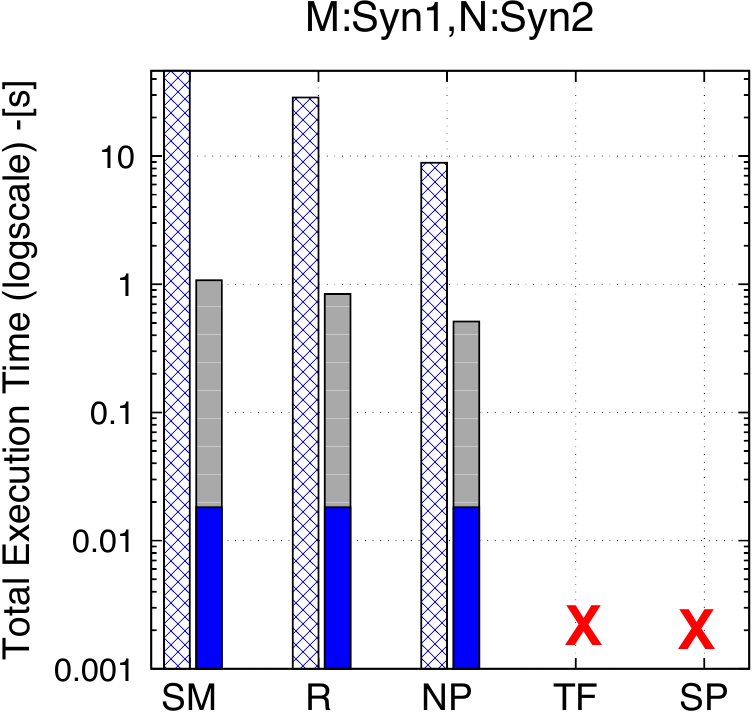}}
 \figspb\figspb\figspb
 \caption{P2.10 evaluation time with and without rewriting}
 \label{fig:la-p2.10}
\end{figure*}

\begin{figure*}[!htbp]
\figspa
  \centering
  \subfigure[\textbf{P2.11}]{\includegraphics[scale=0.25,width=4.2cm,height=3.6cm]{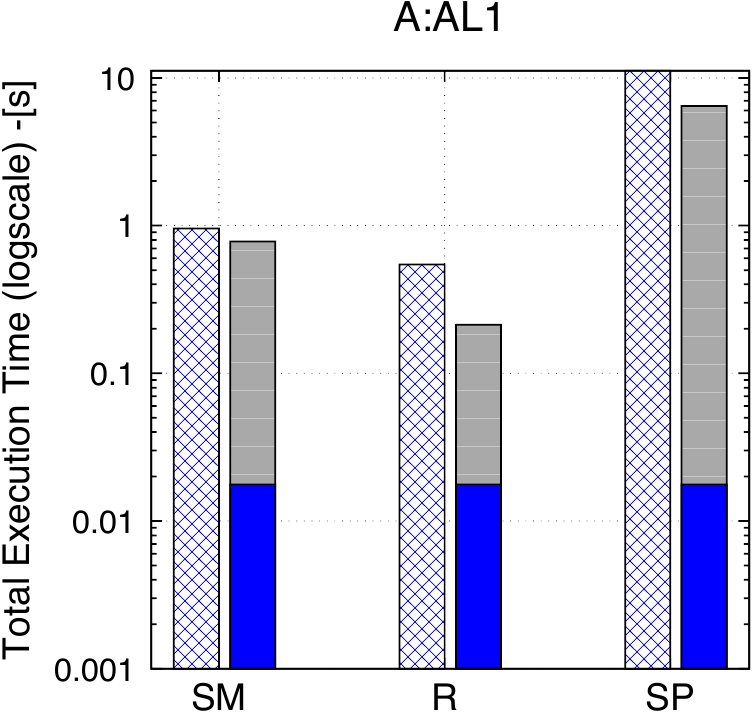}}
  \subfigure[\textbf{P2.11}]{\includegraphics[scale=0.25,width=4.4cm,height=3.6cm]{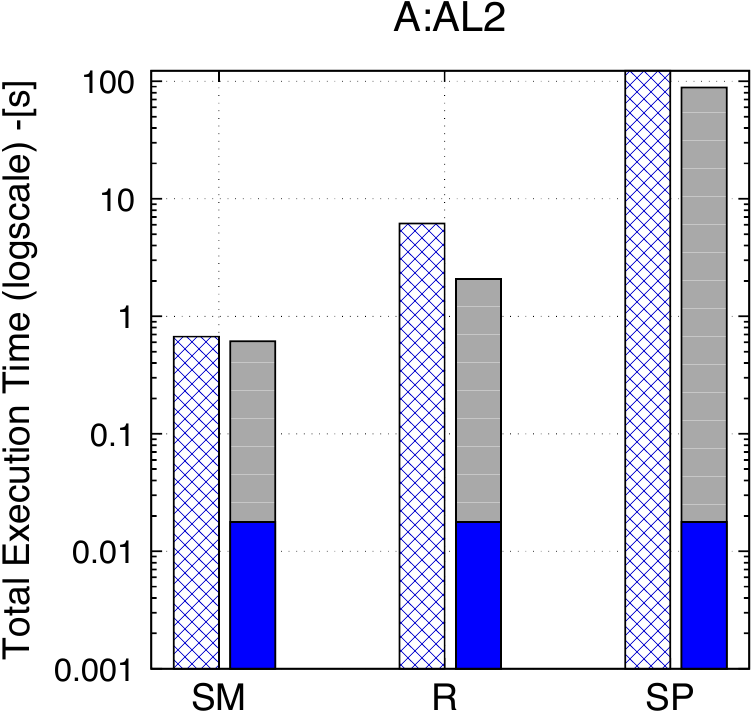}}
    \subfigure[\textbf{P2.11}]{\includegraphics[scale=0.25,width=4.2cm,height=3.6cm]{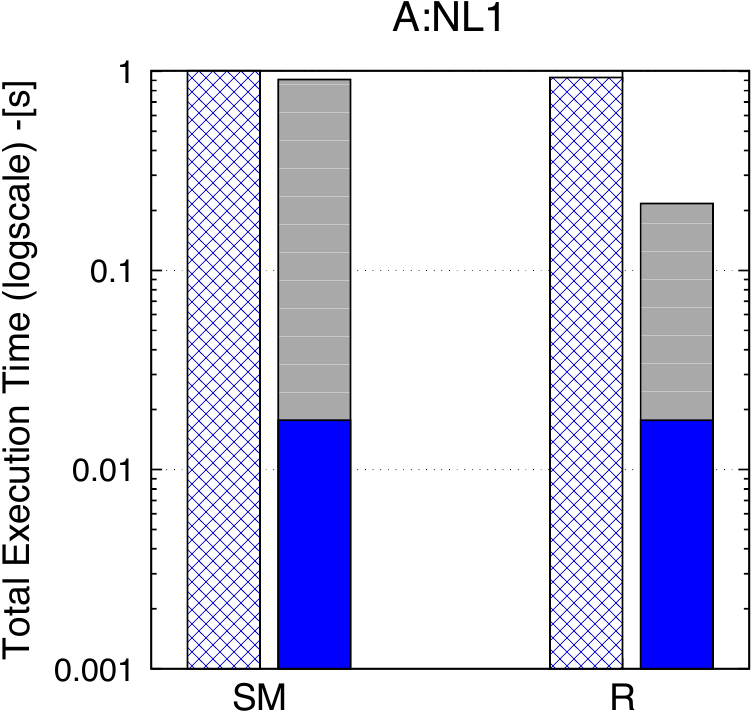}}
 \figspb\figspb\figspb
 \caption{P2.11 evaluation time with and without rewriting}
 \label{fig:la-p2.11}
\end{figure*}

\begin{figure*}[!htbp]
\figspa
  \centering
  \subfigure[\textbf{P2.11}]{\includegraphics[scale=0.25,width=4.2cm,height=3.6cm]{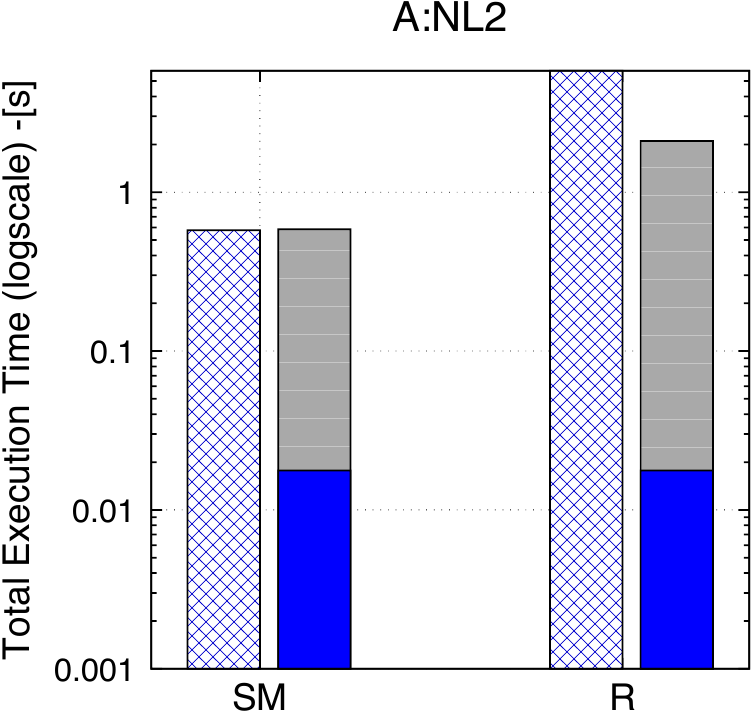}}
  \subfigure[\textbf{P2.11}]{\includegraphics[scale=0.25,width=4.4cm,height=3.6cm]{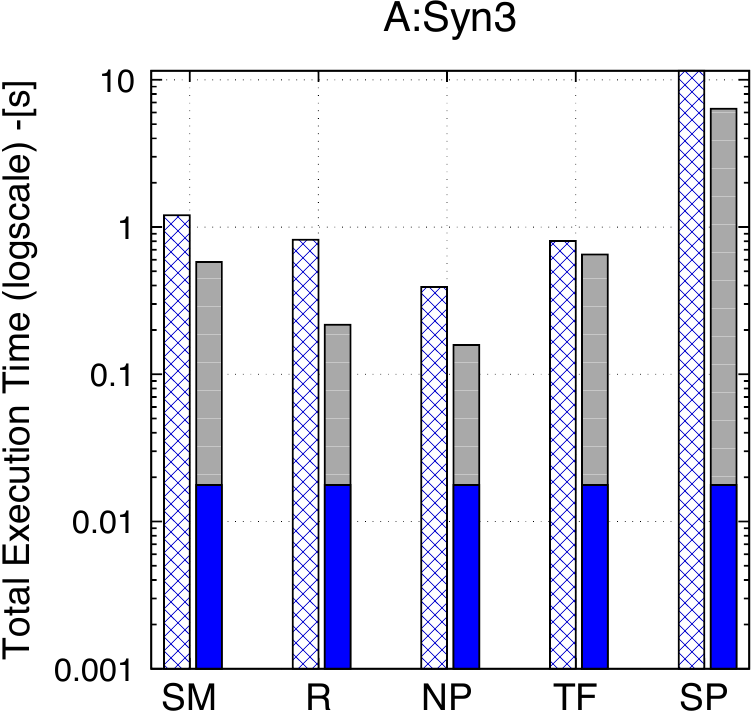}}
    \subfigure[\textbf{P2.11}]{\includegraphics[scale=0.25,width=4.2cm,height=3.6cm]{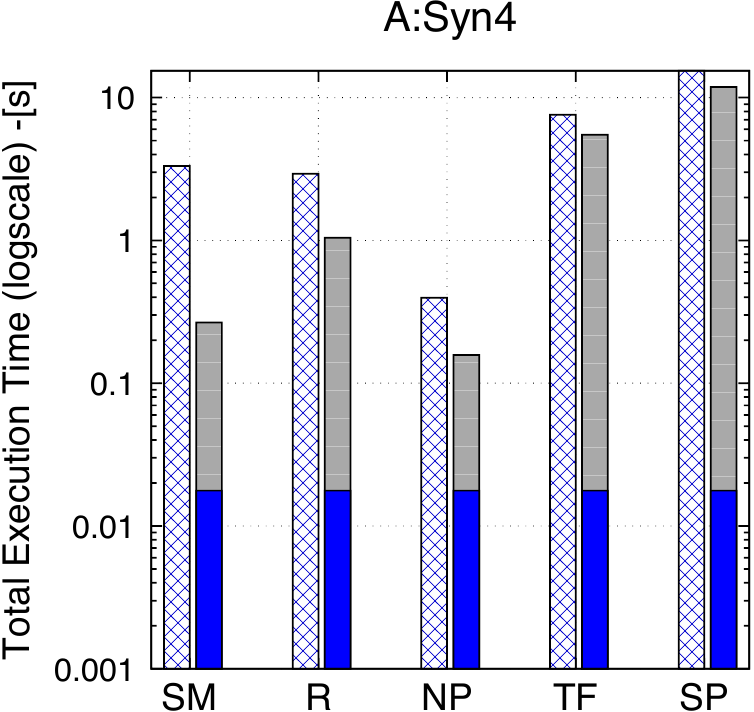}}
 \figspb\figspb\figspb
 \caption{P2.11 evaluation time with and without rewriting}
 \label{fig:la-p2.11}
\end{figure*}

\begin{figure*}[!htbp]
\figspa
  \centering
  \subfigure[\textbf{P2.13}]{\includegraphics[scale=0.25,width=4.2cm,height=3.6cm]{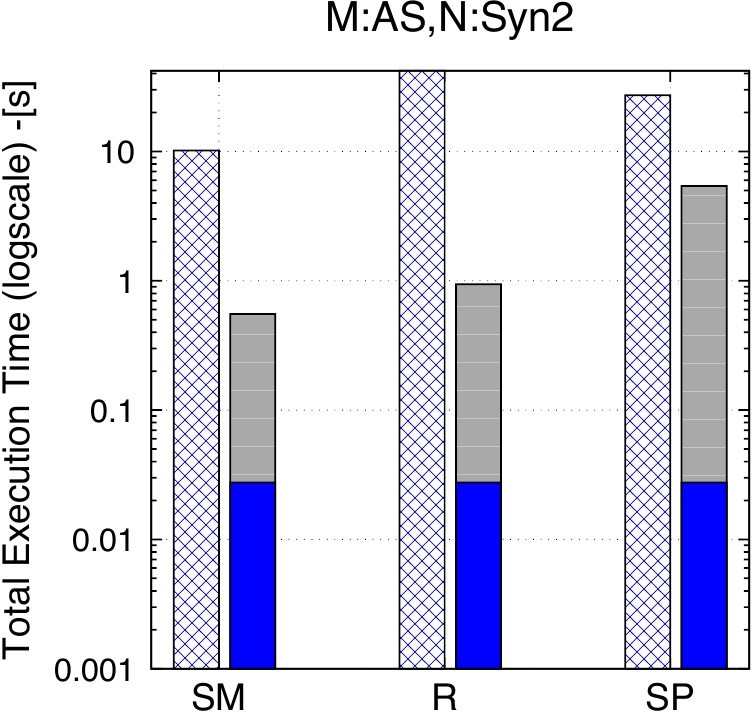}}
  \subfigure[\textbf{P2.13}]{\includegraphics[scale=0.25,width=4.4cm,height=3.6cm]{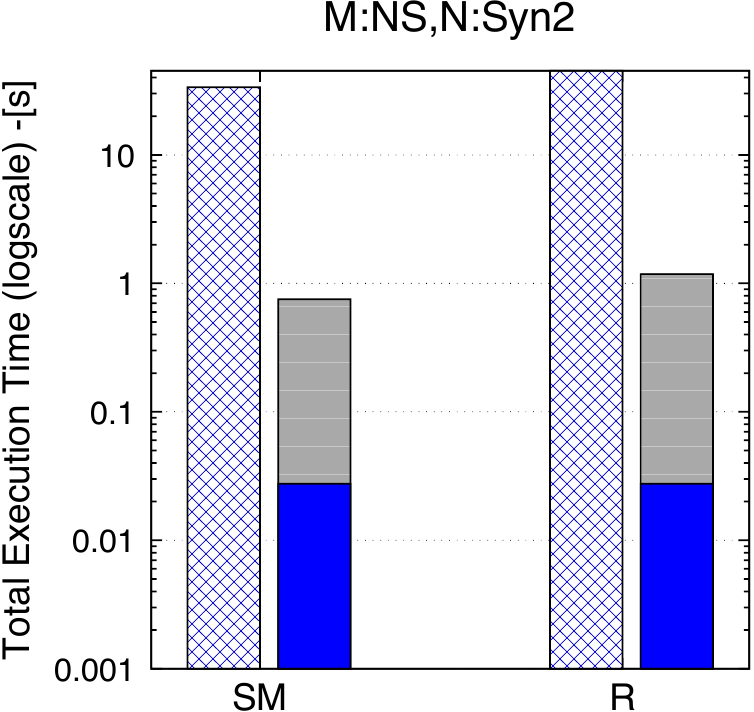}}
   \subfigure[\textbf{P2.13}]{\includegraphics[scale=0.25,width=4.4cm,height=3.6cm]{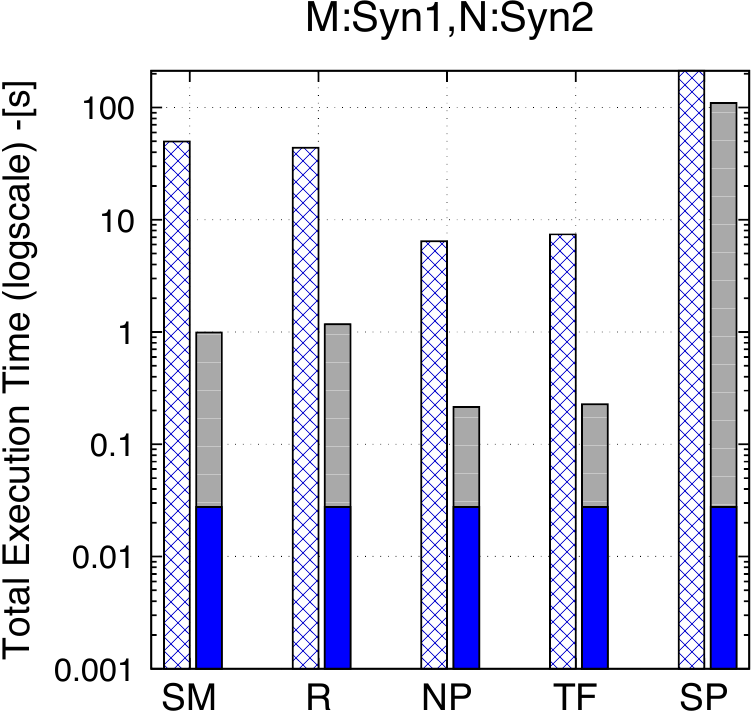}}
 \figspb\figspb\figspb
 \caption{P2.13 evaluation time with and without rewriting}
 \label{fig:la-p2.13}
\end{figure*}

\begin{figure*}[!htbp]
\figspa
  \centering
  \subfigure[\textbf{P2.14}]{\includegraphics[scale=0.25,width=4.2cm,height=3.6cm]{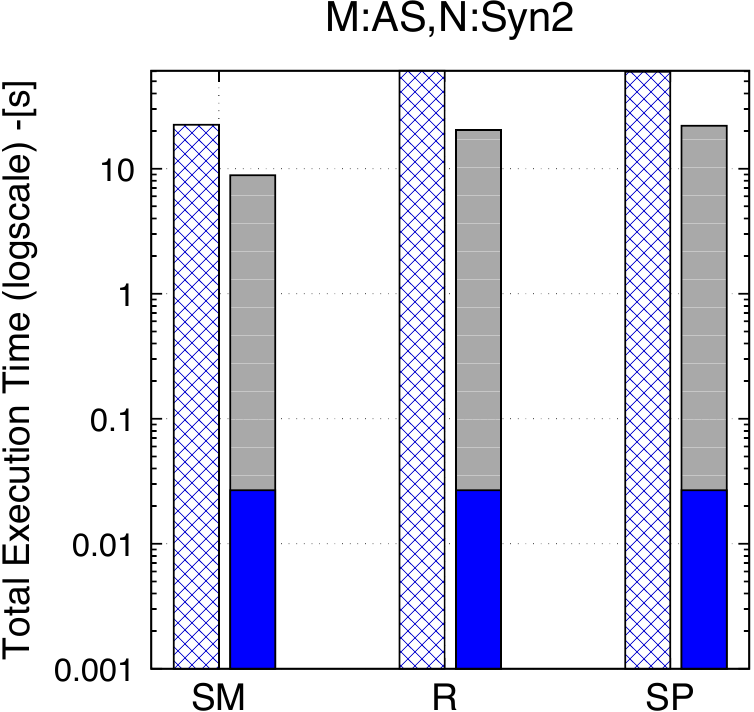}}
  \subfigure[\textbf{P2.14}]{\includegraphics[scale=0.25,width=4.4cm,height=3.6cm]{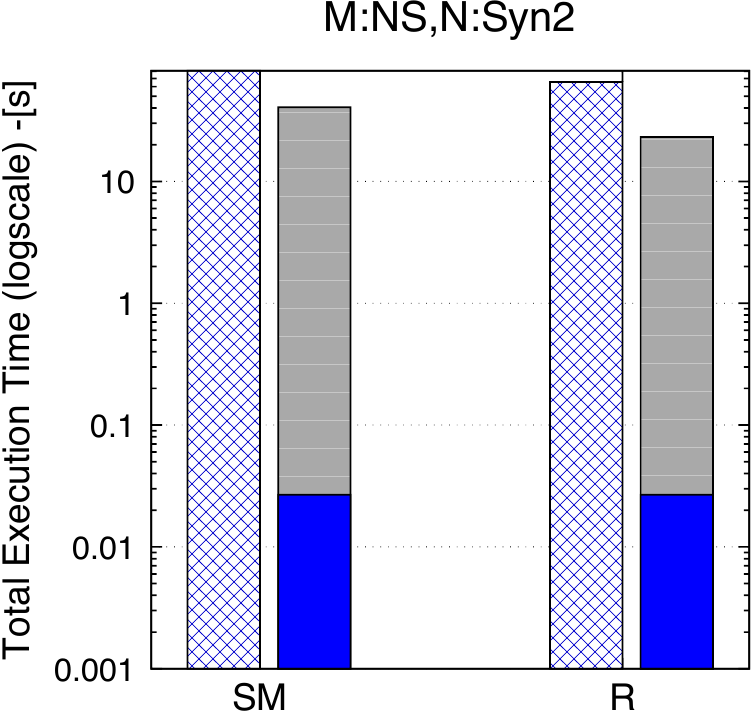}}
 \figspb\figspb\figspb
 \caption{P2.14 evaluation time with and without rewriting}
 \label{fig:la-p2.14}
\end{figure*}

\begin{figure*}[!htbp]
\figspa
  \centering
  \subfigure[\textbf{P2.15}]{\includegraphics[scale=0.25,width=4.2cm,height=3.6cm]{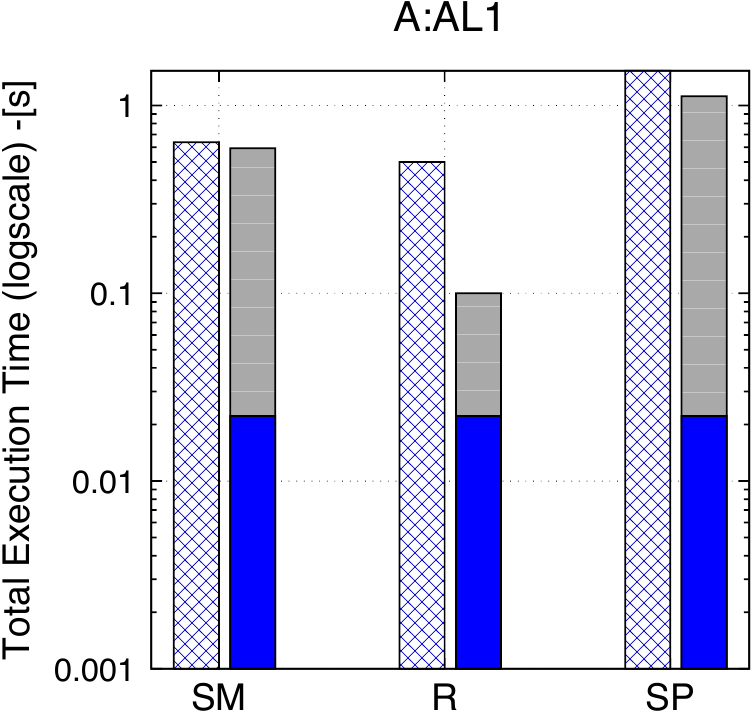}}
  \subfigure[\textbf{P2.15}]{\includegraphics[scale=0.25,width=4.4cm,height=3.6cm]{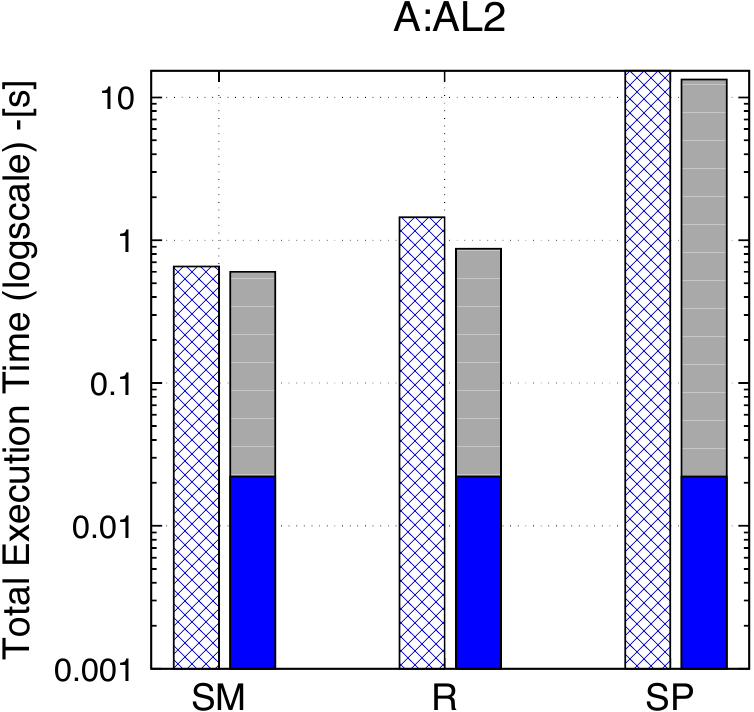}}
    \subfigure[\textbf{P2.15}]{\includegraphics[scale=0.25,width=4.2cm,height=3.6cm]{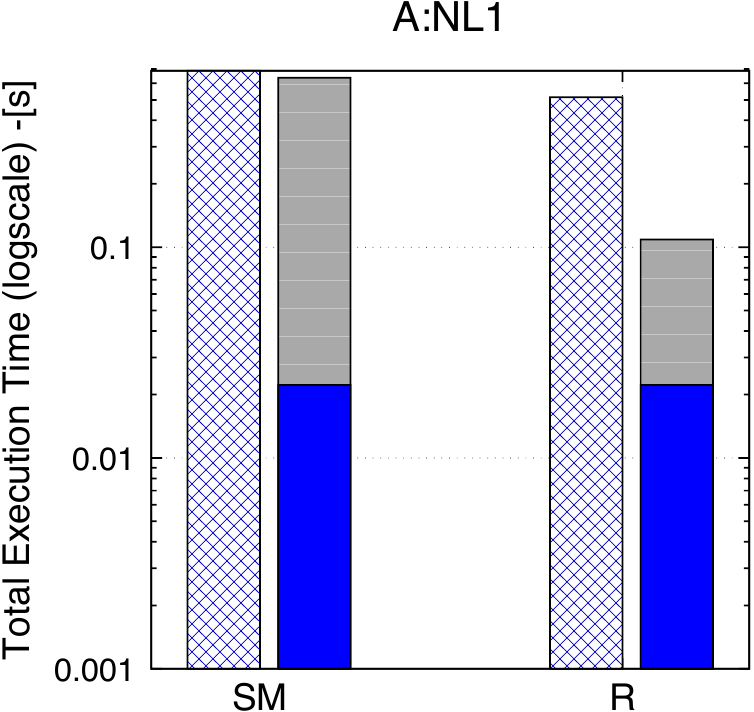}}
 \figspb\figspb\figspb
 \caption{P2.15 evaluation time with and without rewriting}
 \label{fig:la-p2.15}
\end{figure*}

\begin{figure*}[!htbp]
\figspa
  \centering
  \subfigure[\textbf{P2.15}]{\includegraphics[scale=0.25,width=4.2cm,height=3.6cm]{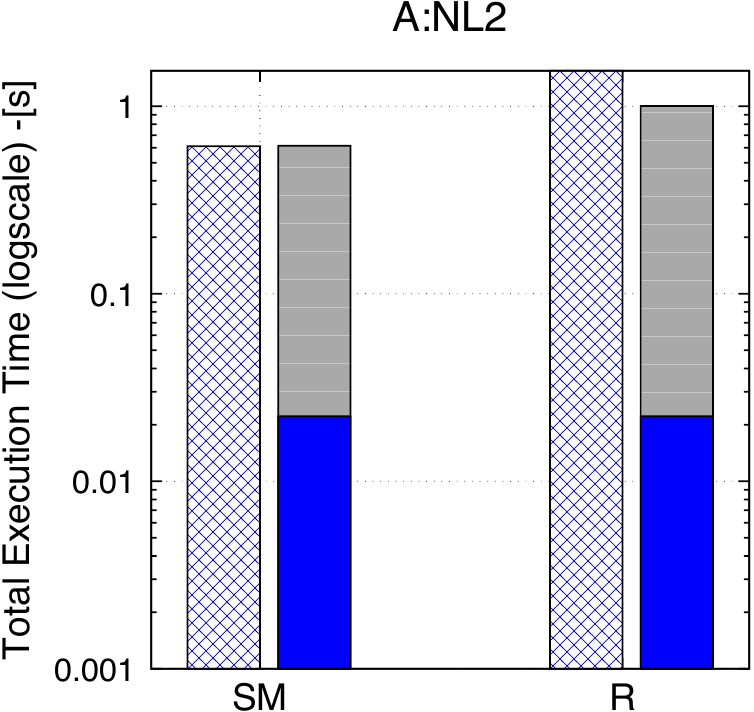}}
  \subfigure[\textbf{P2.15}]{\includegraphics[scale=0.25,width=4.4cm,height=3.6cm]{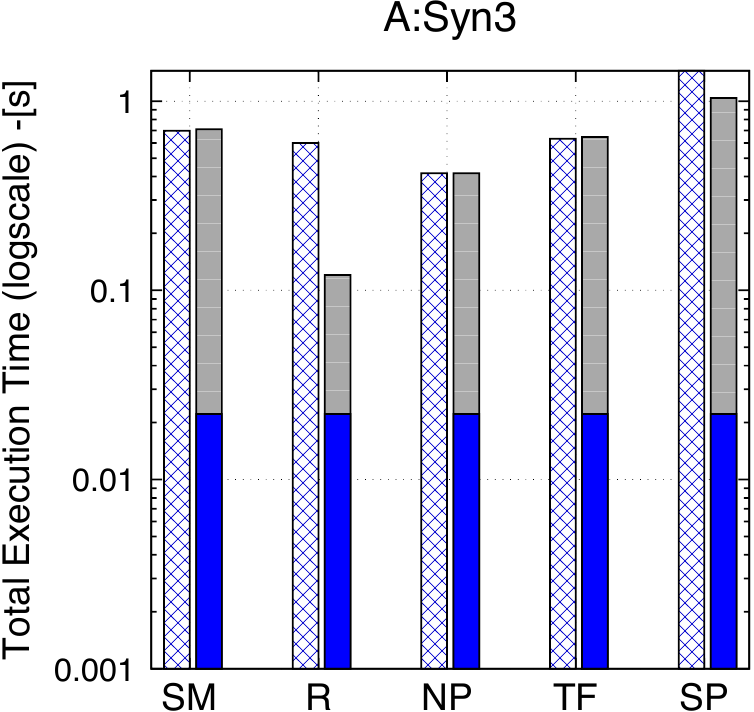}}
    \subfigure[\textbf{P2.15}]{\includegraphics[scale=0.25,width=4.2cm,height=3.6cm]{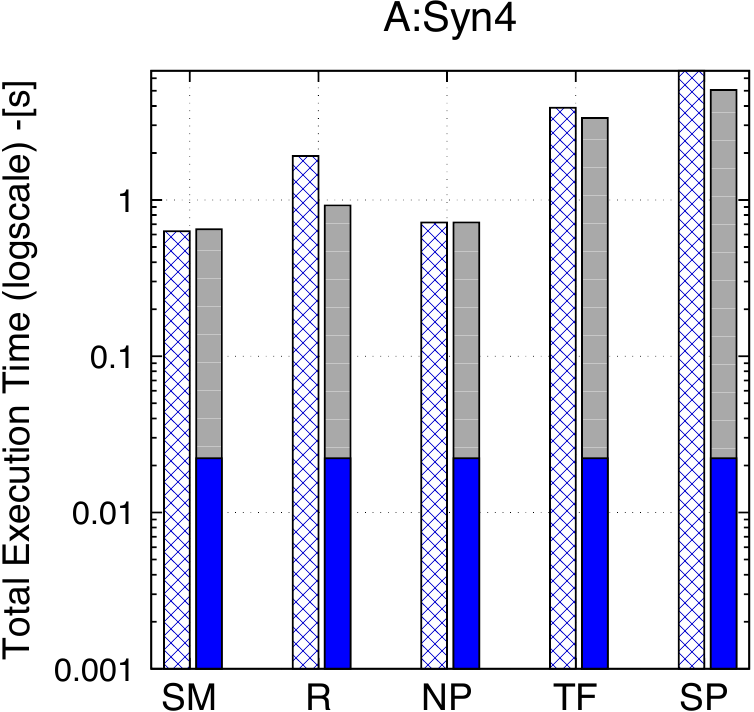}}
 \figspb\figspb\figspb
 \caption{P2.15 evaluation time with and without rewriting}
 \label{fig:la-p2.15}
\end{figure*}

\begin{figure*}[!htbp]
\figspa
  \centering
  \subfigure[\textbf{P2.16}]{\includegraphics[scale=0.25,width=4.2cm,height=3.6cm]{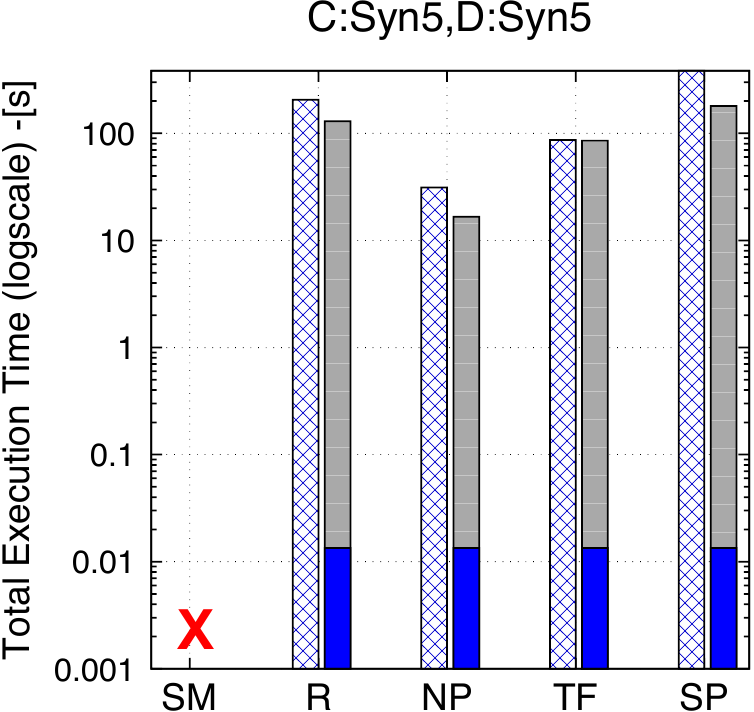}}
  \figspb\figspb\figspb
 \caption{P2.16 evaluation time with and without rewriting}
 \label{fig:la-p2.16}
\end{figure*}

\begin{figure*}[!htbp]
\figspa
  \centering
  \subfigure[\textbf{P2.18}]{\includegraphics[scale=0.25,width=4.2cm,height=3.6cm]{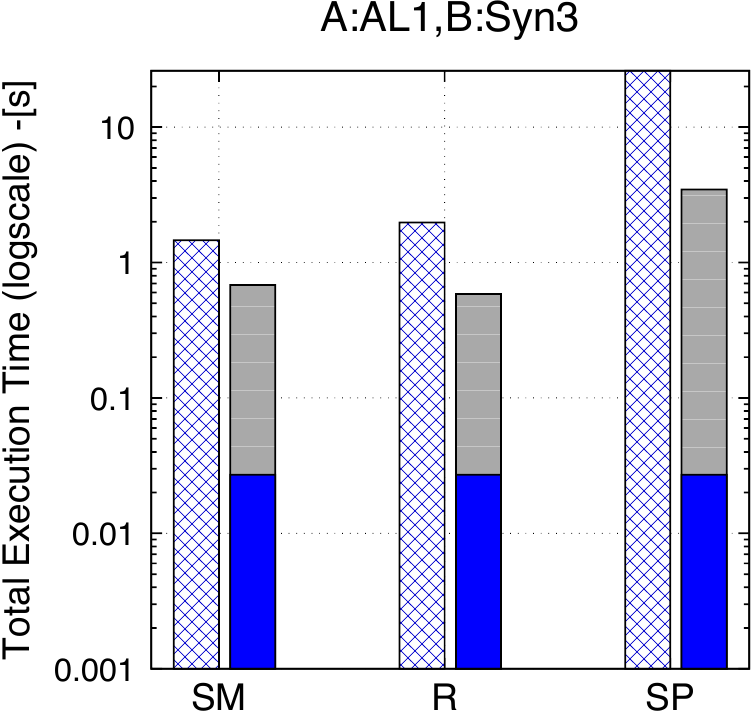}}
  \subfigure[\textbf{P2.18}]{\includegraphics[scale=0.25,width=4.4cm,height=3.6cm]{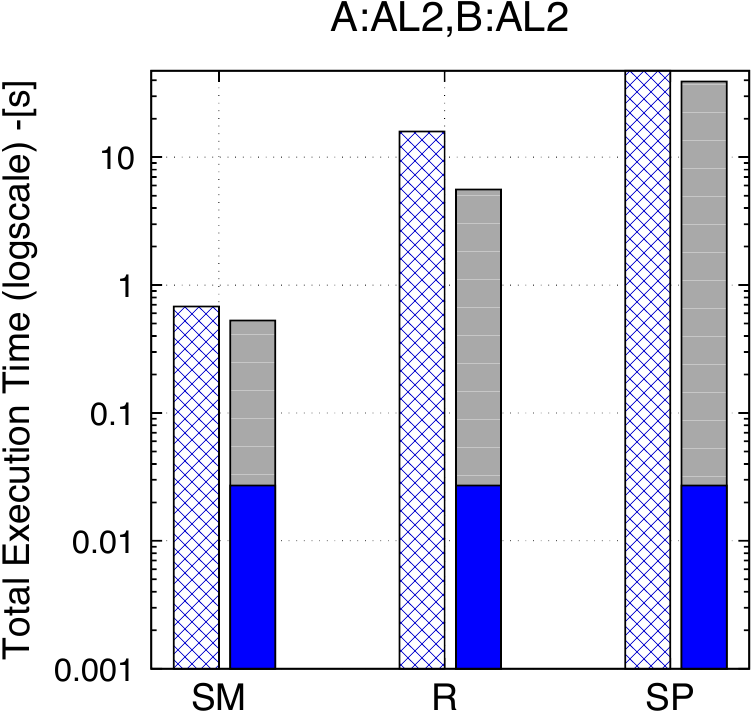}}
    \subfigure[\textbf{P2.18}]{\includegraphics[scale=0.25,width=4.2cm,height=3.6cm]{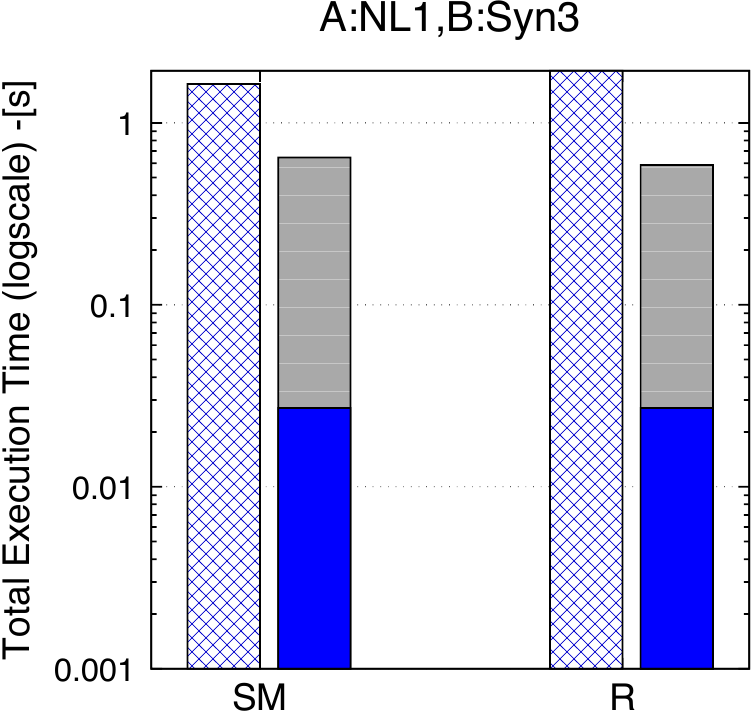}}
 \figspb\figspb\figspb
 \caption{P2.18 evaluation time with and without rewriting}
 \label{fig:la-p2.15}
\end{figure*}

\begin{figure*}[!htbp]
\figspa
  \centering
  \subfigure[\textbf{P2.18}]{\includegraphics[scale=0.25,width=4.2cm,height=3.6cm]{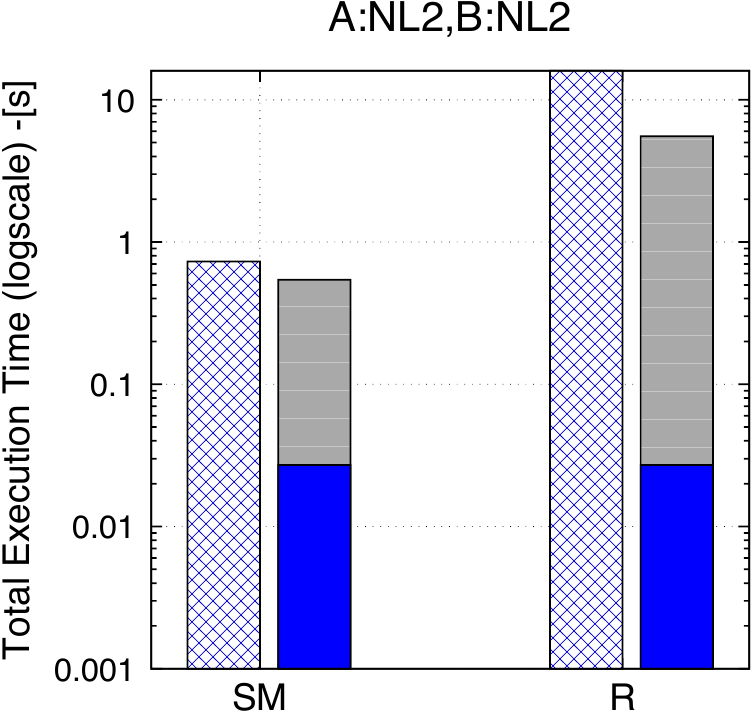}}
  \subfigure[\textbf{P2.18}]{\includegraphics[scale=0.25,width=4.4cm,height=3.6cm]{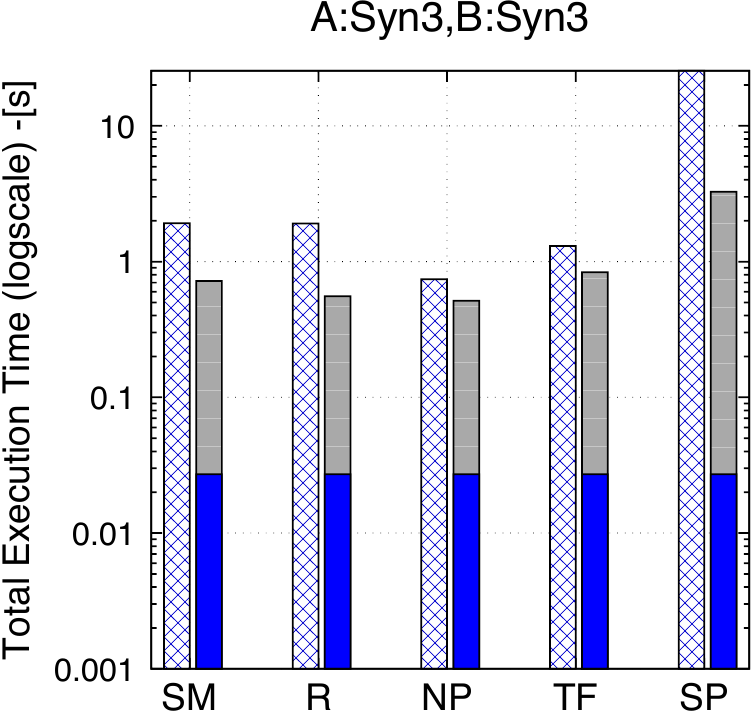}}
    \subfigure[\textbf{P2.18}]{\includegraphics[scale=0.25,width=4.2cm,height=3.6cm]{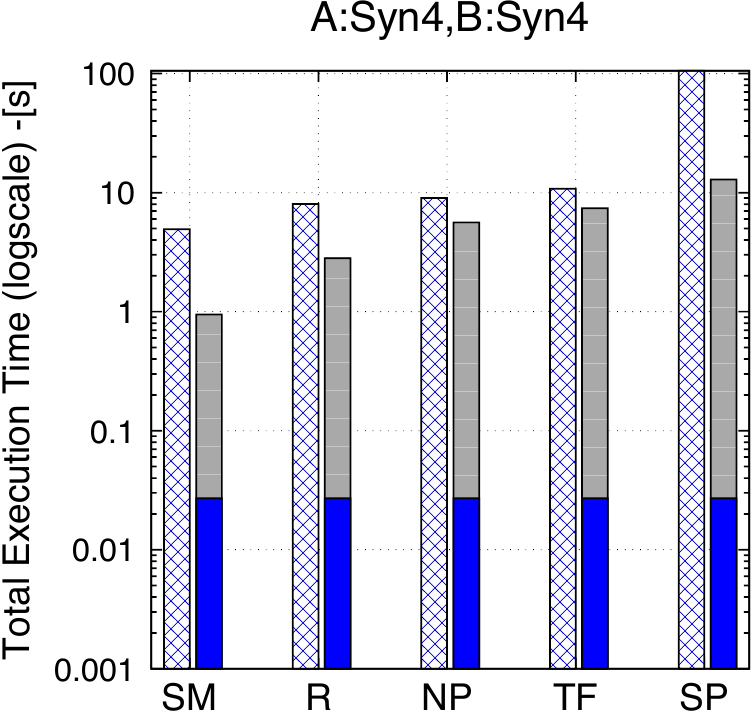}}
 \figspb\figspb\figspb
 \caption{P2.18 evaluation time with and without rewriting}
 \label{fig:la-p2.18}
\end{figure*}

%% file: appendixE.tex
\onecolumn
\section{Additional Results: $\notopt$ Pipelines - MNC-based Cost Model}
\label{appendixE}
\begin{figure*}[!htbp]
\figspa
  \centering
  \subfigure[\textbf{P1.2}]{\includegraphics[scale=0.25,width=4.2cm,height=3.6cm]{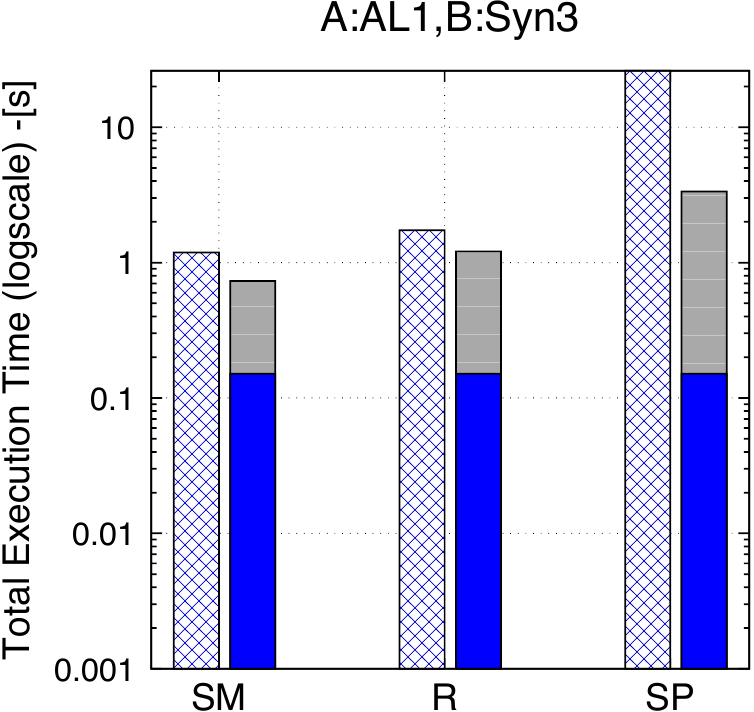}}
    \subfigure[\textbf{P1.2}]{\includegraphics[scale=0.25,width=4.2cm,height=3.6cm]{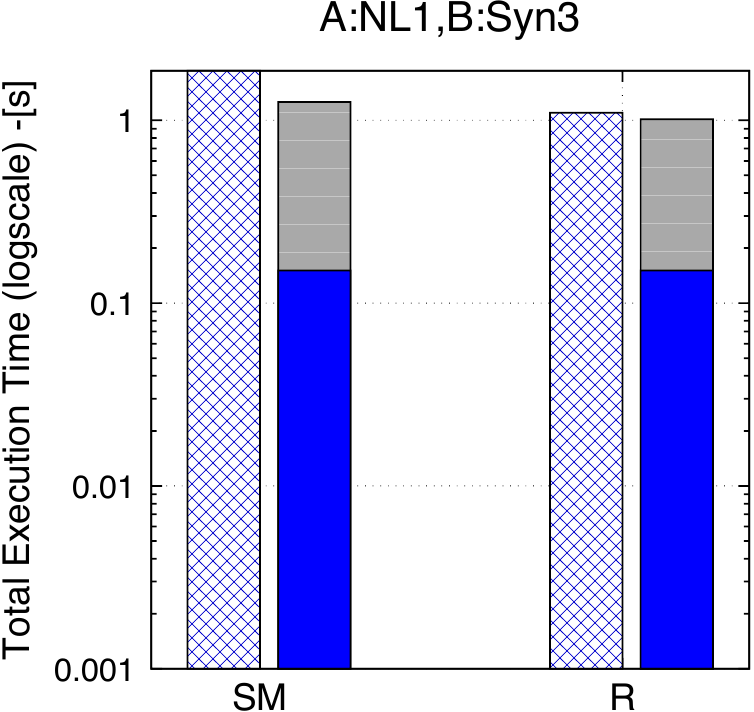}}
      \subfigure[\textbf{P1.2}]{\includegraphics[scale=0.25,width=4.4cm,height=3.6cm]{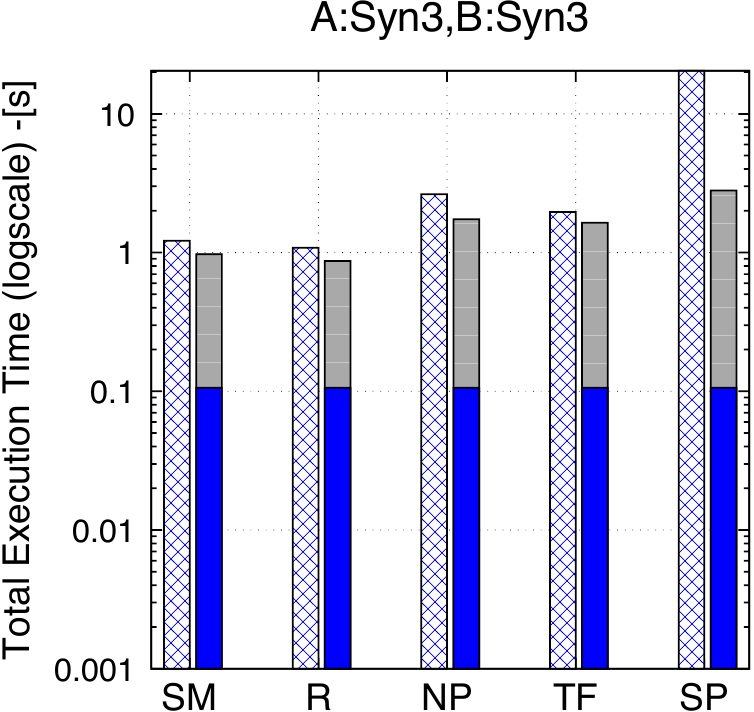}}
 \figspb\figspb\figspb
 \caption{P1.2 evaluation time with and without rewriting}
 \figspc
 \label{fig:la-p1.2}
\end{figure*}

\begin{figure*}[!htbp]
\figspa
  \centering
  \subfigure[\textbf{P1.6}]{\includegraphics[scale=0.25,width=4.2cm,height=3.6cm]{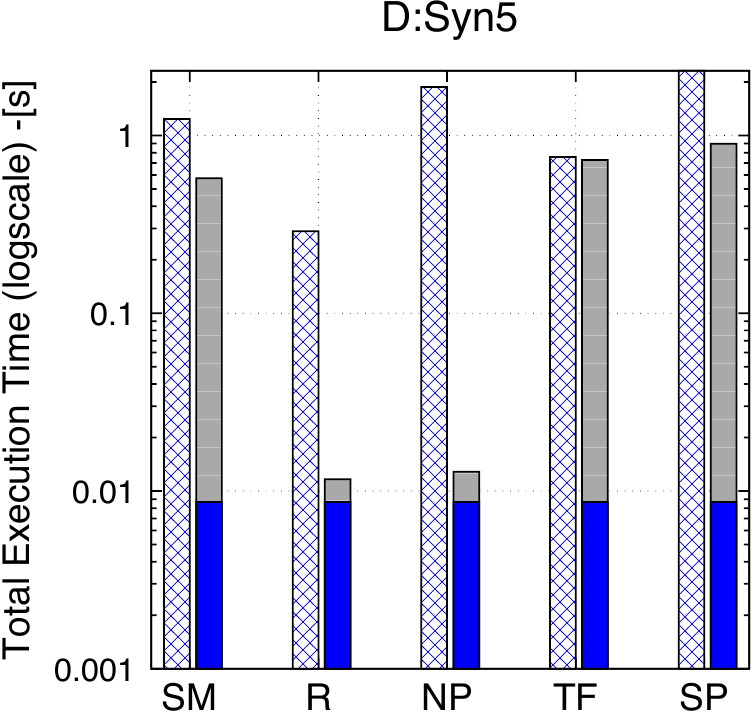}}
 \figspb\figspb\figspb
 \caption{P1.6 evaluation time with and without rewriting}
 \figspc
 \label{fig:la-p1.6}
\end{figure*}

\begin{figure*}[!htbp]
\figspa
  \centering
  \subfigure[\textbf{P1.8}]{\includegraphics[scale=0.25,width=4.2cm,height=3.6cm]{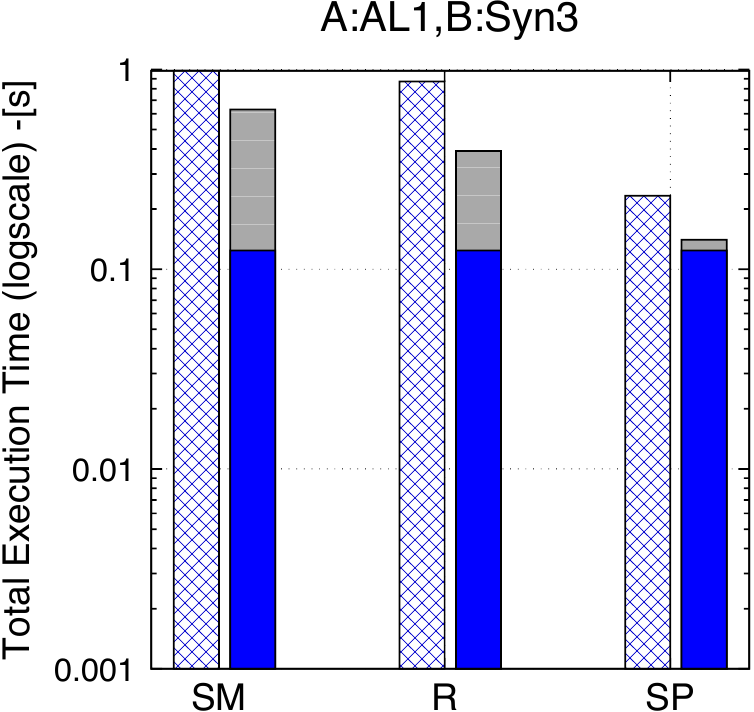}}
    \subfigure[\textbf{P1.8}]{\includegraphics[scale=0.25,width=4.2cm,height=3.6cm]{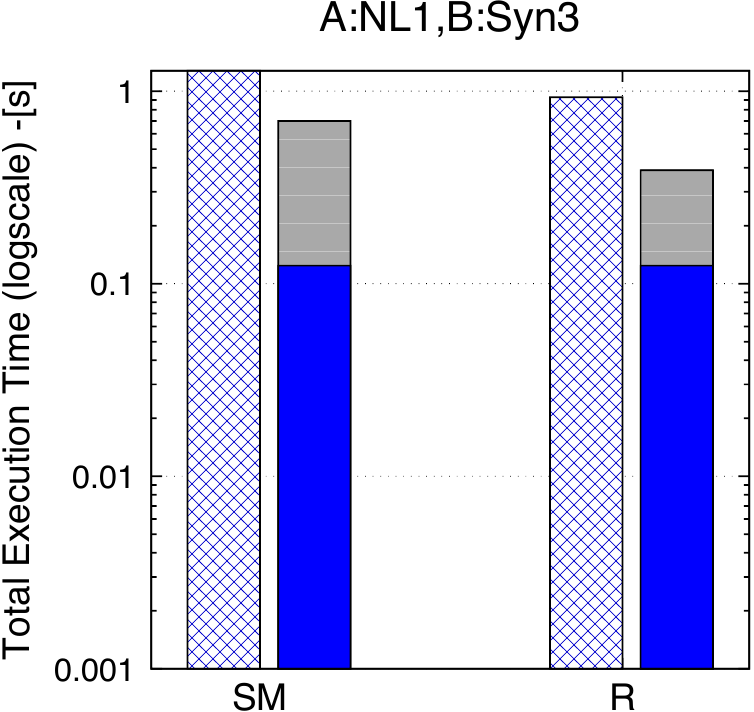}}
      \subfigure[\textbf{P1.8}]{\includegraphics[scale=0.25,width=4.4cm,height=3.6cm]{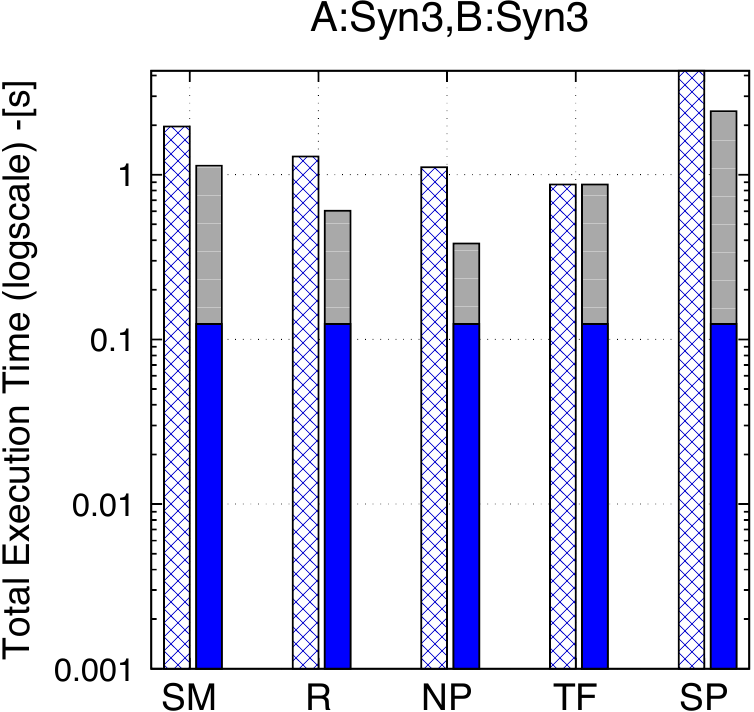}}
 \figspb\figspb\figspb
 \caption{P1.8 evaluation time with and without rewriting}
 \figspc
 \label{fig:la-p1.8}
\end{figure*}

\begin{figure*}[!htbp]
\figspa
  \centering
  \subfigure[\textbf{P1.9}]{\includegraphics[scale=0.25,width=4.2cm,height=3.6cm]{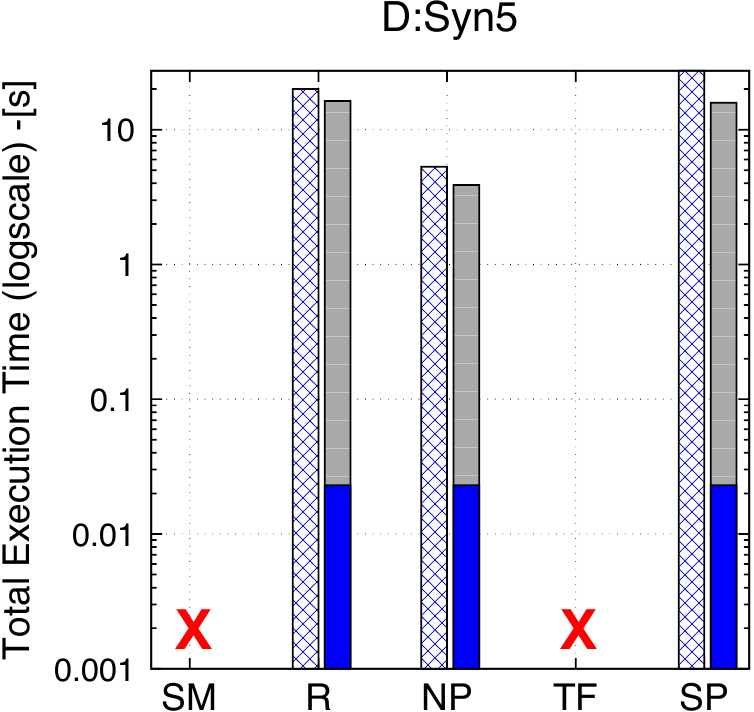}}
 \figspb\figspb\figspb
 \caption{P1.9 evaluation time with and without rewriting}
 \label{fig:la-p1.9}
\end{figure*}

\begin{figure*}[!htbp]
\figspa
  \centering
  \subfigure[\textbf{P1.10}]{\includegraphics[scale=0.25,width=4.2cm,height=3.6cm]{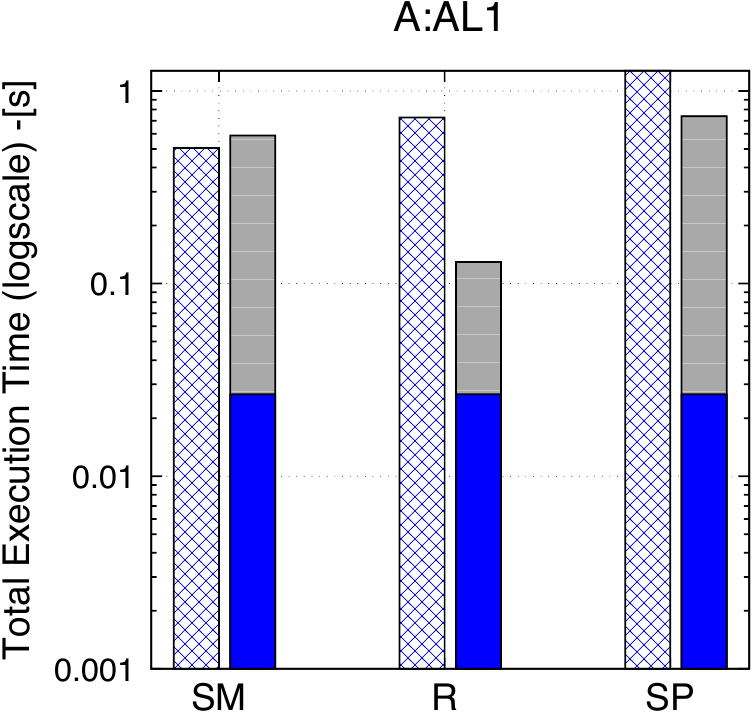}}
    \subfigure[\textbf{P1.10}]{\includegraphics[scale=0.25,width=4.2cm,height=3.6cm]{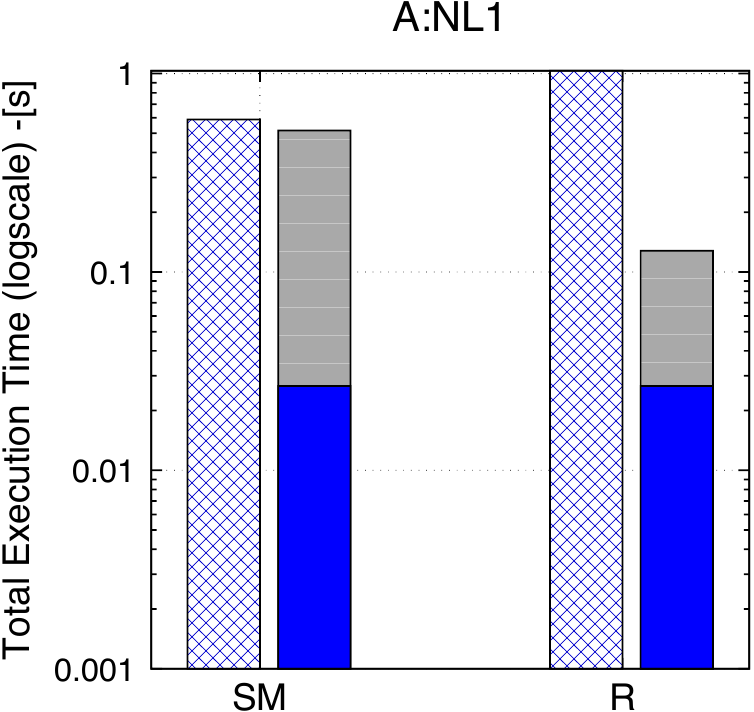}}
    \subfigure[\textbf{P1.10}]{\includegraphics[scale=0.25,width=4.4cm,height=3.6cm]{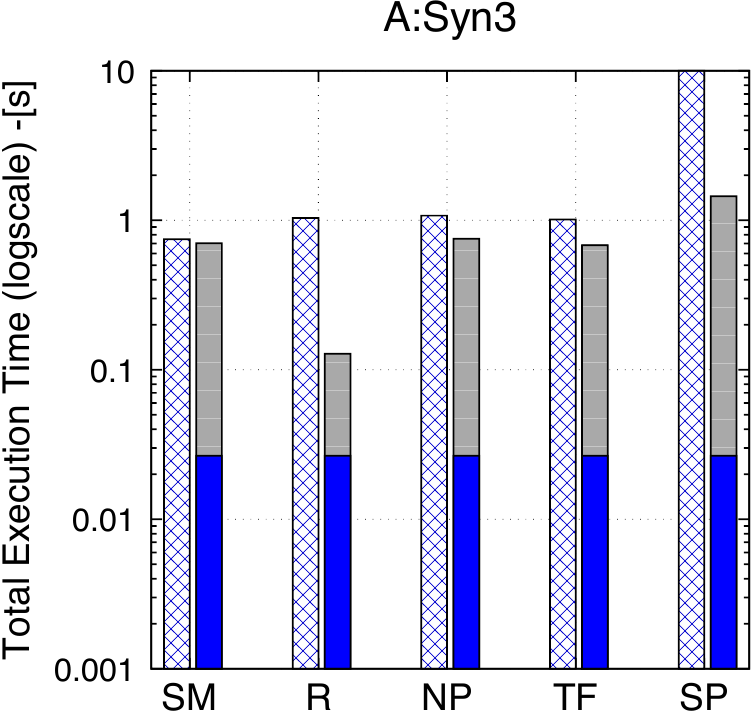}}
 \figspb\figspb\figspb
 \caption{P1.10 evaluation time with and without rewriting}
 \label{fig:la-p1.10}
\end{figure*}

\begin{figure*}[!htbp]
\figspa
  \centering
  \subfigure[\textbf{P1.11}]{\includegraphics[scale=0.25,width=4.2cm,height=3.6cm]{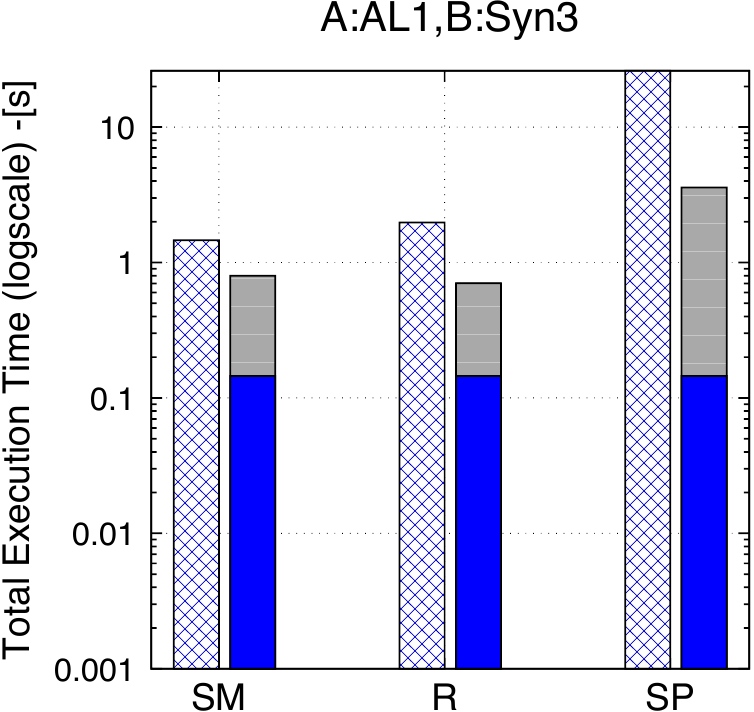}}
    \subfigure[\textbf{P1.11}]{\includegraphics[scale=0.25,width=4.2cm,height=3.6cm]{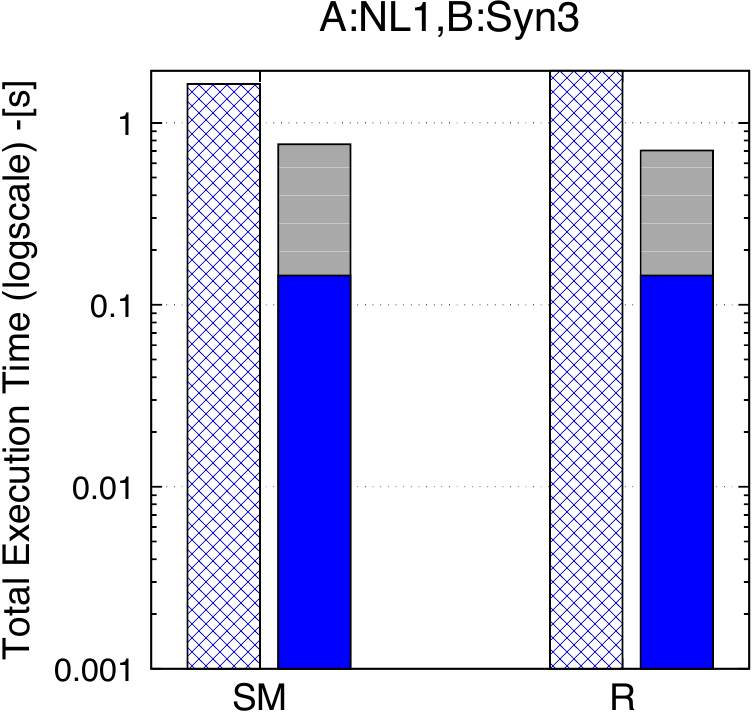}}
      \subfigure[\textbf{P1.11}]{\includegraphics[scale=0.25,width=4.4cm,height=3.6cm]{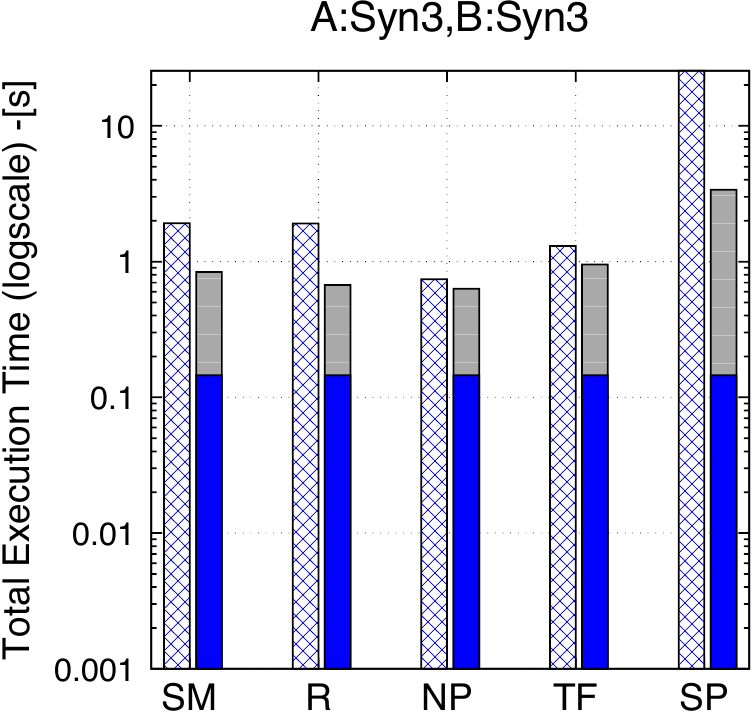}}
 \figspb\figspb\figspb
 \caption{P1.11 evaluation time with and without rewriting}
 \label{fig:la-p1.11}
\end{figure*}

\begin{figure*}[!htbp]
\figspa
  \centering
  \subfigure[\textbf{P1.12}]{\includegraphics[scale=0.25,width=4.2cm,height=3.6cm]{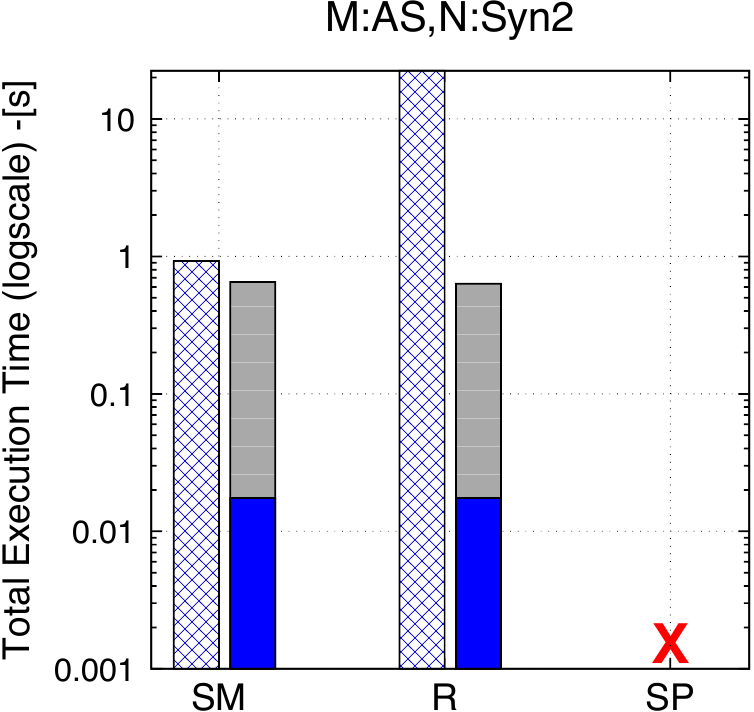}}
  \subfigure[\textbf{P1.12}]{\includegraphics[scale=0.25,width=4.4cm,height=3.6cm]{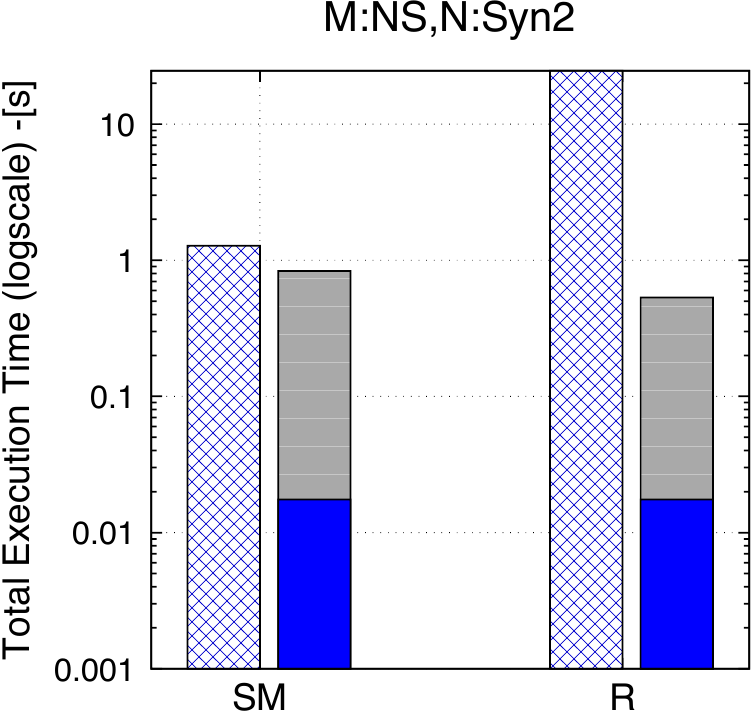}}
    \subfigure[\textbf{P1.12}]{\includegraphics[scale=0.25,width=4.2cm,height=3.6cm]{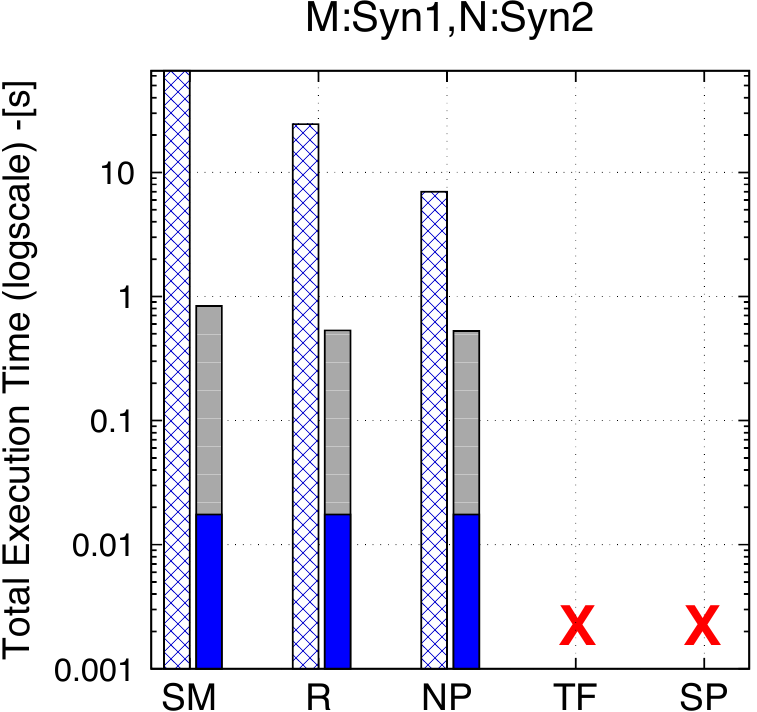}}
 \figspb\figspb\figspb
 \caption{P1.12 evaluation time with and without rewriting}
 \label{fig:la-p1.12}
\end{figure*}

\begin{figure*}[!htbp]
\figspa
  \centering
  \subfigure[\textbf{P1.14}]{\includegraphics[scale=0.25,width=4.2cm,height=3.6cm]{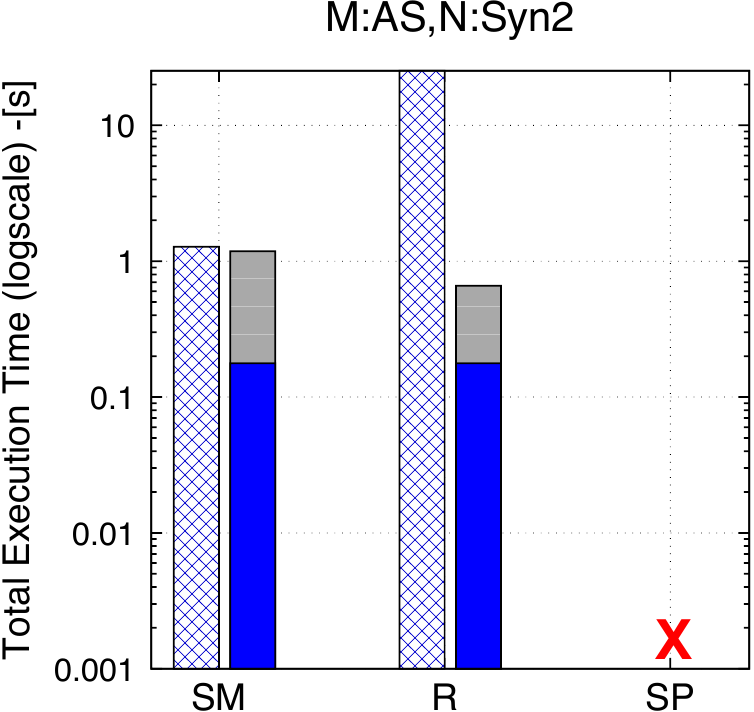}}
  \subfigure[\textbf{P1.14}]{\includegraphics[scale=0.25,width=4.4cm,height=3.6cm]{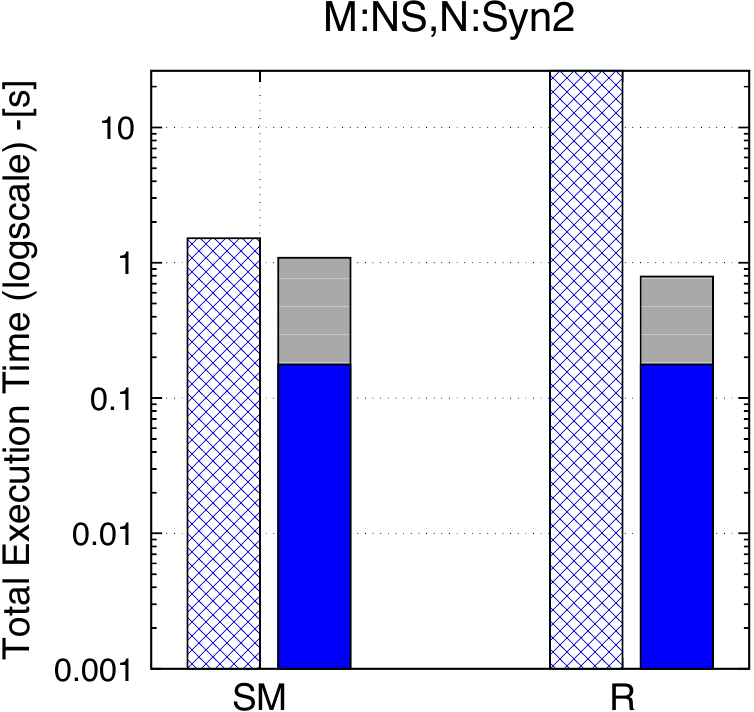}}
 \figspb\figspb\figspb
 \caption{P1.14 evaluation time with and without rewriting}
 \label{fig:la-p1.14}
\end{figure*}

\begin{figure*}[!htbp]
\figspa
  \centering
  \subfigure[\textbf{P1.15}]{\includegraphics[scale=0.25,width=4.2cm,height=3.6cm]{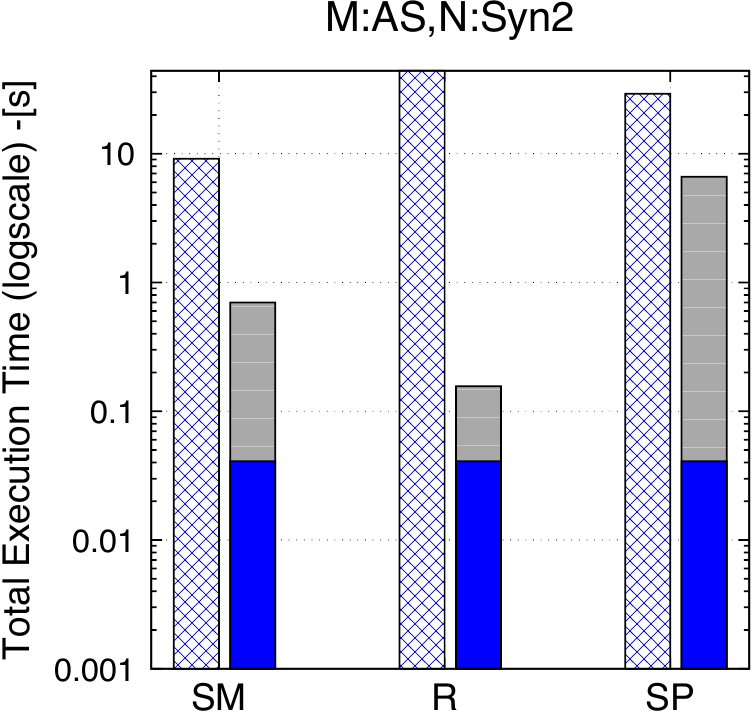}}
  \subfigure[\textbf{P1.15}]{\includegraphics[scale=0.25,width=4.4cm,height=3.6cm]{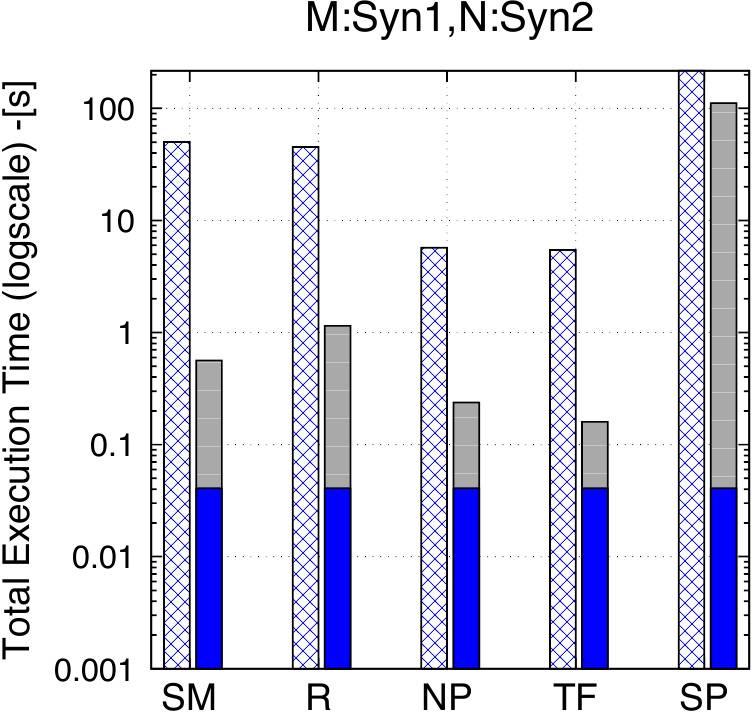}}
 \figspb\figspb\figspb
 \caption{P1.15 evaluation time with and without rewriting}
 \label{fig:la-p1.15}
\end{figure*}

\begin{figure*}[!htbp]
\figspa
  \centering
  \subfigure[\textbf{P1.16}]{\includegraphics[scale=0.25,width=4.2cm,height=3.6cm]{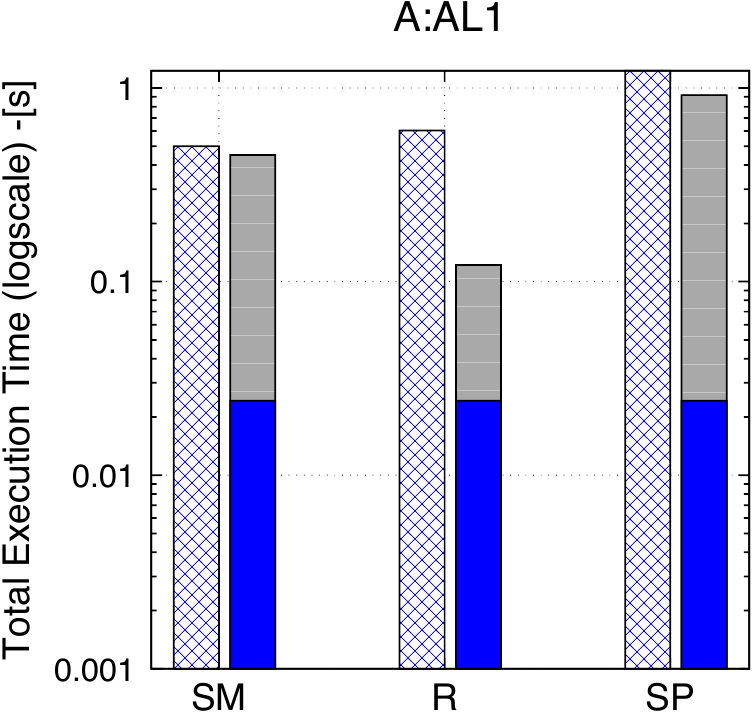}}
    \subfigure[\textbf{P1.16}]{\includegraphics[scale=0.25,width=4.2cm,height=3.6cm]{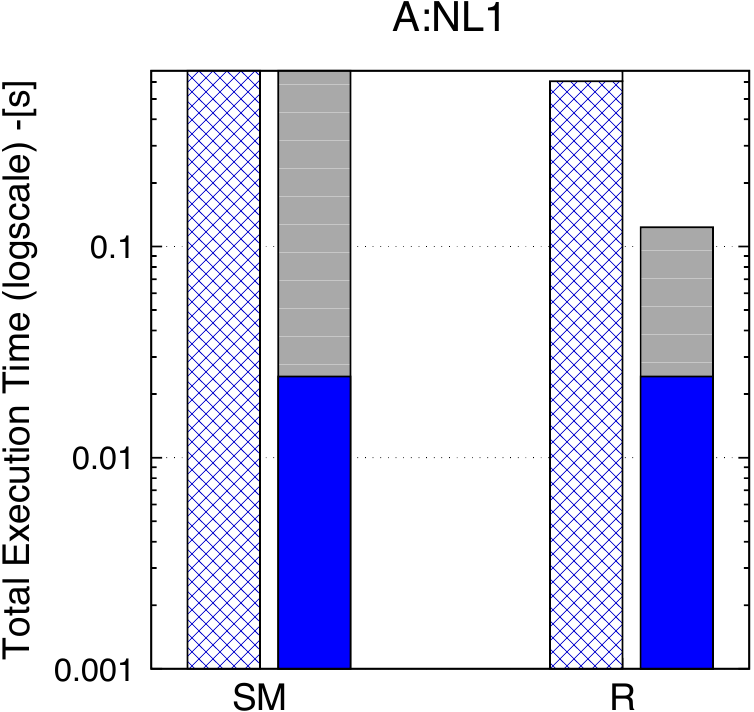}}
      \subfigure[\textbf{P1.16}]{\includegraphics[scale=0.25,width=4.4cm,height=3.6cm]{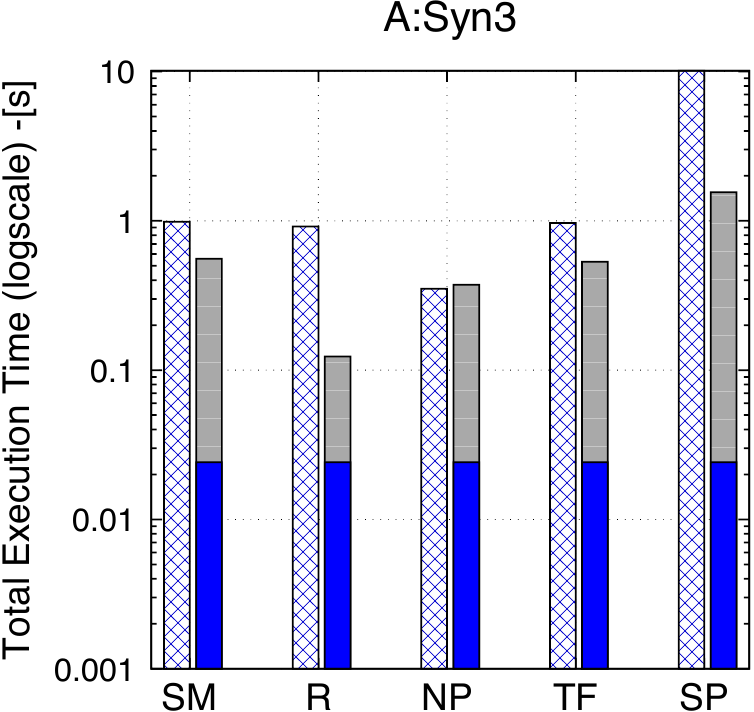}}
 \figspb\figspb\figspb
 \caption{P1.16 evaluation time with and without rewriting}
 \label{fig:la-p1.16}
\end{figure*}

\begin{figure*}[!htbp]
\figspa
  \centering
  \subfigure[\textbf{P1.17}]{\includegraphics[scale=0.25,width=4.2cm,height=3.6cm]{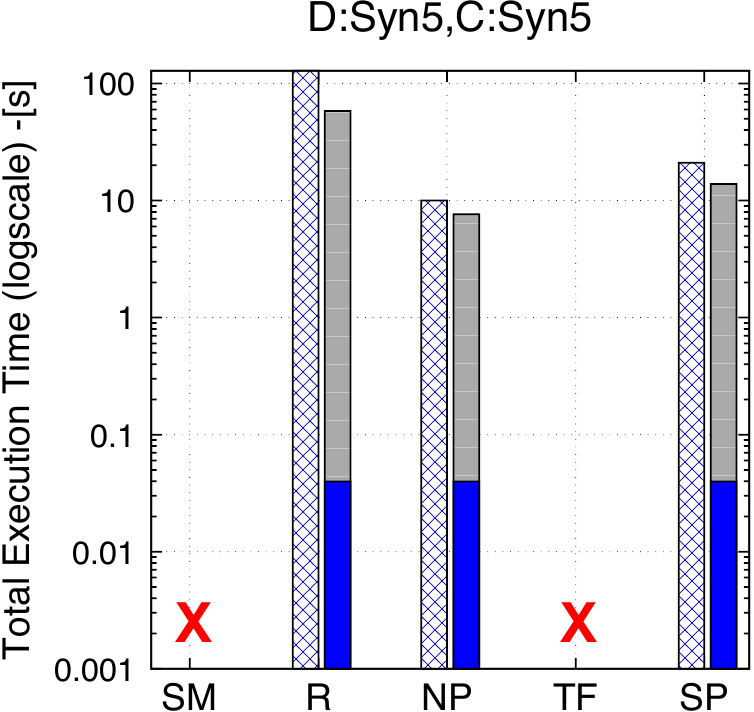}}
  \figspb\figspb\figspb
 \caption{P1.17 evaluation time with and without rewriting}
 \label{fig:la-p1.17}
\end{figure*}

%%%Part2

\begin{figure*}[!htbp]
\figspa
  \centering
  \subfigure[\textbf{P1.18}]{\includegraphics[scale=0.25,width=4.2cm,height=3.6cm]{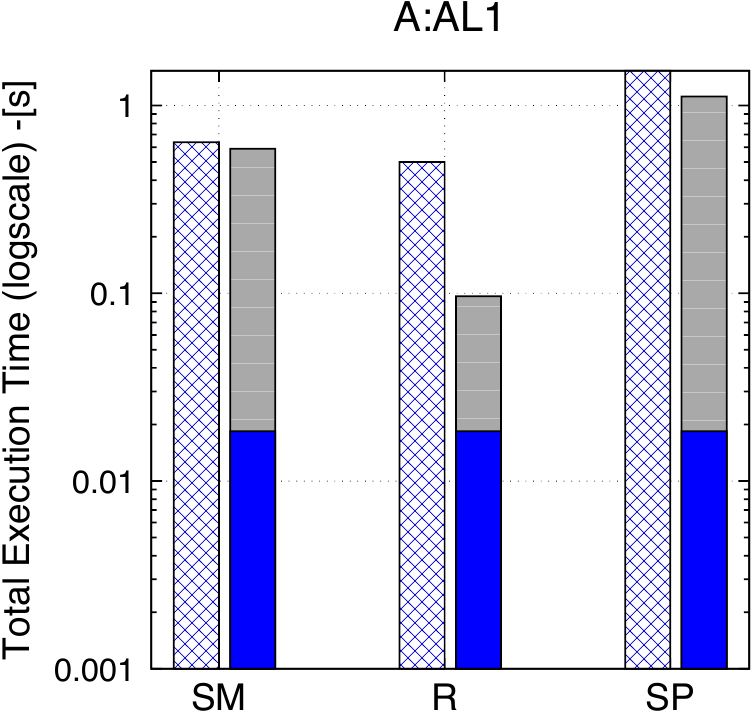}}
    \subfigure[\textbf{P1.18}]{\includegraphics[scale=0.25,width=4.2cm,height=3.6cm]{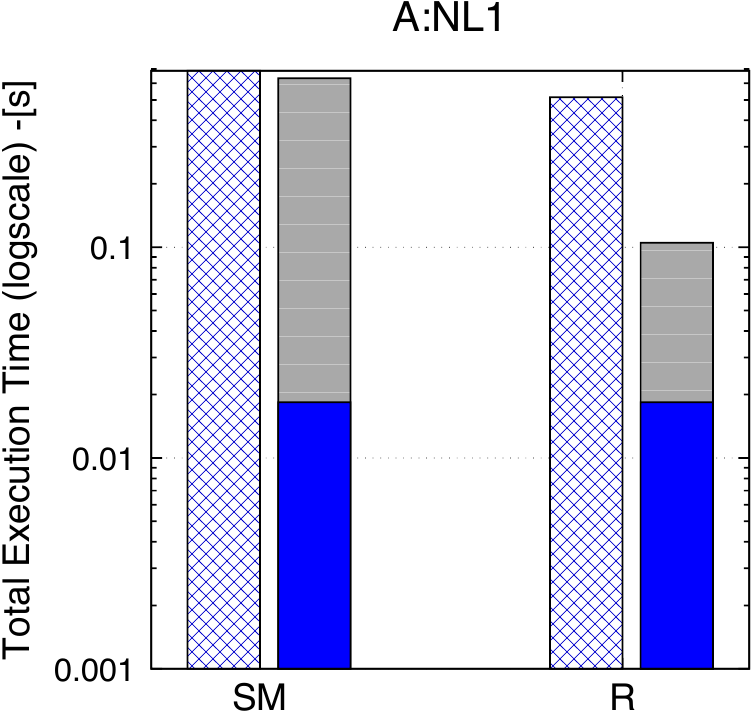}}
      \subfigure[\textbf{P1.18}]{\includegraphics[scale=0.25,width=4.4cm,height=3.6cm]{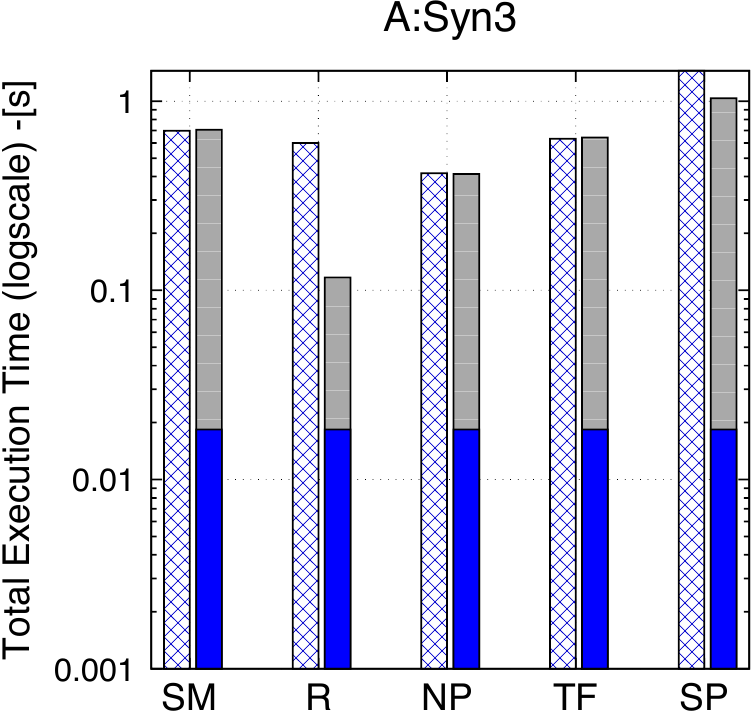}}
 \figspb\figspb\figspb
 \caption{P1.18 evaluation time with and without rewriting}
 \label{fig:la-p1.18}
\end{figure*}

\begin{figure*}[!htbp]
\figspa
  \centering
  \subfigure[\textbf{P1.25}]{\includegraphics[scale=0.25,width=4.2cm,height=3.6cm]{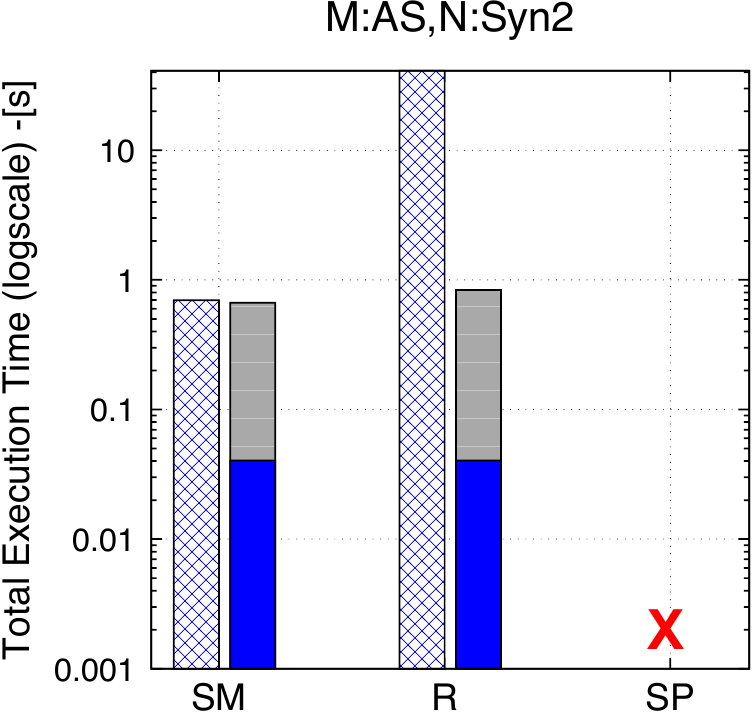}}
  \subfigure[\textbf{P1.25}]{\includegraphics[scale=0.25,width=4.4cm,height=3.6cm]{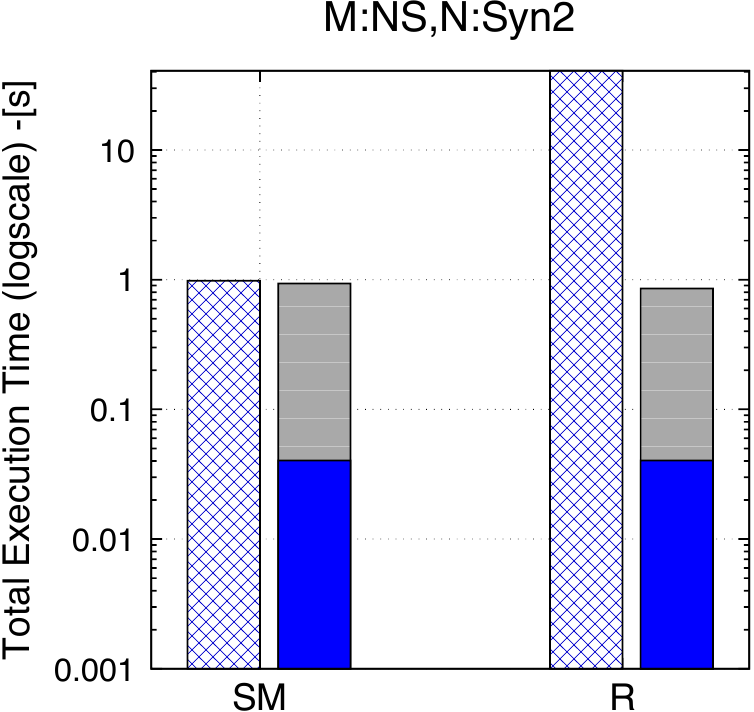}}
 \figspb\figspb\figspb
 \caption{P1.25 evaluation time with and without rewriting}
 \label{fig:la-p1.25}
\end{figure*}

\begin{figure*}[!htbp]
\figspa
  \centering
  \subfigure[\textbf{P2.1}]{\includegraphics[scale=0.25,width=4.2cm,height=3.6cm]{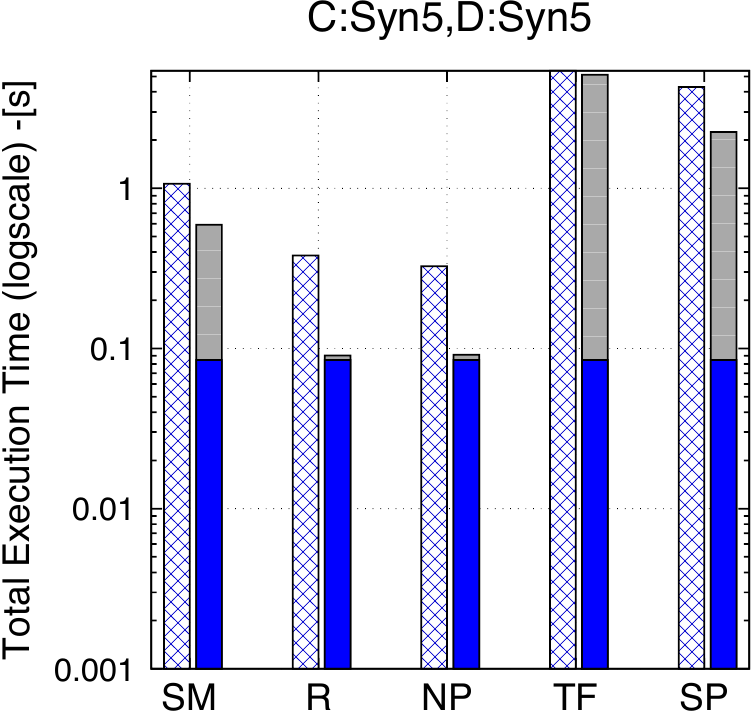}}
    \subfigure[\textbf{P2.2}]{\includegraphics[scale=0.25,width=4.2cm,height=3.6cm]{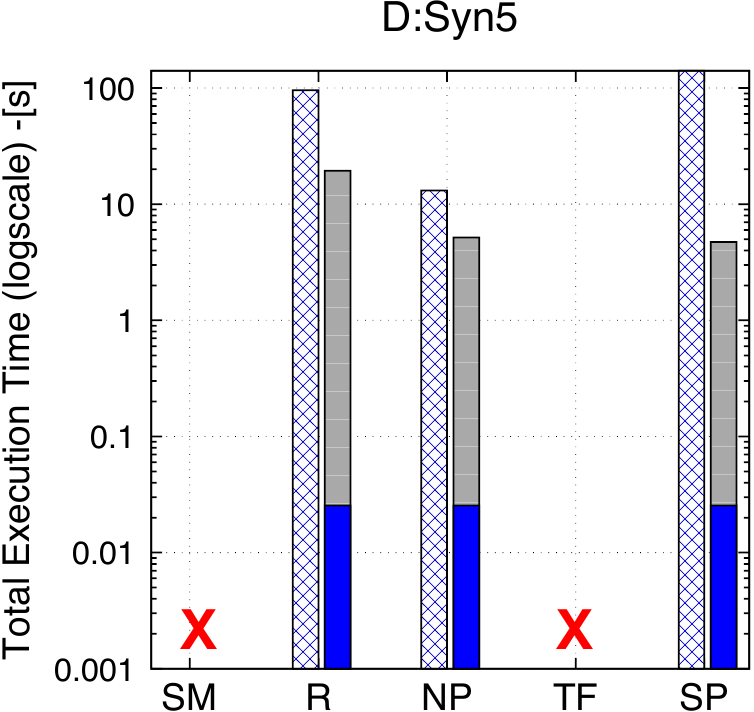}}
      \subfigure[\textbf{P2.3}]{\includegraphics[scale=0.25,width=4.2cm,height=3.6cm]{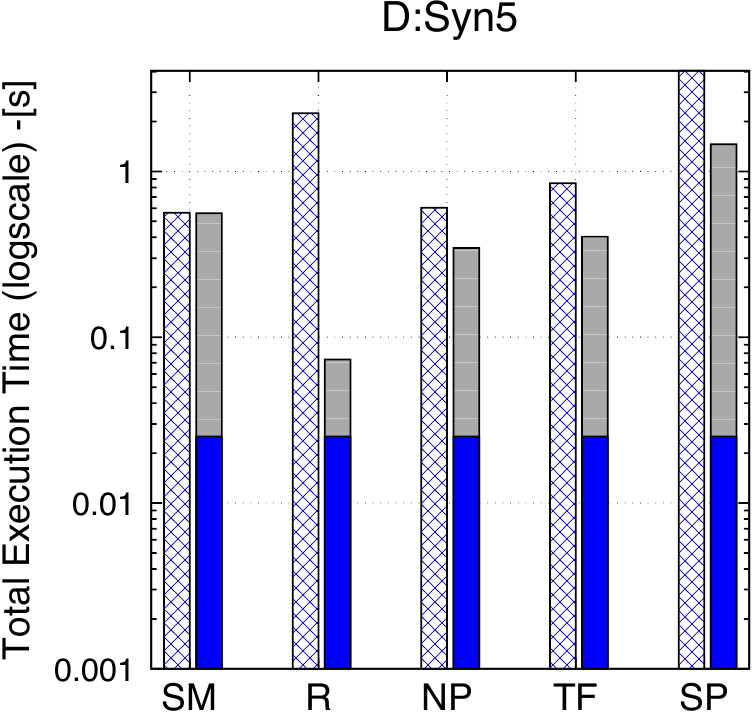}}

  \figspb\figspb\figspb
 \caption{P2.1 evaluation time with and without rewriting}
 \label{fig:la-p2.1}
\end{figure*}

\begin{figure*}[!htbp]
\figspa
  \centering
  \subfigure[\textbf{P2.4}]{\includegraphics[scale=0.25,width=4.2cm,height=3.6cm]{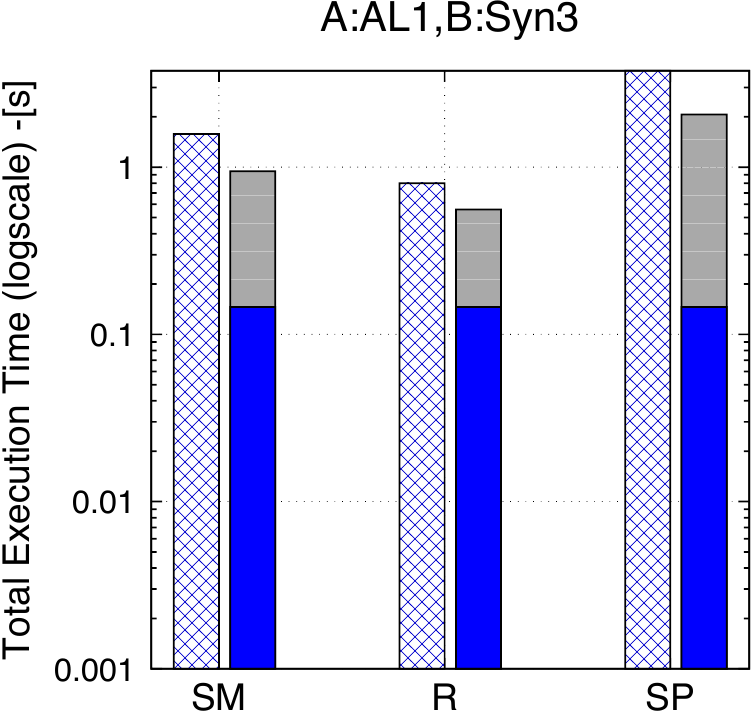}}
    \subfigure[\textbf{P2.4}]{\includegraphics[scale=0.25,width=4.2cm,height=3.6cm]{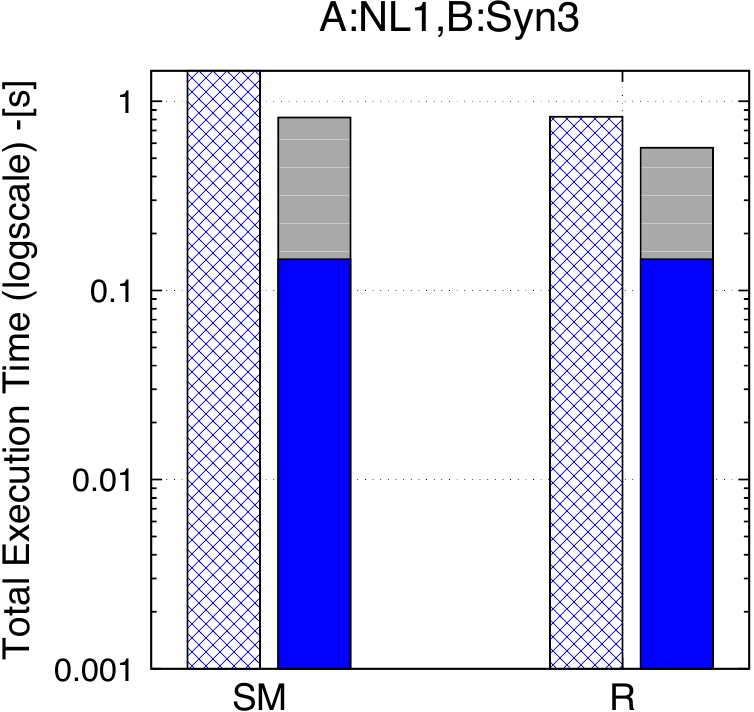}}
      \subfigure[\textbf{P2.4}]{\includegraphics[scale=0.25,width=4.4cm,height=3.6cm]{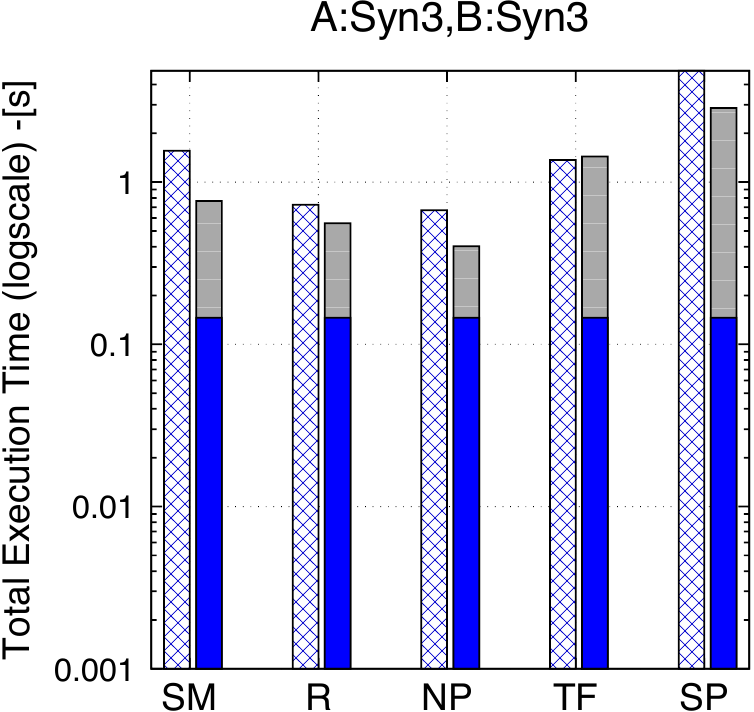}}
 \figspb\figspb\figspb
 \caption{P2.4 evaluation time with and without rewriting}
 \label{fig:la-p2.4}
\end{figure*}

\begin{figure*}[!htbp]
\figspa
  \centering
  \subfigure[\textbf{P2.5}]{\includegraphics[scale=0.25,width=4.2cm,height=3.6cm]{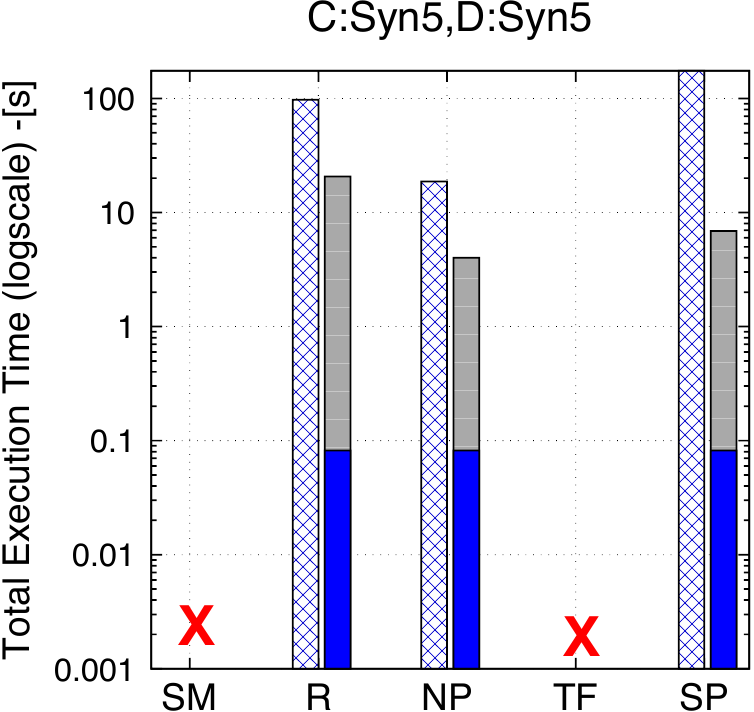}}
  \subfigure[\textbf{P2.6}]{\includegraphics[scale=0.25,width=4.2cm,height=3.6cm]{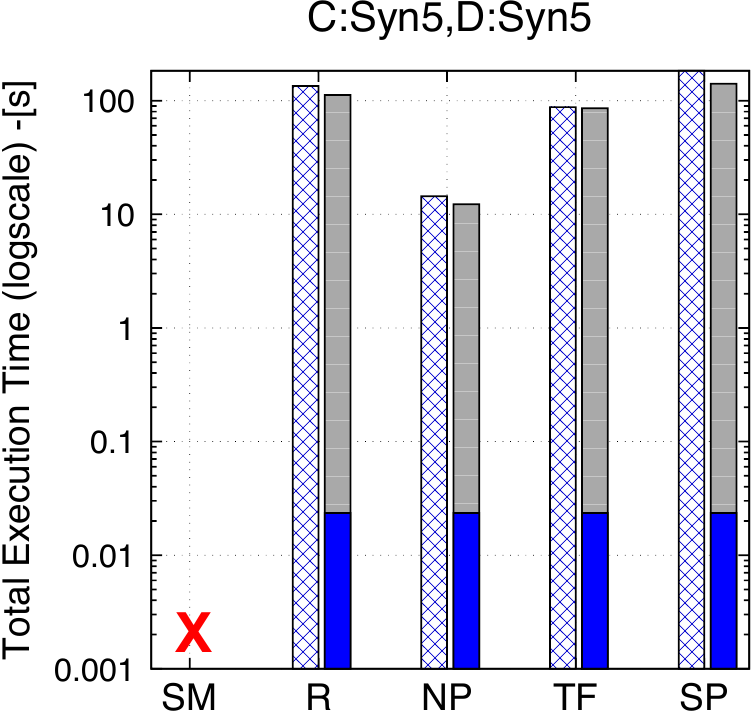}}
  \subfigure[\textbf{P2.8}]{\includegraphics[scale=0.25,width=4.2cm,height=3.6cm]{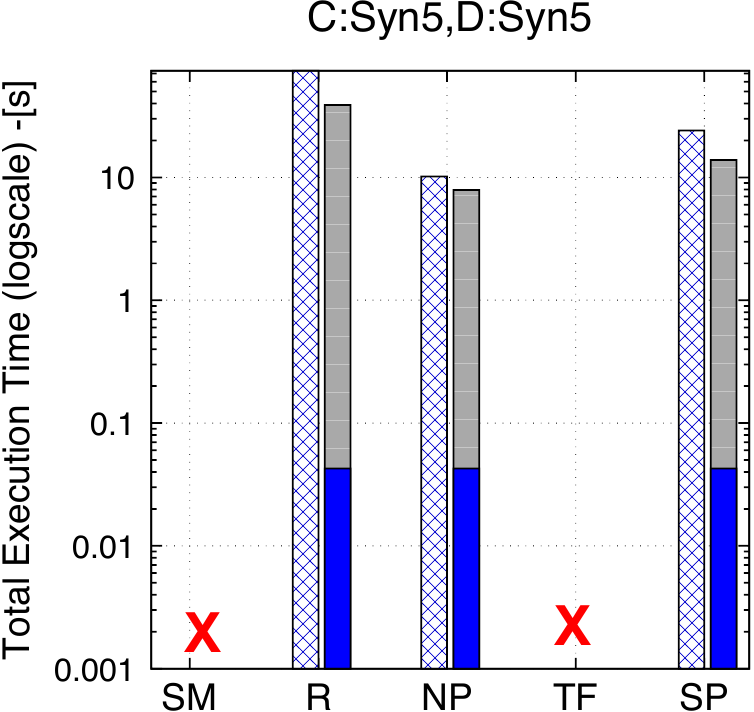}}
    \subfigure[\textbf{P2.9}]{\includegraphics[scale=0.25,width=4.2cm,height=3.6cm]{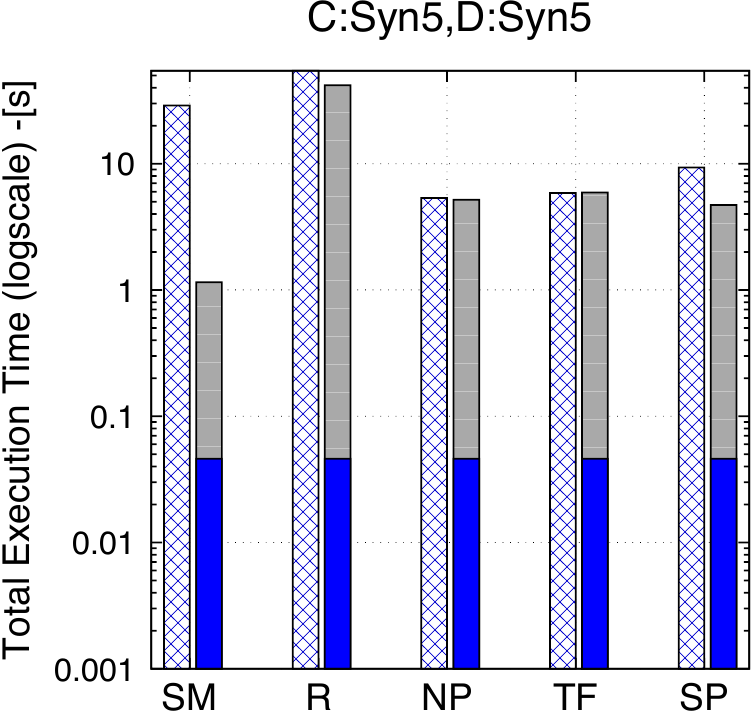}}
  \figspb\figspb\figspb
 \caption{P2.5, P2.6, P2.8 and P2.9evaluation time with and without rewriting}
 \label{fig:la-p2.568}
\end{figure*}

\begin{figure*}[!htbp]
\figspa
  \centering
  \subfigure[\textbf{P2.10}]{\includegraphics[scale=0.25,width=4.2cm,height=3.6cm]{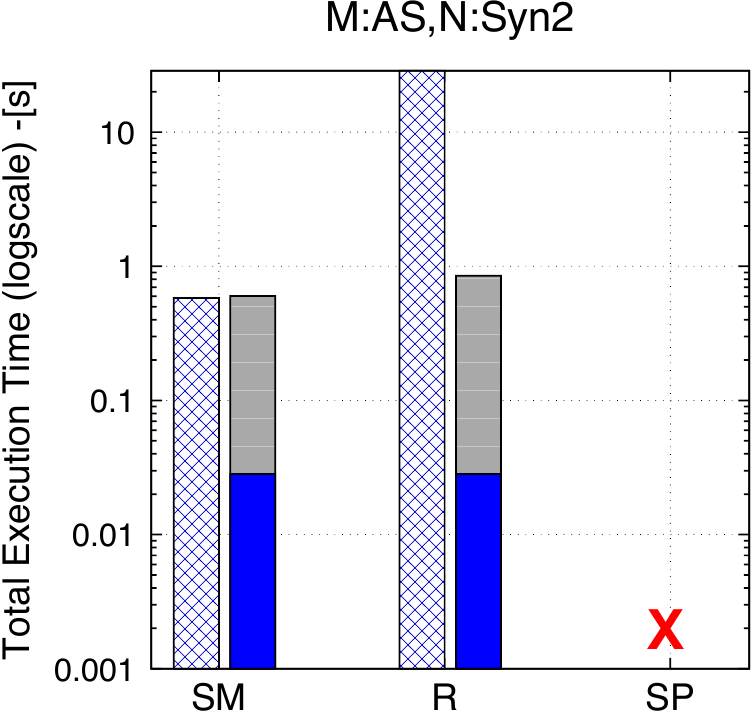}}
  \subfigure[\textbf{P2.10}]{\includegraphics[scale=0.25,width=4.4cm,height=3.6cm]{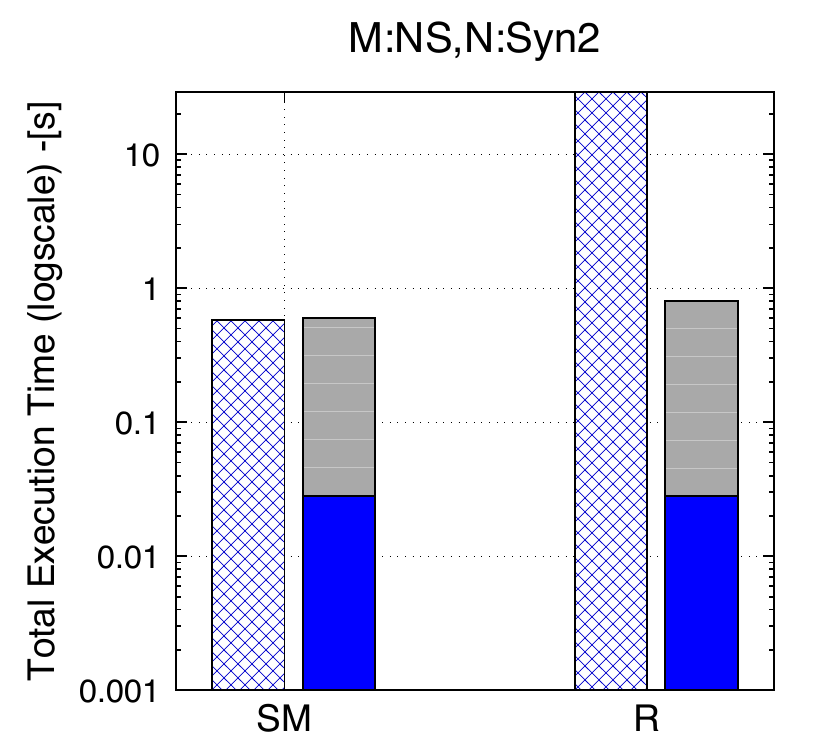}}
    \subfigure[\textbf{P2.10}]{\includegraphics[scale=0.25,width=4.2cm,height=3.6cm]{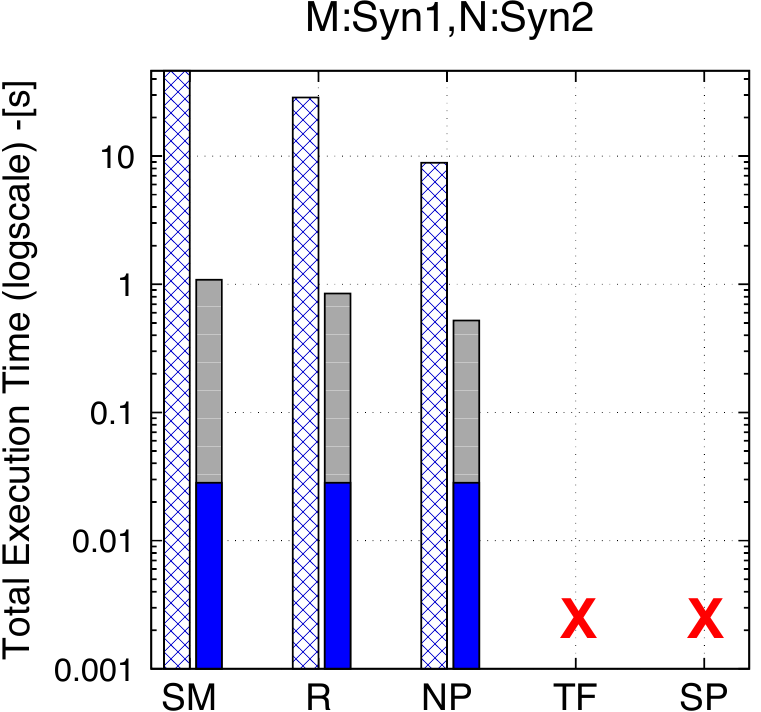}}
 \figspb\figspb\figspb
 \caption{P2.10 evaluation time with and without rewriting}
 \label{fig:la-p2.10}
\end{figure*}

\begin{figure*}[!htbp]
\figspa
  \centering
  \subfigure[\textbf{P2.11}]{\includegraphics[scale=0.25,width=4.2cm,height=3.6cm]{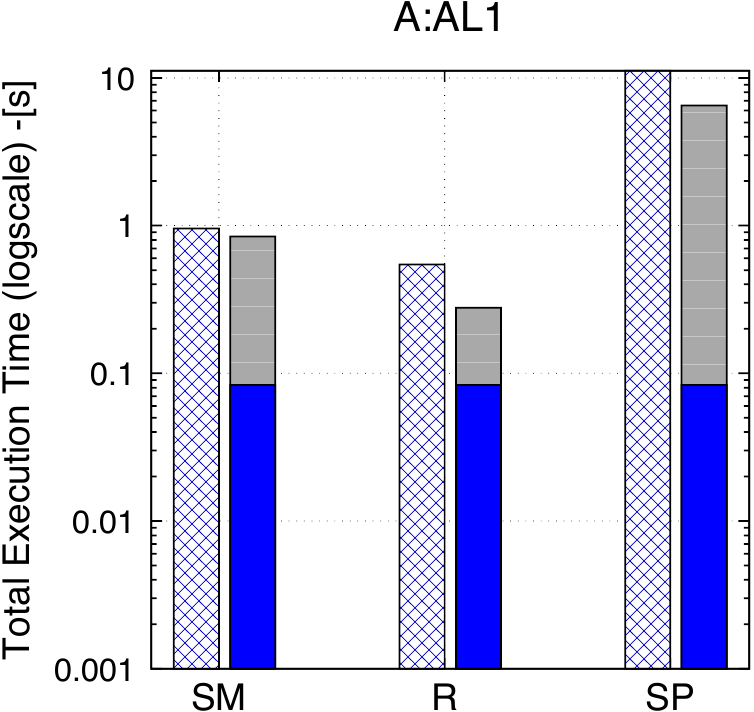}}
    \subfigure[\textbf{P2.11}]{\includegraphics[scale=0.25,width=4.2cm,height=3.6cm]{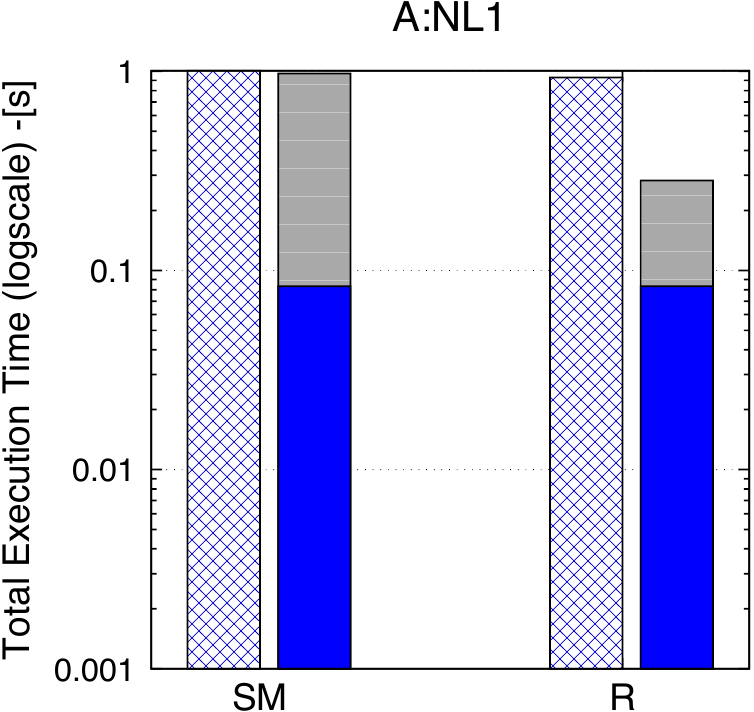}}
      \subfigure[\textbf{P2.11}]{\includegraphics[scale=0.25,width=4.4cm,height=3.6cm]{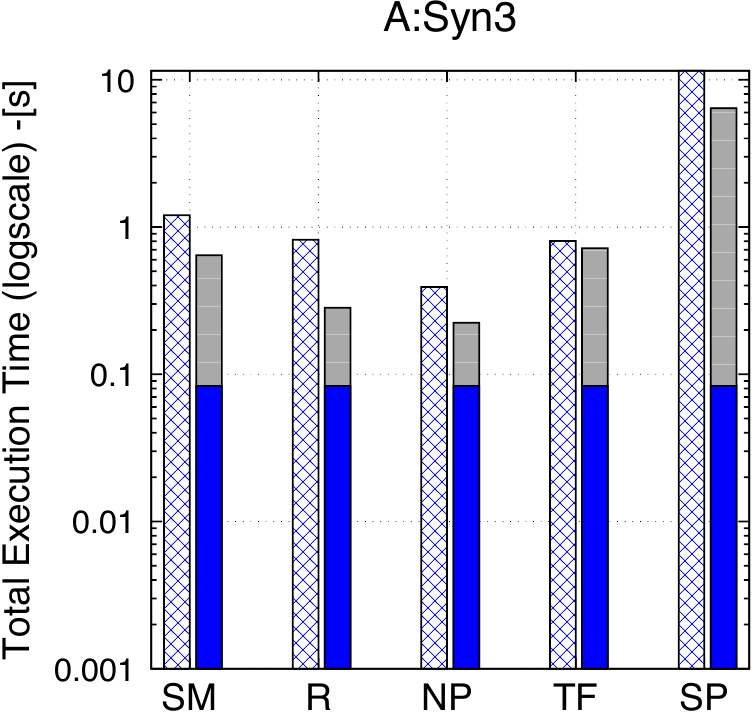}}
 \figspb\figspb\figspb
 \caption{P2.11 evaluation time with and without rewriting}
 \label{fig:la-p2.11}
\end{figure*}

\begin{figure*}[!htbp]
\figspa
  \centering
  \subfigure[\textbf{P2.13}]{\includegraphics[scale=0.25,width=4.2cm,height=3.6cm]{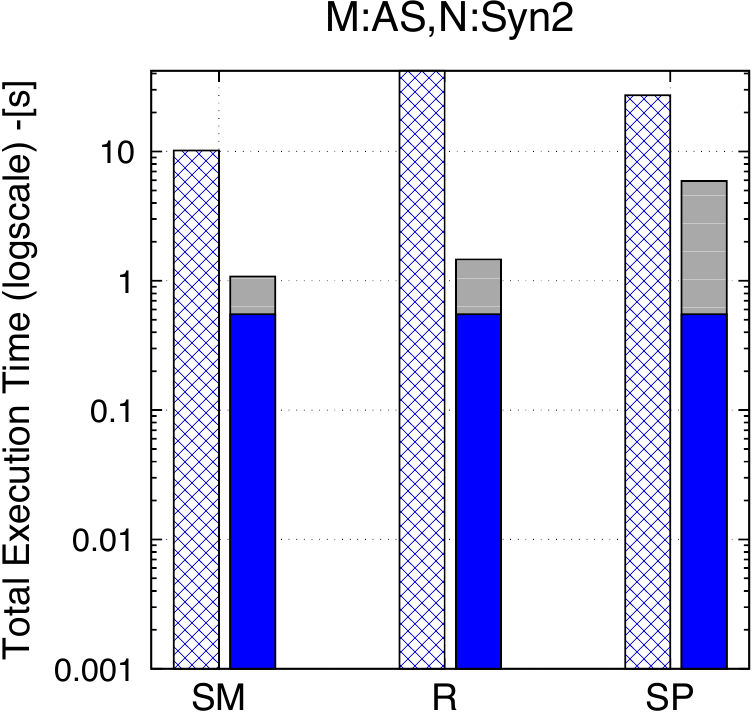}}
  \subfigure[\textbf{P2.13}]{\includegraphics[scale=0.25,width=4.4cm,height=3.6cm]{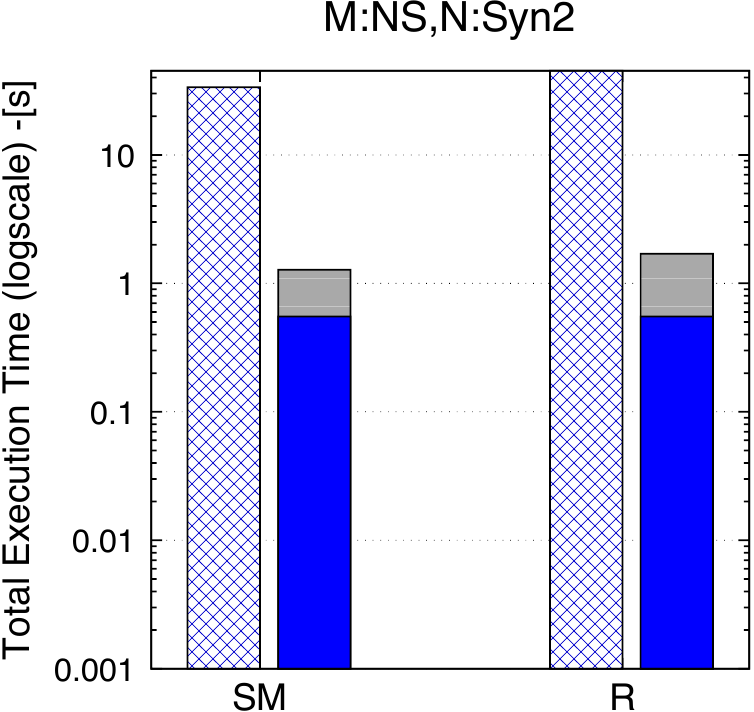}}
   \subfigure[\textbf{P2.13}]{\includegraphics[scale=0.25,width=4.4cm,height=3.6cm]{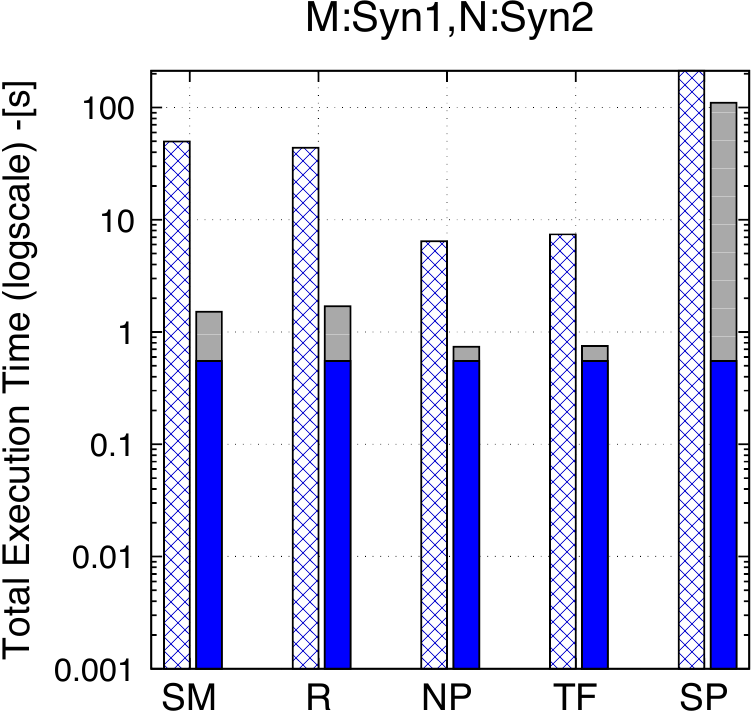}}
 \figspb\figspb\figspb
 \caption{P2.13 evaluation time with and without rewriting}
 \label{fig:la-p2.13}
\end{figure*}

\begin{figure*}[!htbp]
\figspa
  \centering
  \subfigure[\textbf{P2.14}]{\includegraphics[scale=0.25,width=4.2cm,height=3.6cm]{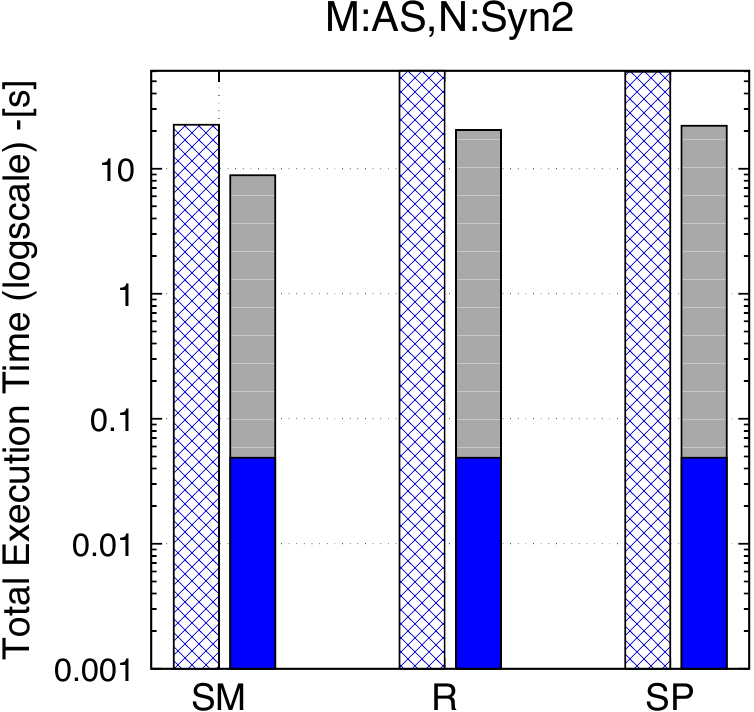}}
  \subfigure[\textbf{P2.14}]{\includegraphics[scale=0.25,width=4.4cm,height=3.6cm]{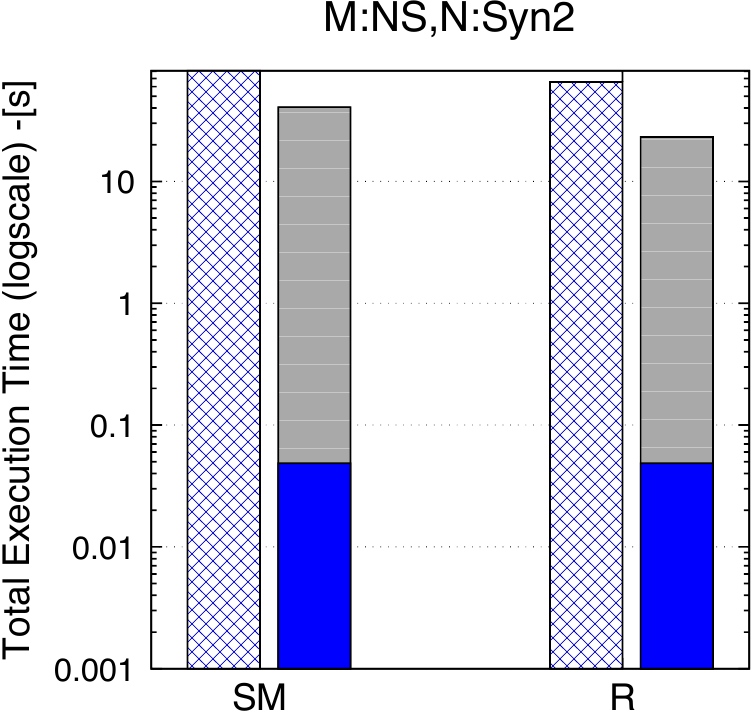}}
 \figspb\figspb\figspb
 \caption{P2.14 evaluation time with and without rewriting}
 \label{fig:la-p2.14}
\end{figure*}

\begin{figure*}[!htbp]
\figspa
  \centering
  \subfigure[\textbf{P2.15}]{\includegraphics[scale=0.25,width=4.2cm,height=3.6cm]{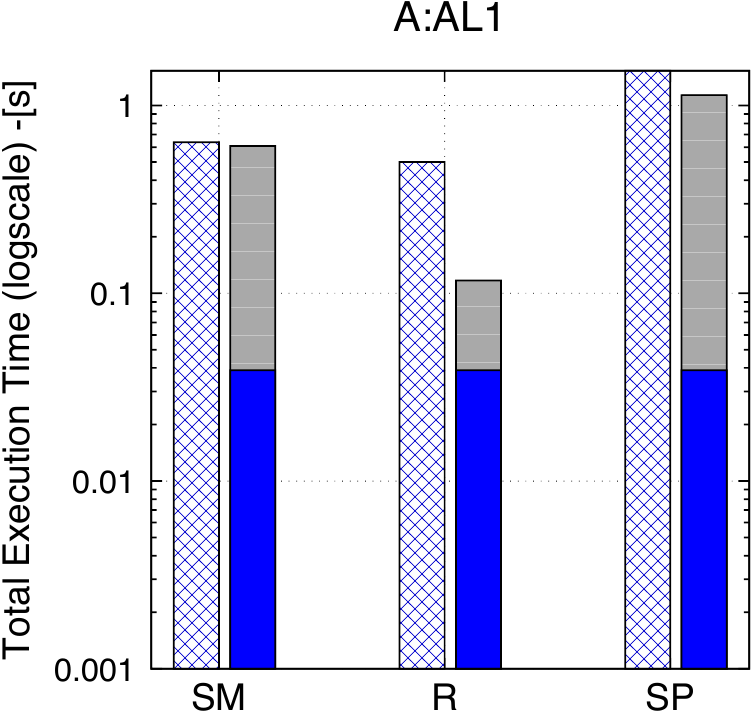}}
    \subfigure[\textbf{P2.15}]{\includegraphics[scale=0.25,width=4.2cm,height=3.6cm]{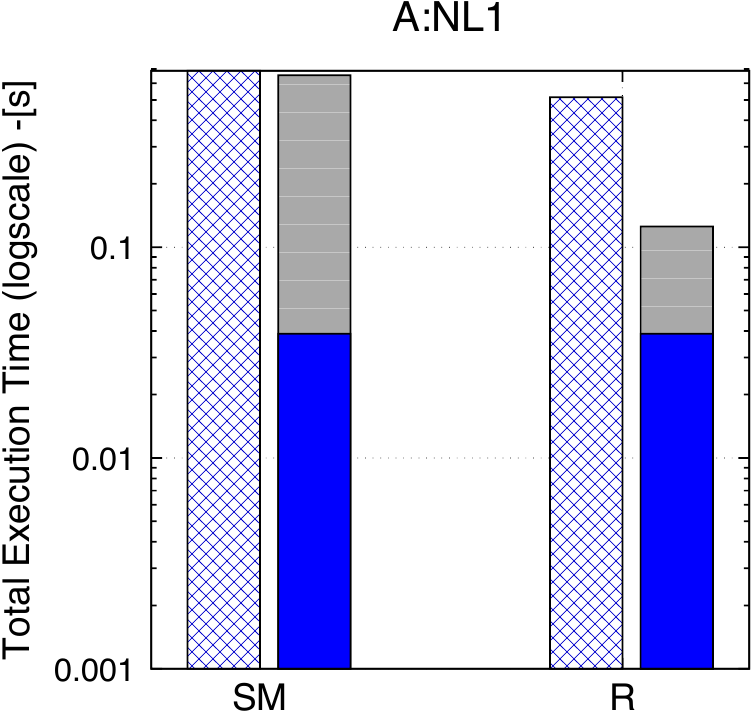}}
      \subfigure[\textbf{P2.15}]{\includegraphics[scale=0.25,width=4.4cm,height=3.6cm]{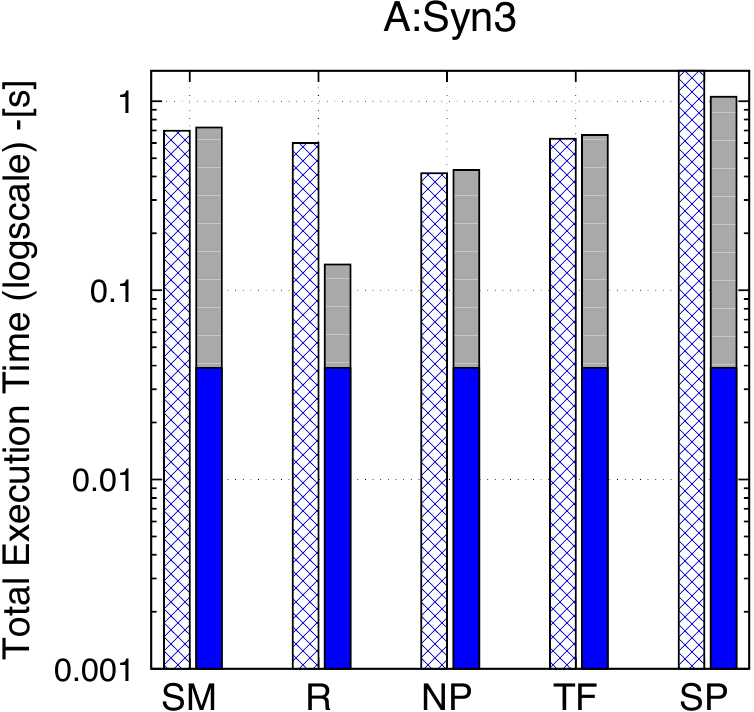}}
 \figspb\figspb\figspb
 \caption{P2.15 evaluation time with and without rewriting}
 \label{fig:la-p2.15}
\end{figure*}

\begin{figure*}[!htbp]
\figspa
  \centering
  \subfigure[\textbf{P2.16}]{\includegraphics[scale=0.25,width=4.2cm,height=3.6cm]{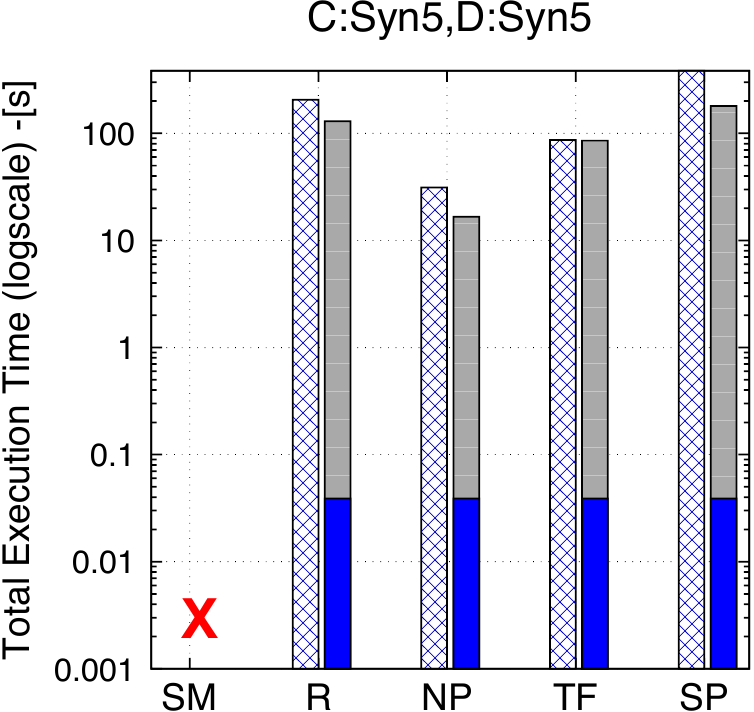}}
  \figspb\figspb\figspb
 \caption{P2.16 evaluation time with and without rewriting}
 \label{fig:la-p2.16}
\end{figure*}

\begin{figure*}[!htbp]
\figspa
  \centering
  \subfigure[\textbf{P2.18}]{\includegraphics[scale=0.25,width=4.2cm,height=3.6cm]{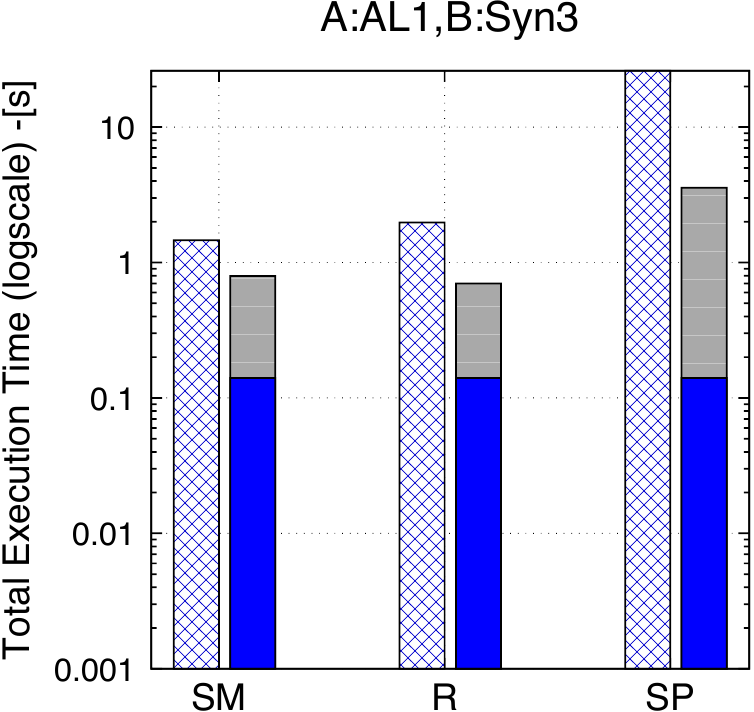}}
    \subfigure[\textbf{P2.18}]{\includegraphics[scale=0.25,width=4.2cm,height=3.6cm]{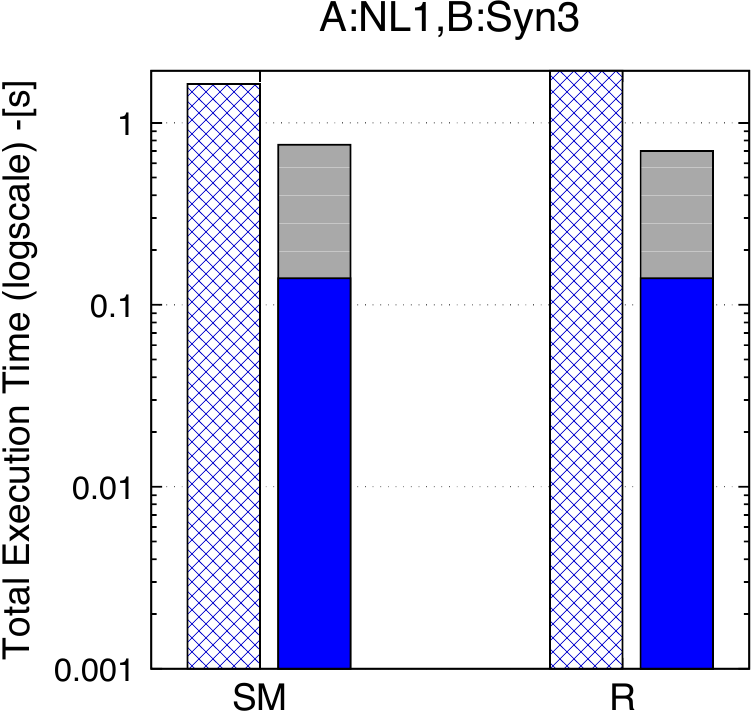}}
      \subfigure[\textbf{P2.18}]{\includegraphics[scale=0.25,width=4.4cm,height=3.6cm]{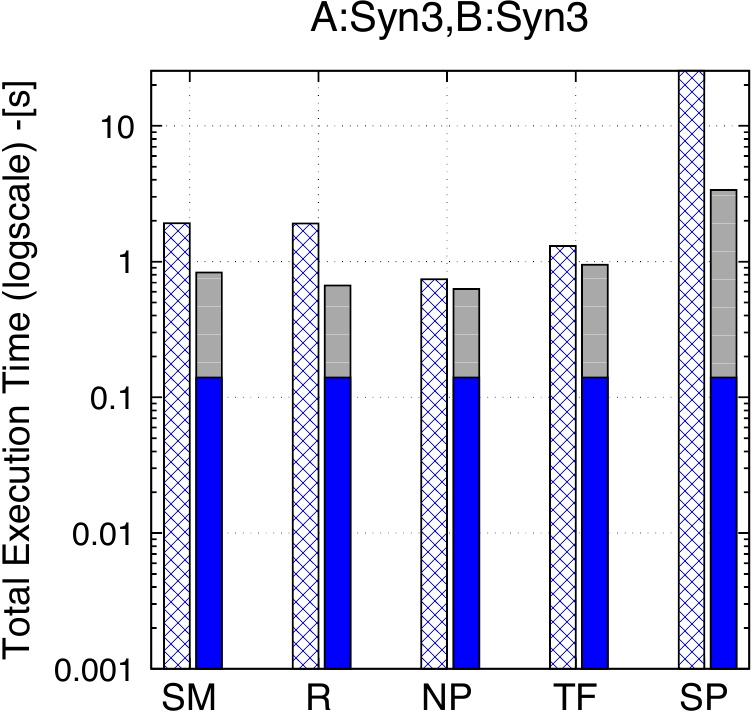}}
 \figspb\figspb\figspb
 \caption{P2.18 evaluation time with and without rewriting}
 \label{fig:la-p2.15}
\end{figure*}

%% file: appendixF.tex
\onecolumn
\section{Additional Results: $\views$ Pipelines}
\label{appendixF}
\begin{figure*}[!htbp]
\figspa
  \centering
      \subfigure[\textbf{P1.2}]{\includegraphics[scale=0.25,width=4.4cm,height=3.6cm]{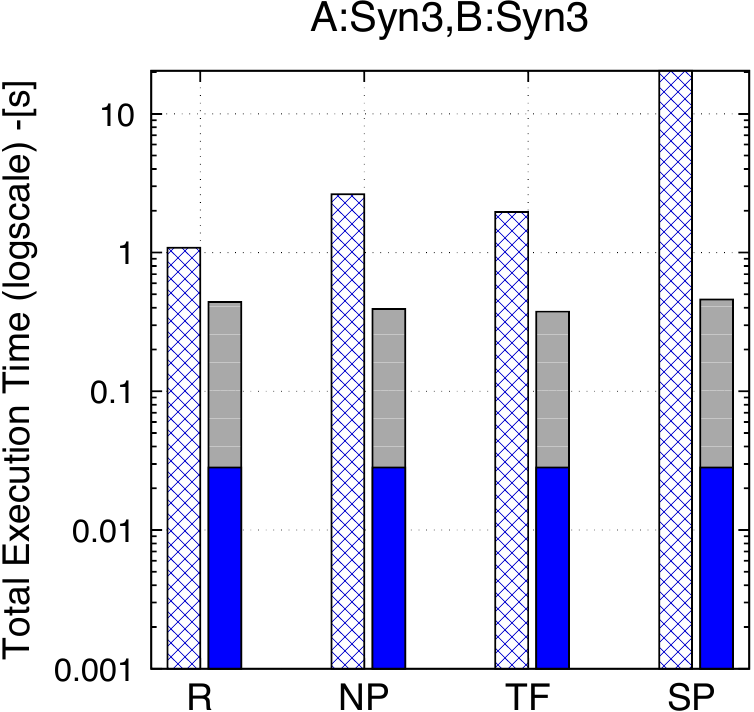}}
      \subfigure[\textbf{P1.2}]{\includegraphics[scale=0.25,width=4.4cm,height=3.6cm]{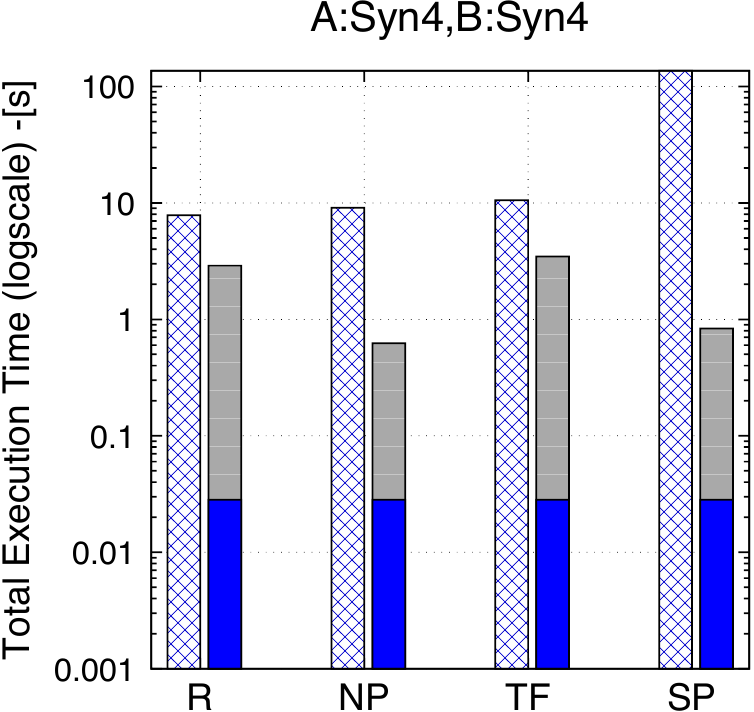}}
      \subfigure[\textbf{P1.3}]{\includegraphics[scale=0.25,width=4.4cm,height=3.6cm]{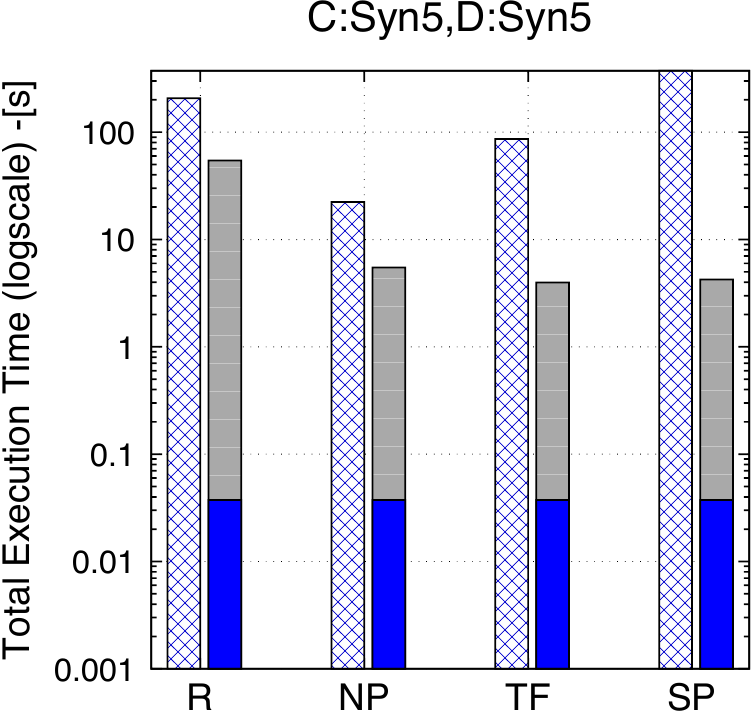}}
 \figspb\figspb\figspb
 \caption{P1.2 and P1.3 evaluation time with and without rewriting}
 \figspc
 \label{fig:la-p1.23}
\end{figure*}

\begin{figure*}[!htbp]
\figspa
  \centering
      \subfigure[\textbf{P1.4}]{\includegraphics[scale=0.25,width=4.2cm,height=3.6cm]{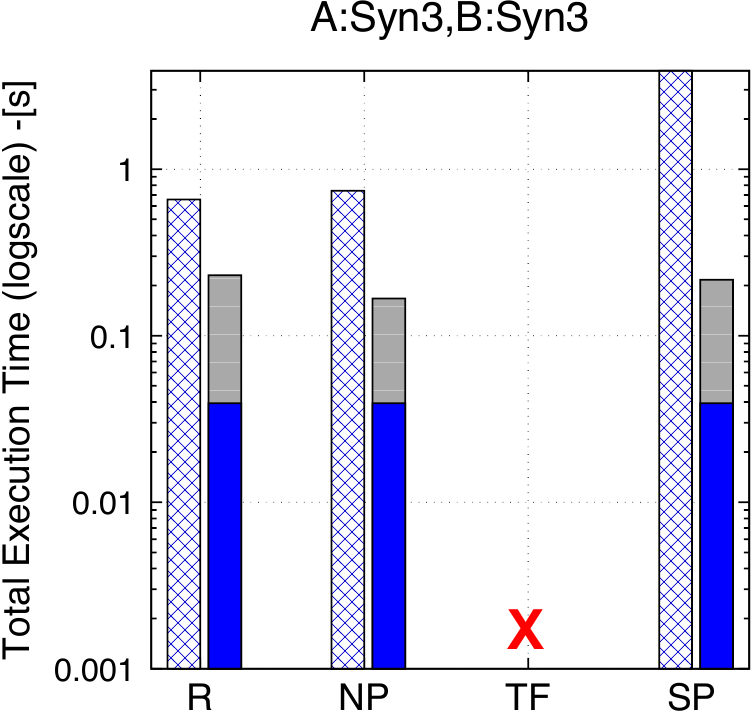}}
      \subfigure[\textbf{P1.4}]{\includegraphics[scale=0.25,width=4.2cm,height=3.6cm]{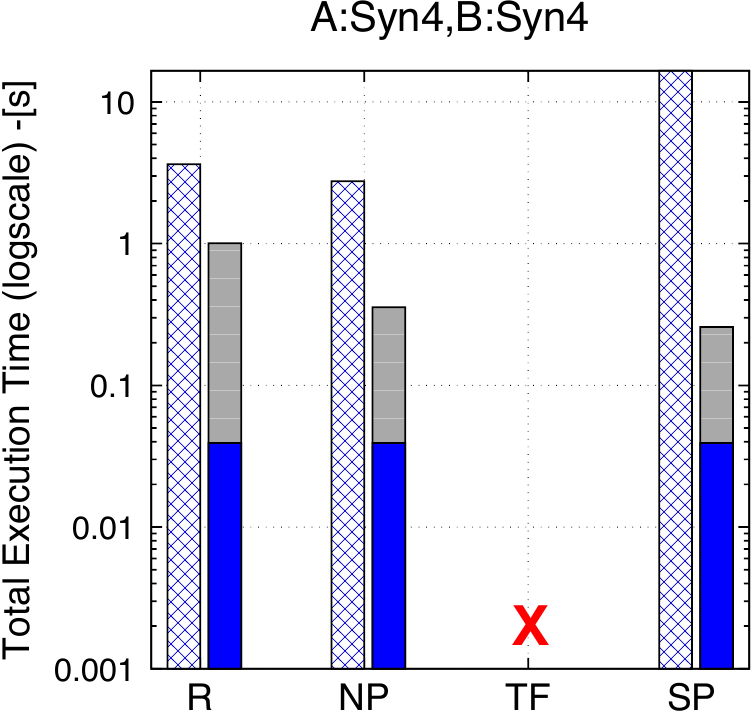}}
      \subfigure[\textbf{P1.11}]{\includegraphics[scale=0.25,width=4.2cm,height=3.6cm]{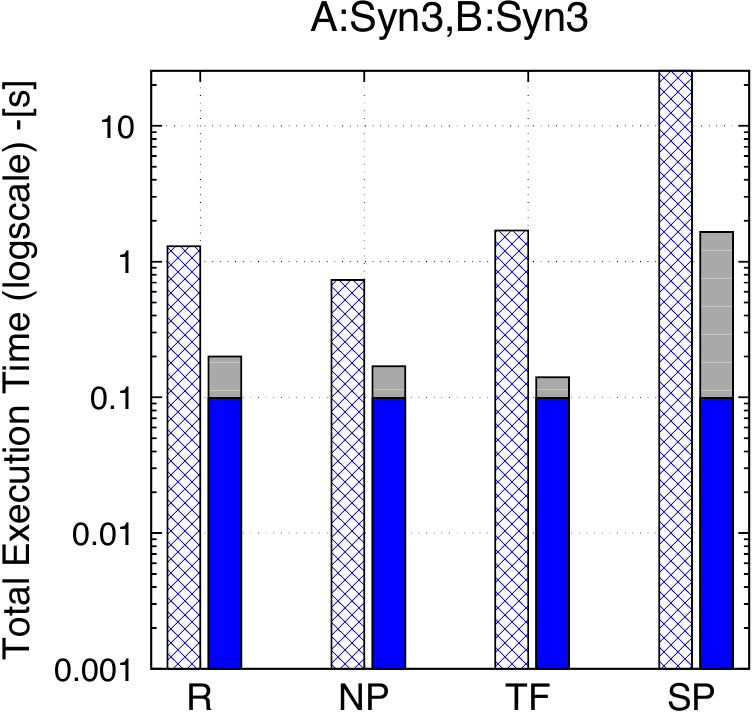}}
      \subfigure[\textbf{P1.11}]{\includegraphics[scale=0.25,width=4.2cm,height=3.6cm]{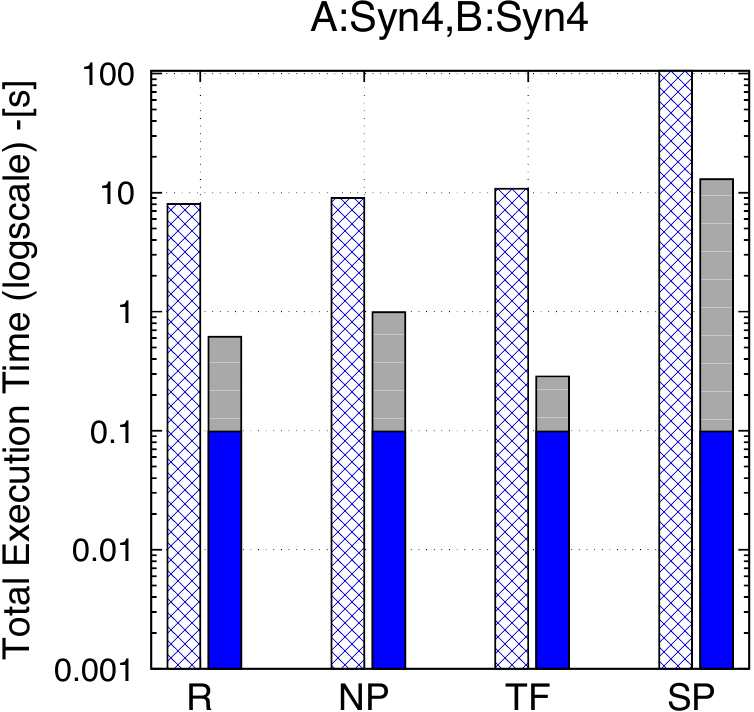}}
 \figspb\figspb\figspb
 \caption{P1.4 and P1.11 evaluation time with and without rewriting}
 \figspc
 \label{fig:la-p1.411}
\end{figure*}

\begin{figure*}[!htbp]
\figspa
  \centering
      \subfigure[\textbf{P1.17}]{\includegraphics[scale=0.25,width=4.2cm,height=3.6cm]{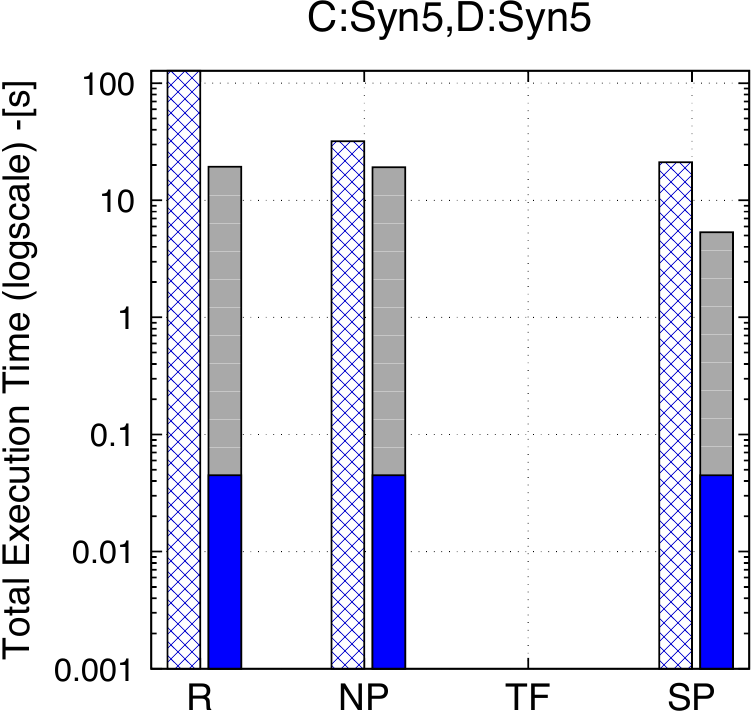}}
      \subfigure[\textbf{P1.19}]{\includegraphics[scale=0.25,width=4.2cm,height=3.6cm]{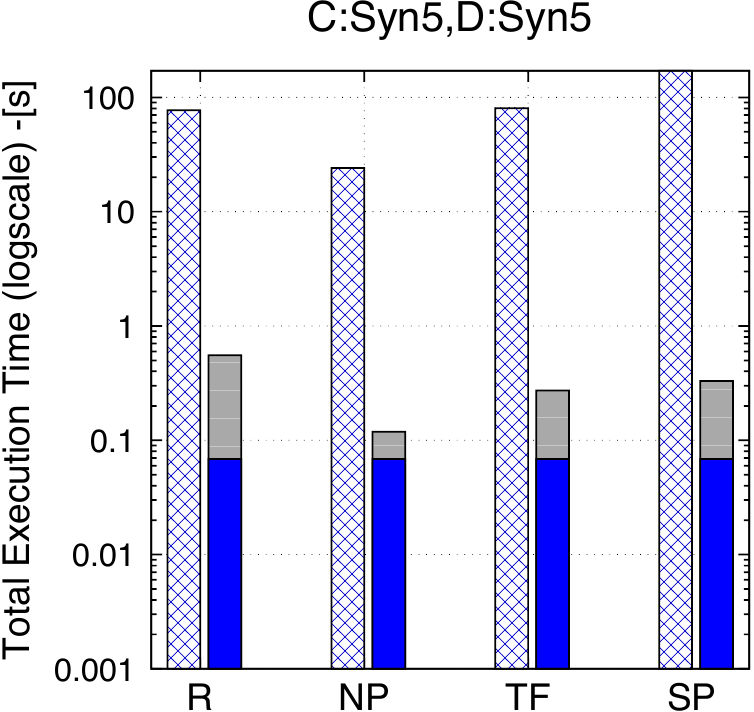}}
      \subfigure[\textbf{P1.20}]{\includegraphics[scale=0.25,width=4.2cm,height=3.6cm]{charts/AdditionalRes/Views/P1-11Syn3-crop.pdf}}
      \subfigure[\textbf{P1.21}]{\includegraphics[scale=0.25,width=4.2cm,height=3.6cm]{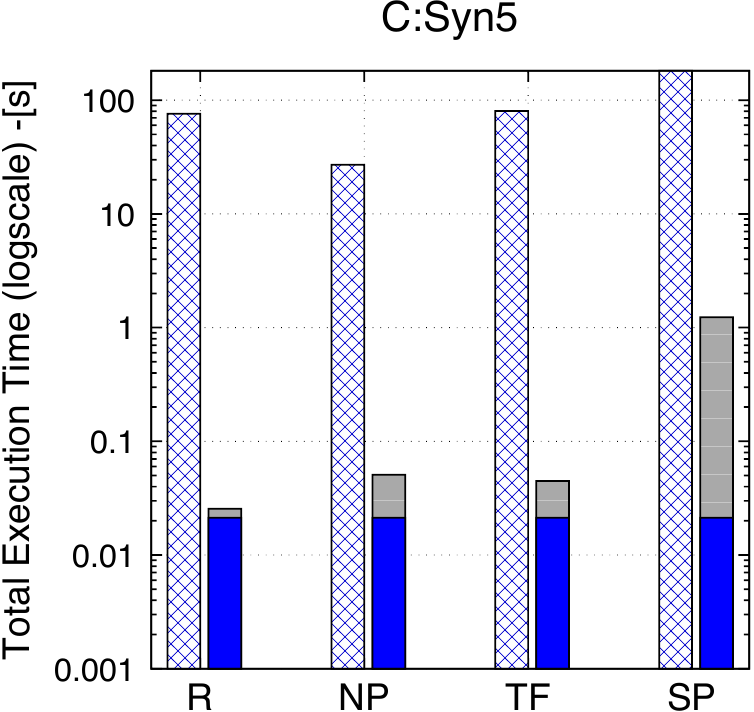}}
 \figspb\figspb\figspb
 \caption{P1.17,P1.19,P1.20 and P1.21 evaluation time with and without rewriting}
 \figspc
 \label{fig:la-p1.17-21}
\end{figure*}

\begin{figure*}[!htbp]
\figspa
 \centering
     \subfigure[\textbf{P1.22}]{\includegraphics[scale=0.25,width=4.2cm,height=3.6cm]{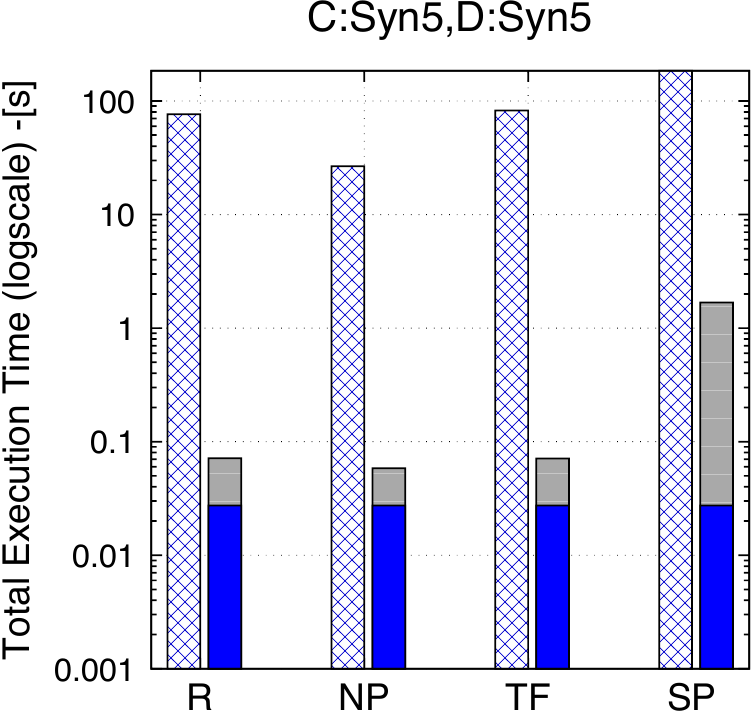}}
     \subfigure[\textbf{P1.23}]{\includegraphics[scale=0.25,width=4.2cm,height=3.6cm]{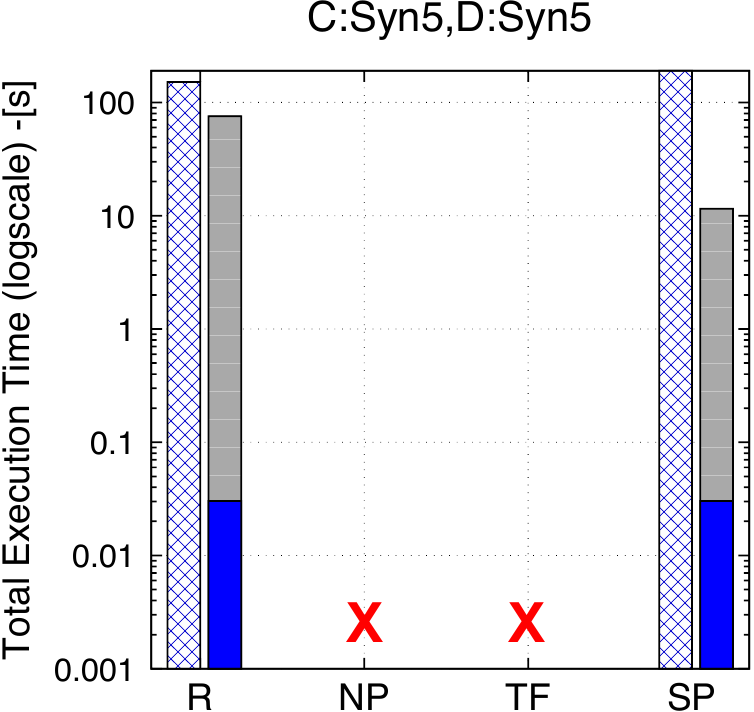}}
     \subfigure[\textbf{P1.24}]{\includegraphics[scale=0.25,width=4.cm,height=3.6cm]{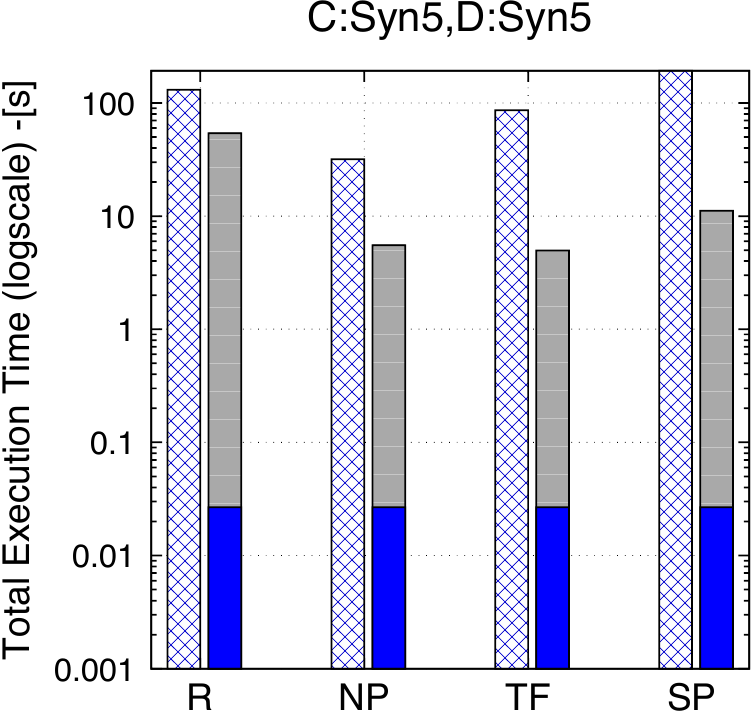}}
     \subfigure[\textbf{P1.29}]{\includegraphics[scale=0.25,width=4.2cm,height=3.6cm]{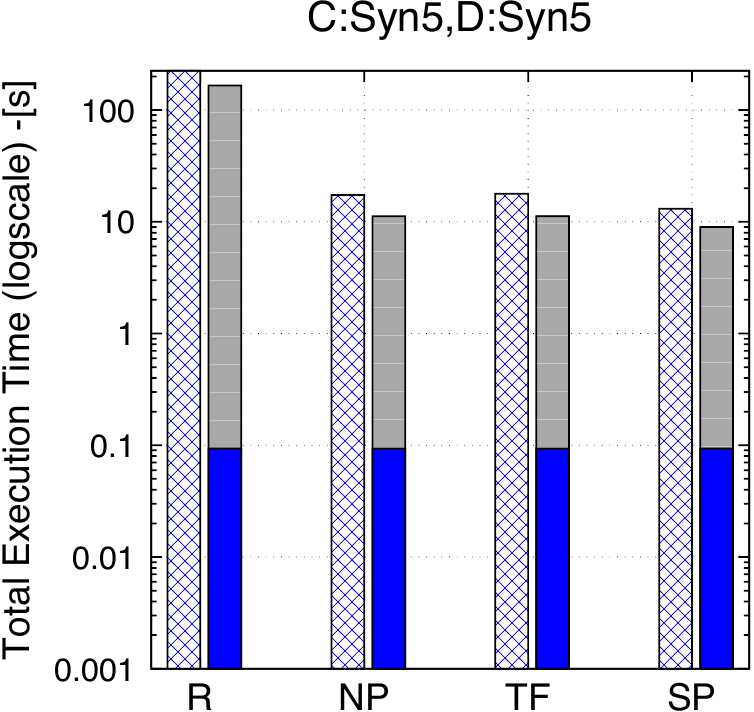}}
\figspb\figspb\figspb
\caption{P1.22,P1.23,P1.24 and P1.29 evaluation time with and without rewriting}
\figspc
\label{fig:la-p1.17-21}
\end{figure*}

 \begin{figure*}[!htbp]
 \figspa
  \centering
      \subfigure[\textbf{P2.2}]{\includegraphics[scale=0.25,width=4.2cm,height=3.6cm]{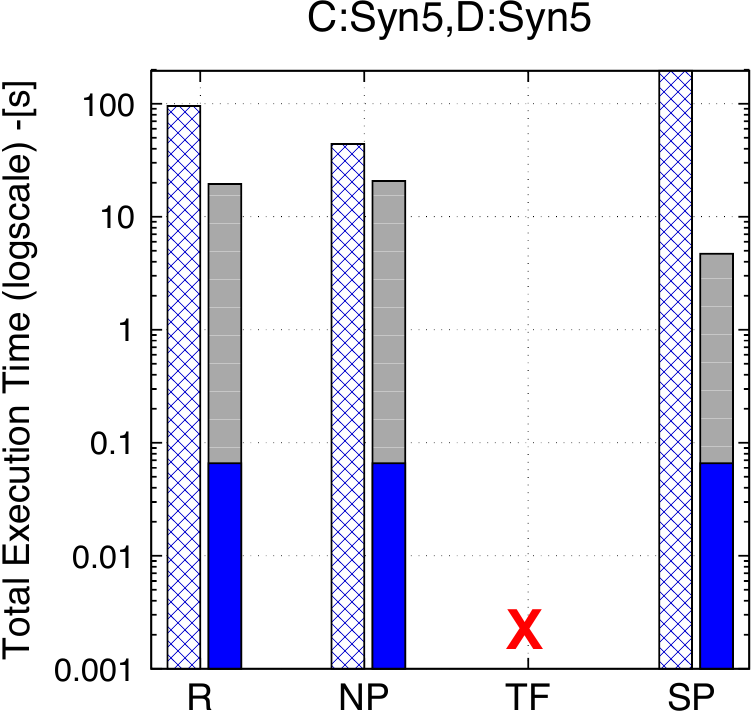}}
      \subfigure[\textbf{P2.4}]{\includegraphics[scale=0.25,width=4.2cm,height=3.6cm]{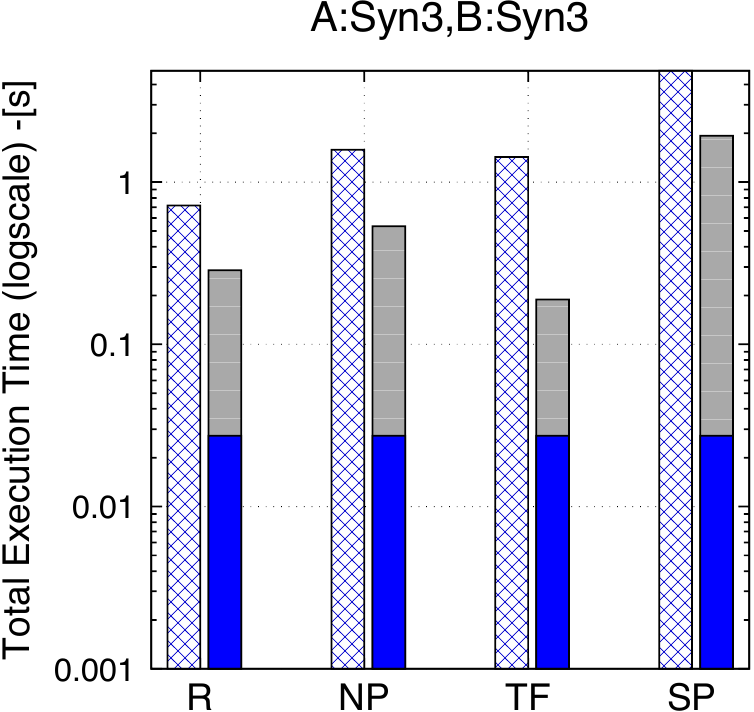}}
      \subfigure[\textbf{P2.4}]{\includegraphics[scale=0.25,width=4.2cm,height=3.6cm]{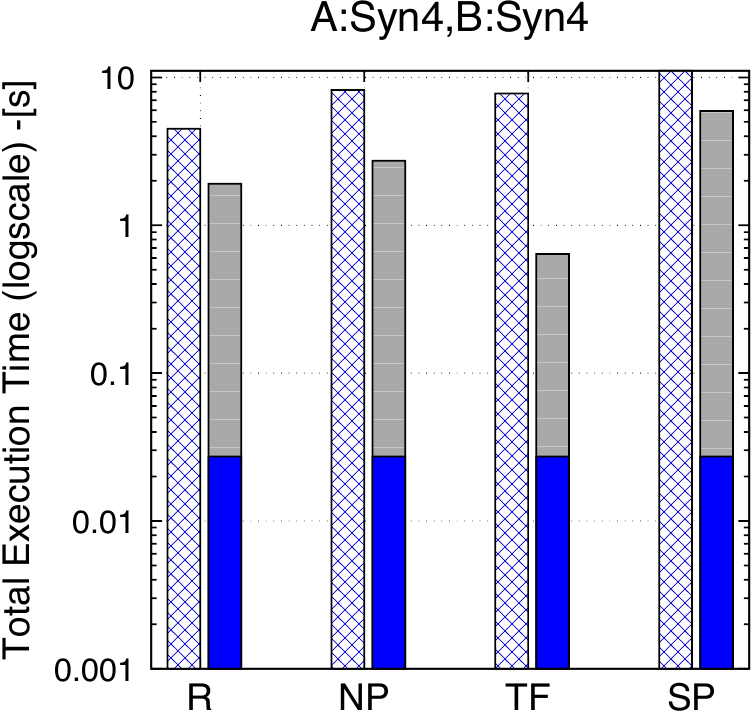}}
      \subfigure[\textbf{P2.5}]{\includegraphics[scale=0.25,width=4.2cm,height=3.6cm]{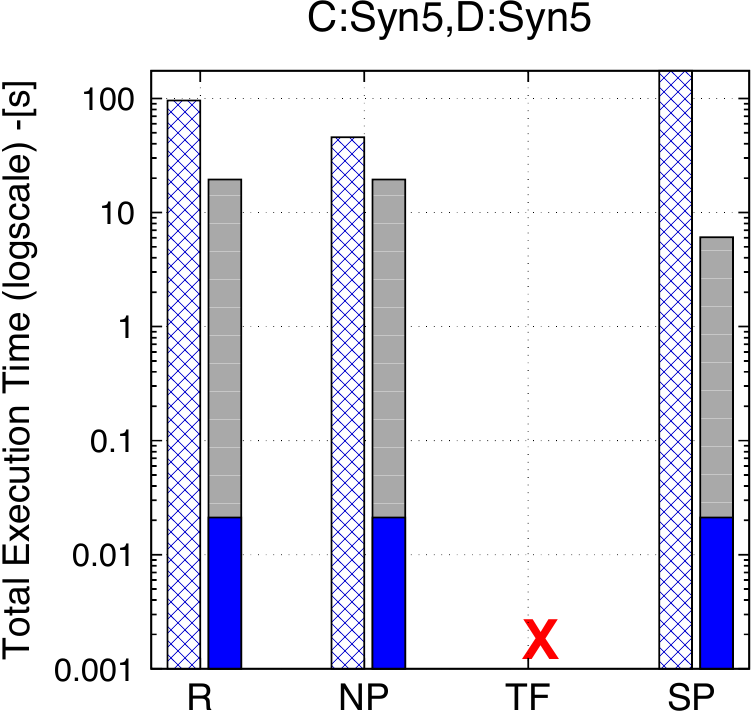}}
 \figspb\figspb\figspb
 \caption{P2.2,P2.4 and P2.5 evaluation time with and without rewriting}

 \label{fig:la-p2-5}
 \end{figure*}

 \begin{figure*}[!htbp]
 \figspa
  \centering
      \subfigure[\textbf{P2.9}]{\includegraphics[scale=0.25,width=4.2cm,height=3.6cm]{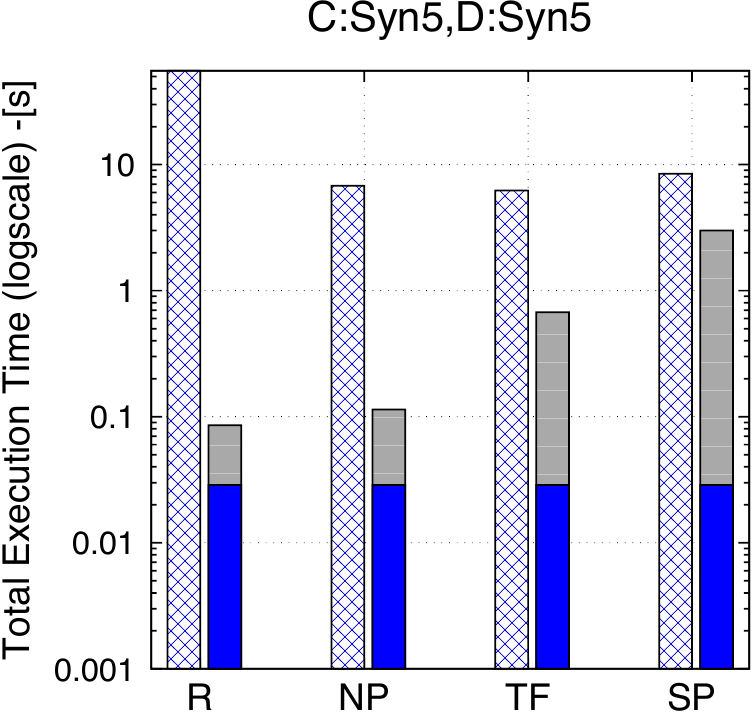}}
      \subfigure[\textbf{P2.11}]{\includegraphics[scale=0.25,width=4.2cm,height=3.6cm]{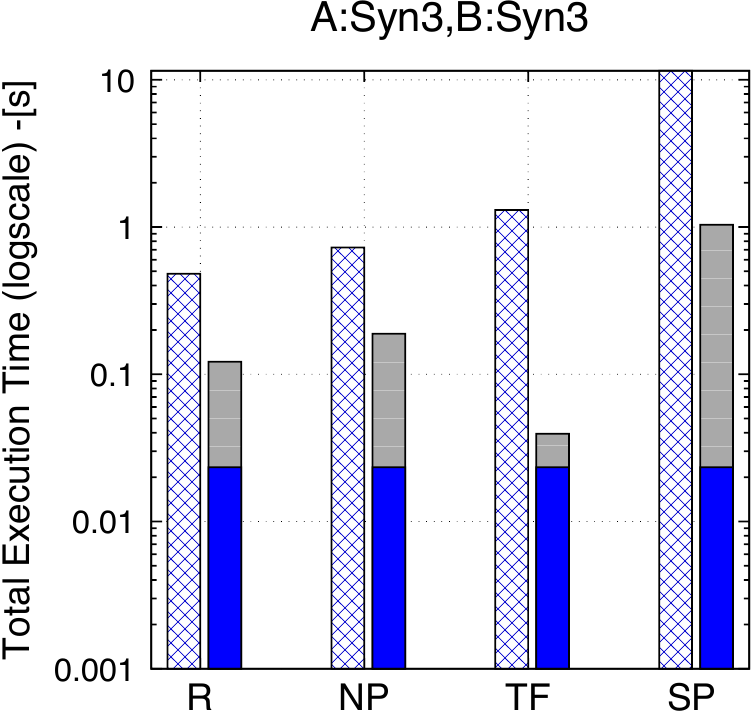}}
      \subfigure[\textbf{P2.11}]{\includegraphics[scale=0.25,width=4.2cm,height=3.6cm]{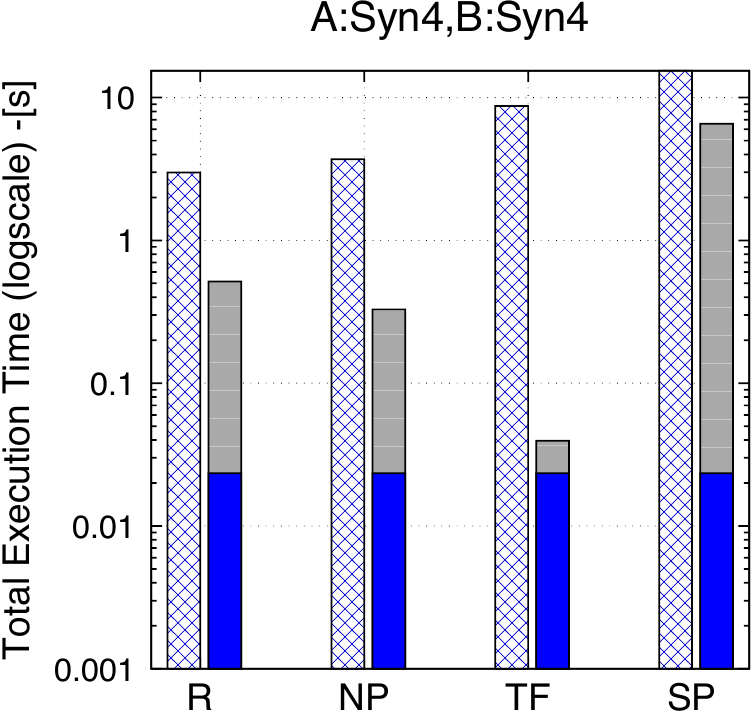}}
      \subfigure[\textbf{P2.16}]{\includegraphics[scale=0.25,width=4.2cm,height=3.6cm]{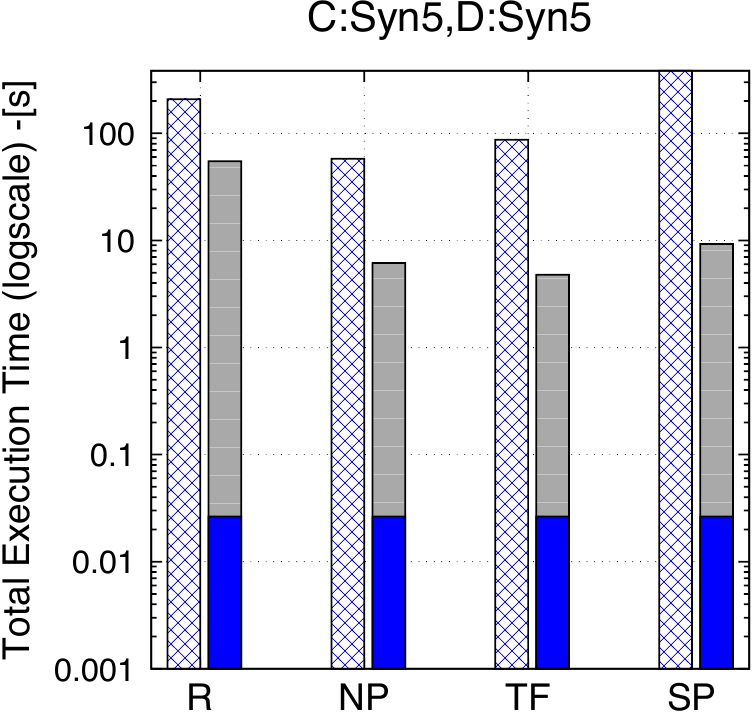}}
 \figspb\figspb\figspb
 \caption{P2.9,P2.11 and P2.16 evaluation time with and without rewriting}
 \figspc
 \label{fig:la-p9-16}
 \end{figure*}

%% file: appendixI.tex
\onecolumn
\section{Hybrid Twitter Benchmark Queries and Views}
\label{appendixI}
\counterwithin{lstlisting}{section}
\eat{\RA{Simplified version}}
\definecolor{dkgreen}{rgb}{0,0.6,0}
\definecolor{gray}{rgb}{0.5,0.5,0.5}
\definecolor{mauve}{rgb}{0.58,0,0.82}
\lstdefinestyle{myScalastyle}{
  frame=tb,
  language=scala,
  aboveskip=3mm,
  belowskip=3mm,
  showstringspaces=false,
  columns=flexible,
  basicstyle={\small\ttfamily},
  numbers=none,
  numberstyle=\tiny\color{gray},
  keywordstyle=\color{blue},
  commentstyle=\color{dkgreen},
  stringstyle=\color{mauve},
  frame=single,
  breaklines=true,
  breakatwhitespace=true,
  tabsize=3,
}
\lstset{basicstyle=\tiny,style=myScalastyle}
\begin{lstlisting}[caption={Q1},captionpos=b]
import scala.math._
import org.apache.sysml.api.mlcontext._
import org.apache.sysml.api.mlcontext.ScriptFactory._
import org.apache.sysml.api.mlcontext.MatrixFormat._
import org.apache.spark.SparkContext
import org.apache.spark.SparkContext._
import org.apache.spark.SparkConf
import org.apache.spark.sql._
import scala.io.Source
import scala.tools.nsc.io._
import scala.collection.immutable._
import org.apache.spark.storage.StorageLevel._

//Preprocessing  M 
val sqlContext = new org.apache.spark.sql.SQLContext(sc)
val User = sqlContext.read.format("csv").option("header", "true").load("User.csv");
User.createOrReplaceTempView("User")
val Tweet = sqlContext.read.format("csv").option("header", "true").load("Tweet.csv");
Tweet.createOrReplaceTempView("Tweet")
val M = spark.sql("""SELECT  followers_count,friends_count, listed_count,protected,verified
										 favorite_count,quote_count, reply_count,retweet_count,favorited,
							 			 possibly_sensitive,retweeted 
                     FROM User as U, Tweet as T 
                     WHERE  T.id = U.id""")
constructMatirx (M2,"M.csv");

//Preprocessing  (N)
val TweetJSON = spark.read.json("tweets/")
TweetJSON.createOrReplaceTempView("TweetJSON")
val N = spark.sql("""SELECT TJ.id, hID, TJ.filter_level
                        FROM TweetJSON AS TJ
                             LATERAL VIEW EXPLOD (entities.hashtags.id) AS hID
                        WHERE text LIKE '%covid%' AND 
                              TJ.place.country_code ='US'""")
constructMTXMatirx (2380000,1000,N,"N.mtx");

//Analysis
val ml = new MLContext(spark)
ml.setConfigProperty("sysml.cp.parallel.ops","true")
ml.setConfigProperty("sysml.optlevel","4")
ml.setConfigProperty("sysml.localtmpdir","systemml")
//DML Query
val dmlTextQuery =
  s"""
    |  M  = read("M.csv", format="csv", header=FALSE, sep=",");
    |  N  = readMM("N.mtx");
    |  V1 = read("V1.csv", format="csv", header=FALSE, sep=",");
    |  U1 = read("U1.csv", format="csv", header=FALSE, sep=",");
    |  X = read("X.csv", format="csv", header=FALSE, sep=",");
    |  NF = ifelse(N<=4,N,0)
    |  print(as.scalar(NF[20000,400]))
    |  res =rowSums(X%*%M) + (((U1%*%t(V1) ) + t(NF))%*%V1)
    |  while(FALSE){}
    |   print(as.scalar(res[500,1]))
    | }
""".stripMargin
val dmlScript = dml(dmlTextQuery)
val result = ml.execute(dmlScript)
\end{lstlisting}
\newpage

\begin{lstlisting}[caption={Q2},captionpos=b]
import scala.math._
import org.apache.sysml.api.mlcontext._
import org.apache.sysml.api.mlcontext.ScriptFactory._
import org.apache.sysml.api.mlcontext.MatrixFormat._
import org.apache.spark.SparkContext
import org.apache.spark.SparkContext._
import org.apache.spark.SparkConf
import org.apache.spark.sql._
import scala.io.Source
import scala.tools.nsc.io._
import scala.collection.immutable._
import org.apache.spark.storage.StorageLevel._

//Preprocessing  M 
val sqlContext = new org.apache.spark.sql.SQLContext(sc)
val User = sqlContext.read.format("csv").option("header", "true").load("User.csv");
User.createOrReplaceTempView("User")
val Tweet = sqlContext.read.format("csv").option("header", "true").load("Tweet.csv");
Tweet.createOrReplaceTempView("Tweet")
val M = spark.sql("""SELECT  followers_count,friends_count, listed_count,protected,verified
										 favorite_count,quote_count, reply_count,retweet_count,favorited,
							 			 possibly_sensitive,retweeted 
                     FROM User as U, Tweet as T 
                     WHERE  T.id = U.id""")
constructMatirx (M2,"M.csv");

//Preprocessing  (N)
val TweetJSON = spark.read.json("tweets/")
TweetJSON.createOrReplaceTempView("TweetJSON")
val N = spark.sql("""SELECT TJ.id, hID, TJ.filter_level
                        FROM TweetJSON AS TJ
                             LATERAL VIEW EXPLOD (entities.hashtags.id) AS hID
                        WHERE text LIKE '%covid%' AND 
                              TJ.place.country_code ='US'""")
constructMTXMatirx (2380000,1000,N,"N.mtx");

//Analysis
val ml = new MLContext(spark)
ml.setConfigProperty("sysml.cp.parallel.ops","true")
ml.setConfigProperty("sysml.optlevel","4")
ml.setConfigProperty("sysml.localtmpdir","systemml")
//DML Query
val dmlTextQuery =
  s"""
    |  M  = read("M.csv", format="csv", header=FALSE, sep=",")
    |  N  = readMM("N.mtx")
    |  X = read("X.csv", format="csv", header=FALSE, sep=",");
    |  U = read("U.csv", format="csv", header=FALSE, sep=",");
    |  NF = ifelse(N<=4,N,0)
    |  print(as.scalar(NF[20000,400]))
    |  res =U%*%colSums(t(X%*%M))+NF
    |  while(FALSE){}
    |   print(as.scalar(res[1,80]))
    | }
""".stripMargin
val dmlScript = dml(dmlTextQuery)
val result = ml.execute(dmlScript)
\end{lstlisting}
\newpage

\begin{lstlisting}[caption={Q3},captionpos=b]
import scala.math._
import org.apache.sysml.api.mlcontext._
import org.apache.sysml.api.mlcontext.ScriptFactory._
import org.apache.sysml.api.mlcontext.MatrixFormat._
import org.apache.spark.SparkContext
import org.apache.spark.SparkContext._
import org.apache.spark.SparkConf
import org.apache.spark.sql._
import scala.io.Source
import scala.tools.nsc.io._
import scala.collection.immutable._
import org.apache.spark.storage.StorageLevel._

//Preprocessing  M 
val sqlContext = new org.apache.spark.sql.SQLContext(sc)
val User = sqlContext.read.format("csv").option("header", "true").load("User.csv");
User.createOrReplaceTempView("User")
val Tweet = sqlContext.read.format("csv").option("header", "true").load("Tweet.csv");
Tweet.createOrReplaceTempView("Tweet")
val M = spark.sql("""SELECT  followers_count,friends_count, listed_count,protected,verified
										 favorite_count,quote_count, reply_count,retweet_count,favorited,
							 			 possibly_sensitive,retweeted 
                     FROM User as U, Tweet as T 
                     WHERE  T.id = U.id""")
constructMatirx (M2,"M.csv");

//Preprocessing  (N)
val TweetJSON = spark.read.json("tweets/")
TweetJSON.createOrReplaceTempView("TweetJSON")
val N = spark.sql("""SELECT TJ.id, hID, TJ.filter_level
                        FROM TweetJSON AS TJ
                             LATERAL VIEW EXPLOD (entities.hashtags.id) AS hID
                        WHERE text LIKE '%covid%' AND 
                              TJ.place.country_code ='US'""")
constructMTXMatirx (2380000,1000,N,"N.mtx");

//Analysis
val ml = new MLContext(spark)
ml.setConfigProperty("sysml.cp.parallel.ops","true")
ml.setConfigProperty("sysml.optlevel","4")
ml.setConfigProperty("sysml.localtmpdir","systemml")
//DML Query
val dmlTextQuery =
  s"""
    |  M  = read("M.csv", format="csv", header=FALSE, sep=",");
    |  N  = readMM("N.mtx");
    |  X = read("X.csv", format="csv", header=FALSE, sep=",");
    |  V = read("V.csv", format="csv", header=FALSE, sep=",");
    |  NF = ifelse(N<=4,N,0)
    |  print(as.scalar(NF[20000,400]))
    |  res = ((NF+X)%*%V)%*% colSums(M)
    |  while(FALSE){}
    |   print(as.scalar(res[500,1]))
    | }
""".stripMargin
val dmlScript = dml(dmlTextQuery)
val result = ml.execute(dmlScript)
\end{lstlisting}
\newpage

\begin{lstlisting}[caption={Q4},captionpos=b]
import scala.math._
import org.apache.sysml.api.mlcontext._
import org.apache.sysml.api.mlcontext.ScriptFactory._
import org.apache.sysml.api.mlcontext.MatrixFormat._
import org.apache.spark.SparkContext
import org.apache.spark.SparkContext._
import org.apache.spark.SparkConf
import org.apache.spark.sql._
import scala.io.Source
import scala.tools.nsc.io._
import scala.collection.immutable._
import org.apache.spark.storage.StorageLevel._

//Preprocessing  M 
val sqlContext = new org.apache.spark.sql.SQLContext(sc)
val User = sqlContext.read.format("csv").option("header", "true").load("User.csv");
User.createOrReplaceTempView("User")
val Tweet = sqlContext.read.format("csv").option("header", "true").load("Tweet.csv");
Tweet.createOrReplaceTempView("Tweet")
val M = spark.sql("""SELECT  followers_count,friends_count, listed_count,protected,verified
										 favorite_count,quote_count, reply_count,retweet_count,favorited,
							 			 possibly_sensitive,retweeted 
                     FROM User as U, Tweet as T 
                     WHERE  T.id = U.id""")
constructMatirx (M2,"M.csv");

//Preprocessing  (N)
val TweetJSON = spark.read.json("tweets/")
TweetJSON.createOrReplaceTempView("TweetJSON")
val N = spark.sql("""SELECT TJ.id, hID, TJ.filter_level
                        FROM TweetJSON AS TJ
                             LATERAL VIEW EXPLOD (entities.hashtags.id) AS hID
                        WHERE text LIKE '%covid%' AND 
                              TJ.place.country_code ='US'""")
constructMTXMatirx (2380000,1000,N,"N.mtx");

//Analysis
val ml = new MLContext(spark)
ml.setConfigProperty("sysml.cp.parallel.ops","true")
ml.setConfigProperty("sysml.optlevel","4")
ml.setConfigProperty("sysml.localtmpdir","systemml")
//DML Query
val dmlTextQuery =
  s"""
    |  M  = read("M.csv", format="csv", header=FALSE, sep=",")
    |  N  = readMM("N.mtx")
    |  X = read("X.csv", format="csv", header=FALSE, sep=",");
    |  U = read("U.csv", format="csv", header=FALSE, sep=",");
    |  v = read("v.csv", format="csv", header=FALSE, sep=",");
    |  NF = ifelse(N<=4,N,0)
    |  print(as.scalar(NF[20000,400]))
    |  res = sum(U+NF%*%colSums(X%*%M)%*%v)
    |  while(FALSE){}
    |   print(res)
    | }
""".stripMargin
val dmlScript = dml(dmlTextQuery)
val result = ml.execute(dmlScript)
\end{lstlisting}
\newpage
\begin{lstlisting}[caption={Q5},captionpos=b]
import scala.math._
import org.apache.sysml.api.mlcontext._
import org.apache.sysml.api.mlcontext.ScriptFactory._
import org.apache.sysml.api.mlcontext.MatrixFormat._
import org.apache.spark.SparkContext
import org.apache.spark.SparkContext._
import org.apache.spark.SparkConf
import org.apache.spark.sql._
import scala.io.Source
import scala.tools.nsc.io._
import scala.collection.immutable._
import org.apache.spark.storage.StorageLevel._

//Preprocessing  M 
val sqlContext = new org.apache.spark.sql.SQLContext(sc)
val User = sqlContext.read.format("csv").option("header", "true").load("User.csv");
User.createOrReplaceTempView("User")
val Tweet = sqlContext.read.format("csv").option("header", "true").load("Tweet.csv");
Tweet.createOrReplaceTempView("Tweet")
val M = spark.sql("""SELECT  followers_count,friends_count, listed_count,protected,verified
										 favorite_count,quote_count, reply_count,retweet_count,favorited,
							 			 possibly_sensitive,retweeted 
                     FROM User as U, Tweet as T 
                     WHERE  T.id = U.id""")
constructMatirx (M2,"M.csv");

//Preprocessing  (N)
val TweetJSON = spark.read.json("tweets/")
TweetJSON.createOrReplaceTempView("TweetJSON")
val N = spark.sql("""SELECT TJ.id, hID, TJ.filter_level
                        FROM TweetJSON AS TJ
                             LATERAL VIEW EXPLOD (entities.hashtags.id) AS hID
                        WHERE text LIKE '%covid%' AND 
                              TJ.place.country_code ='US'""")
constructMTXMatirx (2380000,1000,N,"N.mtx");

//Analysis
val ml = new MLContext(spark)
ml.setConfigProperty("sysml.cp.parallel.ops","true")
ml.setConfigProperty("sysml.optlevel","4")
ml.setConfigProperty("sysml.localtmpdir","systemml")
//DML Query
val dmlTextQuery =
  s"""
    |  M  = read("M.csv", format="csv", header=FALSE, sep=",")
    |  N  = readMM("N.mtx")
    |  U = read("U.csv", format="csv", header=FALSE, sep=",");
    |  X = read("X.csv", format="csv", header=FALSE, sep=",");
    |  NF = ifelse(N<=4,N,0)
    |  print(as.scalar(NF[20000,400]))
    |  res = U%*%colSums(M%*%X) + NF 
    |  while(FALSE){}
    |   print(res)
    | }
""".stripMargin
val dmlScript = dml(dmlTextQuery)
val result = ml.execute(dmlScript)
\end{lstlisting}
\newpage

\begin{lstlisting}[caption={Q6},captionpos=b]
import scala.math._
import org.apache.sysml.api.mlcontext._
import org.apache.sysml.api.mlcontext.ScriptFactory._
import org.apache.sysml.api.mlcontext.MatrixFormat._
import org.apache.spark.SparkContext
import org.apache.spark.SparkContext._
import org.apache.spark.SparkConf
import org.apache.spark.sql._
import scala.io.Source
import scala.tools.nsc.io._
import scala.collection.immutable._
import org.apache.spark.storage.StorageLevel._

//Preprocessing  M 
val sqlContext = new org.apache.spark.sql.SQLContext(sc)
val User = sqlContext.read.format("csv").option("header", "true").load("User.csv");
User.createOrReplaceTempView("User")
val Tweet = sqlContext.read.format("csv").option("header", "true").load("Tweet.csv");
Tweet.createOrReplaceTempView("Tweet")
val M = spark.sql("""SELECT  followers_count,friends_count, listed_count,protected,verified
										 favorite_count,quote_count, reply_count,retweet_count,favorited,
							 			 possibly_sensitive,retweeted 
                     FROM User as U, Tweet as T 
                     WHERE  T.id = U.id""")
constructMatirx (M2,"M.csv");

//Preprocessing  (N)
val TweetJSON = spark.read.json("tweets/")
TweetJSON.createOrReplaceTempView("TweetJSON")
val N = spark.sql("""SELECT TJ.id, hID, TJ.filter_level
                        FROM TweetJSON AS TJ
                             LATERAL VIEW EXPLOD (entities.hashtags.id) AS hID
                        WHERE text LIKE '%covid%' AND 
                              TJ.place.country_code ='US'""")
constructMTXMatirx (2380000,1000,N,"N.mtx");

//Analysis
val ml = new MLContext(spark)
ml.setConfigProperty("sysml.cp.parallel.ops","true")
ml.setConfigProperty("sysml.optlevel","4")
ml.setConfigProperty("sysml.localtmpdir","systemml")
//DML Query
val dmlTextQuery =
  s"""
    |  M  = read("M.csv", format="csv", header=FALSE, sep=",");
    |  N  = readMM("N.mtx");
    |  V1 = read("V1.csv", format="csv", header=FALSE, sep=",");
    |  U1 = read("U1.csv", format="csv", header=FALSE, sep=",");
    |  X = read("X.csv", format="csv", header=FALSE, sep=",");
    |  NF = ifelse(N<=4,N,0)
    |  print(as.scalar(NF[20000,400]))
    |  res = rowSums(t(M%*%X)) + (((U1%*%t(V1) ) + NF)%*%V1) 
    |  while(FALSE){}
    |   print(res[1000,1])
    | }
""".stripMargin
val dmlScript = dml(dmlTextQuery)
val result = ml.execute(dmlScript)
\end{lstlisting}
\newpage

\begin{lstlisting}[caption={Q7},captionpos=b]
import scala.math._
import org.apache.sysml.api.mlcontext._
import org.apache.sysml.api.mlcontext.ScriptFactory._
import org.apache.sysml.api.mlcontext.MatrixFormat._
import org.apache.spark.SparkContext
import org.apache.spark.SparkContext._
import org.apache.spark.SparkConf
import org.apache.spark.sql._
import scala.io.Source
import scala.tools.nsc.io._
import scala.collection.immutable._
import org.apache.spark.storage.StorageLevel._

//Preprocessing  M 
val sqlContext = new org.apache.spark.sql.SQLContext(sc)
val User = sqlContext.read.format("csv").option("header", "true").load("User.csv");
User.createOrReplaceTempView("User")
val Tweet = sqlContext.read.format("csv").option("header", "true").load("Tweet.csv");
Tweet.createOrReplaceTempView("Tweet")
val M = spark.sql("""SELECT  followers_count,friends_count, listed_count,protected,verified
										 favorite_count,quote_count, reply_count,retweet_count,favorited,
							 			 possibly_sensitive,retweeted 
                     FROM User as U, Tweet as T 
                     WHERE  T.id = U.id""")
constructMatirx (M2,"M.csv");

//Preprocessing  (N)
val TweetJSON = spark.read.json("tweets/")
TweetJSON.createOrReplaceTempView("TweetJSON")
val N = spark.sql("""SELECT TJ.id, hID, TJ.filter_level
                        FROM TweetJSON AS TJ
                             LATERAL VIEW EXPLOD (entities.hashtags.id) AS hID
                        WHERE text LIKE '%covid%' AND 
                              TJ.place.country_code ='US'""")
constructMTXMatirx (2380000,1000,N,"N.mtx");

//Analysis
val ml = new MLContext(spark)
ml.setConfigProperty("sysml.cp.parallel.ops","true")
ml.setConfigProperty("sysml.optlevel","4")
ml.setConfigProperty("sysml.localtmpdir","systemml")
//DML Query
val dmlTextQuery =
  s"""
    |  M  = read("M.csv", format="csv", header=FALSE, sep=",")
    |  N  = readMM("N.mtx")
    |  X = read("X.csv", format="csv", header=FALSE, sep=",");
    |  U = read("U,csv", format="csv", header=FALSE, sep=",");
    |  NF = ifelse(N<=4,N,0)
    |  print(as.scalar(NF[20000,400]))
    |   res= X%*%NF%*%U + rowSums(t(M))
    |  while(FALSE){}
    |   print(res[2,30])
    | }
""".stripMargin
val dmlScript = dml(dmlTextQuery)
val result = ml.execute(dmlScript)
\end{lstlisting}
\newpage

\begin{lstlisting}[caption={Q8},captionpos=b]
import scala.math._
import org.apache.sysml.api.mlcontext._
import org.apache.sysml.api.mlcontext.ScriptFactory._
import org.apache.sysml.api.mlcontext.MatrixFormat._
import org.apache.spark.SparkContext
import org.apache.spark.SparkContext._
import org.apache.spark.SparkConf
import org.apache.spark.sql._
import scala.io.Source
import scala.tools.nsc.io._
import scala.collection.immutable._
import org.apache.spark.storage.StorageLevel._

//Preprocessing  M 
val sqlContext = new org.apache.spark.sql.SQLContext(sc)
val User = sqlContext.read.format("csv").option("header", "true").load("User.csv");
User.createOrReplaceTempView("User")
val Tweet = sqlContext.read.format("csv").option("header", "true").load("Tweet.csv");
Tweet.createOrReplaceTempView("Tweet")
val M = spark.sql("""SELECT  followers_count,friends_count, listed_count,protected,verified
										 favorite_count,quote_count, reply_count,retweet_count,favorited,
							 			 possibly_sensitive,retweeted 
                     FROM User as U, Tweet as T 
                     WHERE  T.id = U.id""")
constructMatirx (M2,"M.csv");

//Preprocessing  (N)
val TweetJSON = spark.read.json("tweets/")
TweetJSON.createOrReplaceTempView("TweetJSON")
val N = spark.sql("""SELECT TJ.id, hID, TJ.filter_level
                        FROM TweetJSON AS TJ
                             LATERAL VIEW EXPLOD (entities.hashtags.id) AS hID
                        WHERE text LIKE '%covid%' AND 
                              TJ.place.country_code ='US'""")
constructMTXMatirx (2380000,1000,N,"N.mtx");

//Analysis
val ml = new MLContext(spark)
ml.setConfigProperty("sysml.cp.parallel.ops","true")
ml.setConfigProperty("sysml.optlevel","4")
ml.setConfigProperty("sysml.localtmpdir","systemml")
//DML Query
val dmlTextQuery =
  s"""
    |  M  = read("M.csv", format="csv", header=FALSE, sep=",")
    |  N  = readMM("N.mtx")
    |  C = read("C.csv", format="csv", header=FALSE, sep=",");
    |  V = read("V.csv", format="csv", header=FALSE, sep=",");
    |  X = read("X.csv", format="csv", header=FALSE, sep=",");
    |  NF = ifelse(N<=4,N,0)
    |  print(as.scalar(NF[20000,400]))
    |  res = NF*trace(C+ V%*%colSums(M%*%X)%*%C)
    |  while(FALSE){}
    |   print(res[2,30])
    | }
""".stripMargin
val dmlScript = dml(dmlTextQuery)
val result = ml.execute(dmlScript)
\end{lstlisting}
\newpage

\begin{lstlisting}[caption={Q9},captionpos=b]
import scala.math._
import org.apache.sysml.api.mlcontext._
import org.apache.sysml.api.mlcontext.ScriptFactory._
import org.apache.sysml.api.mlcontext.MatrixFormat._
import org.apache.spark.SparkContext
import org.apache.spark.SparkContext._
import org.apache.spark.SparkConf
import org.apache.spark.sql._
import scala.io.Source
import scala.tools.nsc.io._
import scala.collection.immutable._
import org.apache.spark.storage.StorageLevel._

//Preprocessing  M 
val sqlContext = new org.apache.spark.sql.SQLContext(sc)
val User = sqlContext.read.format("csv").option("header", "true").load("User.csv");
User.createOrReplaceTempView("User")
val Tweet = sqlContext.read.format("csv").option("header", "true").load("Tweet.csv");
Tweet.createOrReplaceTempView("Tweet")
val M = spark.sql("""SELECT  followers_count,friends_count, listed_count,protected,verified
										 favorite_count,quote_count, reply_count,retweet_count,favorited,
							 			 possibly_sensitive,retweeted 
                     FROM User as U, Tweet as T 
                     WHERE  T.id = U.id""")
constructMatirx (M2,"M.csv");

//Preprocessing  (N)
val TweetJSON = spark.read.json("tweets/")
TweetJSON.createOrReplaceTempView("TweetJSON")
val N = spark.sql("""SELECT TJ.id, hID, TJ.filter_level
                        FROM TweetJSON AS TJ
                             LATERAL VIEW EXPLOD (entities.hashtags.id) AS hID
                        WHERE text LIKE '%covid%' AND 
                              TJ.place.country_code ='US'""")
constructMTXMatirx (2380000,1000,N,"N.mtx");

//Analysis
val ml = new MLContext(spark)
ml.setConfigProperty("sysml.cp.parallel.ops","true")
ml.setConfigProperty("sysml.optlevel","4")
ml.setConfigProperty("sysml.localtmpdir","systemml")
//DML Query
val dmlTextQuery =
  s"""
    |  M  = read("M.csv", format="csv", header=FALSE, sep=",")
    |  N  = readMM("N.mtx")
    |  C = read("C.csv", format="csv", header=FALSE, sep=",");
    |  X = read("X.csv", format="csv", header=FALSE, sep=",");
    |  NF = ifelse(N<=4,N,0)
    |  print(as.scalar(NF[20000,400]))
    |  res = X*sum(t(colSums(C)*rowSums(M))+NF
    |  while(FALSE){}
    |   print(res[2,30])
    | }
""".stripMargin
val dmlScript = dml(dmlTextQuery)
val result = ml.execute(dmlScript)
\end{lstlisting}
\newpage

\begin{lstlisting}[caption={Q10},captionpos=b]
import scala.math._
import org.apache.sysml.api.mlcontext._
import org.apache.sysml.api.mlcontext.ScriptFactory._
import org.apache.sysml.api.mlcontext.MatrixFormat._
import org.apache.spark.SparkContext
import org.apache.spark.SparkContext._
import org.apache.spark.SparkConf
import org.apache.spark.sql._
import scala.io.Source
import scala.tools.nsc.io._
import scala.collection.immutable._
import org.apache.spark.storage.StorageLevel._

//Preprocessing  M 
val sqlContext = new org.apache.spark.sql.SQLContext(sc)
val User = sqlContext.read.format("csv").option("header", "true").load("User.csv");
User.createOrReplaceTempView("User")
val Tweet = sqlContext.read.format("csv").option("header", "true").load("Tweet.csv");
Tweet.createOrReplaceTempView("Tweet")
val M = spark.sql("""SELECT  followers_count,friends_count, listed_count,protected,verified
										 favorite_count,quote_count, reply_count,retweet_count,favorited,
							 			 possibly_sensitive,retweeted 
                     FROM User as U, Tweet as T 
                     WHERE  T.id = U.id""")
constructMatirx (M2,"M.csv");

//Preprocessing  (N)
val TweetJSON = spark.read.json("tweets/")
TweetJSON.createOrReplaceTempView("TweetJSON")
val N = spark.sql("""SELECT TJ.id, hID, TJ.filter_level
                        FROM TweetJSON AS TJ
                             LATERAL VIEW EXPLOD (entities.hashtags.id) AS hID
                        WHERE text LIKE '%covid%' AND 
                              TJ.place.country_code ='US'""")
constructMTXMatirx (2380000,1000,N,"N.mtx");

//Analysis
val ml = new MLContext(spark)
ml.setConfigProperty("sysml.cp.parallel.ops","true")
ml.setConfigProperty("sysml.optlevel","4")
ml.setConfigProperty("sysml.localtmpdir","systemml")
//DML Query
val dmlTextQuery =
  s"""
    |  M  = read("M.csv", format="csv", header=FALSE, sep=",")
    |  N  = readMM("N.mtx")
    |  C = read("C.csv", format="csv", header=FALSE, sep=",");
    |  X = read("X.csv", format="csv", header=FALSE, sep=",");
    |  NF = ifelse(N<=4,N,0)
    |  print(as.scalar(NF[20000,400]))
    |  res = NF*sum((X+C)%*%M)
    |  while(FALSE){}
    |   print(res[2,30])
    | }
""".stripMargin
val dmlScript = dml(dmlTextQuery)
val result = ml.execute(dmlScript)
\end{lstlisting}
\newpage
	
\lstdefinestyle{myRstyle}{
  language=R
}
\lstset{style=myRstyle}
\begin{lstlisting}[caption={V3},captionpos=b]
T<-read.csv("Tweet.csv")
U<-read.csv("User.csv")
K <-sparseMatrix(i=1:nrow(U), j=T[,"id"], x=1)
T1 <- as.matrix(subset(T, select = c("favorite_count","quote_count","reply_count", 
									 				"retweet_count","favorited","possibly_sensitive","retweeted" )))
U1 <-as.matrix(subset(U, select = c("followers_count","friends_count", "listed_count","protected","verified" )))
V3<-rowSums(T1)+K%*%rowSums(U1)
write.csv(as.matrix(V3),"V3.csv", row.names = FALSE)
\end{lstlisting}

\lstset{style=myRstyle}
\begin{lstlisting}[caption={V4},captionpos=b]
T<-read.csv("Tweet.csv")
U<-read.csv("User.csv")
K <-sparseMatrix(i=1:nrow(U), j=T[,"id"], x=1)
T1 <- as.matrix(subset(T, select = c("favorite_count","quote_count","reply_count", 
									 				"retweet_count","favorited","possibly_sensitive","retweeted" )))
U1 <-as.matrix(subset(U, select = c("followers_count","friends_count", "listed_count","protected","verified" )))
V4<-cbind(colSums(T1),colSums(K)%*%U1)
write.csv(as.matrix(V4),"V4.csv", row.names = FALSE)
\end{lstlisting}

\lstset{style=myRstyle}
\begin{lstlisting}[caption={V5},captionpos=b]
T<-read.csv("Tweet.csv")
U<-read.csv("User.csv")
C<-read.csv("C.csv")
K <-sparseMatrix(i=1:nrow(U), j=T[,"id"], x=1)
T1 <- as.matrix(subset(T, select = c("favorite_count","quote_count","reply_count", 
									 				"retweet_count","favorited","possibly_sensitive","retweeted" )))
U1 <-as.matrix(subset(U, select = c("followers_count","friends_count", "listed_count","protected","verified" )))
V4<-cbind(C%*%T1,(C%*%K)%*%U1)
write.csv(as.matrix(V5),"V5.csv", row.names = FALSE)
\end{lstlisting}